\documentclass[preprint,authoryear,12pt]{elsarticle}
\usepackage{natbib}
\usepackage{lscape}
\usepackage{ulem}

\date{ }
\usepackage{amssymb}
\journal{Astroparticle Physics}

\newcommand  \gag	{$\gamma-\gamma$}
\newcommand  \eg {$E_{\gamma}$}
\newcommand  \mic {$\mu$m}
\newcommand  \eb {$\epsilon_{\rm b}$}

\newcommand   \gray {$\gamma-$ray}
\newcommand   \taugg {$\tau_{\gamma \gamma}$}
\newcommand   \nwat  {nW~m$^{-2}$~sr$^{-1}$}
\newcommand   \lsun  {L$_{\odot}$}
\newcommand   \msun  {M$_{\odot}$}

\begin{document}

\begin{frontmatter}



\title{The Extragalactic Background Light and the Gamma-ray Opacity of the Universe}


\author{Eli Dwek}

\address{Observational Cosmology Lab., Code 665, NASA Goddard Space Flight Center, Greenbelt, MD 20771; e-mail: eli.dwek@nasa.gov}

\author{Frank Krennrich}
\address{Department of Physics and Astronomy, Iowa State University, Ames, IA 50011; \\ e-mail: krennrich@iastate.edu}
\begin{abstract}
The extragalactic background light (EBL) is one of the fundamental observational quantities in cosmology. All energy releases from resolved and unresolved extragalactic sources, and the light from any truly diffuse background, excluding the cosmic microwave background (CMB), contribute to its intensity and spectral energy distribution. It therefore plays a crucial role in cosmological tests for the formation and evolution of stellar objects and galaxies, and for setting limits on exotic energy releases in the universe. The EBL also plays an important role in the propagation of very high energy \gray s which  are attenuated en route to Earth by pair producing \gag\ interactions with the EBL and CMB. The EBL affects the spectrum of the sources, predominantly blazars, in the $\sim 10$~GeV to 10~TeV energy regime. Knowledge of the EBL intensity and spectrum will allow the determination of the intrinsic blazar spectrum in a crucial energy regime that can be used to test particle acceleration mechanisms and VHE \gray\ production models. Conversely, knowledge of the intrinsic \gray\ spectrum and the detection of blazars at increasingly higher redshifts will set strong limits on the EBL and its evolution.
This paper reviews the latest developments in the determination of the EBL and its impact on the current understanding of the origin and production mechanisms of \gray s in blazars, and on energy releases in the universe. The review concludes with a summary and future directions in Cherenkov Telescope Array techniques and in infrared ground-based and space observatories that will greatly improve our knowledge of the EBL and the origin and production of very high energy \gray s.
\end{abstract}


\end{frontmatter}

%===============================================================
\section{Introduction}
%===============================================================
The extragalactic background light (EBL), defined here as the emission in the 0.1 to 1000~\mic\ wavelength region, is one of the fundamental observational quantities in cosmology. It comprises the integrated light from resolved and unresolved extragalactic sources, and the light from any truly diffuse background, excluding the cosmic microwave background (CMB). It is therefore the repository of all energy released by nuclear and gravitational processes since the epoch of recombination. A significant fraction of this radiation is shifted by cosmic expansion and by absorption and reradiation by dust into infrared (IR) wavelengths. Consequently, its intensity and spectral shape hold  key information about the formation and evolution of galaxies and their stellar and interstellar contents throughout cosmic history.  A strict lower limit on the EBL intensity is provided by the integrated light from resolved galaxies, hereafter referred to as the integrated galaxy light (IGL).

The EBL plays also an important role in the propagation of high energy \gray\ rays that are predominantly emitted by blazars, a subgroup of active galaxies hosting active galactic nuclei (AGN), whose relativistic jet is pointed towards the Earth. High energy photons emitted by blazars are attenuated by photon-photon interactions with the EBL, a process that can be used to set important limits on both, the intrinsic spectra of blazars and the intensity of the EBL in select energy and wavelength regions where these interactions are most prominent.

The EBL is intimately connected to the diffuse X-ray, radio, and supernova neutrino backgrounds. Deep X-ray surveys have resolved the X-ray background into point sources, most of which are dust enshrouded AGNs \citep{mushotzky00}. 
Up to 90\% of the X-ray energy produced in individual AGN can be degraded and reradiated predominantly at mid-IR wavelengths \citep[e.g.][]{franceschini02, ballantyne06}.  Consequently, the X-ray background can be used to predict the EBL intensity at at these wavelengths. Current estimates show that about 15\% of the 24~\mic\ EBL intensity is powered by AGN activity \citep[][and references therein]{treister06a, soifer08}. Conversely, the connection between mid-IR bright sources and AGN can be used to estimate the contribution of obscured AGN to the X-ray background \citep[][and references therein]{gandhi03, soifer08}.

Massive stars that power the IR emission also emit radio free-free emission during the main sequence phase, and radio synchrotron emission during the supernova remnant phase of their evolution. The IR emission from star-forming galaxies is therefore correlated with the radio emission \citep{lisenfeld96, condon91}. This correlation can be used to estimate the contribution of star-forming galaxies to the cosmic radio background  \citep{haarsma98, dwek02, ponente11}.

Most of the EBL intensity is powered by massive stars that end their life as core collapse supernovae. The total EBL intensity can therefore be used to derive an estimate of the supernova rate and the resulting flux of supernova neutrinos  \citep{horiuchi09, beacom10}. The detectability of these neutrinos can be greatly enhanced by the proposed introduction of gadolinium in existing large water Cherenkov detectors (such as Super-Kamiokande) \citep{ beacom04}. Gadolinium has a very high capture cross section for neutrons generated in $\, \bar{\nu}_e + p \rightarrow e^+ + n\, $ reactions, and can be introduced in the form of soluble trichloride (GdCl$_3$). Following the neutron capture, the Gd emits an 8~MeV \gray\ which produces relativistic electrons by Compton scattering. The Cherenkov radiation from these electrons is more easily detected than that produced in the cascade of the 2.2~MeV \gray\ generated by the capture of neutrons by free protons. 
 
Several reviews have appeared in the literature, presenting a historical overview of the importance of the EBL, early estimates of its intensity, the quests for its detection, and its many astrophysical implications \citep{hauser01, kashlinsky05a, lagache05}.  Since these reviews were written, significant advances have been made in studies of the EBL with the launch of UV ({\it Galex}) and IR space observatories ({\it Spitzer}, {\it Herschel}, and {\it Akari}).  These  observatories, together with ground-based telescopes, such as 2MASS, have provided new limits on the EBL ranging from UV to submillimeter wavelengths. Deeper galaxy number counts and new data analysis techniques of stacking astronomical images have narrowed the gap between the contribution of resolved galaxies and the true intensity of the EBL. 

The {\it Fermi} Gamma-ray Space Telescope, operating between 200~MeV and 300~GeV,   and ground-based  air Cherenkov detectors (H.E.S.S., MAGIC, and VERITAS) operating in the $\sim 50$~GeV to 100~TeV range have broadened the energy window for the studies of \gray\ sources. These advances have led to the detection of new GeV and TeV \gray\ sources and provided new data for determining their intrinsic spectra. Reviews of these subjects were presented by \cite{weekes08a} and \cite{hinton09}. More recently, \cite{dermer12a} presented a review of the {\it Fermi} catalog of \gray\ sources and the physics of the production of relativistic particles and \gray s from these sources. Table~1 presents a glossary to the acronyms of the observatories and instruments referred to in this review.  

These developments provide the main impetus for this review. We first present, in \S2, the basic formulae describing the attenuation of photons by pair producing interactions with other photons. We then show how this attenuation will affect $\gamma$ rays traversing a radiation field characterized first by a pure black body, representing the stellar emission component of the EBL, and then by a more realistic EBL that includes the dust emission component. This attenuation can, in principle, be used to determine the intensity of the attenuating radiation field if the intrinsic source spectrum is known. In \S3 we survey the type of \gray\ sources that are used in these studies, their spectral characteristics, the physical mechanisms for generating their spectra, and constraints on their spectral shape imposed by general physical principles. In \S4 we summarize measurements and limits on the EBL intensity determined by direct measurements and by adding the light from resolved galaxies. Models for the EBL intensity and its evolution with redshift are summarized in \S5. In \S6 we summarize the constraints on the EBL intensity derived from \gray\ observations of blazars, emphasizing the different assumptions made on the intrinsic blazar spectra to derive these limits. EBL models predict the \gray\ opacity of the universe at different energies, and in \S7 we compare these model predictions with blazar observation. Throughout this review it was tacitly assumed that the production of \gray s takes place exclusively in the sources. In \S8 we consider alternative scenarios of \gray\ production that could have important implications for EBL limits, namely, that a significant fraction of the observed \gray s could be produced en route to Earth. The role of the EBL in setting limits on exotic energy releases in the universe in briefly discussed in \S9. A  summary and future prospects for the fields of \gray\ and EBL research is given in \S10.      
 
 %===============================================================
\section{The EBL and the Attenuation of Gamma-Ray Photons}
%===============================================================

%--------------------------------------------------------------------------------------------------------
\subsection{The EBL}
%--------------------------------------------------------------------------------------------------------
The differential specific flux at wavelength $\lambda_0$, d$F_{\nu}(\lambda_0)$, received from radiative sources within a comoving volume element d$V_c(z)$ at redshift $z$  at wavelength $\lambda$ is given by \citep[e.g.][]{mo10}:
\begin{equation}
\label{ }
{\rm d}F_{\nu}(\lambda_0) = (1+z)\, {{\cal L}_{\nu}(\lambda, z) \, {\rm d}V_c(z)\over 4\pi\, d_{\small L}(z)^2}
\end{equation} 
where ${\cal L}_{\nu}(\lambda,z)$ is the comoving specific luminosity density of the sources, $d_{\small L}$ is their luminosity distance, and the (1+z) factor arises from the decrease in energy of the emitted photons due to the redshift, and $\lambda_0=(1+z)\lambda$. 
 
The specific comoving intensity of the EBL per unit solid angle, $\delta \Omega$, at redshift $z_0$ and wavelength $\lambda_0$  is given by an integral over all energy releases over cosmic history:
%---------------
\begin{eqnarray}
\label{eq:ebl2}
I_{\nu}(\lambda_0, z_0) &  = &  \int_{z_0}^{\infty}\ (1+z){{\cal L}_{\nu}(\lambda, z) \, \over 4\pi\, d_{\small L}(z)^2}\, 
{{\rm d}V_c(z)\over \delta \Omega} \\ \nonumber
 & = &  \left({1\over 4\pi}\right)\, \int_{z_0}^{\infty}\ {\cal L}_{\nu}(\lambda, z)\, \left|{c\, dt\over dz}\right|\, dz 
\end{eqnarray}
where $c|dt/dz|$ is given by \citep[e.g.][]{mo10}:
%---------------
\begin{eqnarray}
\label{eq:dldz}
c\left|{dt\over dz}\right| & = & {R_H\over (1+z) E(z)}; \qquad R_H\equiv {c\over H_0} \\ \nonumber
E(z) & \equiv & \left[\Omega_R(1+z)^4 + \Omega_m(1+z)^3 + \Omega_k(1+z)^2 +\Omega_{\Lambda}\right]^{1/2} \\
     & = & \left[(1+z)^2\, (\Omega_m z+1) -z(2+z)\, \Omega_{\Lambda}\right]^{1/2}  \\ \nonumber
     & = & \left[\Omega_m (1+z)^3 + \Omega_{\Lambda}\right]^{1/2} \qquad.
\end{eqnarray}
$H_0$ is the Hubble constant, and $\Omega_R$ $\Omega_m$, $\Omega_k$ and $\Omega_{\Lambda}$ are the dimensionless density parameters of the radiation, matter, the curvature, and the cosmological constant $\Lambda$, obeying the relation: $\Omega_R + \Omega_m + \Omega_k + \Omega_{\lambda} = 1$. The second expression for $E(z)$ is for a matter dominated ($\Omega_R << 1$) universe, and the third is for one that is matter dominated and flat ($\Omega_k = 0$).
In the concordance cosmology model: $H_0=70$~km~s$^{-1}$~Mpc$^{-1}$; $\Omega_m = 0.27$, and $\Omega_{\Lambda}=0.73$  \citep{hinshaw09}.
 
%--------------------------------------------------------------------------------------------------------
\subsection{Gamma-ray attenuation by pair production}
%--------------------------------------------------------------------------------------------------------
The interaction between two photons with energies \eg\ and \eb, will lead to the creation of a particle anti-particle pair when the total \gray\ energy in the center of momentum of the system exceeds the rest frame energy of the two particles. The threshold for the creation of an $e^+$+$e^-$ pair is given by:
\begin{equation}
\label{eq:eth}
\epsilon_{th}(E_{\gamma}, \mu, z) = {2\, (m_e\, c^2)^2\over E_{\gamma}\, (1-\mu)}
\end{equation}

%----- figure 1 ----- diagram of gg production
 \begin{figure}[htbp]
  \centering
  \includegraphics[width=1.5in]{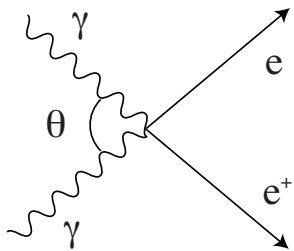} 
%  \hspace{0.1in}
%   \includegraphics[width=2.5in]{sigma_mu.eps}
%  \vspace{0.1in}
  \caption{{\footnotesize Schematic illustration of the \gag\ pair production reaction, showing the definition of the angle $\theta$ between the interacting photons}. \label{fig:theta}}
\end{figure} 
%--------------------
\noindent
where $\mu \equiv \cos\theta$, and $\theta$ is the angle between the two photons, as illustrated in Figure \ref{fig:theta}. 

The cross-section for the \gag\ interaction is given by:
\begin{equation}
\label{eq:sigma}
\sigma_{\gamma \gamma}(E_{\gamma}, \epsilon, \mu, z) = {3 \sigma_T\over 16}\ (1-\beta^2)\left[2\beta\, (\beta^2-2) + (3-\beta^4)\, \ln\left({1+\beta)\over(1-\beta)}\right)\right] \
\end{equation}
\noindent
where
%---------------
\begin{equation}
\label{eq:beta}
\beta \equiv \sqrt{\left(1-{\epsilon_{th}\over \epsilon}\right)}
\end{equation}

%----- figure 2 ----- sigma dependence on angle
 \begin{figure}[htbp]
  \centering
  \includegraphics[width=2.5in]{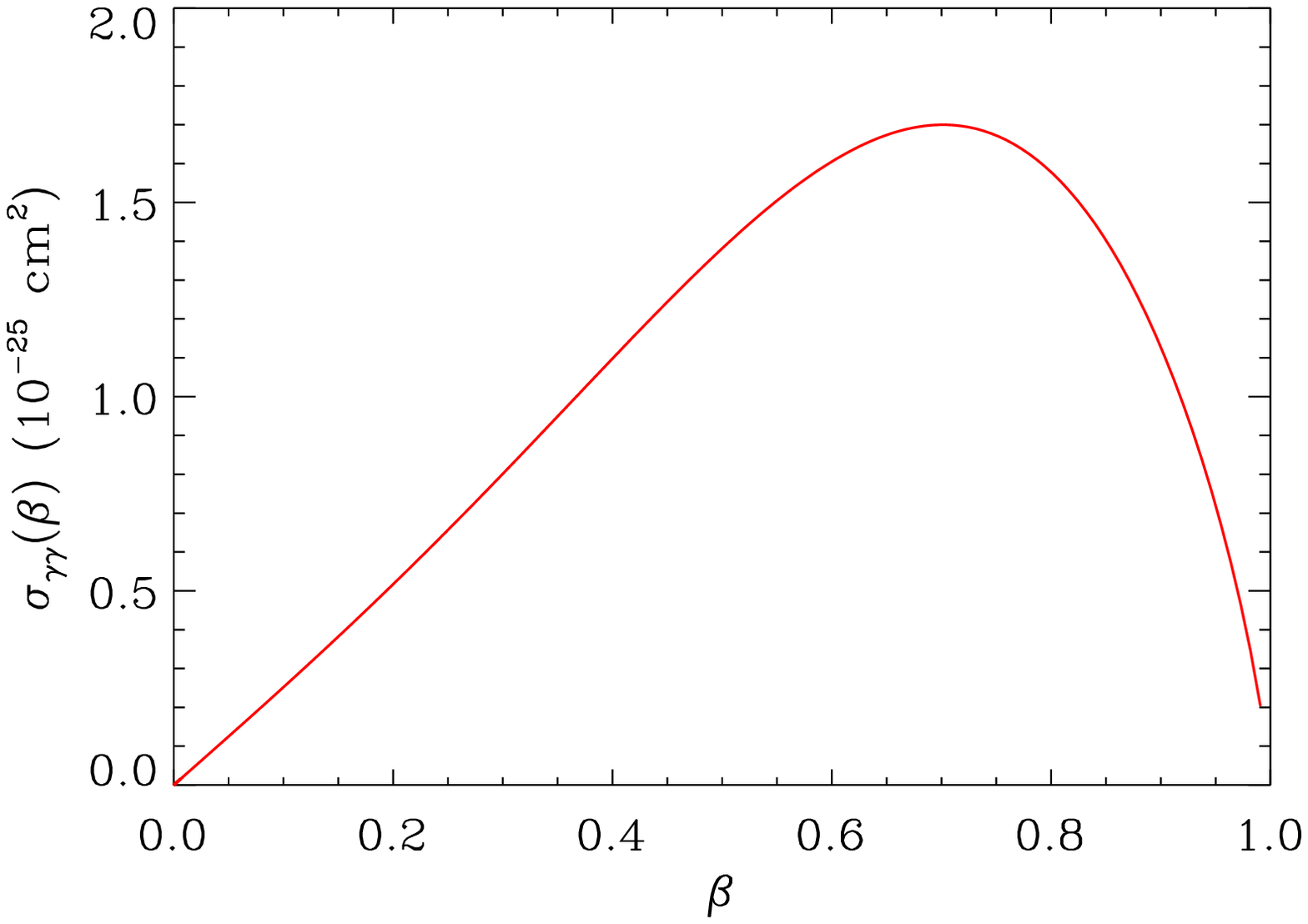} 
  \hspace{0.1in}
   \includegraphics[width=2.5in]{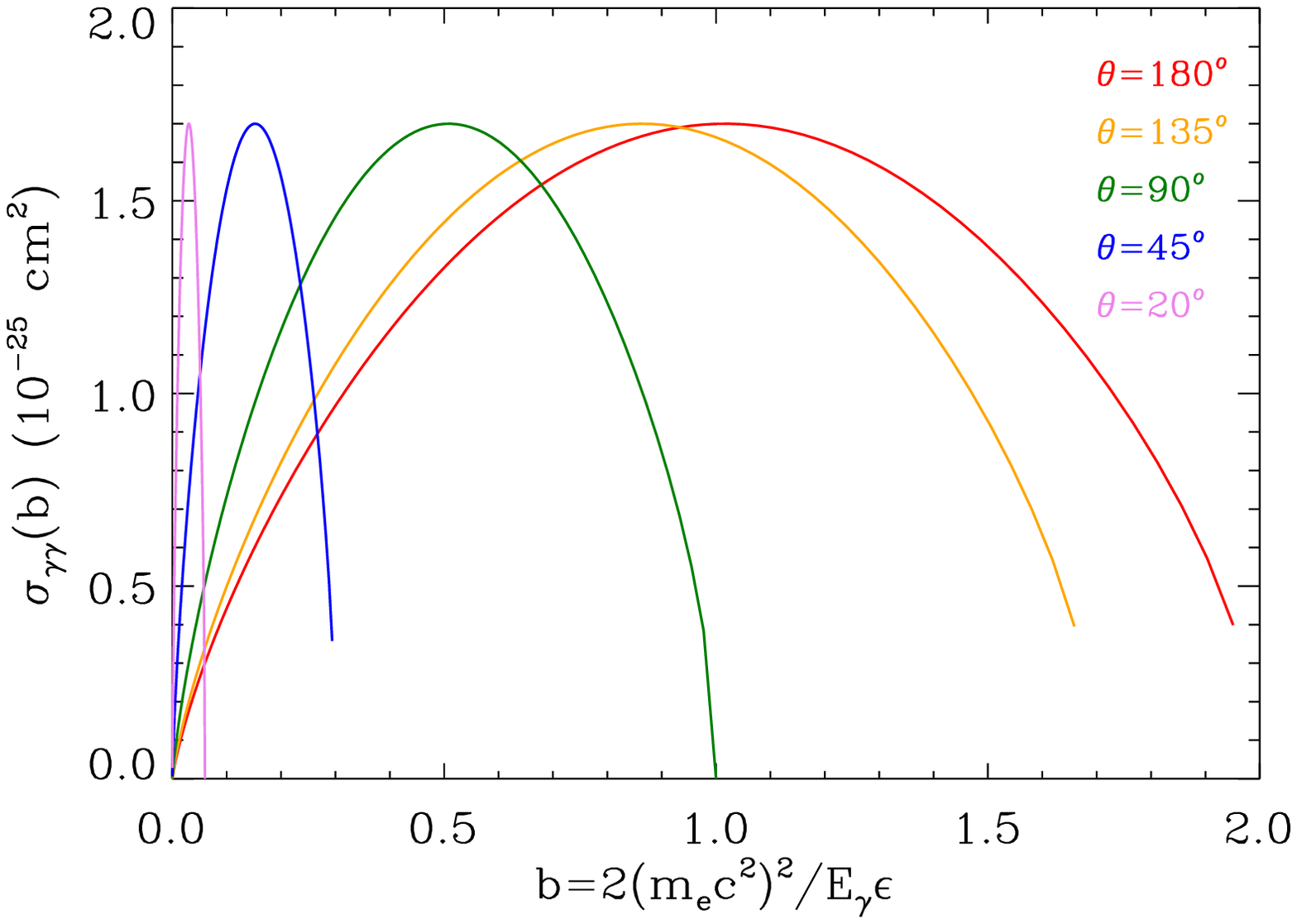}
%  \vspace{0.1in}
	  \caption{{\footnotesize The cross section for the \gag\ interaction. {\bf Left panel}: its dependence on $\beta$ [eq. (7)]; {\bf Right panel}: its dependence on $b$ for different angles of incidence.} \label{fig:sigma} }
\end{figure} 
%--------------------
Figure~\ref{fig:sigma} (left panel) depicts the cross section as a function of $\beta$.
The cross section peaks at a value of $\beta = 0.70$, providing a relation between the energies \eg\ and $\epsilon$ (or wavelength $\lambda$) at the peak, given by:
\begin{equation}
\label{eq:eg}
E_{\gamma}(TeV) = {1.07 \over \epsilon(eV)\, (1-\mu)}= {0.86\, \lambda(\mu m)\over (1-\mu)}
\end{equation}
The right panel of the figure depicts the cross section as a function of $b\equiv 2(mc^2)^2/E_{\gamma} \epsilon$ for different values of the angle $\theta$. When the photons are moving in the same direction ($\theta=0$), the cross section collapses to a delta-function at $b=0$, and the energy threshold becomes infinite.

%--------------------------------------------------------------------------------------------------------
\subsection{The Attenuation of $\gamma-$rays from Cosmological Sources}
%--------------------------------------------------------------------------------------------------------
En route to Earth, $\gamma-$rays from cosmological sources have to pass through the radiation field of the EBL, resulting in their attenuation by pair producing interactions. The optical depth of a \gray\ photon at an observed energy \eg, emitted by  a source at redshift $z$ due to this process is given by:
%---------------
\begin{equation}
\label{eq:taugg}
\footnotesize
\tau_{\gamma \gamma}(E_{\gamma}, z)  = \int_0^z\, dz'\, {d\ell \over dz'}\ \int_{-1}^1\, d\mu\, {1-\mu\over 2}\ \int_{\epsilon'_{th}}^{\infty}\, d\epsilon\, n_{\epsilon}(\epsilon, z')(1+z')^3\, \sigma_{\gamma \gamma}(\beta', z')
\end{equation}
where $n_{\epsilon}(\epsilon, z) \equiv dn(\epsilon, z)/d\epsilon$ is the specific comoving number density (cm$^{-3}$~eV$^{-1}$) of background photons with energy $\epsilon$ at redshift $z$, and the $(1+z)^3$ term represents its conversion to a proper number density. 
The pair-production threshold energy is $\epsilon'_{th}=2(m_ec^2)^2/E_{\gamma}(1-\mu)(1+z)$, where the $(1+z)$ factor takes into account that the observed \gray\ photon had a higher energy at the redshift of the interaction. The parameter $\beta'=(1-\epsilon'_{th}/\epsilon)^{1/2}$, and $d\ell/dz=c|dt/dz|$, where $\ell$ is the proper distance.

Calculating the EBL opacity to \gray s from cosmological distant sources requires knowledge of the evolution of the comoving specific photon number density $n_{\epsilon}(\epsilon, z)$ as a function of redshift. The specific number density of photons with energy $\epsilon$ at redshift $z$ is related to the specific EBL intensity at a given redshift $z$ by:
%---------------
\begin{eqnarray}
\label{eq:e2ne}
\epsilon^2\, n_{\epsilon}(\epsilon, z) & = & {4\pi\over c}\, \nu\, I_{\nu}(\nu, z) \\ \nonumber
 & = & 2.62\times10^{-4}\, \nu\, I_{\nu}(\nu, z)
\end{eqnarray}
where $\epsilon = h \nu$, $I_{\nu}(\nu, z)$ is given by eq. (2), and the coefficient in the second line was calculated for $\epsilon$ in eV, $n_{\epsilon}$ in cm$^{-3}$~eV$^{-1}$, and $\nu\, I_{\nu}$ in \nwat. 

Finally, we point out that the \gag\ cross section is wide, so that in calculating the \gray\ opacity, strong variations in the EBL spectrum are smoothed out over a wide range of \gray\ energies. The EBL intensity at a given wavelength is therefore effecting \taugg\ over a wide range of \gray\ energies around the peak given by eq. (8).
% are given by \citep[e.g.][]{mo10}:
%
%%---------------
%\begin{eqnarray}
%\label{eq:dldz}
%{d\ell \over dz} & = & c\left|{dt\over dz}\right| = {R_H\over (1+z) E(z)}; \qquad R_H\equiv {c\over H_0} \\ \nonumber
%E(z) & \equiv & \left[\Omega_R(1+z)^4 + \Omega_m(1+z)^3 + \Omega_k(1+z)^2 +\Omega_{\Lambda}\right]^{1/2} \\
%     & = & \left[\Omega_m (1+z)^3 + \Omega_{\Lambda}\right]^{1/2} \qquad .
%\end{eqnarray}
%where $\ell$ is the proper distance, $H_0$ is the Hubble constant, and $\Omega_R$ $\Omega_m$, $\Omega_k$ and $\Omega_{\Lambda}$ are the dimensionless density parameters of the radiation, matter, the curvature, and the cosmological constant $\Lambda$, obeying the relation: $\Omega_R + \Omega_m + \Omega_k + \Omega_{\lambda} = 1$. For a flat ($\Omega_k = 0$) and matter dominated ($\Omega_R << 1$) universe, $E(z)$ reduces to the second line in eq. (7).
%In the concordance cosmology model: $H_0=70$~km~s$^{-1}$~Mpc$^{-1}$; $\Omega_m = 0.27$, and $\Omega_{\Lambda}=0.73$  \citep{hinshaw09}. 

%--------------------------------------------------------------------------------------------------------
\subsection{A Simple Example: An EBL given by a diluted blackbody spectrum}
\begin{figure}[t]%[htbp]
  \centering
  \includegraphics[width=2.in]{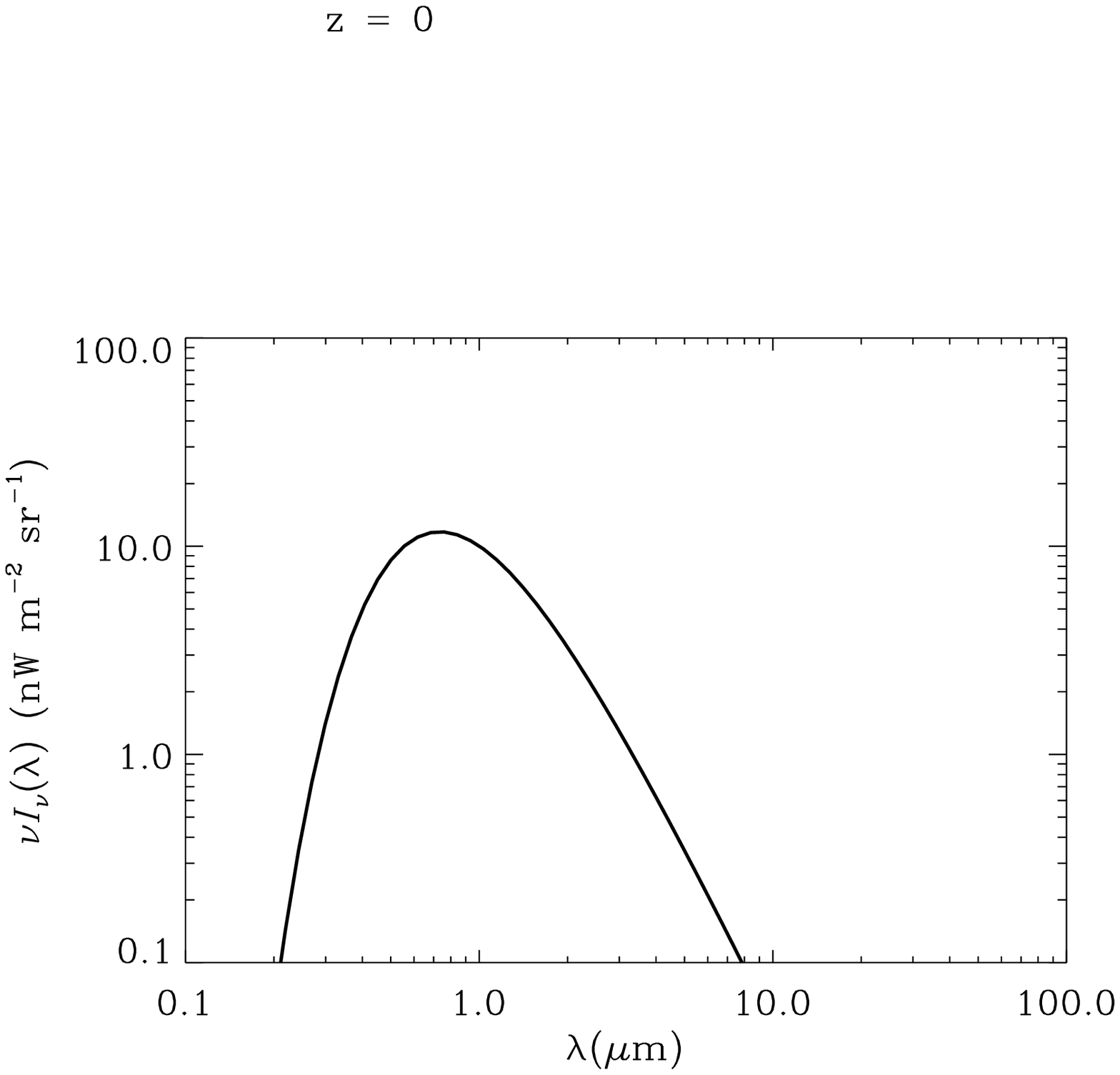} 
  \hspace{0.1in}
   \includegraphics[width=2.in]{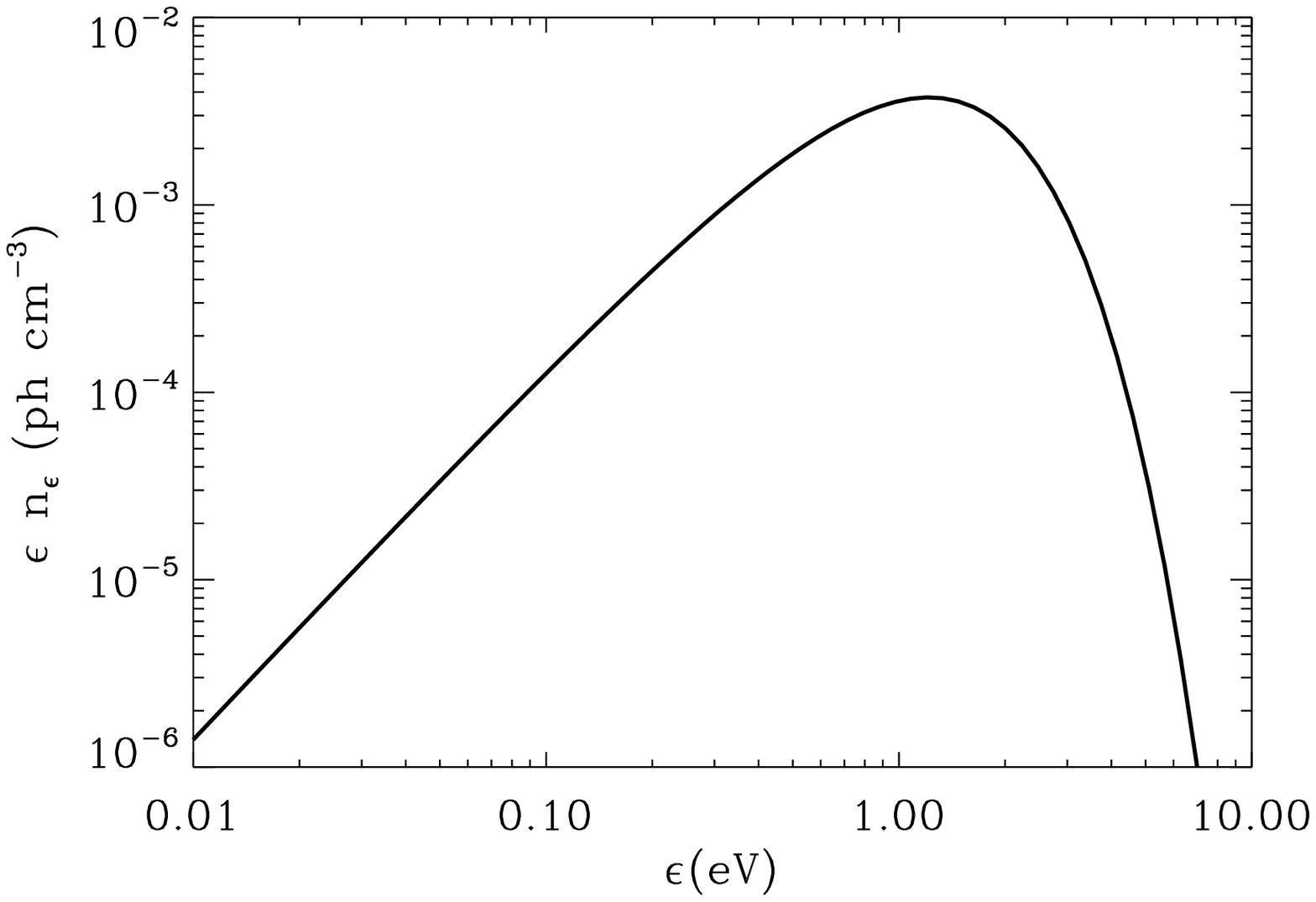} \\
  \includegraphics[width=2.in]{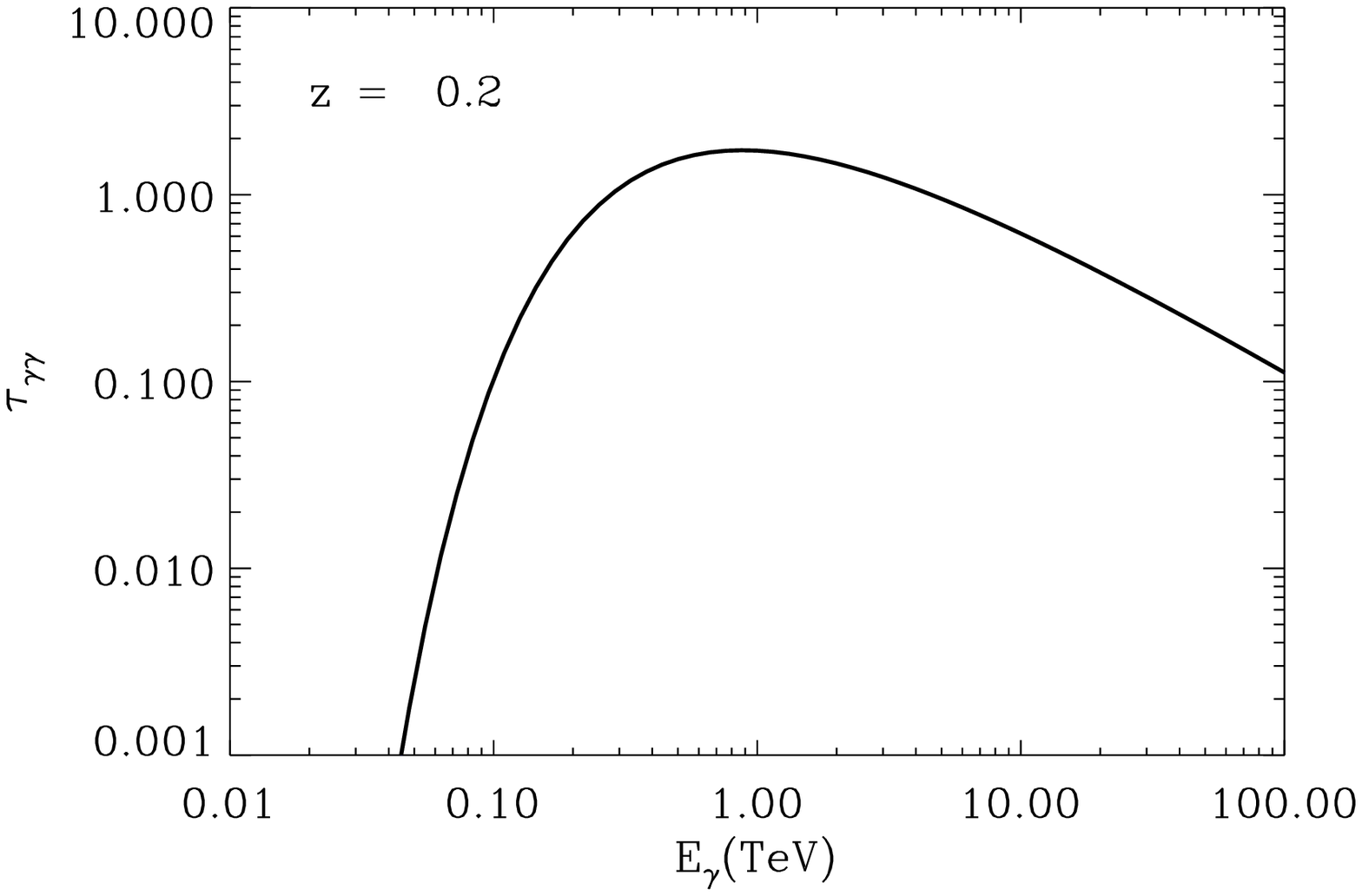} 
  \hspace{0.1in}
   \includegraphics[width=2.in]{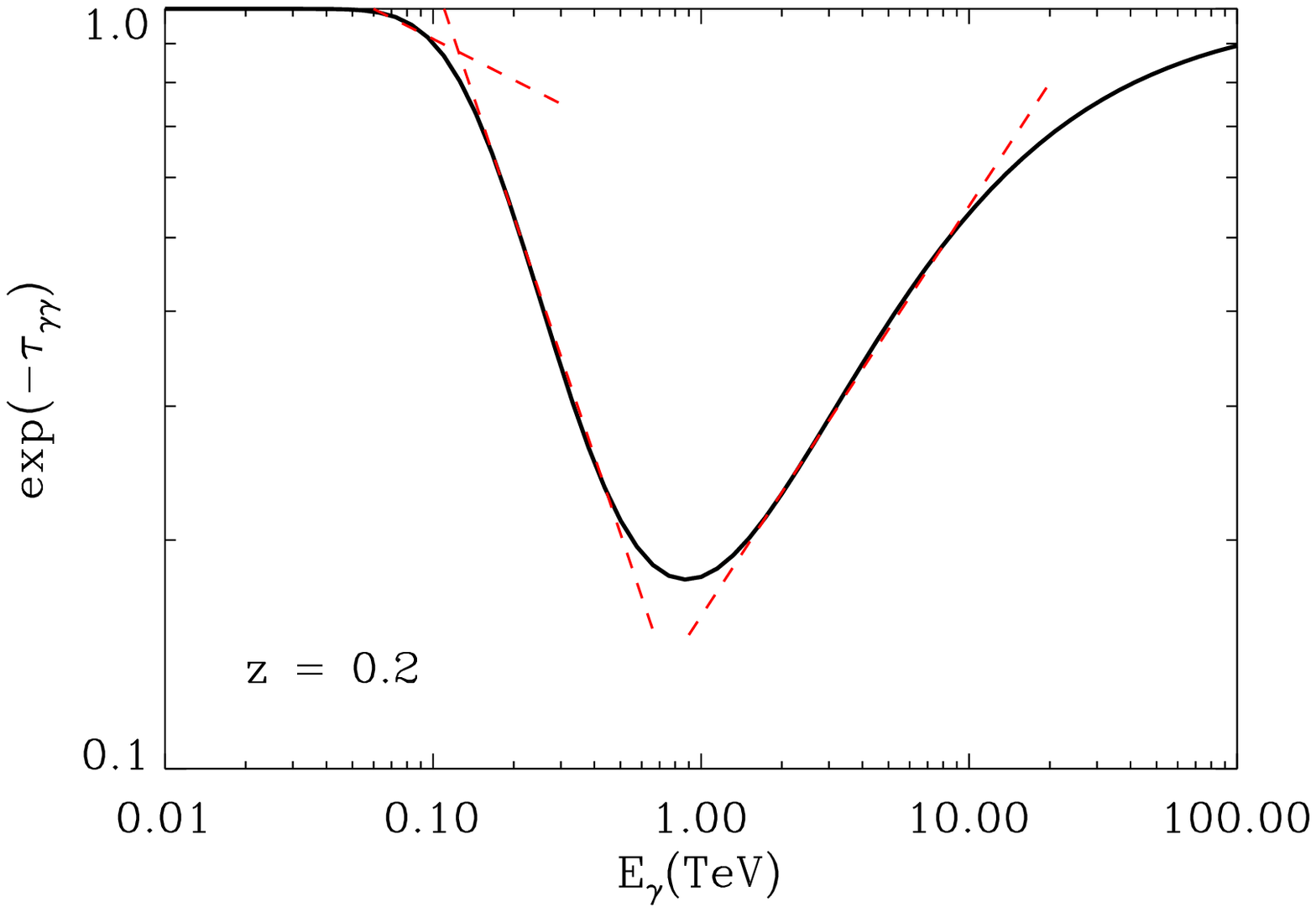} \\
   %  \vspace{0.1in}
  \caption{{\footnotesize {\bf Top left}: A diluted black body representation of the stellar emission component of the EBL; {\bf Top right}: The corresponding proper photon number density versus energy; {\bf Bottom left}: The \gray\ opacity versus energy, \eg; {\bf Bottom right}: The \gray\ attenuation. The figure illustrates the dramatic change in the attenuation at the \gray\ energy that corresponds to the wavelength at which the slope of the EBL spectrum changes. The different slopes are depicted as dashed lines in the figure. } \label{fig:black-body}}
\end{figure}

%--------------------------------------------------------------------------------------------------------
Of particular interest is the behavior of \taugg\ for a background radiation field that is represented by a diluted blackbody. Figure~\ref{fig:black-body} (upper left panel) depicts a local EBL characterized by a Planck function, normalized to an intensity of 10~\nwat\ at 1~\mic. The upper right panel of the figure depicts the photon number density. The bottom left panel shows the \gray\ opacity at redshift $z=0.2$, assuming a non-evolving EBL, and the right panel shows the source attenuation as a function of \gray\ energies. 
Also shown in the figure are the energy regimes in which substantial changes in the slope of the opacity occur (dashed lines).

The rapid rise in the EBL spectrum between 0.5 and 1~\mic\ results in a rise of the \gray\ opacity, and the onset of substantial source attenuation in the 10 to 500~GeV energy region.
This sudden increase in the GeV attenuation creates a break, $\Gamma_{GeV}$, in the spectrum, defined as the difference in power law index between the unattenuated and the attenuated region of the spectrum (see Figure~\ref{fig:gevtevbreak} in this paper).
At higher \gray\ energies, the spectrum of a blazar characterized by an intrinsic power law will exhibit a second spectral break around $\sim 1$~TeV. For an evolving EBL, the magnitude and location of this spectral break are expected to evolve with redshift.  The substantial decrease in the attenuation at a few TeV is a consequence of the particular choice of the EBL spectrum, which decreases rapidly at wavelengths beyond $\sim 2$~\mic.  

%----- figure 3 ----- change in TeV attŽnuation versus EBL spectrum
%--------------------

%--------------------------------------------------------------------------------------------------------
\subsection{A More Realistic Example: An EBL that includes dust emission}
%--------------------------------------------------------------------------------------------------------

\begin{figure}[t]%[htbp]
  \centering
  \includegraphics[width=2.5in]{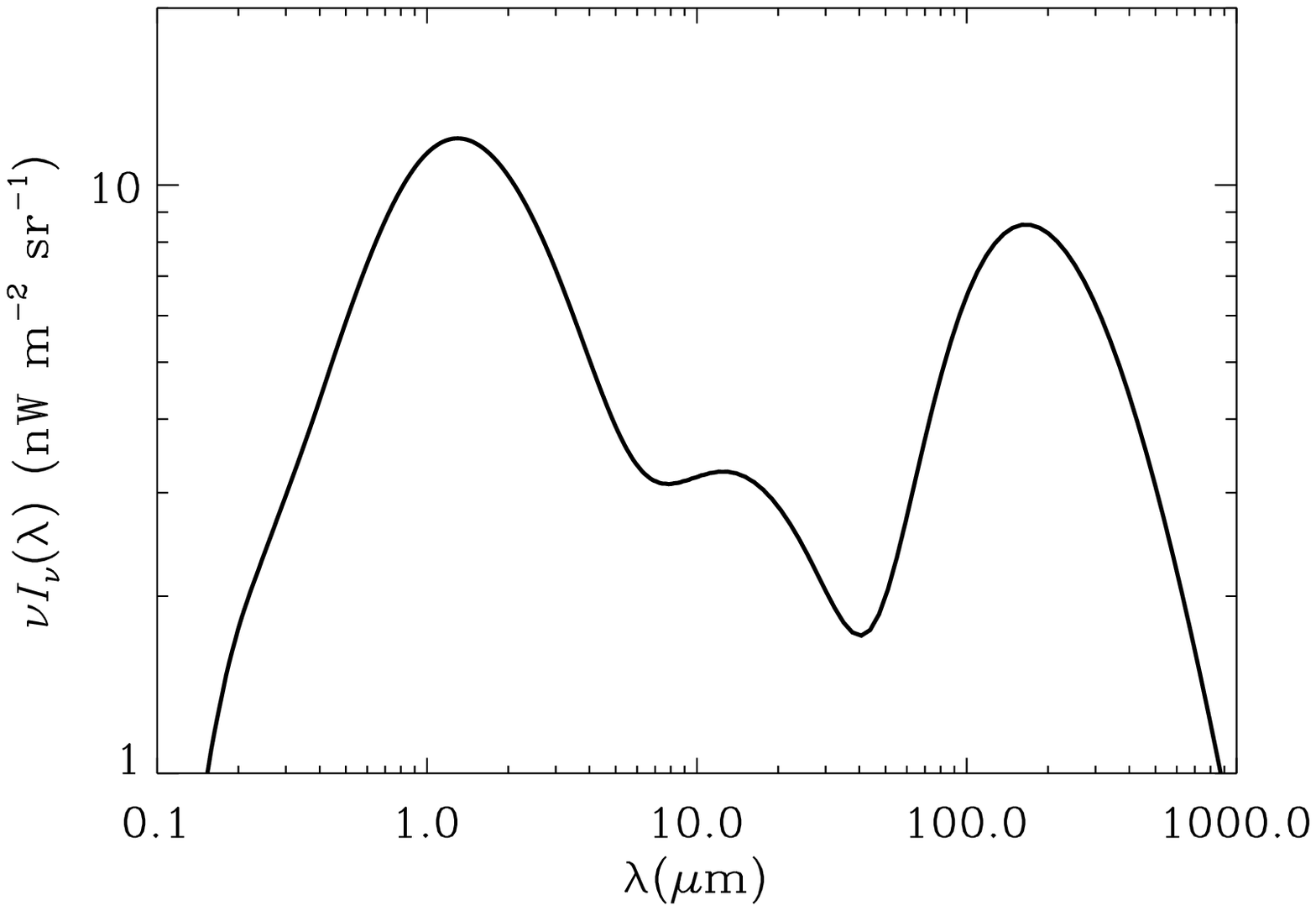} 
  \hspace{0.1in}
   \includegraphics[width=2.5in]{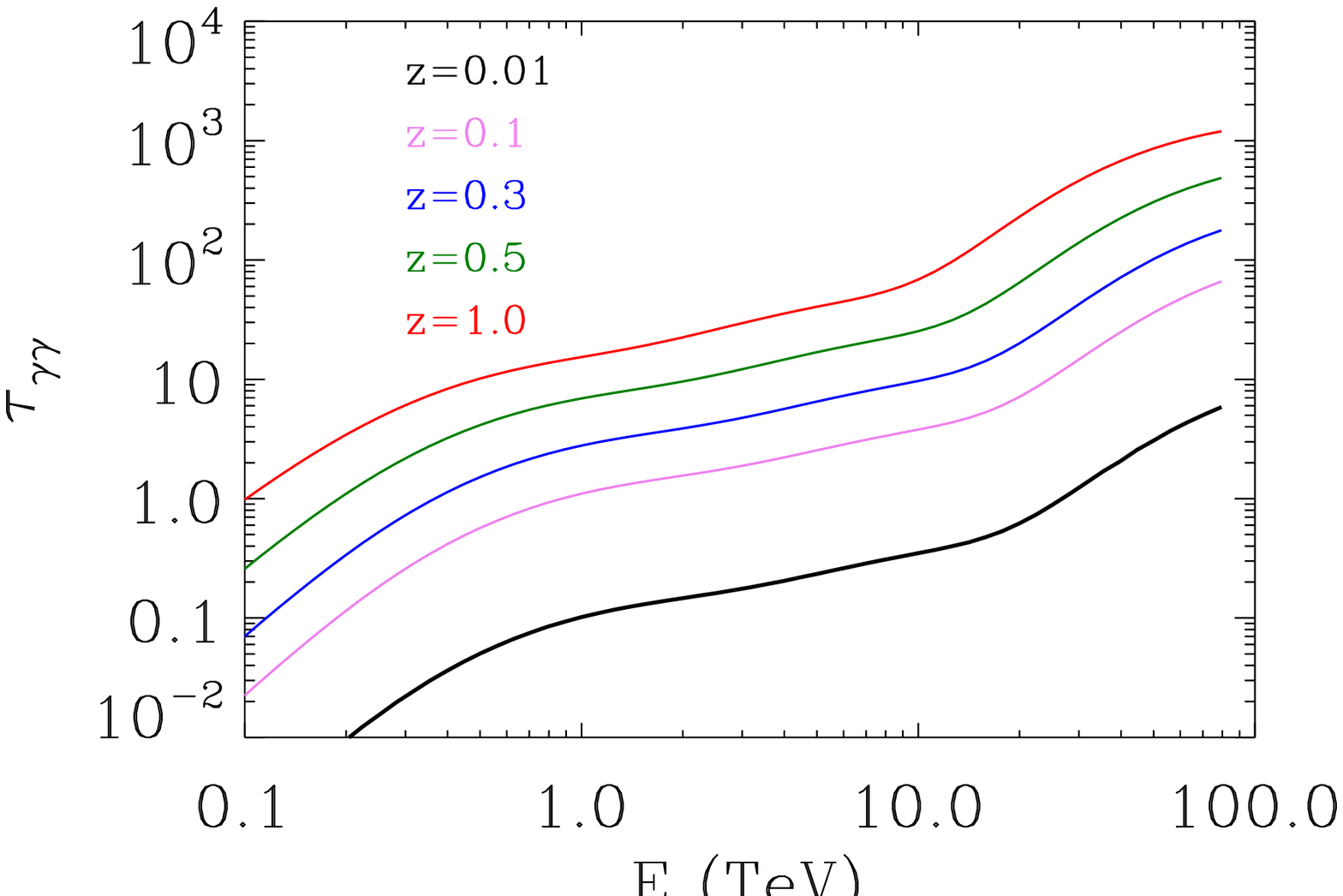} 
   \caption{{\footnotesize {\bf Left panel}: Calculated EBL  intensity versus wavelength at $z=0$ ; {\bf Right panel}: The \gray\ opacity versus energy for sources at different redshift (see labels). The figure illustrates the correlation between the changes in the slope of the EBL intensity with those in \taugg. Model calculations by \cite{finke10}. Details in \S2.5 of the text.}  \label{fig:ebl-tau}}
\end{figure}

Figure~\ref{fig:ebl-tau} depicts a more realistic presentation of the current EBL spectrum (left panel) and the \gray\ opacity for different redshifts (right panel), taken from model calculations of \cite{finke10}.
At wavelengths short wards of $\sim 5$~\mic\ the spectrum represents the stellar and AGN contributions to the EBL. At longer wavelengths the spectrum represents the AGN and starlight energy that was absorbed and reradiated by the dust. The right panel shows the energy dependence of the \gray\ opacity for sources at different redshifts. The opacity calculations took into account the evolution of the EBL with redshift.
The figure illustrates the relation between the EBL spectrum and the energy dependence of the \gray\ opacity. 
The initial rise of the EBL intensity at UV-optical wavelengths causes an increase in the \gag\ opacity between 10 and 500~GeV. The decline in the EBL intensity between $\sim 1$ and 15~\mic\ causes \taugg\ to rise less rapidly between 1 and 10~TeV. The slope of \taugg\ in this region reflects the ratio of the $\sim 1$ to 15~\mic\ intensities of the EBL. The rise in \taugg\ beyond 10~TeV reflects the rise in the EBL towards the peak of the dust emission at $\sim 100 -200$~\mic. 

The energy dependence of \taugg\ will give rise to several breaks in the spectrum of \gray\ sources that reflect the changes in the slope of the opacity. The first spectral break, $\Delta \Gamma_{GeV}$ occurs between 10 and 500~GeV. The second, $\Delta \Gamma_{TeV}$ around 1~TeV, and the third around 10~TeV. 

The first break has been used in most EBL studies to date with various assumptions on the intrinsic source spectra, and the second break has been most recently explored in the analysis of \cite{orr11}.   A review of studies that utilize the first and second break for constraining the EBL is presented in \S6.
A break at $\sim 10$~TeV has yet to be discovered. Starburst galaxies, which have a hard \gray\ spectrum, are the most promising subject for such analysis.

%----- figure 4 ----- change in TeV attŽnuation versus EBL spectrum

%===============================================================
\section{The Types and Spectra of Extragalactic  GeV/TeV Sources}
%===============================================================
Determination of the EBL intensity from GeV--TeV~\gray\ observations requires knowledge of the intrinsic spectrum of the sources. Here we list the  different sources, their spectral characteristics, the different proposed mechanisms for their \gray\ production, and the physical limits on their spectral energy distribution at very high energies. 
%--------------------------------------------------------------------------------------------------------
\subsection{The Types of Extragalactic GeV/TeV Sources}
%--------------------------------------------------------------------------------------------------------
The currently available GeV--TeV \gray\ sources that are being used to derive limits and constraints on the EBL are listed in Table~2.  They  include the accretion-powered relativistic jets of active galactic nuclei (AGNs), namely BL Lacertae objects  (BL Lacs),  flat spectrum radio quasars (FSRQs) and a few radio galaxies.  The list is complemented by the recent detections of two nearby starburst (SB) galaxies. In contrast to AGNs, their \gray\ spectrum is generated by the cumulative effects of cosmic-ray acceleration in shocks generated by a large number of supernova remnants \citep{volk96}.

{\bf \underline{Blazars}}: To date, the most numerous sources used in EBL studies are blazars. Historically, they have been divided into two sub-classes based on their optical properties: FSRQs, characterized by strong emission lines; and BL~Lacs, characterized by weak or lack of emission lines. Because of the weakness of their emission lines, the redshift determination of BL~Lac blazars has proven difficult or even impossible in many cases.  The status of blazars as bright GeV and TeV sources arises from the  fact that their relativistic jets are closely aligned with the observers line of sight.   Consequently, the luminosity of a \gray\ emission region moving relativistically along the jet axis in the direction of the observer  is strongly beamed, enabling its detection at cosmological distances.   Occasional strong flaring activity renders the following BL~Lacs: PKS~2155-304 \citep{aharonian07b}; Mrk~501 \citep{catanese97, abdo11b}, and Mrk~421, \citep{gaidos96, acciari11b, aleksic11c} the brightest TeV sources; and the following FSRQs: 3C~454.3 \citep{donnarumma09}, and 3C~279, \citep{wehrle98}  the brightest GeV emitters in the sky.  The flaring has provided high quality \gray\  spectra and has led to  their detection at redshifts as far as $z\approx$~0.5 at TeV energies with IACTs, and as far as $z \approx$~3.2 at $\sim 10$~GeV energies with {\it Fermi}.

The combined GeV-TeV observations of  blazars make it possible to study their spectra over a larger range of redshifts, thereby enabling the studies of the EBL over a wider range of wavelengths. GeV photons interact mainly with UV/optical photons, whereas TeV photons probe mainly the near- to mid-IR region of the EBL. Since the intensity of the EBL is much lower at UV energies, the universe is transparent to \gray s below 10~GeV, becoming essentially opaque for TeV sources at redshifts of z $\rm > 0.5$.   
The {\it Fermi} Gamma-Ray Space Telescope provides important probes of the UV region of the EBL, and the GeV transparency can be used to test evolutionary models of the EBL to relatively large redshifts ($z > 1$). \\
{\bf \underline{Radio galaxies}}:
The jets in radio galaxies are significantly misaligned with respect to the observer's viewing direction, and thereby provide no relativistic Doppler boosting. This limits the detection of radio galaxies with current generation \gray\ telescopes to the local group and the Perseus galaxy cluster.    Deep \gray\  observations  of radio galaxies with CTA combined with spatially resolved studies in the radio, optical and X-ray will play an important role in understanding the physics of relativistic jets. These observations are likely to yield spectra up to  $\rm \approx$~10s TeV which will provide useful constraints on the EBL in the mid- and far-IR wavelength regions \citep[][for the VERITAS Collaboration, the VLBA 43 GHz M 87 Monitoring Team, the H.E.S.S. Collaboration, and the MAGIC Collaboration]{acciari09a}. 

With sufficiently high spatial resolution, the \gray s produced by IC scattering of CMB and EBL photons off
the relativistic electrons of the lobes of radio galaxies can be used to set limits on their energy density in the immediate vicinity of these objects \citep[][see \S6 below]{georganopoulos08}.\\
{\bf \underline{Starburst galaxies}}:
The detection of starburst galaxies M82 \citep{acciari09d,abdo09} and NGC~253 \citep{acero09}, potentially opened a new wavelength regime for studying the EBL.  The  \gray s in starburst galaxies are generated by cosmic rays that are accelerated by a large number of  supernova remnants, giving rise to hard \gray\   spectra that extend to energies of 10s  of TeV.  The $\sim 10$~TeV opacity to nearby starburst galaxies is quite small,  and about unity at energies of $\sim$~50 - 100~TeV. Nearby starbursts are therefore important probes of the EBL at far-IR ($\sim 100$~\mic) wavelengths that cannot be probed by other \gray\ sources because of the relative softness of their spectra compared to those of SB galaxies.

%------ figure 5 ------ spectral break vs redshift
 \begin{figure}[t]%[htbp]
  \centering
  \includegraphics[width=3.88in]{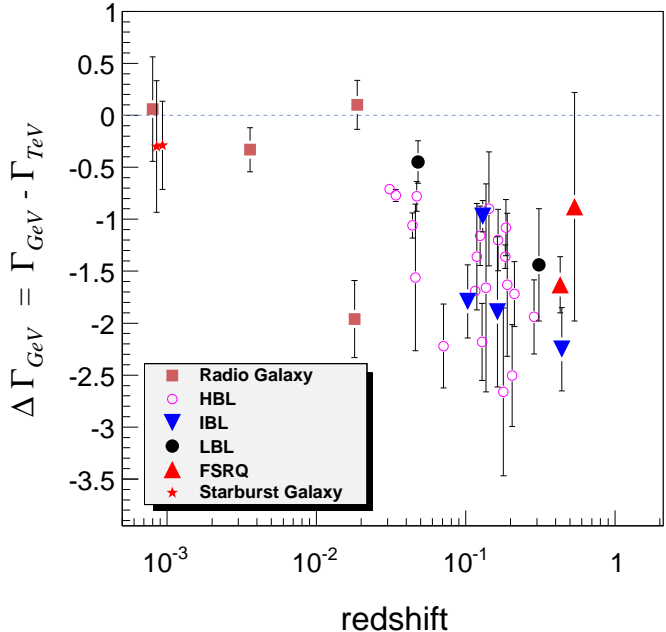}
  \hspace{0.1in}
%  \vspace{0.1in}
  \caption{{\footnotesize The difference between  $\rm  \Gamma_{GeV} $, the spectral index at GeV (Fermi) energies,  and $\rm  \Gamma_{TeV} $, the energy spectral index in the TeV regime (H.E.S.S, MAGIC, VERITAS) is shown as a function of their redshift.  Red squares (radio galaxies), red stars (starburst galaxies), empty circles (HBLs, high-frequency peaked BL Lacs), blue downward triangles (intermediate-frequency peaked BL Lacs), filled circles (LBLs, low-frequency peaked BL Lacs), red upward triangles (FSRQs, flat spectrum radio quasars)  indicate the different types of \gray\ sources.}\label{fig:gevtevbreak} }
\end{figure}
%--------------------

%--------------------------------------------------------------------------------------------------------
\subsection{The Spectra of Extragalactic GeV/TeV Sources}
%--------------------------------------------------------------------------------------------------------
Over a sufficiently small energy range the blazar spectrum can be characterized by a power law, $\rm dN/dE \propto E^{-\Gamma}$, with different indices, $\Gamma_{GeV}$ and $\Gamma_{TeV}$, at GeV and TeV energies, respectively. 
An important characteristic of the observed spectra is the presence of a break, defined as $\rm \Delta \Gamma_{GeV} \equiv \Gamma_{GeV} - \Gamma_{TeV}$, occurring between GeV and TeV energies, the exact location depending on the source's redshift. A source with an intrinsic spectrum characterized by a single power law out to TeV energies will have a value of $\rm \Delta \Gamma_{GeV} = 0$. Without any intergalactic absorption this value will remain constant with redshift. 

The spectral index $\Gamma_{GeV}$ is obtained from a power law fit to the $\sim 1 - 10$~GeV region of the spectrum which is unaffected by EBL absorption. If the intrinsic blazar spectrum is an extension of this power law to energies of $\sim 1$~TeV, then any spectral break ($\Gamma_{TeV} > \Gamma_{GeV}$) in the observed spectrum can be regarded as evidence for EBL absorption.   

A spectral break analysis of the amount of EBL absorption provides therefore a powerful method for studying the EBL. It is a differential method that replaces knowledge of the intrinsic blazar spectrum with a weaker requirement, namely that the power law representing the intrinsic blazar spectrum at GeV energies can be extended to TeV energies as well. 

Table~2 lists the values of $\Gamma_{GeV}$ and $\Gamma_{TeV}$ and the redshifts for all GeV and TeV detected blazars.   
Almost all sources exhibit a spectral break ($\rm \Delta \Gamma_{GeV} < 0$) at energies between 10~GeV and 1~TeV.
Figure~5 depicts the dependence of $\rm \Delta \Gamma_{GeV}$ on redshift. 
The figure  shows a clear trend of increasing $|\rm \Delta \Gamma_{GeV}| $  with redshift, strongly suggesting that the break is the consequence of the attenuation of the source spectrum by the EBL. As the optical depth increases with redshift, the observed \gray\ spectrum becomes softer, the position of the break moves to lower energies, and $\rm \Delta \Gamma_{GeV}$ becomes more negative. The detailed redshift dependence of $\rm \Delta \Gamma_{GeV}$  reflects the evolution of the spectrum and proper photon number density of the EBL with redshift (see \S7). 

Figure~\ref{fig:gevtevbreak} also shows that there is  significant scatter in $\rm \Delta \Gamma_{GeV}$ at any given redshift.  This suggests that some sources have considerable intrinsic hardening/softening in their spectra. Indeed, observations show that  sources exhibit a wide range of spectral trends, i.e.,  spectral softening (3C~279, PKS~1510-08) and spectral hardening (1ES~0502+675) in the GeV regime \citep{abdo10a},  while the spectra of other blazars  \citep[Mrk~421,][]{abdo11a}, \citep[Mrk~501,][]{abdo11b} extend without any cutoff to 300~GeV  \citep{abdo10b}, or multi-TeV energies \citep{krennrich02}, depending on the flaring state of the source. A lower limit in the scatter of $\rm \Delta \Gamma_{GeV}$ at a given redshift may suggest the combined effects of intrinsic spectral softening of the source spectrum and the effects of EBL absorption.  An upper limit in the scatter may simply indicate that the break in the spectral index is  only created by EBL attenuation.
% Under this assumption,  the upper bound corresponds to a lower limit to the EBL, while the lower bound corresponds to an upper limit to the EBL.

Disentangling intrinsic spectral softening from the effects caused by the EBL is complicated, and requires a clear understanding of the physical processes that cause the intrinsic softening or hardening (which is very rare) between the GeV and TeV energy regions of the source spectrum.  

Multiwavelength observations are required to solve this problem. Such observations will provide conclusive tests of non-thermal  \gray\ emission models, and constitute an important step towards achieving  the ultimate goal of using the radio-optical-X-ray and low energy \gray\ spectra as a predictor for the TeV spectrum.  The application of this approach to different source classes, i.e., blazars, radio galaxies and starburst galaxies  could provide additional redundancy to help constrain the EBL.

 The current constraints on the UV/optical to mid-IR regions of the EBL have come predominantly from studies of blazars, since  the vast majority of  the  extragalactic GeV/TeV \gray\  sources are FSRQs and BL Lacs.   AGN population studies with {\it Fermi}  reveal only a small number of non-blazar AGNs \citep{ackermann11}.  Consequently, we will focus in the following sections on EBL limits that were derived from studies of blazar spectra, which have   already provided important limits  on the near- to mid-IR spectra of the EBL.

%--------------------------------------------------------------------------------------------------------
\subsection{Phenomenology of Blazar Spectra and Models for TeV \gray\ Production}
%--------------------------------------------------------------------------------------------------------

\begin{figure}[t]%[htbp]
  \centering
 \includegraphics[width=3.88in]{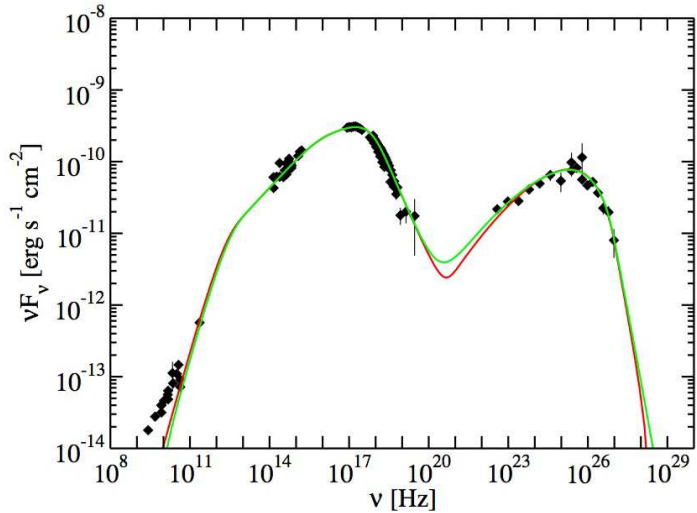}
  \hspace{0.1in}
%  \vspace{0.1in}
  \caption{{\footnotesize The SED of Mrk~421 depicts the synchrotron peak at X-ray energies, and the inverse Compton peak at TeV energies. The figure was taken from \cite[][Figure 11]{abdo11a}.} \label{fig:mrk421spec}}
\end{figure}

The non-thermal emission spectra of blazars generally exhibit  two emission peaks in $\rm \nu F_{\nu}$, the power emitted per unit logarithmic photon energy [see Figure (\ref{fig:mrk421spec}), \citep{abdo11a}].   The peak in the radio-UV-X-ray waveband is unequivocally attributed to synchrotron radiation that is produced by ultra-relativistic electrons. The second peak, located at  X-rays or  \gray\ energies, is commonly believed to stem from soft photons that were upscattered by the inverse Compton process to X- and \gray\ energies by the very same electrons responsible for the synchrotron emission. This mechanism for creating the second peak is often referred to as the synchrotron-self-Compton (SSC) mechanism. If the second peak includes upscattered photons drawn from other ambient radiation fields the mechanism is called the external-Compton (EC) mechanism.

%------ figure 6 ------ Mrk 421 spectrum
%--------------------

Alternatively, the second emission peak can be produced by a hadronic jet containing a significant amount of energy in ultra-high-energy (UHE) protons that subsequently interact with soft photons via $\rm p + \gamma \rightarrow \pi^{0},  \pi^{\pm}, .... $ interactions. The pions decay giving rise to \gray s, electrons, muons, and neutrinos. The strong magnetic fields required to collimate the hadronic jets also lead to the generation of considerable synchrotron radiation by protons and charged leptons in the pair cascade. The electrons from the pair cascade contribute to the lower energy synchrotron peak, whereas the muons together with the secondary photons from neutral pion decay contribute to the higher energy \gray\ peak. For a recent review of blazar models see \cite{boettcher10} and \cite{dermer12a}. 

 BL Lacs are classified loosely by their synchrotron peak position. They are referred to as low-frequency peaked BL Lacs (LBLs) if their synchrotron peak  is at  $ 10^{13}  \le \nu_{peak} \le 10^{14}$~Hz; as intermediate-frequency peaked BL Lacs (IBLs) if $\rm     10^{15}  \le   \nu_{peak}   \le 10^{16}  $Hz; and as high-frequency peaked  BL Lacs (HBLs) if $\rm \nu_{peak} \ge 10^{17}  $Hz \citep{nieppola06}.  The corresponding \gray\ peak  follows a similar pattern, with the peak energy progressively shifting towards higher energy as the synchrotron peak shifts to higher frequencies.  LBLs peak  at  a few GeV (similar to FSQRs), IBLs in the tens of GeV, and HBLs generally peak beyond 100 GeV.

 As the \gray\ peak is found at increasingly higher energy, its luminosity also decreases relative to that of the synchrotron peak, a trend often referred to as the Fossati blazar sequence, \citep{fossati98}.  As a result, the multiwavelength spectra of FSRQs and LBLs exhibit a prominent and large \gray\  luminosity,  whereas in IBLs and in particularly HBLs, the \gray\ peak  is typically dwarfed by the  synchrotron emission.  This makes  both FSRQs and BL Lacs very useful and complementary for EBL studies.   FSRQs are extremely bright in the GeV regime and despite their low peak energy, some are still detectable in the sub-TeV regime with atmospheric Cherenkov telescopes. They are thus becoming increasingly useful for  EBL constraints in the UV/optical/near-IR.  HBLs are at the other extreme, their \gray\ peak can extend well into the multi-TeV regime, while their \gray\  to synchrotron  luminosity ratio is much smaller.  However,  during strong flares some  HBLs  have been found to show  a \gray\  dominated spectral energy distribution. They are therefore useful for  constraining the EBL at near- to mid-IR wavelengths.

In general, HBLs can be well described by basic SSC models where the ultrarelativistic electrons and their target photons are closely linked via synchrotron radiation. IBLs are better described by external Compton (EC) models, that include a strong ambient photon field external to the blazar jet, thereby providing additional target photons for IC scattering \citep{acciari09e}. FSRQs, such as 3C 279, are difficult to fit with either model and may require the addition of a hadronic component.

%In general,  HBLs can be well described by basic SSC models where the ultrarelativistic electrons and their target photons are closely linked via synchrotron radiation.  IBLs are better described by a model that includes the strong external (to the jet) photon fields, that naturally provide additional target photons for IC scattering, which is included in the EC model \citep{acciari09e}.  FSRQs such as 3C~279 are difficult to fit with either model and may require the addition of a hadronic component.

The detailed fitting of blazar spectra with models requires extensive multiwavelength monitoring including optical, X-ray, and  \gray s.     Results from such multi-wavelength campaigns carry the potential to reduce the spread in the $\rm \Delta \Gamma_{GeV}  (z) $ relation in figure~\ref{fig:gevtevbreak}.    While numerous  successful multiwavelength campaigns have been reported in the literature \citep{abdo11b,abdo11a,aleksic10b,fossati08,abramowski11}, the unequivocal interpretation of the spectral energy distributions with a clear prediction for the intrinsic  TeV spectrum has not been possible.

In spite of the fact that blazar spectra are complex, it is possible to set limits on the behavior of their GeV/TeV spectra that are based on fundamental physical limits  imposed  by energy losses in the particle acceleration and radiation processes. The most prominent limitation is the maximum hardness of the \gray\ spectrum which provides an  important constraint on the EBL.    The energy spectrum of electrons produced in models of diffusive shock acceleration in blazar jets \citep{malkov01}, strongly constrains the hardness of the resulting \gray\ spectra produced in SSC and EC models, limiting their power law index to values larger than $\Gamma  \approx 1.5$.   

Even spectra that obey this limit are difficult to produce at higher energies where the Klein-Nishina effect softens the energy spectra substantially.  The detection of relatively hard TeV spectra of blazars with redshift $\rm \approx 0.1-0.2$ therefore came as a surprise,  since  the absorption corrected spectra are already reaching the $\rm \Gamma   \approx 1.5$ limit for the minimal EBL imposed by the IGL \citep{aharonian06, levenson08, krennrich08, ackermann11,abdo10c}.

To which extent a problem of hard TeV spectra persists depends on the theoretical scenarios invoked to explain these  spectra.  A solution to the problem was proposed by \cite{katarzynski06}. In their model, a high low-energy cutoff in the electron distribution could give the appearance of a hard \gray\ spectrum for a given energy regime.    Other ideas by \cite{aharonian08} show that \gag\ absorption in the source due to narrow band emission from the AGN could lead to unusually hard TeV spectra. The  emission produced by  proton synchrotron radiation \citep{aharonian07, zacharopoulou11}  combined with internal absorption at lower energies has been shown to produce spectra that exceed the hardness achievable by diffusive shock acceleration.    Other explanations avoid  substantial EBL absorption by introducing axion-like particles that couple with photons in intergalactic magnetic fields thus reducing the \gray\ opacity of the universe substantially \citep{sanchez-conde09}. Mechanisms that would substantially weaken EBL limits derived from \gray\ observations are discussed in Section 8.

%--------------------------------------------------------------------------------------------------------
\subsection{Unphysical Blazar Spectra}
%--------------------------------------------------------------------------------------------------------

In spite of the difficulties and the complexity associated with modeling the intrinsic \gray\ spectra of  blazars,   models generally predict a  common feature of blazar spectra:   they follow a power law with curvature and/or exponential cutoff and overall can be described by a concave shape  (curvature in energy flux is downward rather than upward).  This means that any absorption corrected \gray\ spectrum showing an exponential rise cannot represent a physical source spectrum, and must have its origin in an over-correction for EBL absorption\footnote{This assumes that no other external effects such as pileups due to  pair halos are playing a major role in the modification of blazar spectra.  The production of pair halos would require a magnetic field intensity between $10^{-7}$ and $10^{-12}$~G, sufficiently large so that electrons are isotropized, and sufficiently small so that IC losses exceed synchrotron losses.  Instead of a power law with curvature term, a power law with exponential cutoff or a broken power law is also consistent with the above statement.}.

The relation between the intrinsic, $\rm (dN/dE)_{int}$, and observed, $\rm (dN/dE)_{obs}$, blazar spectrum is given by:
%------------------  eq. 7
\begin{equation}
\label{eq:dnde-obs}
\left({dN \over dE}\right)_{int} =   \left({dN \over dE}\right)_{obs}    e^{\tau_{\gamma \gamma}(E_{\gamma}, z)}
\end{equation}
%--------------------- 
The equation illustrates that an overestimate of the \gag\ opacity will lead to an exponential rise in the inferred intrinsic spectrum of the blazar. Such exponential rise is unphysical. It runs contrary to our basic understanding of blazar models, and is absent in the \gray\ spectra of blazars at energies below 10~GeV \citep{abdo10c}.  This behavior can therefore be used to exclude EBL models with optical depths that will  result in an exponential rise in the corrected blazar spectrum \citep{guy00, dwek05a}. 
 
Similarly,  EBL scenarios can lead to an absorption correction for which the reconstructed intrinsic spectra follow a power law with an extremely hard spectral slope.   Additional constraints can be derived from the spectral slope itself,   however these are model dependent.
EBL models leading to slopes of $\rm \Gamma  < 1.5$ for the intrinsic spectrum, have  been rejected by  \cite{aharonian06} on the basis of the diffuse shock acceleration model.   Caveats  to this approach have been extensively discussed in the literature \citep{stecker07, katarzynski06, bottcher08, aharonian08, krennrich08, lefa11,zacharopoulou11}, and while some extreme blazars may exhibit harder spectra with  $\rm \Gamma < 1.5$,  most GeV energy spectra of blazars obey the $\rm \Gamma  = 1.5$ limit, with few exceptions \citep{ackermann11}.

These general constraints on the hardness of the intrinsic blazar spectra have been used to derive constraints on the EBL by using few individual objects \citep{dwek05a, aharonian06}, and by using samples of blazars \citep{mazin07,meyer12,orr11}.

%===============================================
\section{The Extragalactic Background Light I: Measurements \\and Limits}
%===============================================

%--------------------------------------------------------------------------------------------------------
\subsection{Spectral Measurements and Limits}
%--------------------------------------------------------------------------------------------------------
The specific intensity of the EBL is usually presented in units of \nwat. The conversion between these units and others that are sometimes used in the literature is given by:
%---------------
\begin{eqnarray}
\label{eq:conv}
\nu \, I_{\nu}(\lambda)\, [\rm {nW~m}^{-2}~\rm {sr}^{-1}]  & = & {3000 \over \lambda(\mu m)} \, I_{\nu}(\lambda)\, [\rm {MJy~sr}^{-1}] \\ \nonumber
 & = & {9.85\times10^{-3}\over \lambda(\mu m)}\, I_{\nu}(\lambda)\, [\rm {mJy~deg}^{-2}] \nonumber
\end{eqnarray}

The EBL intensity can be determined in several ways. The first consists of direct measurements, a method that poses considerable technical and astronomical challenges. Technically, it requires the absolute calibration of the instruments, and the understanding and removal of all measurements uncertainties. Astronomically, it requires the removal of strong foreground emission from interplanetary dust particles (the zodiacal light, ZL) and from stellar and interstellar emission components in the Milky Way. A thorough review of the challenges in determining the EBL was presented by \cite{hauser01}.   

A strict lower limit to the EBL intensity can be obtained by adding the light emitted by resolved galaxies. In principle, the integrated galaxy light (IGL) can converge to the total intensity of the EBL. A necessary condition for convergence is that the spectral index $\alpha$ of the differential galaxy number count versus flux $S$, $dN/dS \sim S^{-\alpha}$, becomes smaller than 2 at lower fluxes, so that the total integrated intensity, $\int S^2\, (dN/dS)\, dS$, is finite. At short wavelengths the intensity of the IGL is limited by the sensitivity of the survey. However, even in deep surveys the convergence of the IGL does not ensure the measurement of the total EBL intensity, since the low surface brightness regions of galaxies may be missed in standard aperture photometry \citep{bernstein02,levenson08}. Furthermore, a truly diffuse background will always remain undetected in such surveys. 

At longer wavelengths and large beam sizes, unresolved galaxies become a source of confusion, limiting the depth of the survey. Below a certain flux (the confusion limiting flux) individual sources become indistinguishable from the variation in the sky brightness caused by statistical fluctuations in the number of faint resolved or unresolved sources \citep[][and references therein]{dole04}. These limitations can be partially circumvented by  stacking analysis. Stacking of astronomical images of sources detected at one wavelength enhances their signal relative to the random background fluctuations at some other wavelength \citep[e.g.][]{dole06}.  The integrated light obtained by this method is thus closer to the EBL intensity than that obtained by integration down to the confusion limit.

Finally, given a model for $dN/dS$, one can extrapolate the differential source count to very faint fluxes, and evaluate the sensitivity of the integrated intensity to the lower flux limit and functional shape of the extrapolation. 

\begin{figure}[t!]%[htbp]
  \centering
  \includegraphics[width=4in]{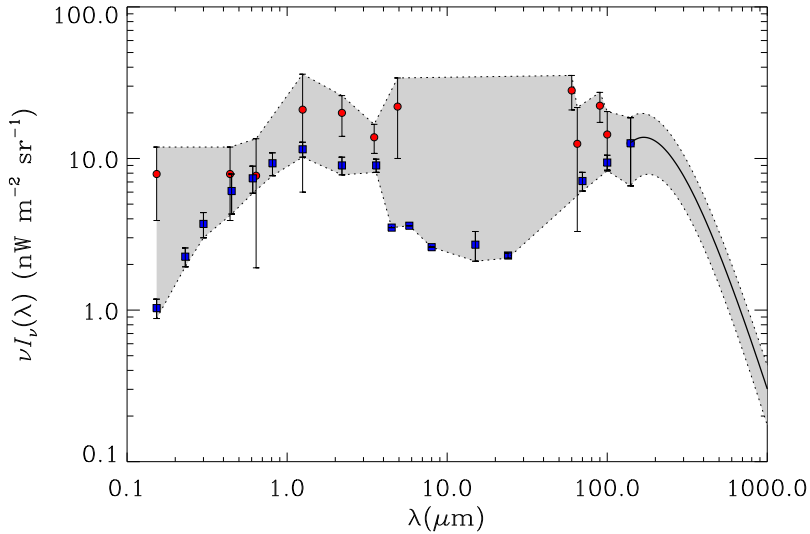} 
%  \vspace{0.1in}
  \caption{{\footnotesize  Limits on the EBL intensity. Lower limits (blue squares) are determined by the intensity of the IGL. Upper limits (red circles) are determined by absolute measurements of the EBL. The data used in the figure are listed in Tables~3-5 in bold. The shaded area depicts the range of the allowed EBL intensity as determined by UV to sub millimeter observations.}  \label{fig:ebl_limits}}
\end{figure}

Tables 3-5 list measurements and limits on the EBL intensity derived by the different methods described above with the different satellites, balloons, and ground observatories.  
Select measurements were used to define the gray area in Figure~\ref{fig:ebl_limits}. Absolute measurements and their 1$\sigma$ uncertainties were used to define the upper limits on the EBL. The integrated light from resolved galaxies and their 1$\sigma$ uncertainty was used to define the lower limit on the EBL. Lower limits derived from stacking analysis were used when available.  At 140~\mic\ the DIRBE detection with the FIRAS calibration  \citep{hauser98}, and at longer wavelengths the  FIRAS detections by \citep{fixsen98} were used to define the limits on the EBL. The measurements used to define the upper and lower limits on the EBL  are shown as bold entries in Tables 3-5. 
The figure shows that the EBL is poorly determined in the $\sim 5 - 60$~\mic\ wavelength region, where the foreground emission from the interplanetary dust cloud is strongest \citep{kelsall98}.

%----- figure 7 ----- observational EBL limits
%--------------------

%--------------------------------------------------------------------------------------------------------
\subsection{Integral Constraints on the EBL Intensity} 
%--------------------------------------------------------------------------------------------------------
The total EBL intensity per unit solid angle is given by:

%---------------
\begin{equation}
\label{eq:ebl1}
I_{EBL}  =  \left({c\over 4\pi}\right)\, \int_{0}^{\infty}\ {\cal L}( z)\, \left|{dt\over dz}\right|\, {dz \over 1+z} 
\end{equation}
where ${\cal L}(z)$ is the luminosity density in a comoving volume element at redshift $z$, and $|{dt\over dz}|$ is given by eq. (\ref{eq:dldz}). 

The comoving luminosity density is dominated by the radiative output from stars. 
On a galactic scale, an AGN can dominate the optical to IR output of a galaxy, however, on a global scale AGN make only a small contribution to the total energy releases in the universe. AGN make up most of the X-ray background \citep{mushotzky00, draper09}, and a significant fraction of the cosmic radio background \citep{dwek02}. However, they make only a small contribution to the total IR background. Accretion onto a central black hole releases about 10\% of the rest mass energy of the accreted matter, significantly more than the 0.7\% released in nuclear processes. However, the mass locked up in BH is only $\sim 0.6$\% of that in stellar objects. 
Rest frame color-color diagrams generated from {\it Spitzer} IRAC and MIPS observations covering the 3.6, 4.5, 5.8, 8.0, 24, and 70~\mic\ bands, of radio-detected submillimeter-selected galaxies with spectroscopic redshifts show that AGN constitute a small fraction, between 13 and 19\%, of the sample dominating its mid-IR spectrum \citep{hainline09}.
So with these limits in mind, the total comoving luminosity density is a direct measure of the cosmic star formation rate (CSFR) at a given redshift.

% is AGNs contributing only a small fraction of the one can use  the integrated light released in the process of star formation can be used to put an integral constraint on the EBL 
%
%The cosmic star formation rate (CSFR) gives an integral constraint on the EBL intensity, provided that AGNs make a negligible contribution to the  tracers used to determine the CSFR, and to the integrated intensity of the EBL.

%----- figure 8 ----- CSFR versus redshift
 \begin{figure}[t]%[htbp]
  \centering
 \includegraphics[width=3.88in]{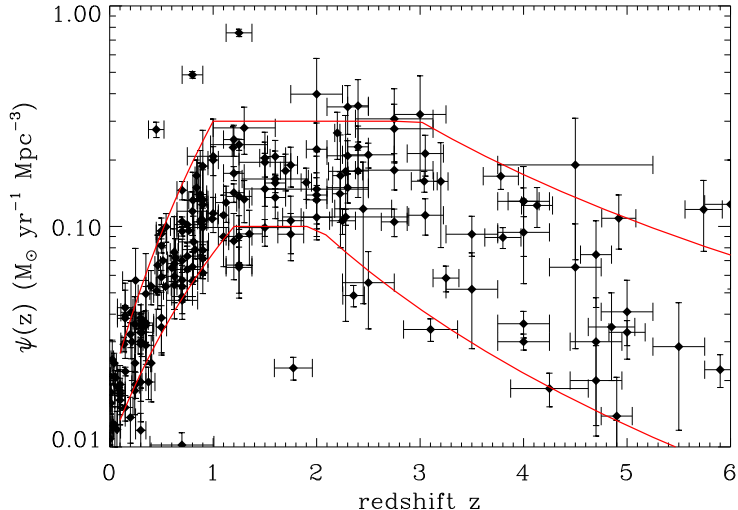} 
  \hspace{0.1in}   
%  \vspace{0.1in}
  \caption{{\footnotesize A compilation of the cosmic star formation rate as inferred from UV, H$_{\alpha}$, mid-IR, submillimeter, radio, and Ly$\alpha$ observations. Lower limits were excluded from the figure. The red curves are analytical approximations representing the upper and lower limits to the CSFR.} \label{fig:csfr} }
\end{figure} 
%--------------------

Figure~\ref{fig:csfr} presents the redshift dependence of the CSFR as determined from UV-optical emission lines, [O~II], [O~III], Ly$\alpha$, H${\alpha}$, H${\beta}$, and from mid-IR, submillimeter, and radio observations \citep {madau96,hopkins06,michalowski10c}. The data in the figure were taken from the compilation of \cite{michalowski10c}, who used the standard $\Lambda$CDM parameters to calculate volume densities and the \cite{kennicutt98b} law with a Salpeter stellar initial mass function (IMF) to convert luminosity densities to star formation rates.  
The two red lines represent a broken power law approximation to the upper and lower limits lines of the observations.
For consistency with the observational determination of the CSFR, the CSFR was converted to a bolometric intensity using the Salpeter IMF and a starburst age of 100~Myr: $L_{bol} = 7.5\times 10^9 \, \Psi$~\msun~yr$^{-1}$ \citep{dwek11b}. The integrated intensity of the CSFR is then bounded by:
\begin{equation}
I_{\small EBL}   =  21 - 66 \qquad \rm{nW~m}^{-2}~\rm{sr}^{-1}
\end{equation}

Table~6 presents the current limits on the EBL intensity, separated into its stellar and dust components, and compares them to this predicted by the various EBL models presented in \S5. 
%--------------------------------------------------------------------------------------------------------  
\subsection{Constraints on the EBL from Fluctuation Measurements}
%--------------------------------------------------------------------------------------------------------
Most of the EBL is generated by discrete galactic or primordial stellar sources. Fluctuations in their number and their clustering properties will give rise to spatial fluctuations in the EBL intensity. Studies of the optical region of the EBL via the fluctuations method were first conducted by \citep{shectman73, shectman74}, and of the near-IR region using the {\it COBE}/DIRBE data by \cite{kashlinsky96}. The fluctuations do not provide a direct measurement of the EBL intensity. The derivation of the EBL intensity from these measurements will require detailed knowledge of the galaxy source counts, their luminosity function, and their clustering properties as a function of redshift. However, spatial fluctuations in the EBL provide a different means of setting limits on its intensity. Fluctuation measurements of the EBL do not require absolute measurements of its intensity, since the removal of foreground emission components is done on the basis of their distinct spatial properties rather than their absolute intensities. At IR wavelengths, these fluctuations can have a spatial signal that is distinctly different from that generated by the interplanetary dust cloud or by interstellar dust.
After the removal of all known resolved sources, any residual fluctuations will measure the EBL contribution
from sources that may represent a yet unknown or unresolved population of stars or galaxies. Recent fluctuation measurements in the Spitzer 3.6 and 4.5~\mic\ bands \citep{kashlinsky12} show that they are in excess of those that can be attributed to known galaxy populations \citep{helgason12a}, suggesting an origin in a very faint, yet unknown, population of highly clustered sources. More details and references on the use of fluctuations analysis to constrain the EBL intensity can be found in \cite{hauser01} and \cite{kashlinsky05a}.  

EBL fluctuations at far-IR wavelengths were detected at 170~\mic\ \citep[][{\it ISO} data]{lagache00b}, at 160~\mic\ \citep[][{\it Spitzer} data]{lagache07} and by \cite{shang12}, using the {\it Planck} data. Fluctuation measurements and EBL colors have been used by \cite{penin12} to derive limits on the EBL at 100 and 160~\mic\ (see Tables 4 and 5).
 
%=======================================================
\section{The Extragalactic background Light II: Models}
%=======================================================
The EBL intensity only provides an integral constraint on all the radiative energy releases over cosmic time. 
%Its specific comoving intensity per unit solid angle at redshift $z_0$ and wavelength $\lambda_0$  is given by:
%%---------------
%\begin{equation}
%\label{eq:ebl2}
%I_{\nu}(\lambda_0, z_0)  =  \left({c\over 4\pi}\right)\, \int_{z_0}^{\infty}\ {\cal L}_{\nu}(\lambda, z)\, \left|{dt\over dz}\right|\, dz 
%\end{equation}
%where ${\cal L}_{\nu}(\lambda, z)$ is the specific luminosity density in a comoving volume element at redshift $z$, and
%$\lambda = \lambda_0/(1+z)$ is the wavelength in the rest frame of the radiating sources.
%The specific intensity of the EBL is usually presented in units of \nwat. The conversion between these units and others that are sometimes used in the literature is given by:
%%---------------
%\begin{eqnarray}
%\label{eq:conv}
%\nu \, I_{\nu}(\lambda)\, [\rm {nW~m}^{-2}~\rm {sr}^{-1}]  & = & {3000 \over \lambda(\mu m)} \, I_{\nu}(\lambda)\, [\rm {MJy~sr}^{-1}] \\ \nonumber
% & = & {9.85\times10^{-3}\over \lambda(\mu m)}\, I_{\nu}(\lambda)\, [\rm {mJy~deg}^{-2}] \nonumber
%\end{eqnarray}
%
In a dust-free universe, it represents the intrinsic stellar or AGN spectra. The EBL intensity and spectral shape depends then on the star formation history, the stellar initial mass function, the evolution of metallicity, the energy released by AGN, and the relative importance of energy releases by nuclear and gravitational processes. In a dusty universe, the comoving luminosity density, ${\cal L}(z)$, and total EBL intensity, $I_{\small EBL}$ remain unchanged, however the energy is redistributed by absorption and reemission processes over a large spectral range. The resulting spectrum depends on many factors, ranging from the size distribution, composition, and optical properties of the dust grains, the evolution of their abundance and properties over time, and on the morphology of the galaxy which determines the spatial distribution of the dust relative to the radiative sources. 

Several distinct approaches have been used to model the intensity and spectral distribution of the EBL at $z=0$. They all represent different approaches for calculating the evolution of ${\cal L}_{\nu}(\lambda, z)$ with redshift [see eq. (2)]. Backward evolution models start from the local determination of ${\cal L}_{\nu}(\lambda, z=0)$, evolving it with redshift using observed galaxy number counts at different wavelengths. Forward evolution models use the CSFR to determine ${\cal L}_{\nu}(\lambda,z)$ as a function of redshift, and population synthesis and radiative transfer models to determine the distribution of the energy over wavelengths. Cosmic chemical evolution models are similar to previous models, except that their system is the universe as a whole. Semi-analytic models calculate ${\cal L}_{\nu}(\lambda, z)$ by including the appropriate physical processes in a more general model for the formation and evolution of structure in the universe.  In the following we discuss these different models in somewhat more detail. For a more extensive discussion and description we refer the reader to the review by \cite{hauser01}.

%--------------------------------------------------------------------------------------------------------
\subsection{Backward Evolution (BE) Models}
%--------------------------------------------------------------------------------------------------------
BE models start with the construction of a library of the galactic spectral energy distributions (SEDs) representing those of galaxies in the local universe and evolve them back in time in order to fit observed number counts. The EBL intensity and spectral shape is used as an integral constraints on their evolution. The galaxies in such library should represent the range of observed galactic morphologies (spiral elliptical, irregular) and activities (AGN, normal, starburst, mergers) in the local universe. The SEDs of the galaxies comprising such library should also satisfy observed statistical properties of the ensemble of local galaxies, such as: the trend of increasing $S(60\, \mu m)/S(100\, \mu m)/)$ and decreasing $S(12\, \mu m)/S(25\, \mu m)/)$ flux ratios with increasing IR luminosities (\citep{soifer91}; and the number density of galaxies in the $L+dL$ luminosity interval, represented by the luminosity function (LF), $\Phi(L) dL$. 

%The specific local luminosity density, ${\cal L}_{\nu}(\lambda, 0)$ is given by the integral of the LF in the different wavebands. References to different libraries are: Chary \& Elbaz 2001; Xu et al. Rowan-Robinson et al. Lagache et al. Franceschini et al.  ;  

Many functional forms have been adopted to characterize the local LF. The most commonly one used at optical and near-IR wavelengths is the Schechter (1976) LF. At wavelengths above $\sim 10$~\mic\ the galaxies' SED is dominated by thermal emission from dust, and their LF cannot be adequately represented by the Schechter LF. Several distinct functional forms have been used to characterize the LF in the different IR wavebands. They are usually characterized by three parameters: a normalization parameter, $\Phi_{\star}$, a characteristic luminosity, $L_{\star}$, that determined the transition point between the low and high luminosity behavior of the L; and a power law index, $\alpha$ that determines the behavior of the LF at low luminosities. A list of references to the functional forms derived from  the {\it IRAS} survey can be found in Hauser \& Dwek (2001). Of those, the parameters of the \cite{saunders90} LF have been recently updated to fit the differential 24~\mic\ number counts obtained by deep {\it Spitzer} surveys \citep{rodighiero10}. 

If neither the galaxies' SED nor their comoving number density evolved with time, the spectral luminosity density, ${\cal L}_{\nu}(\lambda, z)$ would be independent of redshift. However, the recent deep surveys with the {\it Spitzer} and {\it Herschel} satellites show strong evolution in number counts, compared to predictions made with no evolution models \citep{lagache05, rodighiero10}. Evolution in the LF or, equivalently, the spectral luminosity density can be inferred directly from observations if the redshift of the sources is known, and their number counts are complete \citep[e.g.][]{rodighiero10,dunne00}. Alternatively, evolution in the LF can be introduced by adding a redshift dependence, usually of the form $(1+z)^{\gamma}$, to the basic parameters, $\Phi_{\star},\, L_{\star}$, and $\alpha$, that characterize the LF. The value of $\gamma$ is then derived by fitting model prediction to the observed galaxy number counts in a given waveband. Evolution in the LF can also be modeled by evolving the relative number of the different type galaxies: quiescent star forming galaxies, starbursts, AGNs, and ellipticals with redshift \citep{rowan-robinson01, rowan-robinson09, lagache03, dominguez11}, or by simply evolving the relative number of ultraluminous infrared galaxies (ULIRGs), characterized by IR luminosities in excess of $\sim 10^{12}$~\lsun, relative to the rest of the galaxy population [e.g. \cite{chary01}].
Having determined the evolution of the spectral luminosity density with redshift, the EBL is obtained by a simple integration of ${\cal L}_{\nu}(\lambda, z)$ over redshift. 

\cite{helgason12b} adopted a more empirical approach than standard BE models, using number-count derived luminosity functions at UV (0.1~\mic) to mid-IR (25~\mic) bands, to construct  the luminosity density, ${\cal L}_{\nu}(\lambda, z)$, as a function of redshift. 
  
BE models are relatively simple, and their predictions can be easily compared to observed galaxy number counts, their magnitude-color and magnitude number density relations, and their redshift distribution. They are only loosely constrained by physical processes. The predicted dependence of the comoving bolometric luminosity density with redshift should be consistent with that inferred from limits and observations of the cosmic star formation rate.  

%--------------------------------------------------------------------------------------------------------      
\subsection{Forward Evolution (FE) Models}
%--------------------------------------------------------------------------------------------------------
FE models use the redshift dependence of the cosmic star formation rate (CSFR), inferred from a variety of wavebands and line observation \citep{madau96,hopkins06}, as a starting point in their calculations. The CSFR is determined  from the UV-optical line and continuum, and IR and radio emission in the various wavebands using statistically determined conversion factors derived from galaxies in the local universe  \citep{kennicutt98b, Haarsma00}. Determination of the CSFR is complicated by extinction effects at UV and optical wavelengths, and by the implicit assumption that the IR luminosity is powered by stars and representative of the total bolometric luminosity of the galaxies. Even if the total bolometric luminosity of a given galaxy is determined, its conversion to a star formation rate requires knowledge of the stellar IMF, a poorly determined quantity at high redshifts, and the duration of the starburst activity at each redshift. Once the CSFR is determined, FE models use population synthesis models such as P\'EGASE \cite{fioc97}, Starburst99 \cite{leitherer99}, \cite{bruzual03} or \cite{maraston11} to calculate the stellar bolometric and spectral luminosity density as a function of redshift.

The most difficult part of this approach is determining the fraction of starlight that is absorbed by dust, and the spectrum of the reradiated IR emission. The SED of a galaxy can be determined with radiative transfer models, such as GRASIL \citep{silva98}, DIRTY \citep{gordon01}, or DUSTY \citep{nenkova00}. Alternatively, one can use a parametric approach  in which the fraction of UV-optical light absorbed by the dust, and the reradiated infrared spectrum are statistically determined from observations. EBL spectra derived from FE evolution models were presented by \cite{dwek98a}, \cite{razzaque09}, and \cite{finke10}. 

Population synthesis models, combined with simple radiative transfer calculations, are useful for determining the UV to radio SED of individual galaxies. However, the cosmological application of such models assumes that star formation is a monolithic process, in which in all galaxies star formation commenced at the same redshift and evolved quiescently until the current epoch. The models do not allow for galaxy interactions, stochastic star formation histories associated with merger events, or any morphological evolution of galaxies. Any discrepancies between model predictions and galaxy number counts must be introduced in an ad hoc fashion by evolving the stellar IMF, or by introducing a new population of galaxies at the appropriate redshifts.  

%--------------------------------------------------------------------------------------------------------
\subsection{Cosmic Chemical Evolution (CCE) Models}
%--------------------------------------------------------------------------------------------------------
CCE models treat the universe as a closed system in which all galaxies within a large comoving volume element are represented by their basic ingredients:  stars, interstellar gas,  metallicity, and radiation. Chemical evolution equations, analogous to those used to follow the chemical evolution of the Galaxy \citep[e.g.][]{audouze76,tinsley81,pagel01}, are used to follow the  evolution of the average stellar, gaseous, and radiative contents in each comoving volume in a self consistent manner. CCE models were pioneered by \cite{pei95}, and most recently updated by \cite{pei99}. Inputs parameters for their model are the mean rest frame UV luminosity density as a function of redshift, and the mass of the ISM gas as determined from H~I column densities derived from studies of quasar absorption lines through damped Ly$\alpha$ systems. The decrease in the ISM gas with redshift and the UV luminosity density were used to derive a solution for the evolution of the CSFR with redshift which is consistent with that determined from the extinction-corrected H$\alpha$, and with SCUBA~850 and {\it ISO} 15~\mic\ surveys. Similar to FE models, population synthesis models were then used to calculate the stellar SED at each redshift, and an LMC extinction law was adopted to calculate the fraction of starlight absorbed by the dust. A power-law distribution in dust temperature was used to calculate the spectrum of the reradiated IR emission. The model reproduced various observational constraints, including the comoving rest-frame 0.44, 1.0, and 2.2~\mic\ spectral luminosity densities in the $\sim 0-2$ redshift interval, the 12, 25, 60, and 100~\mic\ local luminosity densities; and the mean abundance of metals in damped Ly$\alpha$ systems in the $\sim 0.4 - 3.5$ redshift interval.   

%--------------------------------------------------------------------------------------------------------
\subsection{Semi-analytical (SA) Models}
%--------------------------------------------------------------------------------------------------------
SA models follow the formation and evolution of galaxies in a cold dark matter Lambda dominated ($\Lambda$CDM) universe using the cosmological parameters derived from the 5-year{\it Wilkinson Microwave Anisotropy Probe} ({\it WMAP5}) observations \citep{hinshaw09} as the initial conditions. SA models then follow the growth and merger of dark matter halos, and the emergence of galaxies which form as baryonic matter falls into the potential wells of these halos. The fate of the infalling gas is determined by many different processes: the formation of stars in a multiphase interstellar medium, AGN and supernovae feedback processes that quench their formation, the evolution of the stellar radiation field, the heating and cooling of the interstellar medium and its chemical enrichment, the exchange of material with the intergalactic medium through infall and galactic winds, and the growth of the central black hole. A description of recent developments and references to previous work can be found in \cite{somerville11}.  
Model prediction are compared to a basic set of observational constraints such as the observed characteristics of galaxies: their morphology, colors, and spectral energy distribution, and morphology; and their integrated cosmological properties: their number counts and luminosity function in different wavebands and redshifts, their mass function, the cosmic star formation rate, and the EBL generated by them.
As in all EBL models, determination of the galaxies' SED is complicated by the detailed microscopic and large scale parameters needed to calculate the amount of starlight that is absorbed by dust, and the spectrum of the reradiated emission. Recent SA models have combined the models for galaxy formation with radiative transfer models to determine the galaxies' SED \citep{fontanot09, fontanot11, somerville11, younger11}. 

SA models are inherently complex, incorporating a large number of physical processes, some poorly known, to derive galaxy properties. However, they are the most physically motivated models, and quite successful in reproducing a large number of observational constraints.

\subsection{Comparison of Model Predictions with Observations}
\begin{figure}[htbp]
  \centering
  \includegraphics[width=4in]{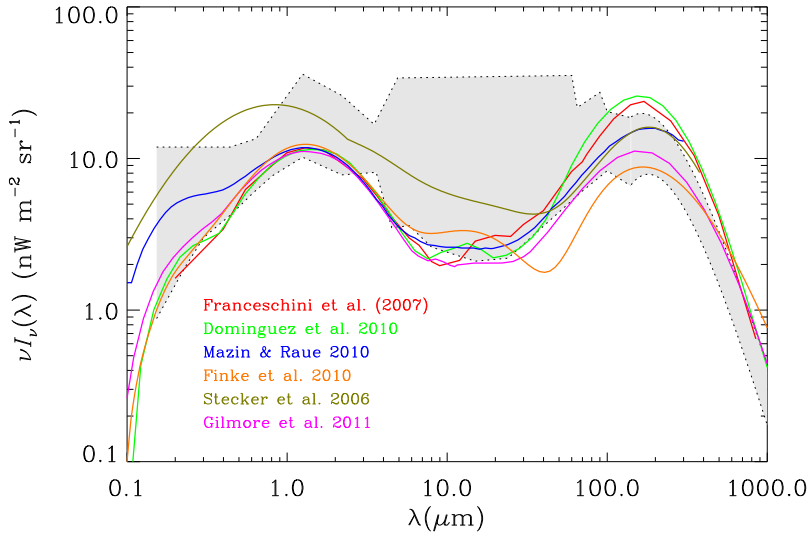} 
  \caption{{\footnotesize Models of the EBL are compared to observational limits on the EBL.}  \label{fig:ebl-models}}
\end{figure}

A detailed comparison of all model types with EBL limits and observations was presented by \cite{hauser01}. Here we will represent mostly the models that have been developed since then: BE models by \cite{stecker06}, \cite{franceschini08}, and \cite{dominguez11}; The FE model of \cite{finke10}; and the SA model of \cite{gilmore11}. Figure~\ref{fig:ebl-models} compares the various models to the current limits and observations of the EBL.
In general, all models, except for the BE models of Stecker et al. provide adequate fits to the EBL. 

%----- figure 9 ----- EBL models versus observational limits
%--------------------

%===============================================================
 \section{EBL Constraints from \gray\  Observations of Blazars}
%===============================================================

\begin{figure}[htbp]
  \centering
    \includegraphics[width=4in]{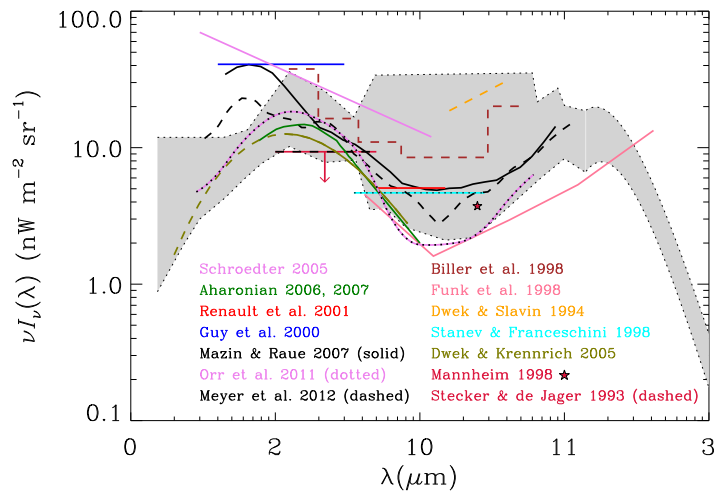} \\
%  \vspace{0.1in}
  \caption{{\footnotesize  Limits on the EBL as determined from \gray\ observations of blazars. Details in text.} \label{fig:tev-limits} }
\end{figure}

The attenuation of \gray s by the EBL can in principle be used to
determine the EBL intensity at wavelengths corresponding to the \gray\
observations. Neglecting the possible scattering or production of second
generation \gray\ photons along the line of sight to the blazar,  the
intrinsic \gray\ flux from the blazar, $F_{Int}(E_{\gamma})$, can be related
to the observed one, $F_{obs}(E_{\gamma})$ by:
\begin{equation}
\label{eq:fgamma}
F_{obs}(E_{\gamma}) = F_{Int}(E_{\gamma})\, \exp[-\tau_{\gamma
\gamma}(E_{\gamma})]
\end{equation}
where the optical depth, \taugg, is given by eq. (\ref{eq:taugg}).
Determination of the EBL assumes that all the attenuation is caused by
interaction with the EBL, instead of  photons in or around the vicinity
of the blazar. Furthermore, it requires knowledge of the intrinsic blazar
spectrum.
Assuming that all the attenuation is attributed to the EBL, several upper 
limits have been derived on the EBL intensity by making various
assumptions on the intrinsic blazar spectrum. The \gray\ derived EBL limits are compared to those derived from UV to sub millimeter observations in Figure~\ref{fig:tev-limits}, and described below. \\
%---------------
{\bf \underline {Fixed power law}}: Early observations of the blazars Mrk~421  and 3C~279  suggested that their $\sim$~GeV--TeV spectrum could be approximated
by a single power law \citep[][respectively]{punch92,stecker92}. If so, then any deviations of the observations
from the extrapolated power law to higher energies should be attributed to
EBL attenuation.
\cite{stecker93} derived  an upper limits at 1-5~\mic\ of 10 \nwat, assuming that a straight power law extrapolation of the  spectrum of Mrk~421 from  GeV energies obtained from the EGRET  \citep{lin92}  with an index of  $\Gamma = 1.96 \pm 0.14$ holds up to the TeV regime,  where the index was  measured by the Whipple collaboration to be $\Gamma = 2.25 \pm 0.19$.   \cite{biller95} included the statistical uncertainties of the GeV spectrum, and demonstrated vastly different extrapolations with significantly higher upper limits,  thereby yielding conservative upper limit to the EBL in the mid-IR at 10$\mu$m.

%----- figure 10 ----- TeV limits on EBL
%--------------------
Strong flaring activity of Mrk~501 provided a well measured TeV spectrum from 0.2 - 24 TeV \citep{aharonian99b, samuelson98, djannati-atai99}.   \cite{stanev98} calculated limits by fitting the energy spectrum of Mrk~501, with a range of possible absorption scenarios assuming intrinsic power law spectra and varying levels of  EBL intensity by scaling the lower limits from galaxy counts.   This provided  strong EBL limits in the near- and mid-IR.  \cite{funk98} followed  a similar approach, except that they used an EBL model by \cite{macminn96} as the basis for scaling the EBL intensity, yielding similar results in the mid-IR.
\cite{mannheim98} argued that the observed spectra would deviate from a power law, if the primary \gray\ spectrum
were substantially attenuated by the EBL.
An upper limit in the mid-IR, based on this hypothesis and on an energy spectrum of Mrk~501 from the HEGRA collaboration, is also shown in Figure~\ref{fig:tev-limits}.  Later on, \cite{vassiliev00} demonstrated that the absence of deviations from a power law does not preclude the presence of substantial absorption in the observed spectra.    

The obvious drawback of the method is that   the assumption of a  single power law for  the intrinsic blazar spectrum does not hold true over a wide range in energy. Blazar spectra  generally exhibit a concave \gray\  peak over a sufficiently large energy range.  Most blazar spectra measured by {\it Fermi} or by Cherenkov telescopes can be represented by a power law over the energy range covered by the instrument.\\
%---------------
 {\bf \underline{Synchrotron Self-Compton (SSC) Spectrum}}: The SSC model is a popular
model explaining the existence of the two peaks in the blazar spectrum: the synchrotron peak at radio-UV-X-ray energies and the inverse Compton (IC) peak at \gray\ energies. 
 The spectrum of the IC peak can be modeled using 
parameters that produce the synchrotron peak and the unabsorbed part ($E < 10$~GeV) of the IC spectrum \citep[see review by][]{dermer12a}.
 
Such models for the intrinsic  blazar spectrum have been used by \cite{guy00} to determine the intensity of  the EBL in the  1-5 \mic\ and 20-80 \mic\ wavelength region.   They applied a multiwavelength fit to the X-ray and TeV data of Mrk~501 in the framework of a standard homogeneous SSC model to derive the  level of absorption present in the TeV spectrum. As a result, they obtained an absolute upper limit on the EBL of 60~\nwat\, and a most likely value of 20~\nwat\ at 1~\mic.  They also pointed out that the lack of an absorption signature in the spectrum of Mrk~501, as suggested by the HEGRA telescopes, does not necessarily imply a lack of EBL absorption.  They emphasized that in the transition region from the near-IR to the mid-IR EBL, the opacity could be nearly constant. This is a consequence of the large width of \gag\ cross section (see Figure 2). So when $\sigma_{\gamma \gamma}$ is convolved with the number density of background photons, any strong wavelength variations in the EBL are smoothed out.  As a result, the observed TeV spectrum at  1 - 10 TeV would corresponds to the intrinsic blazar spectrum since the observed spectrum is now described by $\rm (dN/dE)_{int} \times e^{\tau}$  with $\tau$ a slowly varying function of energy. 

The drawback of using \gray\ emission models to constrain the EBL is the uncertainty in the many parameters  that determine the IC spectrum.  Furthermore, while HBLs  generally can be well  fit by SSC models, IBLs require the inclusion of additional ambient radiation fields that make a contribution to the \gray\  IC component. 

Additional complications arise from the fact that basic one-zone SSC models are not applicable for sources exhibiting ''orphan flares'', where only the TeV flux is enhanced while the synchrotron emission remains unchanged \citep{Krawczynski04}.  Finally, the biggest challenge for the SSC/multi-wavelength approach to constraining the EBL is to get simultaneous measurements for large sets of blazars. \\
%----------------------------------
 {\bf \underline{The $\Gamma > 1.5$ limit on the hardness of the blazar
spectrum}}: A more relaxed assumption on the intrinsic blazar spectrum
is that it cannot produce too many hard photons, so that the \gray\
spectrum, expressed as $dN/dE \sim E^{-\Gamma}$ cannot be flatter
than one with $\Gamma = 1.5$ \citep{malkov01}.    In the spirit of this limit to the spectral index,  \cite{renault01} explored a range of EBL scenarios based on measurements with the minimal assumptions that the intrinsic power of Mrk~501 is concave, effectively requiring a decreasing energy flux distribution above 4~TeV  ($\Gamma > 2.0$).  They derived an upper limit of 5 $\rm nW/m^2/sr$  at 10 \mic.

The strict assumption of $\Gamma > 1.5$ , was used by \cite{aharonian06} to derive upper limits on the 1--5~\mic\ on the EBL which are close to the lower limits determined by the IGL, suggesting that the EBL has been largely resolved at these wavelengths. A comprehensive study by \cite{mazin07} is based on eleven blazars over a redshift range from 0.03 - 0.18, and explores a large number (8 million) of hypothetical EBL scenarios to set upper limits on the EBL, again with the requirement that the source spectra cannot be harder than $\Gamma = 1.5$ or $\Gamma = 2/3$. The lower value arises from the extreme scenario of a mono-energetic energy distribution of ultra-relativistic electrons in which the resulting IC \gray\ spectrum could be as hard as $\gamma =2/3$, leading to two conditional upper limits.  The first condition yielded limits that are slightly above  that of  \cite{aharonian06}. The second, more relaxed condition, yielded limits that were higher by about 30\%. 

The theoretical validity of a strict hardness limit of $\Gamma > 1.5$ has been discussed by a number of authors, with no unanimous verdict \citep{katarzynski06,stecker07,bottcher08, aharonian08, lefa11,zacharopoulou11}.   Observational evidence, e.g. \cite{levenson08}, have provided lower EBL limits from galaxy counts that are higher than previous ones derived by \cite{madau00}.   If these new limits are correct,  they  imply \gray\   spectra that are slightly harder  than $\Gamma = 1.5$ \citep{krennrich08}. \\
%----------------------
 {\bf \underline{Unphysical exponential rise of the blazar spectrum}}: Less model dependent, and therefore more robust limits on the hardness of the intrinsic blazar spectra arise from the notion that an exponential increase of their luminosity with energy is unphysical.  All current  blazar models produce a concave spectrum, rendering intrinsic blazars spectra with an exponential rise in energy flux theoretically unfeasible.
The paradigm of concave intrinsic energy spectra was used by \cite{dwek05a} to reject many different realization of the EBL. Furthermore,  \cite{dwek05b} ruled out the extragalactic origin of the near-IR sky brightness observed by \cite{matsumoto05a}, since it would lead to an exponential rise  in the spectrum of the blazar PKS~2155-304, which is ruled out by observations \citep{aharonian05a}.   An EBL spectrum close to the IGL limits yielded a blazar spectrum consistent with the SSC model, suggesting that the EBL  was mostly resolved at near-IR wavelengths \citep{dwek05b, dwek05c}.  \\
%-----------------------
 {\bf \underline{Spectral break analysis due to EBL spectrum}}:  \cite{orr11} developed a novel approach to set limits on the EBL intensity by studying the effects of the spectral shape of the EBL on the sub-- to multi--TeV spectra of blazars.  The \gray\ opacity around 1 TeV is sensitive to the EBL intensity at $\sim 1$~\mic\ (see Fig. 3). Its subsequent energy dependence hinges on the rate at which the stellar UV, optical, and near-IR emission decreases towards mid-IR wavelengths. The near- to mid-IR intensity ratio of the EBL determines the relative $\sim 1$ to 10~TeV opacity. This is illustrated in Figure~\ref{fig:ebl-tau} which shows the energy dependence of $\tau_{\gamma \gamma}$. A large near- to mid-IR ratio in the EBL intensity would cause the \gray\ opacity to be relatively flat in the $\sim 1-10$~TeV region, resulting in a hard \gray\ spectrum. Conversely, a low  near- to mid-IR ratio would result in an increase in the $\sim 10$~TeV opacity relative to that at $\sim 1$~TeV, resulting in a softer blazar spectrum. 
 
The spectral shape of the EBL in the $\sim 1 - 15$~\mic\ region can be related to the break $\Delta \Gamma_{TeV} \equiv \Gamma(<1~TeV)-\Gamma(>1~TeV)$ in the blazar spectra. 
A study of 12 blazars spanning a redshift range from 0.03 to 0.186, showed a trend of increasing $|\Delta \Gamma_{TeV}|$ with redshift  with a statistical significance of  3.6$\sigma$  \citep{orr11}. This strongly suggests that the trend is caused by EBL absorption, providing strong constraint on the near-IR to mid-IR EBL intensity ratio. Combined with \gray-derived upper limits on the EBL, \cite{orr11} derived correlated constraints between the near-IR and mid-IR intensity ratio of the EBL and the EBL intensity at mid-IR wavelengths. \\
%------------------------
\noindent {\bf \underline{\gray\ inverse Compton emission from radio lobes}}:
A new method for determining the local EBL intensity was described
by \citep{georganopoulos08}. The method relies on the detection of
$\gamma-$rays produced by IC scattering of CMB and EBL photons off
the relativistic electrons of the lobes of radio galaxies. Since the lobes have to be clearly resolved with the \gray\ telescope, the method is presently limited to the nearby radio galaxy
Fornax~A. With the normalization and maximum electron energy in
the lobes determined by their synchrotron spectra (measured by WMAP),
the $\gamma-$ray spectrum of the lobes is determined by the intensity of
the radiation field in the radio lobes, which is
dominated by the CMB and EBL. The resulting $\gamma-$ray spectrum 
comprises distinct contributions by the CMB and EBL photons: The CMB
contribution peaks at energies $\sim 40$~MeV, while that of the
EBL  appears as an excess above steeply dropping CMB contribution at
higher energies. 
This excess emission is relatively flat, and extends to energies of
$\sim 50$~GeV. Its  magnitude provides  a direct measurement of the EBL
intensity at mid- and far-IR wavelengths.

\begin{figure}[htbp]
  \centering
  \includegraphics[width=2.65in]{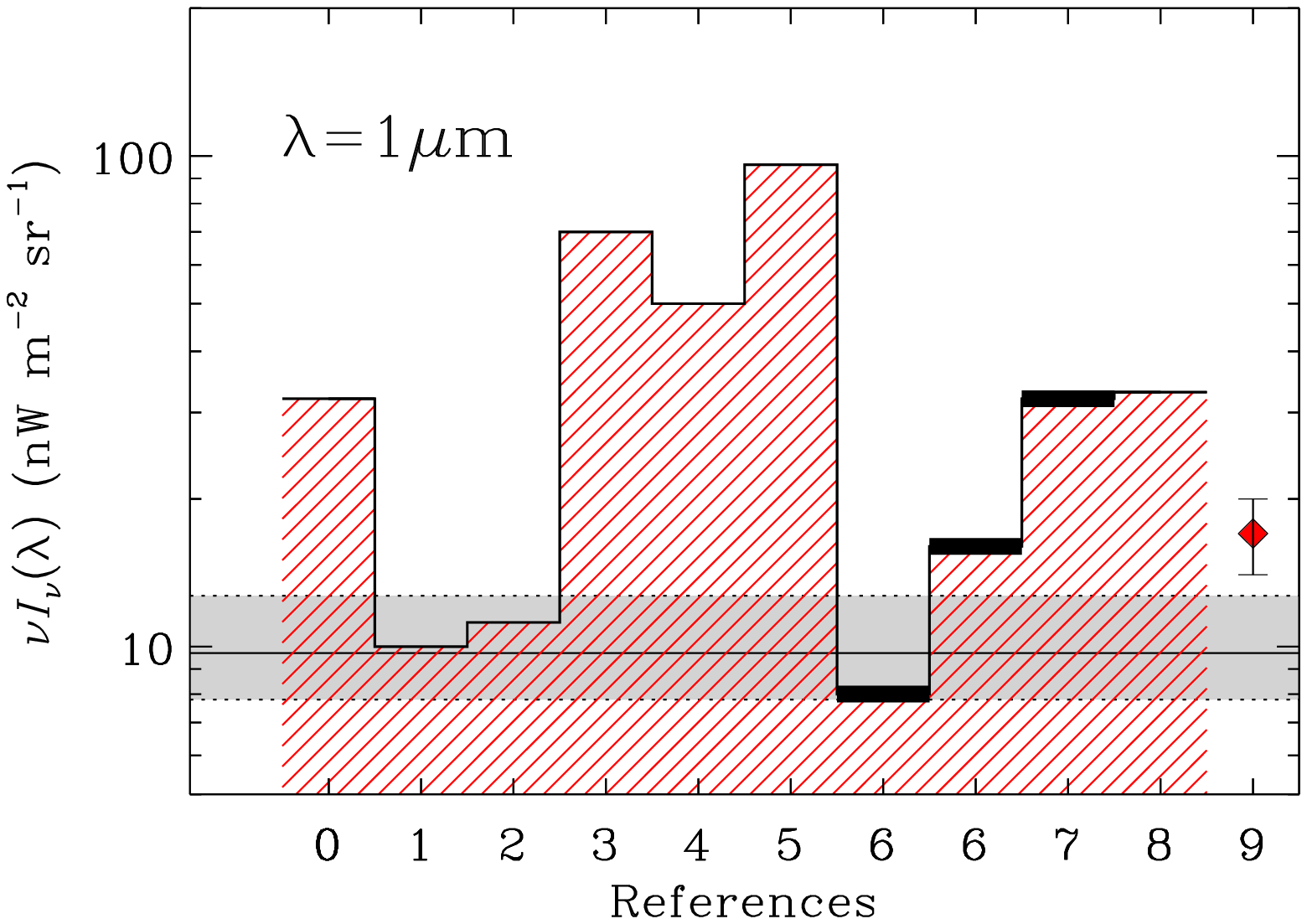}
 % \hspace{0.1in}
 %\hspace{0.1in}
  \includegraphics[width=2.65in]{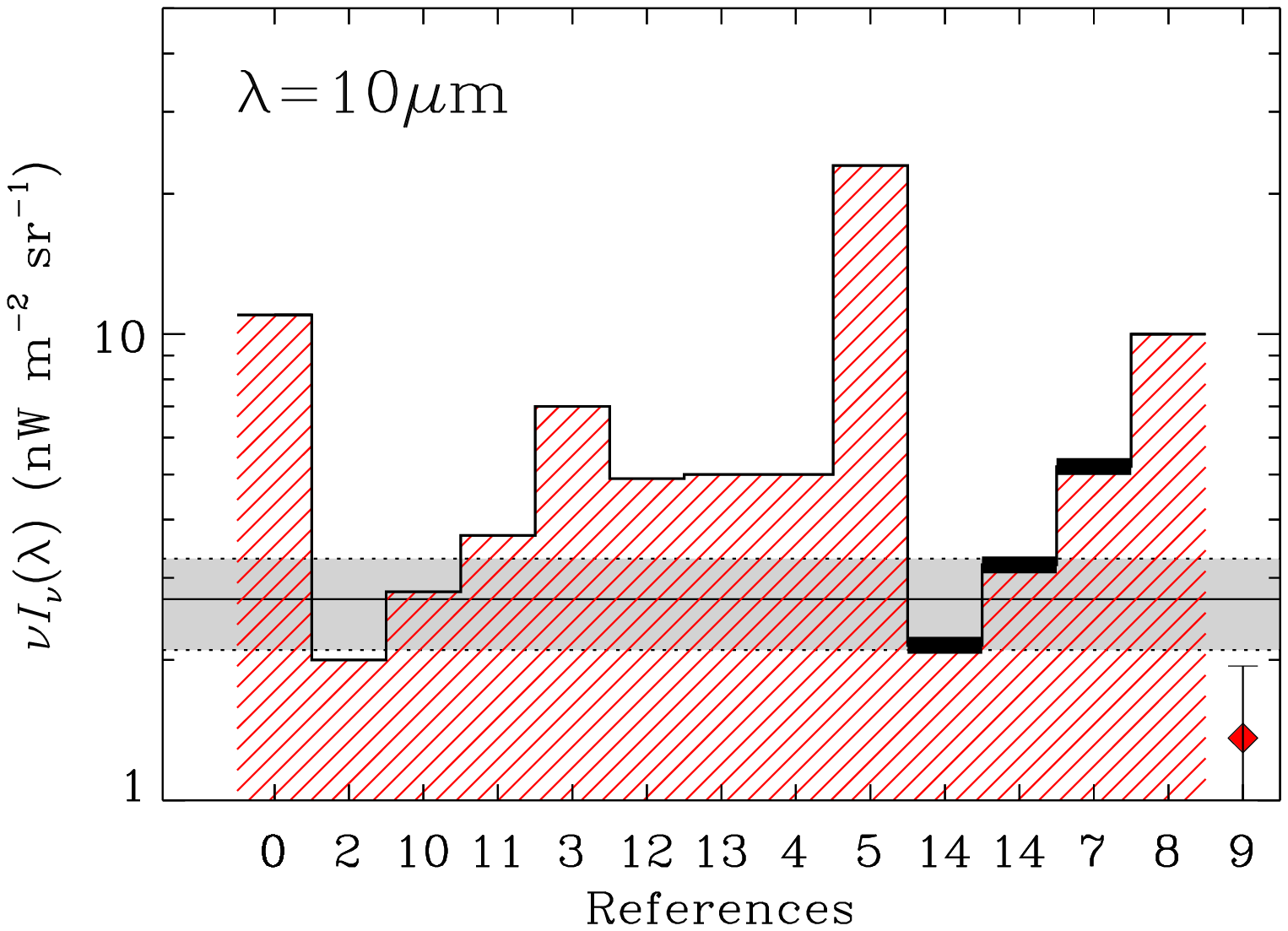}
%  \vspace{0.1in}
  \caption{{\footnotesize The EBL constraints  as derived from \gray\ observations to the EBL are shown for 1~\mic\ (left) and for 10   \mic\  (right). The horizontal line represents the intensity of the integrated galaxy light (IGL), which provides a strict lower limit on the EBL intensity. The most recent limits are represented by bold lines. The shaded area represents the 1$\sigma$ uncertainty in the IGL intensity. References to the \gray\ limits are: (0) \cite{stecker93}; (1) \cite{stanev98}; (2) \cite{biller98}; (3) \cite{guy00}; (4) \cite{dwek05b}; (5) \cite{schroedter05}; (6) \cite{aharonian06}; (7) \cite{mazin07}; (8) \cite{finke09}; (9) \cite{orr11}; (10) \cite{funk98}; (11) \cite{mannheim98}; (12) \cite{dwek94}; (13) \cite{renault01}; (14) \cite{aharonian07}.}\label{fig:ebl-1-10-limits}}
\end{figure}

%---------------------------
\noindent {\bf \underline{Combined \gray\ limits}}:
\noindent Figure \ref{fig:ebl-1-10-limits} shows select limits on the EBL derived from TeV observations.
All limits were adjusted to a common Hubble constant of $70$~km~s$^{-1}$~Mpc$^{-1}$.
Strict convergence  of the different limits is not expected, since each limit was derived
from different assumptions on the intrinsic blazar spectra and  the EBL spectrum. The bold horizontal lines and the \cite{orr11} result represent the most recent constraints on the EBL. It is important to emphasize that different methods were used, i.e., the  $\Gamma > 1.5$ limit on the hardness of the blazar spectrum and the spectral break analysis due to EBL spectrum,  yet  they reached similar conclusions.

In general, much improved upper limits are now available  from the observations with the new generation of atmospheric
Cherenkov telescopes (H.E.S.S., MAGIC, VERITAS).   Both the near-IR and the mid-IR  intensity levels  are now constrained to
much lower values than was possible with early results from the first few TeV blazars, and make only minimal assumptions on the intrinsic blazar spectra.    Extreme EBL scenarios,  with 70 \nwat at 1.5 \mic,  as suggested by \cite{matsumoto05a} are clearly ruled out.    Minimal assumptions on the blazar spectra
such as the absence of exponential rises or applying the  $\Gamma > 1.5$ limit, reject such high EBL intensities in the near-IR \citep{dwek05b, aharonian06}.

%-------figure 11 ---------------- EBL 1 and 10 micron limits from TeV
%--------------------

\cite{mazin07} provide upper limits that are  comparable to the former two  in the near-IR, while in the mid-IR their upper limits are much higher.   For this
work the spectrum of 1ES~0229+200, a blazar with a redshift of z=0.129 and a spectrum up to  10~TeV was not available, while it is the basis for strong mid-IR limits in  \cite{aharonian07b} and \cite{orr11}.  This emphasizes the importance of extending  the TeV energy spectra of distant (z$\ge$ 0.1) blazars into the multi-TeV regime, where \gray s reach their maximum cross-section with photons in the mid-IR.   Furthermore, energy spectra between 100~GeV and 10 TeV are sensitive to the spectral  shape of the EBL, thereby linking the upper limits in the  near- and mid-IR;   a given   mid-IR intensity level combined with the EBL spectral shape limits the range of near-IR EBL intensities,  i.e., by excluding EBL intensities that are either too high or too low in the near-IR based on the spectral shape constraint.

There are now in the order of 30 extragalactic objects with  redshifts up to  $z \approx 0.5$
 available for  constraining the EBL.   Limits derived from the recent generation of experiments (H.E.S.S., MAGIC, VERITAS) suggest a low mid-IR,  such as the results from \cite{aharonian07b} and \cite{orr11}.  

In summary, it should also be noted that all of the most recent TeV constraints are well within the boundaries set by direct measurements and their uncertainties (see shaded area in Figure \ref{fig:tev-limits}).    However, given the debate about some of  the assumptions about the blazar spectra, namely,   the  $\Gamma > 1.5$ limit on the hardness of the blazar spectrum, claims that the EBL has been resolved are premature.

%=======================================================
\section{The \gray\ Opacity of the Universe}
%=======================================================
With blazars being detected at increasingly large redshifts, it becomes possible to use them to discriminate between different EBL models. 
The \gray\ opacity to a blazar at redshift $z$, is given by:
% $\tau_{\gamma \gamma}(E_{\gamma}, z) = -\log\left[{F(E_{\gamma}, z)_{obs} \over F(E_{\gamma})_{int}}\right] $
%------------------  eq. 7
\begin{equation}
\label{eq:dnde-obs}
\tau_{\gamma \gamma}(E_{\gamma}, z) = -\log\left[{F(E_{\gamma}, z)_{obs} \over F(E_{\gamma})_{int}}\right] 
\end{equation}
%---------------------
The opacity can be determined from models of the EBL if its evolution with redshift is known and, independently, from \gray\ observations if the intrinsic blazar spectrum is known. Concordance between these two independent determinations of \taugg\ can serve as a test for the validity of the underlying assumptions in each method. 
  
\begin{figure}[t]%[htbp]
  \centering
  \includegraphics[width=1.8in]{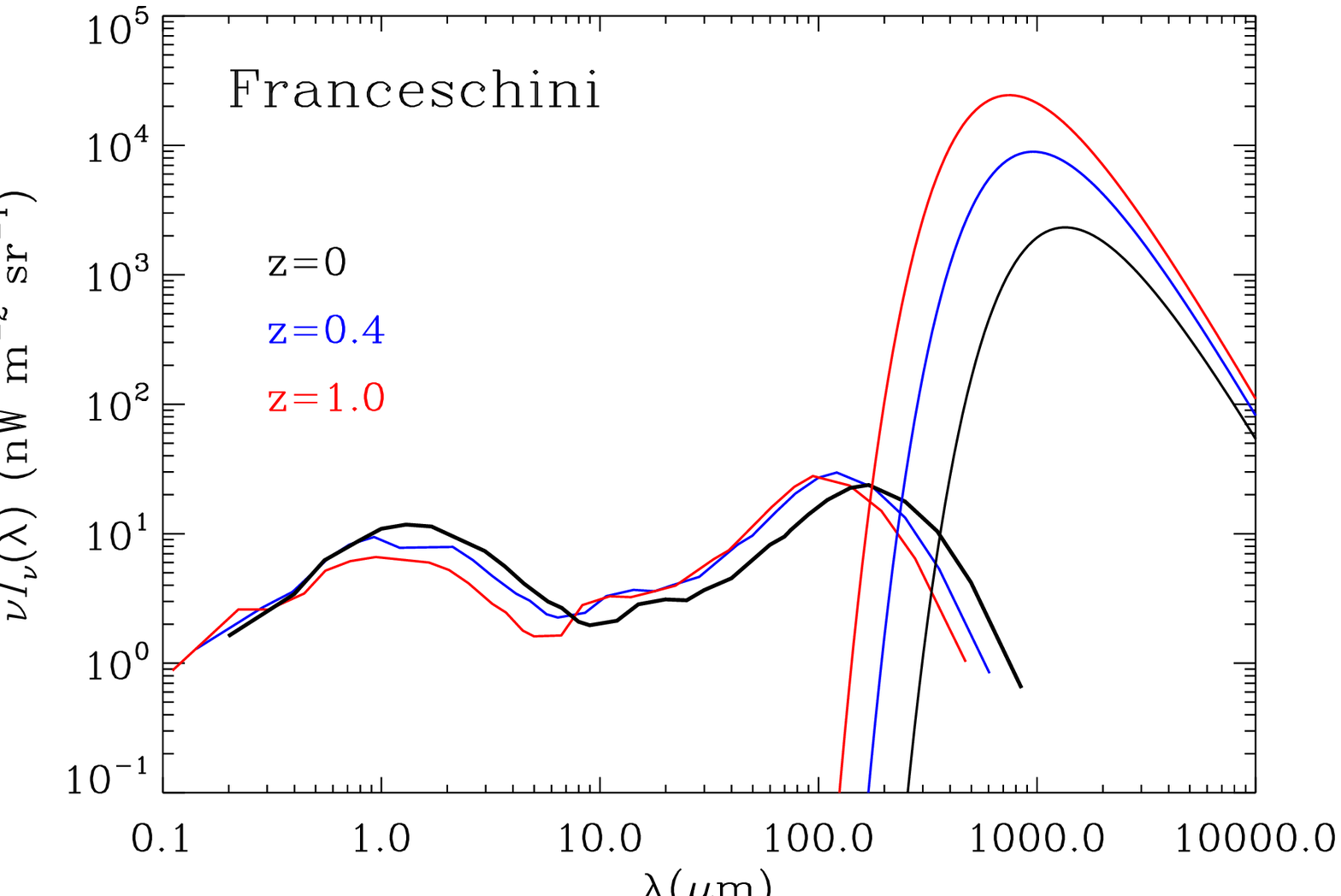} 
  \hspace{0.1in}
   \includegraphics[width=1.8in]{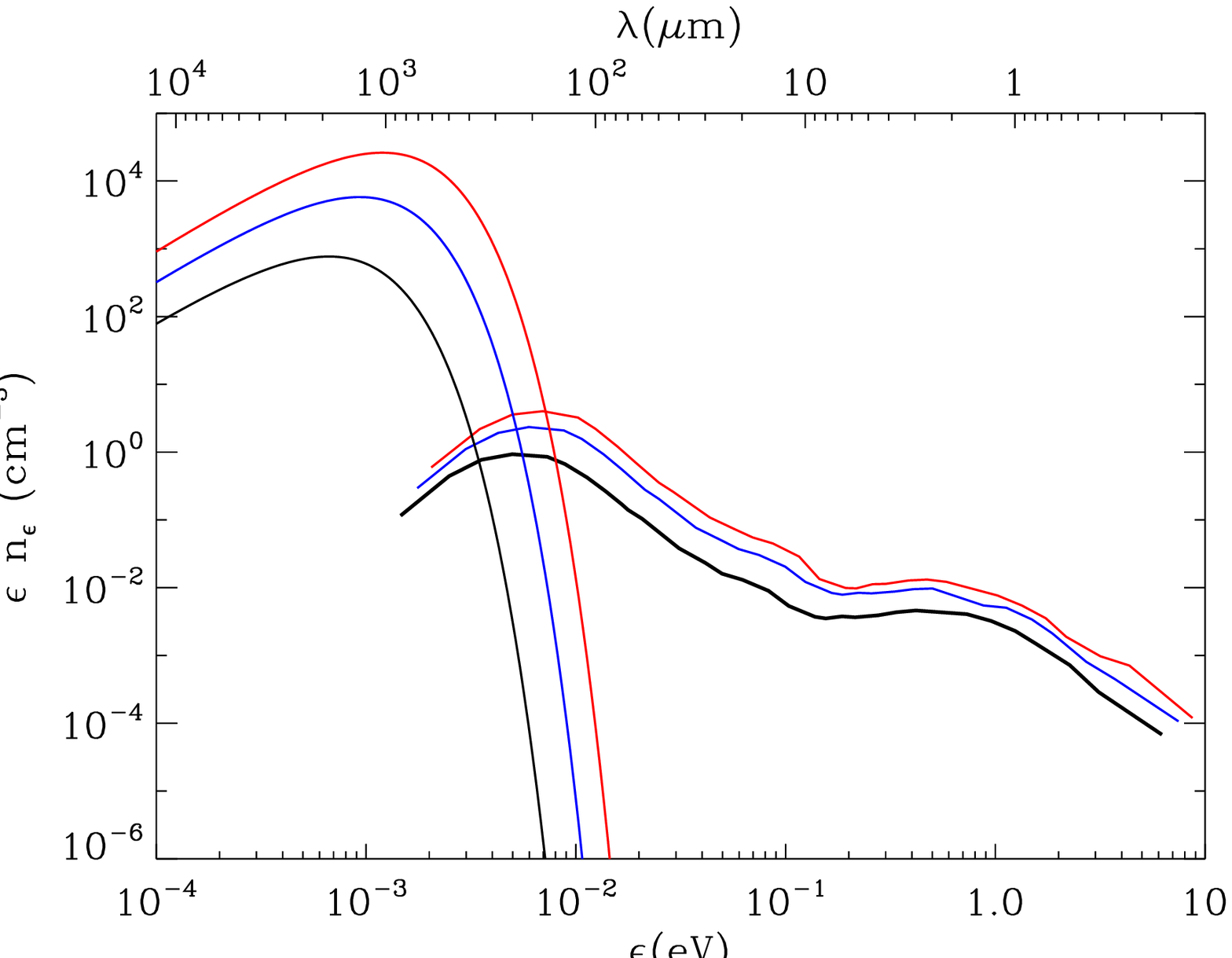} \\
  \includegraphics[width=1.8in]{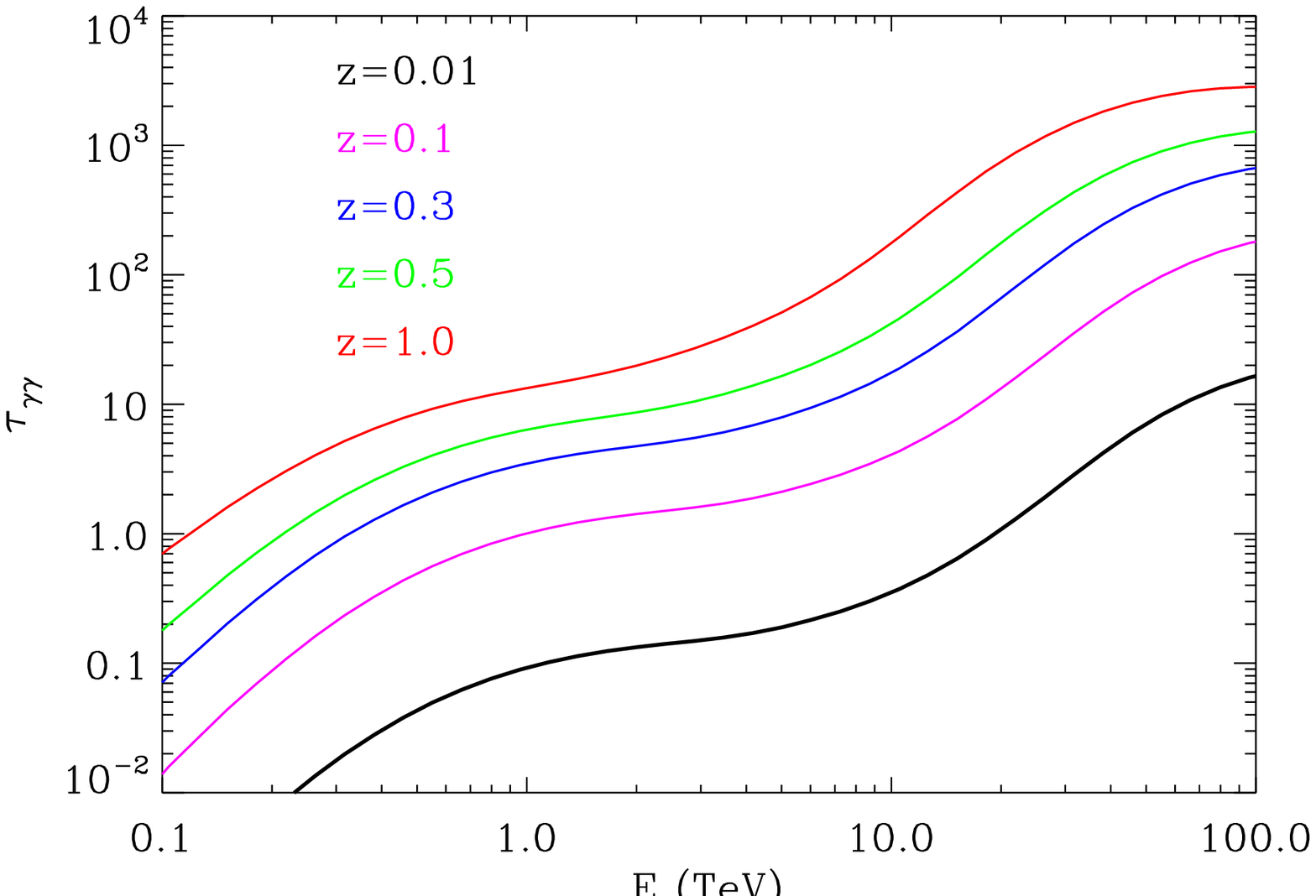} 
  \hspace{0.1in}
   \includegraphics[width=1.8in]{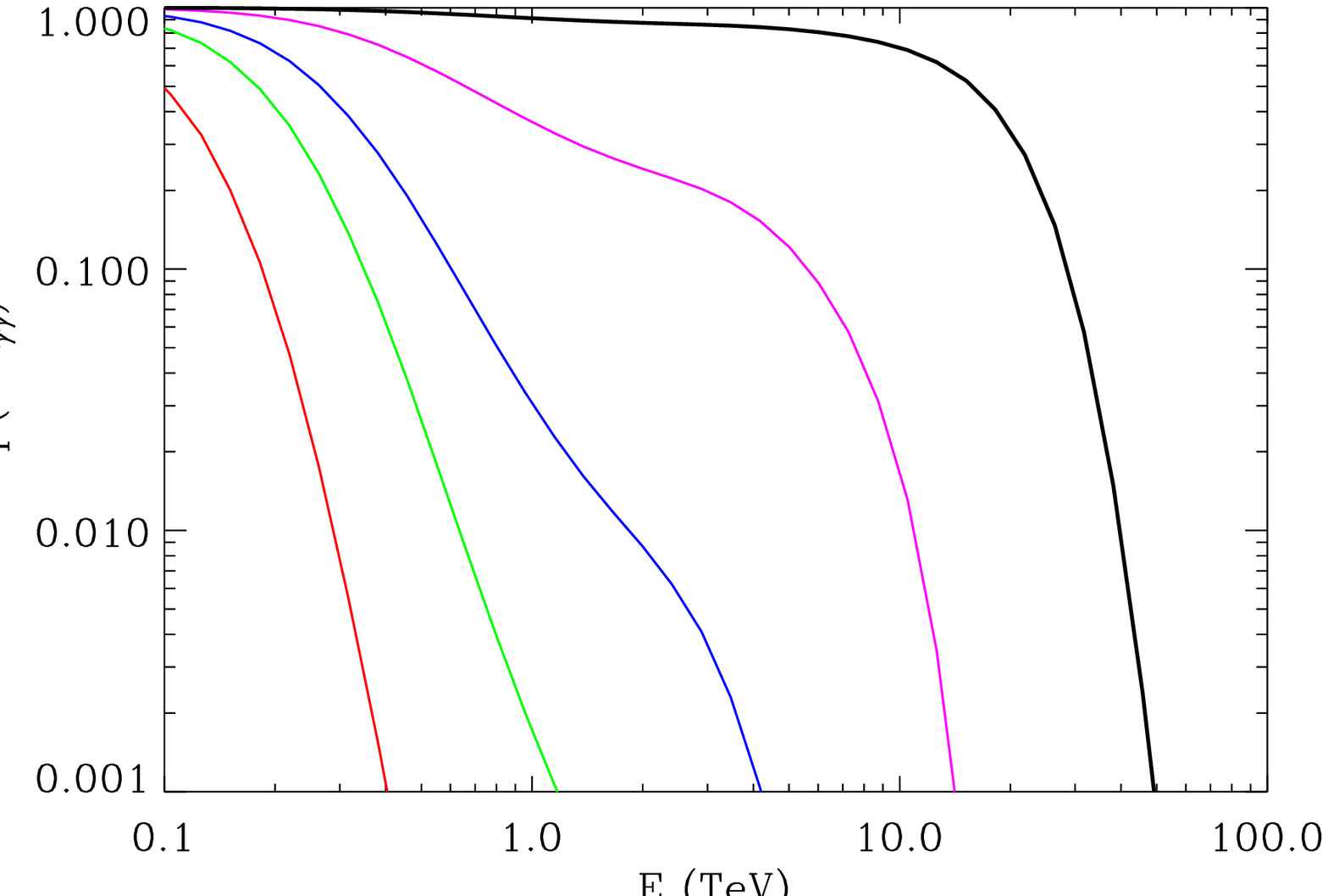} \\
   %  \vspace{0.1in}
 \caption{{\footnotesize Basic EBL model results by \cite{franceschini08}: {\bf Top left}: The comoving EBL and CMB intensities versus wavelength for different redshifts; {\bf Top right}: The proper number density of EBL and CMB photons versus energy for the same grid of redshifts as the previous panel; {\bf Bottom left}: The \gray\ opacity versus energy, \eg\ for different redshifts; {\bf Bottom right}: The amount of attenuation versus energy for the same grid of redshifts as the previous panel. The figure illustrates the change in the slope of \taugg\ at energies corresponding to the wavelength at which the slope of the EBL spectrum changes.}  \label{fig:franceschini}}
 \end{figure}

\begin{figure}[b!]%[htbp]
  \centering
  \includegraphics[width=1.8in]{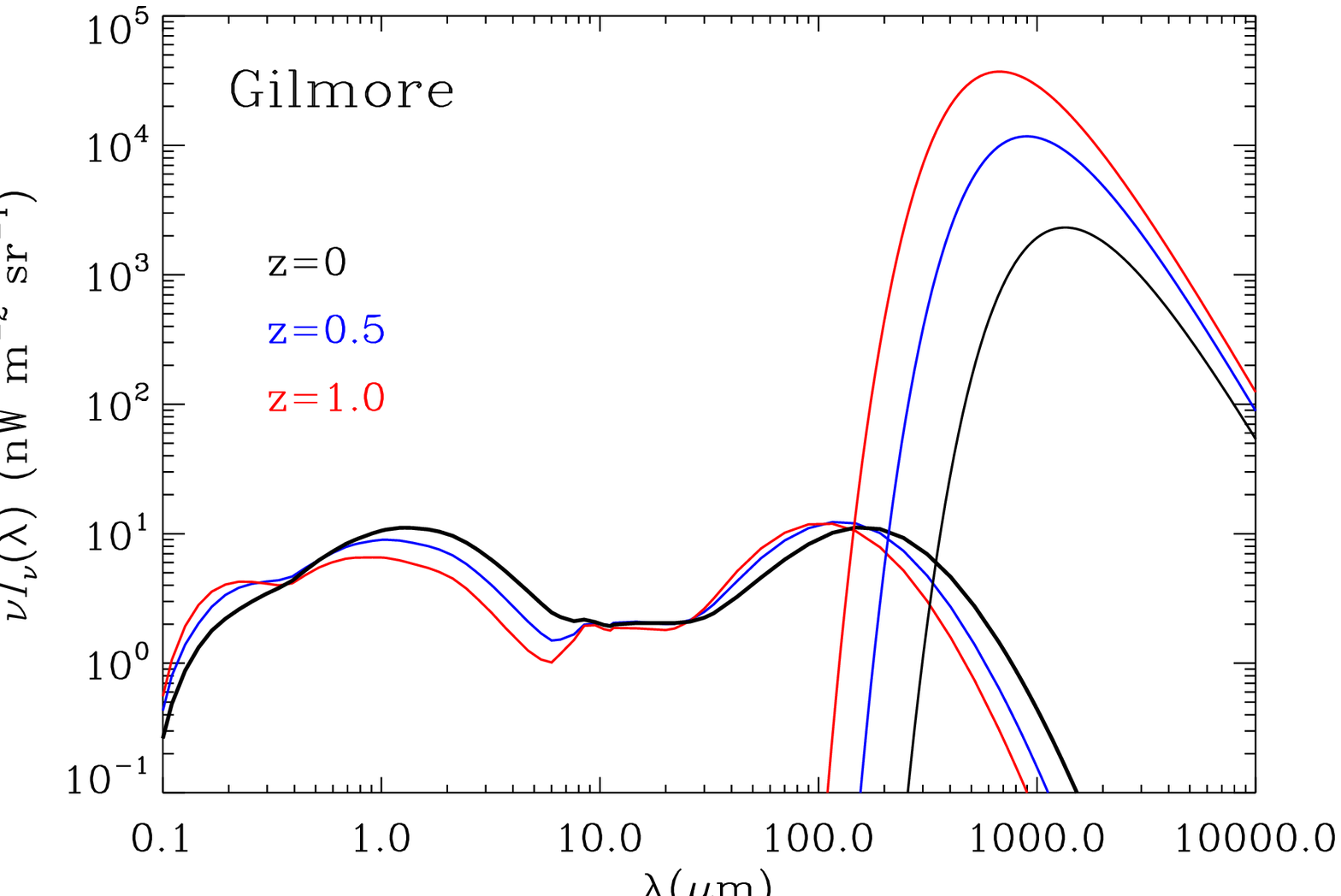} 
  \hspace{0.1in}
   \includegraphics[width=1.8in]{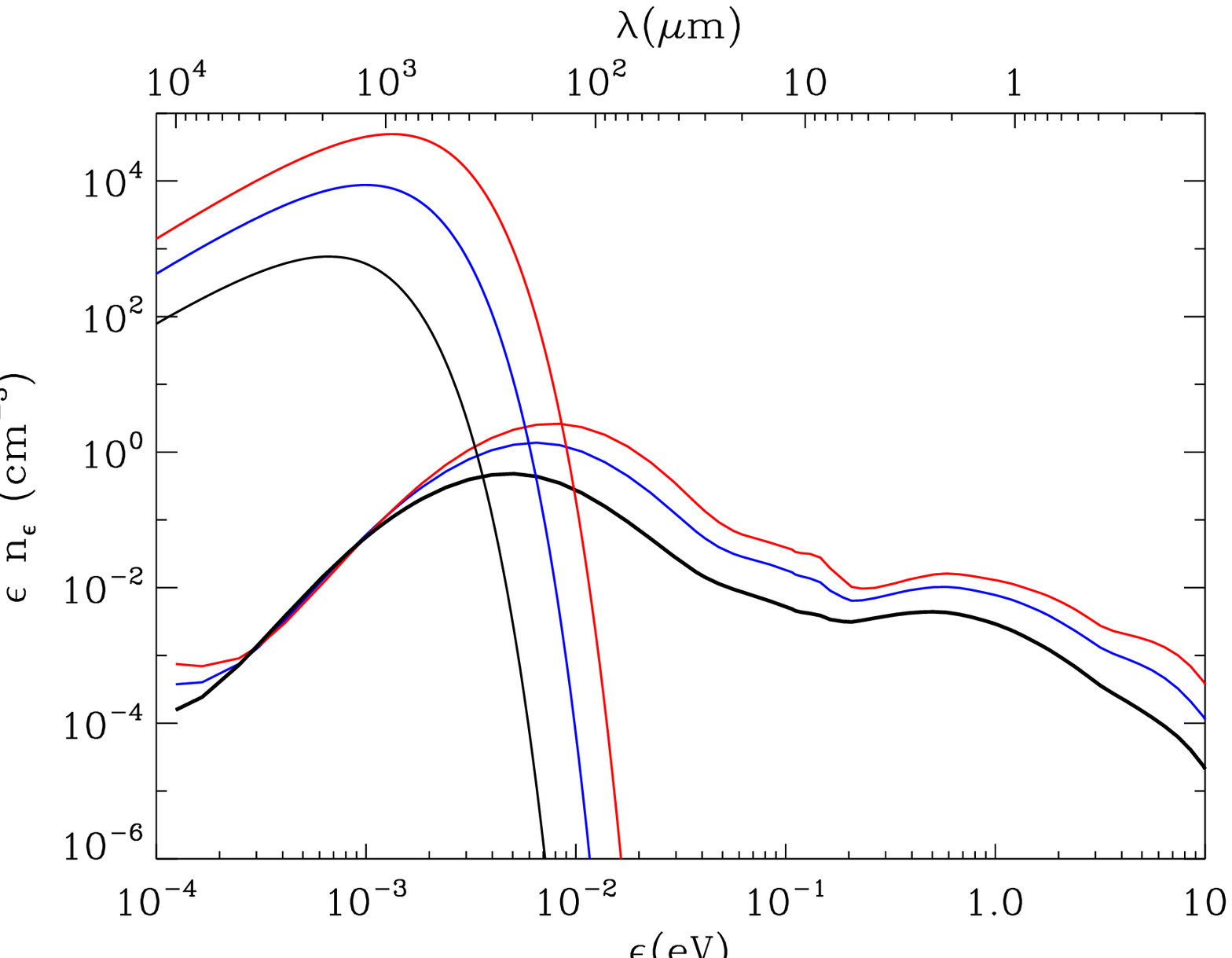} \\
  \includegraphics[width=1.8in]{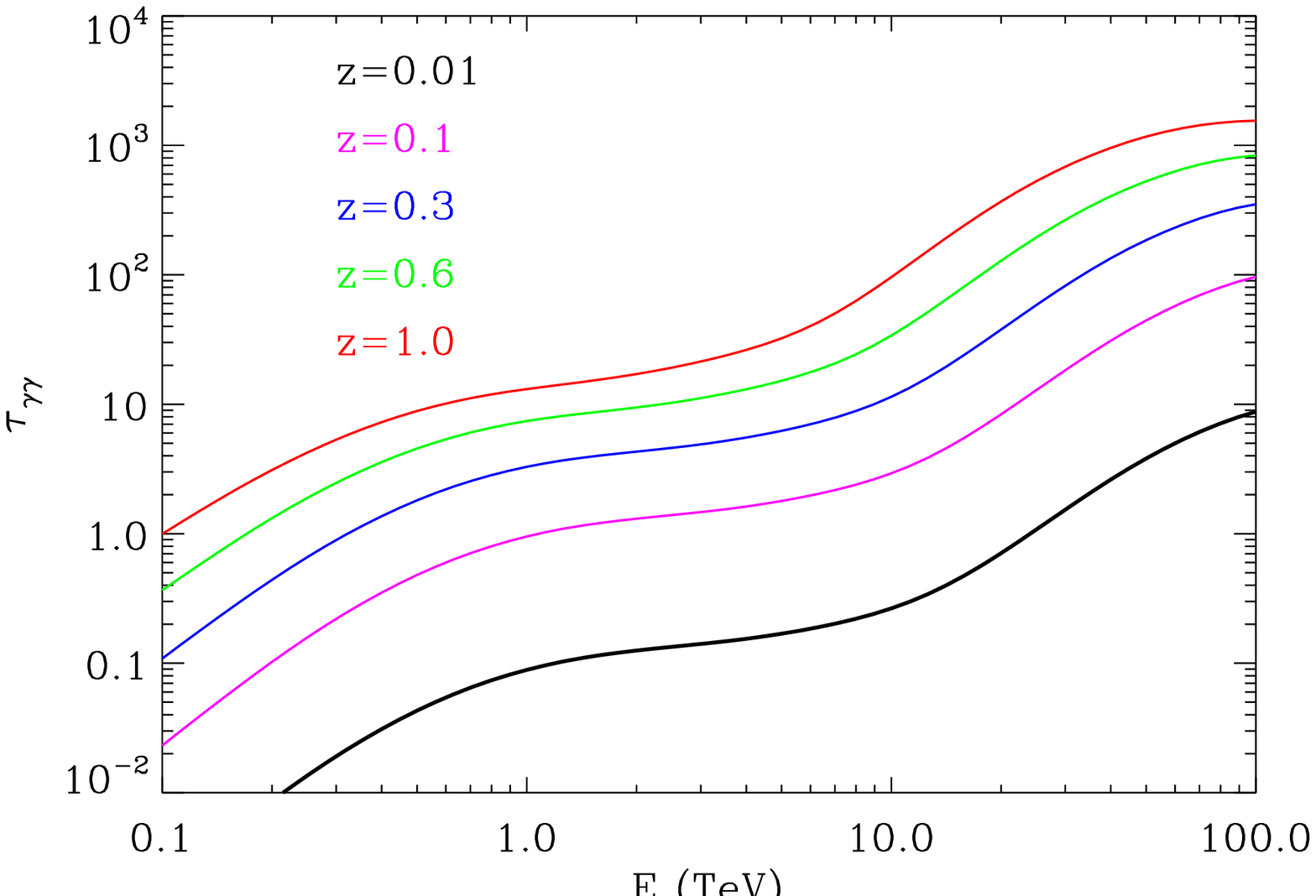} 
  \hspace{0.1in}
   \includegraphics[width=1.8in]{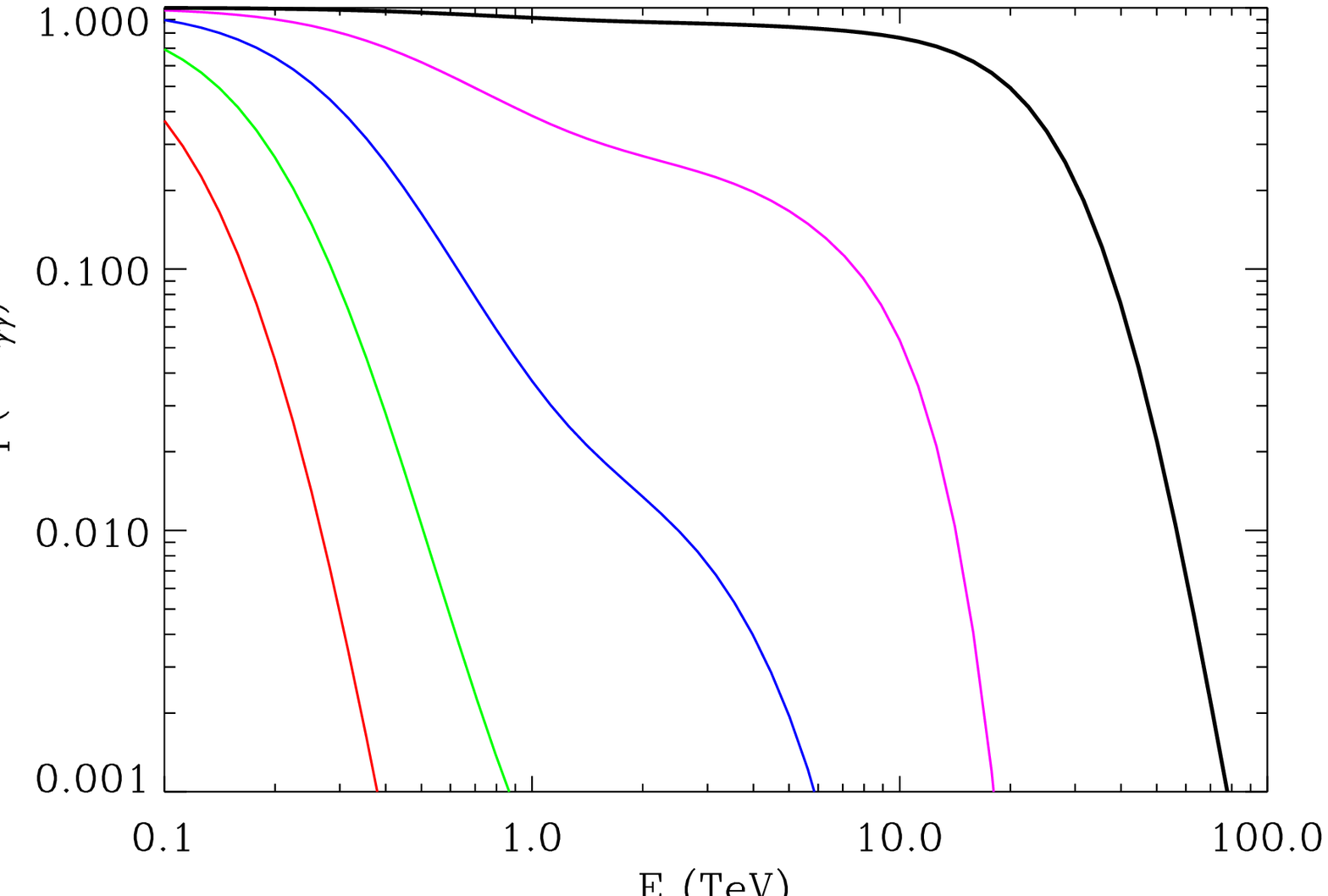} \\
   %  \vspace{0.1in}
  \caption{{\footnotesize Same as Figure~\ref{fig:franceschini} for the \cite{gilmore11} model.} \label{fig:gilmore} }
\end{figure}

Figures~\ref{fig:franceschini} and \ref{fig:gilmore} depict the evolution of the comoving intensity of the EBL, the corresponding evolution of the proper number density of background photons, the optical depth to blazars at various redshift, and the corresponding attenuation factor. Results are plotted for the BE evolution model of \cite{franceschini08} and the BE evolution model of \cite{dominguez11}.  

%\newpage
%\clearpage

%----- figure 12 ----- 
%--------------------

%----- figure 13 ----- 
%--------------------

Determining the \gray\ opacity from observations requires knowledge of the intrinsic blazar spectrum. Differences between the observed and expected flux at a given energy \eg\ would then be simply attributed to EBL attenuation. 
Figures~\ref{fig:franceschini} and \ref{fig:gilmore} show that the sharp drop of the EBL intensity at UV and shorter wavelengths renders the universe almost transparent to GeV photons. Consequently, the observed $\sim 1-50$~GeV spectrum is very likely the intrinsic blazar spectrum. 
So instead of assuming a theoretical limit on the spectral index, one can use the GeV - 10s of GeV energy spectral slope from Fermi data as a proxy for the intrinsic spectra at TeV energies. 

Assuming that this power law can be extrapolated from GeV to TeV energies,  one can derive the TeV optical depth to the observed blazar.  This approach was used by \cite{georganopoulos10a} and in method~1 in \cite{orr11} to set firm upper limits on EBL models using the GeV to TeV spectra of PKS~2155-304 ($z = 0.116$) and 1ES~1218+304 ($z=0.182$). Assuming that the GeV spectrum is unattenuated by the EBL, \cite{mankuzhiyil10} used optical, X-ray and GeV data to model the TeV flux of PKS~2155-304 using a one-zone SSC model. Comparison of the model results with observations, they derived the TeV opacity to this blazar, and found it to be consistent with most EBL models. 

\begin{figure}[t!]%[htbp]
  \centering
    \includegraphics[width=4in]{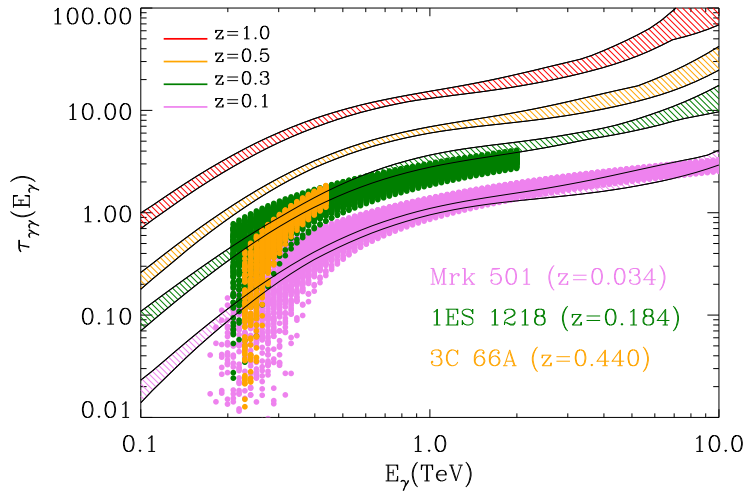} 
   \caption{{\footnotesize Limits on the optical depth at $z$ = 0.1, 0.3, 0.5, and 1.0 as determined by the EBL models are compared to observationally determined optical depths for select blazars. The optical depth to 3C~66A is still uncertain. See \S7 for more details.}\label{fig:horizon}}  
\end{figure}

Figure~\ref{fig:horizon} compares the dependence of the optical depth derived from EBL models (hatched curves) to that derived for select blazars: Mrk 501, 1ES~1218+304, and 3C~66A. Each hatched band spans the range of optical depths predicted by the EBL models of \cite{franceschini08}, \cite{finke10}, \cite{dominguez11}, and \cite{gilmore11}. The colored dots represent the optical depths derived from the \gray\ observations of the three blazars. The intrinsic blazar spectrum was assumed to be a power law determined by the observed flux at 1~GeV and the spectral index, $\Gamma_{GeV}$. The observed flux in the TeV range was assumed to be a power law with a spectral index $\Gamma_{TeV}$ (see Table 2). The \gray\ opacity in the TeV range was then derived from eq.~(9). The band of opacities for each blazar was obtained by performing 100 Monte Carlo simulations of the intrinsic and observed spectra using the uncertainties in the spectral indices and \gray\ energies into account.    

The figure shows that the \gray\ derived optical depths of Mrk~501 and 1ES~1218-304 are in general agreement with model prediction. The discrepancy between the EBL and the \gray\ derived optical depth for 3C~66A is typical of most blazars listed in Table 2. We note that the redshift to 3C~66A is still somewhat uncertain \citep{abdo11d,abdo11c}. The convergence between observational limits on the EBL and models suggests that the origin of the discrepancy can be mostly attributed to our still incomplete knowledge of the intrinsic spectra of blazars.

The EBL not only affects the \gray\ spectra of individual \gray\ sources, but also the spectrum of the extragalactic  \gray\ background (EGRB) which consists of the cumulative contribution of resolved and unresolved sources and a possible truly diffuse emission component.

Recently, the {\it Fermi} Large Area Telescope (LAT) provided a new measurement of the diffuse \gray\ background (DGB) at energies between 0.2 and 100~GeV \citep{abdo10d}, obtained by the subtraction of resolved {\it Fermi} sources and theoretical estimates of the contribution of diffuse Galactic emission from the total sky intensity.

In a recent paper, \cite{inoue11} estimated the contribution of unresolved \gray-loud radio galaxies (GRLRG) to the GeV DGB. Their contribution is given by an integral similar to that used for calculating the EBL [see eq. (\ref{eq:ebl2})], except that it requires the addition of an opacity term which takes the propagation of the \gray\ photons through the EBL into account.

Calculating the contribution of GRLRG to the DGB requires knowledge of their intrinsic spectrum and \gray-luminosity function (LF).  Using a sample of 10 radio galaxies from the {\it Fermi} catalog, \cite{inoue11} derived a correlation between their 5~GHz and 0.1-100~GeV luminosities. Using this correlation and the 5~GHz LF, he then derived the \gray-LF for these objects. 

The results show that GRLRG can account for about 25\% of the intensity of the DGB above 100~MeV. Their cumulative spectrum is characterized by a power law to energies of $\sim 10$~GeV, after which it exhibits a decline resulting from EBL absorption. However, the observed {\it Fermi}--LAT spectrum does not exhibit such decline. The lack of this absorption signature in the DGB spectrum therefore suggests that nearby \gray\ sources or distant sources with hard \gray\ spectrum must be the major contributors to the DGB at energies above $\sim 30$~GeV.  

The derived GRLRG spectrum was derived using a sample of only 10 galaxies to calculate their \gray\ LF. The model also contains some additional free parameters that characterize their intrinsic spectra and their overall contribution to the population of all radio galaxies. Clearly more data are required to resolve the origin of the DGB and detect any signature of EBL absorption in its spectrum.

%----- figure 14 ----- 
%--------------------

%------------------------
 
%------------------------------------------------------
\section{Is the Blazar Spectrum Determined in the Source?}
%------------------------------------------------------
In deriving upper limits on the EBL from TeV observations, it was tacitly assumed that  blazar spectra are produced in the sources and attenuated en route to Earth. 
The detection of  medium redshift blazars at $z \sim 0.2$ with hard \gray\ spectra has revived alternate models for the origin of  their spectra. Since hard photons from these redshifts are easily attenuated, one of these models proposes that the observed blazar spectra  contains secondary photons that are produced close to the observer, and therefore more likely to survive. 

High resolution optical spectra of the blazars PKS~0447-439 and PMN~J0630-24 obtained with the CTIO and NTT observatories show a clear absorption line around 6280~\AA\ which, when attributed to the Mg~{\small II} 2795.5~\AA\ and 2802.7~\AA\ doublet, places them at redshifts $z \gtrsim 1.246$ and $z \gtrsim 1.238$, respectively \citep{landt12}. PKS~0447-439 was detected as a very-high energy \gray\ source with H.E.S.S., showing an energy spectrum from 200 GeV - 1.5 TeV \citep{zech11}. 

The optical depth of 1~TeV photons at such large redshift is about 20 (see Figures~\ref{fig:franceschini} and \ref{fig:gilmore}), so that any flux of $\sim 1$~TeV photons is attenuated by a factor of $\sim 10^{-9}$.  This would suggest that the intrinsic 1~TeV luminosity of this blazar is about $10^{10}$ larger than that of Mrk~421!  Any attempts to solve this problem by lowering the \gag\ opacity of the intervening intergalactic medium will require the unrealistic reduction of the EBL intensity below the lower limits determined by the IGL (see Figure~\ref{fig:ebl_limits}).   

It is therefore unlikely that primary \gray s could have reached the Earth from such distance, suggesting that most of the \gray s from this object must be secondary photons created by the interaction of cosmic-ray protons at relatively close distances to the observer \citep{kusenko12,aharonian12}. 
The secondary photons are generated by the interactions of the protons with the CMB and the EBL. For the secondary photons to be  detected, they have to be produced within a distance $\lambda_{\gamma}$, the mean-free-path of the secondary \gray s, of the Earth, and be deflected into the viewing angle of the telescope \citep[e.g][]{stanev00}. Their angular deviation from the primary source direction must therefore be smaller than the point spread function of the telescope ($\rm \sim 0.1^{\circ}$), requiring the protons to travel on a "straight" trajectory until the interaction region. This requires their gyro-radius to be much larger than the source distance, which sets a range $\sim 10^{-17} - 10^{-15}~$G on the line-of-sight intensity of the intergalactic magnetic field \citep{aharonian12}. While this value is significantly smaller than commonly accepted upper limits for the intergalactic magnetic field \citep{kronberg94,kronberg10}, weak  fields are not ruled out. 

Hadronic particle acceleration is at the heart of the cascade model. This acceleration mechanism may hold the key to understanding blazar jets, and has far reaching consequences in high energy astrophysics. 
Hadronic jet models were brought forward by \citep{mannheim93} around the time of the first detection of TeV photons from a blazar \citep{punch92}.   Protons at
energies of $\rm 10^{18} - 10^{19}$~eV are capable of reaching the threshold for photopion production. The subsequent pion-induced cascade inside the jet generates primary  \gray\ photons which make a substantial contribution to the blazer spectrum.   An additional contribution to the \gray\ spectrum comes from  proton synchrotron radiation, and the synchrotron emission from the secondary pions and muons.

The viability of hadronic \gray\ emission models has been questioned, since they seemed to require exceedingly large magnetic fields to accelerate the protons and to confine them to the small emission region inferred from the observations of short duration TeV flares. 
 However, hadronic  \gray\ emission models  have  been put to test through TeV \gray\  observations of nearby blazars for which EBL absorption is negligible.
 The acceleration of UHE protons along relativistic jets requires large magnetic fields, typically several 10~G. The emission of gamma rays through a hadron induced cascade has to occur within a small emission region because of the observed variability of TeV flares.  The first big flare reported from Mrk~421 exhibited sub-hour scale flux variability \citep{gaidos96}. Assuming a Doppler factor of 10, the short flux variations set a limit of $\rm  \le 10^{13}$~m, in the comoving frame of the relativistic jet, on the size of the gamma-ray emission region \citep{gaidos96}. A magnetic field strength of at least $\rm  \sim 50 $~G is required to  
prevent $\rm   10^{19} $~eV ions and protons from escaping the small emission region \citep{dermer12a}.   Such large magnetic fields are still consistent with
the substantial, but plausible, power requirements for the magnetic field energy, which is less than 0.1\% of the Eddington luminosity. Consequently, the observation of short duration TeV flares does not pose a major obstacle to the viability of hadronic models.

Finally, hadronic jet emission models are also attractive for explaining the origin of ultra-high energy  (UHE: $\rm  E \ge 10^{17} $~eV)  cosmic rays that are detected by air shower arrays such as AUGER and HiRes   \citep{kampert12, sokolsky10}. 
  
However, the need to resort to hadronic acceleration model to specifically explain the observed flux of PKS~0447-439 is probably premature. The redshift determination to this blazar has recently been challenged by \cite{fumagalli12} who independently obtained the PKS~0447-439 spectrum using the 6.5~m Magellan Telescopes with a high S/N ratio of $\sim 150$ in the 6270--6300~\AA\ region of interest. They point out the existence of an atmospheric telluric absorption line at the wavelength of the claimed Mg~{\small II} absorption line, which they showed was also present in the spectrum of two standard stars. \cite{pita12} also called into question the redshift measurement of PKS~0447-439, providing clear evidence that the conclusions in \cite{landt12} are incorrect. As a result, the redshift of PKS~0447-439 is still undetermined.  

Hadronic jet models can play an important role in explaining the origin of the  DGB. The featureless DGB power spectrum derived from the {\it Fermi}--LAT  observations \citep{abdo10d} does not exhibit the  EBL absorption signature around $\gtrsim 30$~GeV, that is expected  from the contribution of cosmologically distant \gray\ sources such as GRLRG, starburst, or active galaxies \citep[for a discussion see][and references therein]{inoue11,inoue12}.  Hadron induced pair cascades interacting with the CMB can generate secondary \gray s that will contribute to the GeV DGB,  filling in the "missing" flux at energies $\gtrsim 30$~GeV \citep{aharonian12}. Further measurements extending the DGB to $\sim 1$~TeV will provide important constraints on the contribution of secondary \gray s from blazars.
 
%------------------------------------------------------
\section{EBL Limits on "Exotic" Energy Releases in the Universe}
%------------------------------------------------------
The intensity and spectral shape of the EBL contain the memory of all energy releases in the universe since the epoch of recombination. The {\it COBE}/DIRBE limits on the UV to near-IR regions of the EBL proved useful in ruling out various "exotic" sources of energy in the early universe, such as decaying particles, exploding stars, or very massive objects \citep[][and references therein]{dwek98a}. 

%----- figure 15 ----- observational EBL limits
 \begin{figure}[t!]%[htbp]
  \centering
  \includegraphics[width=4in]{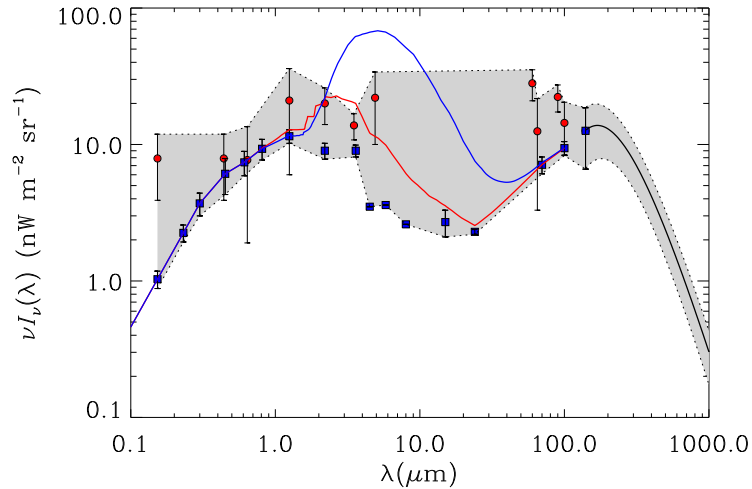} 
%  \vspace{0.1in}
  \caption{{\footnotesize  The contribution of dark stars to the EBL for two different Limits on the EBL intensity. Lower limits, blue squares) are determined by the intensity of the IGL. Upper limits (red circles) are determined by absolute measurements of the EBL. The data used in the figure are listed in Tables~3-5 in bold. The shaded area depicts the range of the allowed EBL intensity as determines by UV to sub millimeter observations.} \label{fig:ebl-x-limits} }
\end{figure} 

Primordial (Population~III) stars were suggested by \cite{salvaterra03} as the source of the excess 1-5~\mic\ diffuse emission above the IGL intensity detected by \cite{matsumoto05a}. However, \citep{dwek05b} ruled out an extragalactic origin for the excess emission, since it would have produced a physically unrealistic intrinsic \gray\ spectrum of the blazar PKS~2155-304. Furthermore, \cite{dwek05c} showed that such origin would have required a Pop~III star formation rate and energy output to be significantly  higher than that predicted by hierarchical models for structure formation in a $\Lambda$CDM universe \citep{bromm02}. Using theoretical limits on their formation rate, \citep{dwek05c} concluded that Pop~III stars can contribute only a fraction of the EBL intensity.

A more detailed study of the contribution of Pop~III stars to the EBL intensity was conducted by \cite{raue09}. Using \gray\ derived constraints on the EBL intensity they set a limit of 0.3 to 3~\msun~Mpc$^{-3}$~yr$^{-1}$ on the formation rate of Pop~III stars in the $z = 7-14$ redshift interval.

More recently, \citep{maurer12} used EBL limits to constrain the properties of dark stars (DS) in the early universe. Dark stars are objects that have either accreted or captured weakly interacting massive particles (WIMPs, a dark matter candidate), which by annihilating inject energy into the stars before their radiative output is dominated by standard nuclear fusion processes. The formation rate of these stars, their luminosity and spectrum, and their effective lifetime are all free parameters of the model. Figure~\ref{fig:ebl-x-limits} depicts the EBL, to which the contribution of dark stars was added to the IGL intensity, for two different sets of parameters. The blue curve represents the EBL with a contribution of 106~\msun\ DS with a surface temperature of 5,000~K, and the red curve that with the added contribution of 690~\msun\ DS with a surface temperature of 7,500~K. The colder dark stars are obviously ruled out, but the hotter ones are marginally consistent with current EBL limits, dominating the intensity of the IGL in the $\sim 2-10$~\mic\ wavelength region.

%------------------------------------------------------
\section{Summary and Future Direction}
%------------------------------------------------------
Very high energy \gray s emitted from extragalactic sources are attenuated en route to earth by \gag\ interaction with EBL photons. \gray\ observations can therefore be used to set limits on the EBL intensity, provided that the intrinsic \gray\ spectrum of the sources is known. Conversely, knowledge of the EBL can be used to determine the intrinsic \gray\ spectrum of the different sources, thereby provide important constraints on mechanisms for their production. The main issues and results discussed in this review can be briefly summarized as follows:   
\begin{enumerate}
%------
\item The EBL spectrum consists of two broad  peaks, one at $\lambda \approx 1$~\mic, representing the cumulative gravitational and nuclear energy releases by stars and AGNs over cosmic history. Their energy has been partially absorbed by dust and reradiated at IR wavelengths. This thermal dust emission component generates a second peak at $\sim 100 - 200$~\mic. Current limits and detections of the EBL were presented in Figure~\ref{fig:ebl_limits}, and the partitioning of its total intensity into the different emission components was presented in Table~6;  
%------
\item The cross section for the \gag\ interaction is broad and peaks at energies  $E_{\gamma}(TeV)  \approx  0.86\, \lambda$(\mic). 
%$E_{\gamma}(TeV)  \approx 0.86\  \lamda (\mu)$m 
%(see Figure 
%\ref{fig:sigma}. 
Consequently, the energy dependence of the \gray\ opacity, a product of the cross-section with the number density of EBL photons, reflects the spectral variation in the EBL. We identified three potential breaks in the spectrum of \gray\ sources, one occurring between 10 and 500~GeV, a second break at 1~TeV and third at 10~TeV (see Figures~\ref{fig:black-body} and \ref{fig:ebl-tau});
%------
\item The spectral breaks at GeV and TeV energies have been used to set limits on the EBL intensity at near-IR wavelengths, and on the relative intensities of the peak of the stellar emission and the trough between the stellar and dust emission components of the EBL;
%------
\item \gray\ derived limits on the EBL vary as different studies used different assumptions on the intrinsic \gray\ source spectra (see Figure~\ref{fig:tev-limits}). The strictest limits on the EBL are around the 1 and $\sim 10$~\mic\ wavelength regions of the EBL, with some approaching the lower limit on the EBL intensity set by the IGL (see Figure~\ref{fig:ebl-1-10-limits});
%------
\item Recent IR space and ground-based observations have resulted in closer agreement between the EBL limits and detections derived from measurements of the absolute sky brightness and lower limits set by the integrated light from galaxies. A summary of the recent observational status of the EBL is presented in Tables 3--5. The major gap in our knowledge of the EBL is in the $\lambda \approx 10 - 70$~\mic\ wavelength region, where the thermal emission from interplanetary dust (the zodiacal light) dominates the brightness of the sky. This wavelength region is best probed by nearby radio and starburst galaxies at energies $\gtrsim 10$~TeV; 
%------
\item Models of the EBL, employing various methods for determining the evolution of the galaxies' spectral energy density with redshift seem to agree on the intensity and spectral shape of the UV to near-IR component of the EBL. However, there are considerable differences in their treatment of the redistribution of the intrinsic stellar and AGN output at IR wavelengths (see Figure~\ref{fig:ebl-models}); 
%------
\item The discovery of TeV blazars with alleged redshifts $z \gtrsim 1$ has revived an alternative model for the creation of these high energy photons. In this model  hadronic jets produce a cascade of $\sim$~TeV \gray\ photons en route to earth, circumventing the attenuation problem from such high redshift sources (see \S8). However, the redshift of these sources has been disputed, so the need to resort to such hadronic model is highly premature;
%------
\item Dark matter has been invoked to postulate the existence of a new class of primordial stars, powered by dark matter annihilation, instead of nuclear fusion. The energy release from these so-called dark stars could, in principle, lead to an observable signature in the EBL spectrum (see Figure~\ref{fig:ebl-x-limits}); and finally
%------
\item Observations suggest that the universe is essentially transparent to \gray s with energies $\lesssim 2$~TeV  up to $z\approx 0.2$, and energies of $\lesssim 400$~GeV up to $z \approx 0.4$ (see Figure~\ref{fig:horizon}).         
\end{enumerate}

The future prospects of intensified EBL studies with \gray s are promising, especially when considering the progress made over the last 5 years through the operation of {\it Fermi} and the current generation of atmospheric Cherenkov telescopes.
These instrument yielded combined energy spectra for at least 3 dozen extragalactic sources that cover up to 5 orders of magnitude in energy.  These data have provided a first glimpse of the transition region from a transparent universe at 1~GeV, to TeV energies at which the universe gradually turns opaque with increasing redshift.   

Furthermore, {\it Fermi} detected a large set of blazars \citep{ackermann11} including 310 FSRQs, 395 BL Lacertae objects and 156 candidate blazars, raising the specter for EBL studies with sizable source samples and different sources classes with CTA.  Large samples of blazars have the potential to constrain the EBL in the optical/near-IR and mid- IR through a better understanding of the blazar subclasses (FSRQs, LBL, IBL, HBL) and their intrinsic spectra, and better photon statistics for the measurement of the redshift dependence of any spectral feature attributable to the EBL.    

Recent discoveries of TeV emission from nearby radio- and starburst galaxies also gives rise to future prospects of extending the reach of  \gray\  constraints  up to the  far-IR through energy spectra spanning a few GeV up to 100~TeV.  These will be important to extend measurements of the \gray\ opacity imposed by the EBL all the way from UV/optical/near-IR/mid-IR to the far-IR and thus provide additional constraints to the intensity ratios between the different wavelength regimes of the EBL.

Limitations of precision EBL studies with \gray s arise from technical reasons, but are not unsurmountable.  First, systematic uncertainties in the measurements of \gray\  spectra with atmospheric Cherenkov telescopes arise from uncertainties in modeling the Earth's atmosphere, which translates into uncertainties in the attenuation of Cherenkov light from electromagnetic cascades.  To first order this affects the absolute energy scale of the measured \gray\ energies and is typically quoted at a level of 15\% to 20\%.  Effects on the spectral index and shape are generally of second order but require detailed studies.    Additional uncertainties lie in the absolute instrument calibration of Cherenkov telescopes, e.g., mirror reflectivity, light losses in the focal plane, and uncertainties in the quantum efficiency of the photodetectors, affecting  the light throughput and absolute energy scale.  Uncertainties in the spectral indices and spectral shape are typically quoted at the level of $\Delta \Gamma = 0.05 - 0.1$ or better, and depend mostly on the \gray\ selection efficiency derived from Monte Carlo simulations and detailed detector modeling.  Currently most energy spectra published (except for few flaring sources) are dominated by statistical uncertainties and leave much room for improvement through a more sensitive instrument such as CTA with a collection area of $\sim  km^{2}$ for 100~GeV to a few TeV energies, and several $km^2$ for higher energies.   At the same time calibration techniques are continually improving to provide a better handle on systematics as well.

Astrophysical challenges facing EBL studies with \gray s arise from the fact that currently only 55\% of all Fermi detected BL~Lacs have reliable redshift measurements \citep{ackermann11}.  Attempts to increase the redshift identifications through dedicated optical follow-up observations of the host galaxies of Fermi detected blazars during low flaring states are promising to increase the fraction of blazars with known redshift. The redshift distribution for FSRQs detected by {\it Fermi} peaks at z = 1 extending to z = 3.1, while BL Lacs peak at z = 0.2 reaching up to a redshift of z = 1.5. As already indicated, GeV/TeV detections of FSRQs are promising to allow constraints to the EBL in the UV/optical. Furthermore, the first detection of FSRQs by atmospheric Cherenkov telescopes out to redshifts of $z = 0.5$ provides the potential to complement EBL studies with a different class of sources, with different intrinsic spectral properties than BL~Lacs and,  most importantly, whose redshifts are known.

Detection and limits on the EBL have been obtained by direct absolute 
measurements, and by galaxy number counts that provided lower limits on its 
intensity. Direct measurements are aided by ground-based or space-based observations 
that resolve the foreground emission from Galactic stars. 
Future observations can considerably improve direct absolute measurement 
of the EBL by determining the absolute brightness of the zodiacal light from high 
resolution observations of reflected solar Fraunhofer lines, or by making 
absolute sky measurements from the outer solar system, thus effectively removing 
any foreground contribution from the zodiacal cloud.  Lower limits 
on the EBL can converge to the EBL itself with observations of sufficient 
depth and resolution to resolve all the galaxies that contribute to its intensity.

The James Webb Space Telescope (JWST) is a large (6.6 m) IR 
space observatory, passively cooled to temperatures below 50 K, 
that will be launched into orbit at the second Earth-Sun Lagrange 
point (L2). The wide band observations with NIRCam at 2 - 5 $\mu$m 
and MIRI at 6 - 25 $\mu$m will fill in crucial 
gaps in our knowledge of the EBL intensity at these wavelengths. 
NIRCam will cover a 10 arcmin$^2$ field of view (FOV) 
and will resolve galaxies down to a 10$\sigma$ flux limit of 11 nJy at 2 $\mu$m
in 10,000 s, making it about 10$^3$ more sensitive than the Spitzer/IRAC 
3.6 $\mu$m filter.  
With unparalleled resolution and sensitivity, the JWST \citep{gardner06} will resolve the EBL 
at near-IR wavelengths. The MIRI instrument operating in its 
broadband imaging mode at 10 and 21~\mic\ will have a smaller 
FOV of $\sim$2.6 arcmin$^2$, 
and sensitivities of 700 nJy and 9 $\mu$Jy, respectively. It will 
resolve galaxies down to the confusion limit which, because of 
its large telescope size, will be significantly fainter than
that of the Spitzer/MIPS instrument at 24 $\mu$m.

Absolute measurements of solar absorption line profiles in the zodiacal
light, for example the Fraunhofer Mg I line at 1.182819~\mic\, or 
the Si I line at 1.210354~\mic\, can be used to determine and 
remove the contribution of the ZL to the total sky brightness 
in the J (1.25~\mic) band \citep{kutyrev04,kutyrev08}. High 
resolution (R $\sim$ 20,000) observations of these lines using the 
ground-based spectrometer ZEFIR (Zodiacal Emission determination 
through Fraunhofer IR lines) will determine their equivalent
width (EW) in the ZL spectrum. The continuum contribution of 
the ZL in the J-band can then be determined from knowledge of 
the EW of these lines in the solar spectrum. The DIRBE 
instrument provided absolute measurements of the sky 
brightness in this band, and ground-based sky surveys (2MASS) 
have removed most of the contribution of Galactic stellar 
emission from this band. Measurements of the absolute brightness 
of the ZL will thus provide the absolute intensity of the EBL 
in the DIRBE 1.25~\mic\ waveband. Similar measurement could also be 
made from the ground in the 2.2~\mic\ band.

Finally, the contribution of the ZL to the observed sky brightness 
can be largely eliminated by mapping the absolute intensity of the 
diffuse emission from the sky from beyond the zodiacal cloud. A 
satellite mission conducting such EBL measurements from the outer 
solar system will be an important step in that direction (Cooray, 2011).
%Need not be a dedicated mission -- you could combine or piggyback 
% with an outer solorsytem planetary mission.

The next decade will see significant improvement in $\gamma$-ray technology, 
and measurements of the EBL, providing new insights into the 
spectrum of the different $\gamma$-ray sources and the mechanism that 
generate them, and into the spectrum of the EBL which will 
provide new constraints on the history of all nuclear and 
gravitational energy releases in the universe. 
%This strikes me as a very broad and sweeping statement. Though perhaps 
% it fits if it really is the penultimate sentance of the article.
The synergy 
between TeV $\gamma$-ray astronomy and EBL research will ensure 
that any development in each field will greatly benefit both. \\

\noindent
{\bf Acknowledgements} \\
ED and FK gratefully acknowledge the support from the NASA Fermi Guest Investigator grant NNX11AO38G. FK also acknowledges the support from the U.S. Department of Energy Office of Science.
We thanks C.D. Dermer, J.D. Finke, A. Dominguez, R.C. Gilmore, A. Franceschini, and J.R. Primack for providing their model results in digital format.
We thank D. Kazanas and R.G. Arendt for their comments on parts of the manuscript, and an anonymous referee for his/her thorough reading of and helpful comments on the manuscript. 

\clearpage
%=====================
% TABLES
%=====================

%=========================
% Table 1
%=========================
\begin{table}
\caption{Glossary of Abbreviations of Spacecrafts$^1$, Telescopes, and Instruments} 
\begin{tabular}{ll} \hline
%\tablewidth{0pt}
%\tabletypesize{\scriptsize}
%\tablecaption{Total EBL Intensity of Different Models}

 Abbreviation  &  Full Name   \\ \hline  \hline
% \tablehead{
%\colhead{Model &
%\colhead{Stars} &
%\colhead{Dust} & 
%\colhead{Total} & 
%}
%\startdata
{\it Akari} & Infrared imaging satellite (ASTRO-F) \\
BLAST & Balloon-borne Large-Aperture Submillimeter Telescope \\
 {\it COBE} & Cosmic Background Explorer \\
~~~DIRBE & Diffuse Infrared Background Experiment \\
~~~FIRAS & Far Infrared Absolute Photometer \\
CTIO & Cerro Tololo Inter-American Observatory \\
{\it GALEX} &  Galaxy Evolution Explorer  \\
 {\it Herschel} & Herschel Space Observatory\\
~~~PACS & Photodetector Array Camera \\
~~~SPIRE & Spectral and Photometric Imaging Receiver \\
{\it HST} & Hubble Space Telescope  \\
 ~~~WFPC2 & Wide Field Planetary Camera \\
~~~NICMOS & Near IR Camera and Multi-Object Spectrometer\\
{\it IRTS} & Infrared Telescope in Space \\
 {\it ISO} & Infrared Space Observatory \\
 ~~~ISOCAM & ISO Camera \\
 JCMT & James Clerk Maxwell Telescope \\
~~~SCUBA & Submillimeter Common User Bolometer Array \\
NTT & New Technology Telescope \\
{\it Spitzer} & Spitzer Space Telescope \\
~~~IRAC & Infrared Array Camera \\
~~~MIPS & Multiband Imaging Photometer \\ 
Subaru & Optical, near-IR telescope \\
{\it Pioneer} &  Interplanetary spacecraft\\ 
{\it WMAP} & Wilkinson Microwave Anisotropy Probe \\
2MASS & Two Micron All Sky Survey \\
\hline
CTA & Cherenkov Telescope Array \\
{\it Fermi} & Fermi Gamma-Ray Space Telescope \\
H.E.S.S. & High Energy Stereoscopic System \\
IACT & Imaging Air Cherenkov Telescope \\
MAGIC & Major Atmospheric Gamma-Ray Imaging Cherenkov Telescope \\
Milagro & Gamma-ray and cosmic-ray telescope \\
VERITAS & Very Energetic Radiation Imaging Telescope Array \\ 
 \hline
%     \enddata
\label{table:ebl}
\end{tabular}\\
$^1$ Spacecraft names are presented in italics, and their instruments are indented. 
\label{table:acronyms}
\end{table}

%OLD TABLE
%==================
% Table 2
%==================
\begin{table}[t]
\scriptsize{
\caption{Extragalactic \gray\  sources with GeV and TeV spectral information.}
%\noindent{\small
 \begin{tabular}{l l c  c  c c c } \hline
%\begin{tabular}{||l|l|p{0.01795in}|c|c||}
\hline \hline
%DAW Institute & PI & Role & Heritage$^\ast$ & Agency & Budget\\
%DAW      &    &      &          &        & \$k/yr\\
Name    &  Class   & redshift & $\Gamma_{GeV}$   &   $\Gamma_{TeV}$   &   Range  [TeV]  & References \\    \hline   \hline
Centaurus~A   & Radio         &     0.0008   &      2.76$\pm 0.05$    &     2.7$\pm 0.5$  &       0.2 - 5     &      [1], [2]     \\
M82        &   SB          &     0.00085   &      2.2$\pm 0.2$    &     2.5$\pm 0.6$  &        0.7 - 4   &    [3], [4]      \\
NGC253 &   SB          &     0.00093   &      1.95$\pm 0.4$    &     2.24$\pm 0.14$  &  0.3 - 50   &    [3], [5]    \\
 M87   & Radio          &     0.0036   &      2.17$\pm 0.07$    &     2.5$\pm 0.2$  &   0.2 - 10   &     [6], [7], 8], [9]   \\
  NGC~1275   & Radio           &     0.018   &      2.00$\pm 0.02$    &     3.96$\pm 0.37$  &  0.1 - 0.3    &     [10], [11]   \\
  IC~310   & Radio         &     0.0188   &      2.10$\pm 0.19$    &     2.0$\pm 0.14$  &  0.1 - 7   &     [12], [13], [14]     \\
   Markarian~421   & HBL        &     0.031   &      1.77$\pm 0.01$    &     2.48$\pm 0.03^{\star}$  &  0.1 - 5    &  [15]   \\
    Markarian 501   & HBL         &     0.034   &      1.74$\pm 0.03$    &     2.51$\pm 0.05^{\triangle}$  &  0.1 - 10    &    [16]    \\
    1ES~2344+514   & HBL         &     0.044   &      1.72$\pm 0.08$    &     2.78$\pm 0.09^{\triangle}$  &   0.3 - 2    &   [6],  [17]    \\
 Markarian~180   & HBL         &     0.046   &      1.74$\pm 0.08$    &     3.3$\pm 0.70$  &   0.2 - 1   &  [6]  [18]    \\
 1ES~1959+650   &   HBL          &   0.047    &   1.94$\pm 0.03$    &     2.72$\pm 0.14$  & 0.2 - 2   &  [6], [19]  \\
 AP~Lib$^{*}$  &     LBL        &   0.048      &    2.05$\pm 0.04$   &   2.5$\pm 0.2$  &  0.3 - 2   &   [6], [20]  \\
 BL Lacertae   &     LBL        &  0.069     &    2.11$\pm 0.04$   &  3.6$\pm 0.5$  & 0.2 - 1   &  [6], [21]  \\
 PKS~2005-489 & HBL         &  0.071    &    1.78$\pm 0.05$ &   4.0$\pm 0.4$  &  0.2 - 2   &  [6], [22]  \\
W~Comae            & IBL          &  0.103     &  2.02$\pm 0.03$ &    3.81$\pm 0.35$  &  0.3 - 1   &   [6], [23]   \\
PKS~2155-304   & HBL        &  0.116    &    1.84$\pm 0.02$ &   3.53$\pm 0.05$  &  0.4 - 5   &  [6], [24]  \\
B3~2247+381     & HBL         &  0.119    &     1.84$\pm 0.11$ &  3.2$\pm 0.5$  & 0.2 - 1   &  [6], [25]   \\
 RGB~J0710+591     &   HBL       &  0.125    &    1.53$\pm 0.12$   &   2.69$\pm 0.26$  &  0.3 - 4.6   &  [6], [26]    \\
 H~1426+428         &    HBL       &    0.129      &   1.32$\pm 0.12$   &   3.50$\pm 0.35$   &  0.3 - 10   &  [6],  [27] \\
 1ES~1215+303        &   IBL       &  0.13$^{\heartsuit}$      &  2.02$\pm 0.02$  &   2.99$\pm 0.15$  &  0.1 - 1    &  [6], [28]   \\
 1ES~0806+524        &   HBL       &  0.137      &  1.94$\pm 0.06$  &   3.6$\pm 1.0$  & 0.3 - 0.7    &  [6],  [29]  \\
 1RXS J101015.9-311909& HBL &   0.143   &  2.24$\pm 0.14$  &            3.14$\pm 0.53$  &  0.3 - 1   &  [6], [20]   \\
 1ES~1440+122     &   IBL    &   0.163    &  1.41$\pm 0.18$  &    3.3$\pm 0.7$  & 0.3 - 1 & [6], [30]  \\
H~2356-309       &   HBL       &   0.165    &  1.89$\pm 0.17$   &    3.09$\pm 0.24$    &  0.3 - 2   &  [6], [31]  \\
VER~J0648+152 &   HBL       & 0.179   &  1.74$\pm 0.11$   &    4.4$\pm 0.8$    &   0.3 - 0.8 &   [6], [32]   \\
1ES~1218+304  &   HBL        &   0.184    &  1.71$\pm 0.07$   &    3.07$\pm 0.09$    &   0.2 - 2 &  [6], [33]  \\
1ES~1101-232  &   HBL       &   0.186    &  1.80$\pm 0.21$   &    2.88$\pm 0.17$    &   0.16 - 3.3 & [6], [31]   \\
RBS~0413    &   HBL        &   0.19    &  1.55$\pm 0.11$   &    3.18$\pm 0.68$    &  0.25 - 1 &  [6], [33] \\
PKS-0447-439 &   HBL       &   0.205    &  1.86$\pm 0.02$   &    4.36$\pm 0.49$    &   0.25 - 1 & [6], [34]  \\
1ES~1011+496 &   HBL      &   0.212    &  1.72$\pm 0.04$   &    4.0$\pm 0.50$    &   0.25 - 0.6 & [6], [35]  \\
1ES~0414+009 &  HBL       &   0.287      &  1.98$\pm 0.16$  &    3.44$\pm 0.27$    &  0.25 - 1.2 & [6],[36]   \\
S5~0716+714  &  LBL       &   0.31      &  2.01$\pm 0.02$  &    3.45$\pm 0.54$    &   0.25 - 1.2  &  [6], [37]  \\
% PKS~1510-089  &  FSRQ       &   0.36      &  2.29$\pm 0.01$  &    ????$\pm 0.??$    &   0.3 - 10,  0.25 - 1.2 & [6],   \\
1ES~0502+675  &  HBL       &   $0.416^{\clubsuit}$      &  1.49$\pm 0.07$  &    3.92$\pm 0.35$    &    0.25 - 1 & [6], [38]   \\
4C~21.35         &  FSRQ       &   0.43      &  2.12$\pm 0.02$  &    3.75$\pm 0.27$    &   0.07 - 0.4 & [6], [39]   \\
3C~66A       &  IBL       &    $0.44^{\clubsuit}$    &  1.85$\pm 0.02$  &    4.1$\pm 0.4$    &   0.22 - 0.45 &  [6], [40]   \\
3C~279     &  FSRQ       &       0.536    &  2.22$\pm 0.02$  &    3.03$\pm 0.9$    &   0.1 - 0.35 &       [6], [41]   \\     \hline  \hline
  \hline
\end{tabular}
}
% \begin{center}
\scriptsize{
$\star$  Spectrum is not well fit by a powerlaw.  
$\triangle$  Spectrum shows variations.   
$\clubsuit$  redshift uncertain.   
$\heartsuit$  redshift recently given by Abdo et al. arXiV:1108.1420v1, different redshift was considered viable by Colin et al. 2011 \\
References: [1]\cite{abdo10};  [2] \cite{aharonian09}; [3] \cite{abdo10b}; [4] \cite{acciari09d};
[5] \cite[][for the H.E.S.S. collaboration]{abramowski12};   [6] \cite{ackermann11}; [7] \cite{berger11a};
[8] \cite{acciari10a}; [9] \cite{aharonian06a};  [10] \cite{abdo09}; [11] \cite{hildebrand11};
[12]  \cite{neronov10} ; [13]  \cite{aleksic10};  [14] \cite{ackermann11}; [15] \cite{abdo11a}; [16] \cite{abdo11b};  [17] \cite{acciari11}; [18] \cite{albert06a};  [19] \cite{albert06b}; [20]\cite{cerruti11}; [21] \cite{albert07};  [22] \cite{aharonian05b}; [23] \cite{acciari08}; [24]  \cite{aharonian07b}; [25] \cite{berger11b};  [26] \cite{acciari10b}; [27] \cite{petry02}; [28] \cite{colin11};  [29] \cite{acciari09b};   [30] \cite{benbow11};  [31] \cite{aharonian06};  [32]  \cite{errando11} [33] \cite{acciari10c}; [33] \cite{senturk11}; [34] \cite{zech11}; [35] \cite{albert07b}; [36] \cite{volpe11};  [37] \cite{anderhub09}; [38]   \cite{benbow11b};  [39] \cite{aleksic11}; [40]  \cite{acciari09c}; [41]  \cite{aleksic11b}.
}
\label{table:sources}
\end{table}

%================
% Table 3
%================
\begin{table}
\vspace{-1.1in}
\footnotesize{
\caption{Limits and detection of the extragalactic background light (EBL).}
\begin{tabular}{cccccc}\hline
$\lambda$    &   & $ \nu\, I_{\nu}$(nW~m$^{-2}$~sr$^{-1}$) &  & Comment  &   Reference   \\ 
(\mic) & IGL & EXT & ABS &  &   \\ \hline
\hline
0.1530 &  $0.68\pm0.10$ 	             & ${\bf1.03\pm0.15}$	  & 	                       & {\it Galex}     &  [1] \\
0.1595 &  $ 3.75\pm1.25$ 	             & 	                   & 	                           & {\it HST}/STIS      &  [2] \\
0.2    &  $0.6$                       &                      &                            & FOCA/balloon       &  [3] \\
0.2310 &  $0.99\pm0.15$ 	             & ${\bf2.25\pm0.32}$	  & 	                       & {\it Galex}     &  [1] \\
0.2365 &  $3.6^{+0.7}_{-0.5}$         &                      &                            & {\it HST}/WFPC2    &  [2] \\
0.30   &                              &                      &     $ 18\pm12 $            & {\it HST}/WFPC2    &  [4] \\
       &  $ 2.7\pm0.3 $               &   ${\bf 3.7\pm0.7}$  &                            & {\it HST}+ground    &  [5] \\
0.36   &  $2.9^{+0.6}_{-0.4}$         &                      &                            & {\it HST}+ground   &   [6]  \\
0.40   &  				             &                      & $<36\, (26\pm10$)          & dark cloud             &   [7]\\
0.44   &  				             &                      & \textbf {\textit{ 7.9$\pm$4.0}}    & Pioneer 10/11      &  [8] \\
0.45   &  $ 4.6^{+0.7}_{-0.5}$        &                      &                            & {\it HST}+ground   &   [6]  \\
       &  $4.4\pm0.4$                 &  ${\bf 6.1\pm1.8}$   &                            & {\it HST}+ground   &  [5]  \\
0.5115 &                              &                      & $<39\, (30\pm9)$           & ground             &   [9] \\
0.55   &                              &                      &      $ 55\pm27$             & {\it HST}/WFPC2    &  [4] \\
0.61   &  $6.0\pm0.6 $                &   ${\bf 7.4\pm1.5}$  &                            & {\it HST}+ground   &  [5]  \\
0.64   &  				             &                      & \textbf {\textit{ 7.7$\pm$5.8}}          & Pioneer 10/11      &  [8] \\
0.67   &  $6.7^{+1.3}_{-0.9}$         &                      &                            & {\it HST}+ground   &   [6]  \\
0.81   &  $ 8.0^{+1.6}_{-0.9}$        &                      &                            & {\it HST}+ground   &   [6]  \\
       &  $8.1\pm0.8$                 &  ${\bf 9.3\pm1.6}$   &                            & {\it HST}+ground   &  [5]  \\
0.814  &                              &                      &      $ 57\pm32 $           & {\it HST}/WFPC2    &  [4] \\
1.1    &  $9.7^{+3.0}_{-1.9}$         &                      &                            & {\it HST}+ground   &   [6]  \\
1.25   &                              &                      & \textbf {\textit{ 21$\pm$15}}             & {\it COBE}/DIRBE   &   [10] \\
       &                              &                      & $54\pm17$                  & {\it COBE}/DIRBE   &   [11]  \\
       &  $10.9\pm1.1$                &  ${\bf 11.5\pm1.3}$  &                            & {\it HST}+ground   &  [5]  \\
       &  $11.7^{+5.6}_{-2.6}$        &                      &                            & Subaru             &   [12]   \\
1.4--4 &                              &                      & $\sim 60 - 15$             & {\it IRTS}         &   [13] \\
1.6    &  $ 9.0^{+2.6}_{-1.7}$        &                      &                            & {\it HST}+ground   &   [6]  \\
       &  $11.5^{+4.5}_{-1.5}$        &                      &                            & Subaru             &   [12]   \\
2.12   &  $10.0^{+2.8}_{-0.8}$        &                      &                            & Subaru             &   [12]  \\
2.2    &  $7.9^{+2.0}_{-1.2}$         &                      &                            & {\it HST}+ground   &   [6]  \\
       &  $8.3\pm0.8$                 &  ${\bf 9.0\pm1.2}$   &                            & {\it HST}+ground   &  [5]  \\
       &                              &                      & \textbf {\textit{ 20$\pm$6}}           & {\it COBE}/DIRBE   &   [10] \\
       &                              &                      & $28\pm7$                   & {\it COBE}/DIRBE   &   [11]  \\
3.5    &                              &                      & $ 13.3\pm2.8$              & {\it COBE}/DIRBE   & [10]\\
       &                              &                      & \textbf {\textit{13.8$\pm$3}}              & {\it COBE}/DIRBE   &  [15]\\
3.6    &  $5.4$                       &                      &                            & {\it Spitzer}/IRAC &    [16]\\
       &                              & ${\bf 9.0^{+1.7}_{-0.9}}$ &                       & {\it Spitzer}/IRAC &    [14] \\
4.5    &  ${\bf 3.5}$                 &                      &                            & {\it Spitzer}/IRAC &    [16]\\
4.9    &                              &                      & \textbf {\textit{ 22$\pm$12}}          & {\it COBE}/DIRBE   &    [17]\\
5.8    &  ${\bf 3.6}$                 &                      &                            & {\it Spitzer}/IRAC &    [16]\\
8.0    &  ${\bf 2.6}$                 &                      &                            & {\it Spitzer}/IRAC &    [16]\\
\\
 
\hline
\end{tabular}
}
\scriptsize{ABS=absolute measurement; IGL=integrated galactic light; STK=lower limits from stacking analysis; EXT=extrapolated intensity from $dN/dS$ \\
$^1$ Calculated for a 2.2~\mic\ intensity of 20.0~\nwat\ and the Kelsall et al. (1998) ZL model.\\
% References: [1] Xu et al. (2005); [2] Gardner et al. (2000); [3] Milliard et al. (1992); [4] Bernstein (2007); [5] Totani et al. (2001);
% [6] Madau \& Pozzetti (2000);  [7] Mattila (1990), Leinert et al. (1998);  [8] Matsuoka et al. (2011); [9] Dube et al. (1979), Leinert et al. (1998);  
% [10] Levenson et al. (2007);  [11] Cambr\'esy et al. (2001); [12] Keenan et al. (2010); [13] Matsumoto et al. (2001);
%  [14] Levenson \& Wright (2008);  [15] Dwek \& Arendt (1998);   [16] Fazio et al. (2004);   [17] Arendt \& Dwek (2003)
  References:[1] \cite{xu05}; [2] \cite{gardner00}; [3] \cite{milliard92}; [4] \cite{bernstein07}; [5] \cite{totani01};
 [6] \cite{madau00};  [7] \cite{mattila90},\cite{leinert98};  [8] \cite{matsuoka11}; [9] \cite{dube79},\cite{leinert98};  
 [10] \cite{levenson07};  [11] \cite{cambresy01}; [12] \cite{keenan10}; [13] \cite{matsumoto01};
  [14] \cite{levenson08};  [15] \cite{dwek98b};   [16] \cite{fazio04};   [17] \cite{arendt03}

} 
\end{table}

\newpage
	
\thispagestyle{empty}

%================
% Table 4
%================
\begin{table}
\vspace{-1.0in}
\scriptsize{
\caption{Limits and detection of the extragalactic background light (EBL).}
\begin{tabular}{ccccccc}\hline
  & \multicolumn{4}{c}{$\nu\, I_{\nu}$ (nW~m$^{-2}$~sr$^{-1}$)}\\ 
\cline{2-5}
%$\lambda$    &  & $\nu\, I_{\nu}$ (nW~m$^{-2}$~sr$^{-1}$)   &  & Comment  &   Reference   \\ 
$\lambda$(\mic) & IGL & STK  & EXT & ABS & Comment & Reference \\ \hline
\hline
15       & $2.4\pm0.5$            &                        &                       &                         & {\it ISO}/ISOCAM     & \cite{elbaz02} \\
         &  ${\bf 2.7\pm0.6}$      &                        &                       &                         & {\it ISO}/ISOCAM     & \cite{metcalfe03} \\
         & $1.9\pm0.5$            &                        &                       &                         & {\it Akari}  & \cite{hopwood10} \\
\hline
16       & $2.2\pm0.2$            &                        &                       &                         & {\it Spitzer} & \cite{teplitz11} \\
\hline
24       & $1.9\pm0.6$            &                        &  $2.7^{+1.1}_{-0.7}$  &                         & {\it Spitzer}/MIPS    & \cite{papovich04} \\
         & $1.8\pm0.2$            &                        &  $2.0\pm0.2$          &                         & {\it Spitzer}/MIPS    & \cite{chary04} \\
         & ${\bf 2.29\pm0.09}$    &                        & $2.86^{+0.19}_{-0.16}$&                         & {\it Spitzer}/MIPS    & \cite{bethermin10a}\\
\hline
60       &                        &                        &                       & \textbf {\textit{ 28.1$\pm$1.8$\pm$7}} & {\it COBE}/DIRBE       & \cite{finkbeiner00} \\
\hline
65       &                        &                        &                       & \textbf {\textit{ 12.5$\pm$1.4$\pm$9.2}} & {\it Akari} & \cite{matsuura11b} \\
\hline
70       &                        &                        & $7.4\pm1.9$           &                         & {\it Spitzer}/MIPS    & \cite{frayer06}\\
         &                        &  ${\bf7.1\pm1.0}$       &                       &                         & {\it Herschel}/PACS    & \cite{dole06}\\
         & $5.4\pm0.4$            &                        & $6.6^{+0.7}_{-0.6}$   &           			    & {\it Spitzer}/MIPS    & \cite{bethermin10a}\\
         & $4.52\pm1.18$          &                        &                       &                         & {\it Herschel}/PACS    & \cite{berta11}\\	
\hline
90       &                       &                       &                         & \textbf {\textit{ 22.3$\pm$1.7$\pm$4.7}}  & {\it Akari}  & \cite{matsuura11b} \\
\hline
100      &                        &                        &                       & $<34\,( 22\pm6)$  & {\it COBE}/DIRBE (D, KZL) & \cite{hauser98}\\
         &                        &                        &                       & $12.5\pm5$       & {\it COBE}/DIRBE (D, WZL)     & \cite{wright04}\\
         &                        &                        &                       & $24.6\pm2.5\pm8$         & {\it COBE}/DIRBE       & \cite{finkbeiner00} \\
         &                        &                        &                       & $23.4\pm6.3$      & {\it COBE}/DIRBE (D, KZL)      & \cite{lagache00a} \\
	     &                        &                        &                       & \textbf {\textit{ 14.4$\pm$6.0}} & DIRBE (F, WZL)      & \cite{dole06} \\
		 & $8.35\pm0.95$          &   ${\bf9.4\pm1.1}$    &         				&             			& {\it Herschel}/PACS   & \cite{berta10} \\
         &                       &                        &                        & $6.6\pm1.8\pm2.1$       & {\it Spitzer}/MIPS           & \cite{penin12} \\        
\hline
140      &                        &                        &                       & $25.0\pm6.9$    & {\it COBE}/DIRBE (D, KZL)     & \cite{hauser98}\\
         &                        &                        &                       & $15.0\pm5.9$    & {\it COBE}/DIRBE  (F, KZL)    & \cite{odegard07}\\
         &                        &                        &                       & $32\pm13$               & {\it COBE}/DIRBE      & \cite{schlegel98}\\
         &                        &                        &                       & $24.2\pm11.6$   & {\it COBE}/DIRBE (D, KZL)   & \cite{lagache00a} \\
         &                        &                        &                       & $22\pm7$        & {\it COBE}/DIRBE (D, WZL)     & \cite{wright04}\\
         &                       &                       &                         & $20.1\pm3.4\pm1.1$       & {\it Akari}  & \cite{matsuura11b} \\
	     &                        &                        &                       & $12.4\pm6.9$    & {\it COBE}/DIRBE (F, WZL)     & \cite{dole06} \\
         &                        &                        &                       & \textbf {\textit{ 12.6$\pm$6.0}}   & {\it COBE}/FIRAS  & \cite{fixsen98}\\
\\
 \hline
\end{tabular}
}
\end{table}

%================
% Table 5
%================
\begin{table}
\vspace{-1.0in}
\scriptsize{
\caption{Limits and detection of the extragalactic background light (EBL).}
\begin{tabular}{ccccccc}\hline
  & \multicolumn{4}{c}{$\nu\, I_{\nu}$ (nW~m$^{-2}$~sr$^{-1}$)}\\ 
\cline{2-5}
%$\lambda$    &  & $\nu\, I_{\nu}$ (nW~m$^{-2}$~sr$^{-1}$)   &  & Comment  &   Reference   \\ 
$\lambda$(\mic) & IGL & STK  & EXT & ABS & Comment & Reference \\ \hline
\hline
160 	     &                        &  $13.4\pm1.7$          &                       &                         & {\it Herschel}/PACS    & \cite{dole06}\\
         &                        &                        &                       & $13.7\pm3.9\pm0.8$       & {\it Akari}    & \cite{matsuura11b} \\
	     & $9.49\pm0.59$          &  $11.4\pm0.7$           &                       &                         & {\it Herschel}/PACS   & \cite{berta10} \\
	     & $8.9\pm1.1$            &                        & $14.6^{+7.1}_{-2.9}$  &                         & {\it Spitzer}/MIPS    & \cite{bethermin10a}\\	
         &                        &                        &                       & $14.4\pm0.8\pm2.3$       & {\it Spitzer}/MIPS           & \cite{penin12} \\ 
                  &                        &                        &             & \textbf {\textit{ 13.7$\pm$6.1}}       & {\it COBE}/FIRAS      & \cite{fixsen98}\\
\hline
170      &  $19.1\pm5.6\pm5.3$    &                        &                       &                         & ISOPHOT         & Juvela et al. (2009) \\
\hline
240      &                        &                        &                       & $13.6\pm2.5$    & {\it COBE}/DIRBE (D, KZL)     & \cite{hauser98} \\
         &                        &                        &                       & $13\pm2.5$      & {\it COBE}/DIRBE (D, WZL)     & \cite{wright04} \\
         &                        &                        &                       & $12.7\pm1.6$     & {\it COBE}/DIRBE (F, KZL)     & \cite{odegard07} \\
         &                        &                        &                       & $17\pm4$               & {\it COBE}/DIRBE      & \cite{schlegel98}\\
         &                        &                        &                       & $11.0\pm6.9$      & {\it COBE}/DIRBE (D, KZL)      & \cite{lagache00a} \\       
	     &                        &                        &                       & $12.3\pm2.5$    & {\it COBE}/DIRBE (F, WZL)     & \cite{dole06} \\ 
	               &                        &                        &             & \textbf {\textit{ 10.9$\pm$4.3}}       & {\it COBE}/FIRAS  & \cite{fixsen98}\\        
\hline
250      &                        &  $8.6\pm0.6$           &                       &                         & BLAST           & \cite{marsden09} \\
	     & $0.24^{+0.18}_{-0.13}$  &  $5.0^{+2.5}_{-2.6}$    &                       &                         &  BLAST           & \cite{bethermin10b}\\	
         & $1.73\pm0.33$          & 			             &                       &                         & {\it Herschel}/SPIRE  & \cite{oliver10} \\
         & $1.55\pm0.30$          & $7.40\pm1.42$           &$10.13^{+2.60}_{-2.33}$ &                         & {\it Herschel}/SPIRE  & \cite{bethermin12} \\
         &                        &                        &                       & \textbf {\textit{ 10.3$\pm$4.0}}       & FIRAS      & \cite{fixsen98}\\
\hline
350      &                        &   $4.93\pm0.34$        &                       &                          & BLAST          & \cite{marsden09} \\
	     & $0.06^{+0.05}_{-0.04}$  &  $2.8^{+1.8}_{-2.0}$    &                       &                         &  BLAST       &  \cite{bethermin10b} \\
         & $0.63\pm0.18$          &                        &                       &                          & {\it Herschel}/SPIRE  & \cite{oliver10} \\
         & $0.77\pm0.16$          & $4.50\pm0.90$           &$6.46^{+1.74}_{-1.57}$ &                         & {\it Herschel}/SPIRE  & \cite{bethermin12} \\
         &                        &                        &                       & \textbf {\textit{ 5.6$\pm$2.1}}       & FIRAS      & \cite{fixsen98}\\         
\hline
500      &                        &  $2.27\pm0.20$         &                       &                          & BLAST         & \cite{marsden09} \\
	     & $0.01^{+0.01}_{-0.01}$  &  $1.4^{+2.1}_{-1.3}$    &                       &                         &  BLAST         &  \cite{bethermin10b}\\
         & $0.15\pm0.07$          &                        &                       &                          & {\it Herschel}/SPIRE  & \cite{oliver10} \\
         & $0.14\pm0.03$          & $1.54\pm0.34$           &$2.80^{+0.93}_{-0.81}$ &                         & {\it Herschel}/SPIRE  & \cite{bethermin12} \\         
         &                        &                        &                       & \textbf {\textit{ 2.4$\pm$0.9}}       & FIRAS      & \cite{fixsen98}\\
\hline
850      & $0.12\pm0.03$          &                        &                       &                         & SCUBA            & \cite{coppin06} \\
         & $0.24\pm0.03$          &                       &                        &                          & SCUBA           & \cite{zemcov10} \\
         &                        &                        &                       & \textbf {\textit{ 0.5$\pm$0.21}}       & {\it COBE}/FIRAS      & \cite{fixsen98}\\
\hline
 & & & & & & \\
200-1000 &       &          &      & $a\left({\lambda_0\over\lambda}\right)^k\nu B_{\nu}(T)$                 & FIRAS      & \cite{fixsen98}$^1$ \\ 
\\
 \hline
\end{tabular}
}
\footnotesize{$^1\ a=(1.3\pm=0.4)\times10^{-5};\ k=0.64\pm0.12;\ T=(18.5\pm1.2)~K;\ \lambda_0=100$~\mic }
\end{table}

%=========================
% Table 6
%=========================
\begin{table}
\caption{Total EBL Intensity of Different Models (\nwat)} 
\begin{tabular}{lllll} \hline
%\tablewidth{0pt}
%\tabletypesize{\scriptsize}
%\tablecaption{Total EBL Intensity of Different Models}

 Model&  & Stars$^1$ & Dust$^2$ & Total \\ \hline  \hline
% \tablehead{
%\colhead{Model &
%\colhead{Stars} &
%\colhead{Dust} & 
%\colhead{Total} & 
%}
%\startdata
  \cite{franceschini08} & & 25  & 40 &  65  \\
   \cite{dominguez11}  & & 25   &  44  &    69  \\
    \cite{gilmore11}&  & 25 &  23  &  48  \\
     \cite{mazin07} & & 30   & 26 &  56  \\
     \cite{finke10}  & & 27   & 20  &  47  \\
    \cite{stecker06b}&  & 61 &  35  &    96  \\
     Observations$^3$  & & 23--93 & 20--110&  42--202  \\
     From CSFR$^4$ & & -- & --  &  21--66  \\ \hline
%     \enddata
\label{table:ebl}
\end{tabular}\\
$^1$ Integrated intensity from 0.1 to 10~\mic.\\
$^2$ Integrated in tensity from 10 to 1000~\mic. \\
%$^1$ \cite{franceschini08}; $^2$ \cite{dominguez11}; \\ $^3$ \cite{gilmore11}; 
%$^4$ \cite{mazin07};\\ $^5$ \cite{finke10}; $^6$ \cite{stecker06b} \\
$3$ From limits and detections of the EBL (Figure \ref{fig:ebl_limits})\\
$4$ Total integrated intensity inferred from the CSFR (Fig.\ref{fig:csfr})
\label{table:iebl}
\end{table}


%\bibliographystyle{/Users/edwek/Library/texmf/tex/latex/misc/aastex52/aas.bst}

%  \bibliographystyle{/Users/edwek/Library/TeXshop/aastex52/apj.bst}
%\bibliography{/Users/edwek/Dropbox/science/00-Bib_Desk/Astro_BIB.bib}

\clearpage

  \bibliographystyle{/Users/edwek/Library/TeXshop/aastex52/apj.bst}
\bibliography{/Users/edwek/Dropbox/science/00-Bib_Desk/Astro_BIB.bib}

\begin{thebibliography}{242}
\expandafter\ifx\csname natexlab\endcsname\relax\def\natexlab#1{#1}\fi

\bibitem[{{Abdo} {et~al.}(2010{\natexlab{a}}){Abdo}, {Ackermann}, {Agudo},
  {Ajello}, {Aller}, {Aller}, {Angelakis}, {Arkharov}, {Axelsson}, {Bach}, \&
  et~al.}]{abdo10c}
{Abdo}, A.~A., {Ackermann}, M., {Agudo}, I., {Ajello}, M., {Aller}, H.~D.,
  {Aller}, M.~F., {Angelakis}, E., {Arkharov}, A.~A., {Axelsson}, M., {Bach},
  U., \& et~al. 2010{\natexlab{a}}, \apj, 716, 30

\bibitem[{{Abdo} {et~al.}(2011{\natexlab{a}}){Abdo}, {Ackermann}, {Ajello},
  {Allafort}, {Baldini}, {Ballet}, {Barbiellini}, {Baring}, {Bastieri},
  {Bechtol}, \& et~al.}]{abdo11b}
{Abdo}, A.~A., {Ackermann}, M., {Ajello}, M., {Allafort}, A., {Baldini}, L.,
  {Ballet}, J., {Barbiellini}, G., {Baring}, M.~G., {Bastieri}, D., {Bechtol},
  K., \& et~al. 2011{\natexlab{a}}, \apj, 727, 129

\bibitem[{{Abdo} {et~al.}(2009){Abdo}, {Ackermann}, {Ajello}, {Asano},
  {Baldini}, {Ballet}, {Barbiellini}, {Bastieri}, {Baughman}, {Bechtol},
  {Bellazzini}, {Blandford}, {Bloom}, {Bonamente}, {Borgland}, {Bregeon},
  {Brez}, {Brigida}, {Bruel}, {Burnett}, {Caliandro}, {Cameron}, {Caraveo},
  {Casandjian}, {Cavazzuti}, {Cecchi}, {Celotti}, {Chekhtman}, {Cheung},
  {Chiang}, {Ciprini}, {Claus}, {Cohen-Tanugi}, {Colafrancesco}, {Cominsky},
  {Conrad}, {Costamante}, {Dermer}, {de Angelis}, {de Palma}, {Digel},
  {Donato}, {do Couto e Silva}, {Drell}, {Dubois}, {Dumora}, {Farnier},
  {Favuzzi}, {Finke}, {Focke}, {Frailis}, {Fukazawa}, {Funk}, {Fusco},
  {Gargano}, {Georganopoulos}, {Germani}, {Giebels}, {Giglietto}, {Giordano},
  {Glanzman}, {Grenier}, {Grondin}, {Grove}, {Guillemot}, {Guiriec},
  {Hanabata}, {Harding}, {Hartman}, {Hayashida}, {Hays}, {Hughes},
  {J{\'o}hannesson}, {Johnson}, {Johnson}, {Johnson}, {Kadler}, {Kamae},
  {Kanai}, {Katagiri}, {Kataoka}, {Kawai}, {Kerr}, {Kn{\"o}dlseder}, {Kuehn},
  {Kuss}, {Latronico}, {Lemoine-Goumard}, {Longo}, {Loparco}, {Lott},
  {Lovellette}, {Lubrano}, {Madejski}, {Makeev}, {Mazziotta}, {McEnery},
  {Meurer}, {Michelson}, {Mitthumsiri}, {Mizuno}, {Moiseev}, {Monte},
  {Monzani}, {Morselli}, {Moskalenko}, {Murgia}, {Nakamori}, {Nolan}, {Norris},
  {Nuss}, {Ohsugi}, {Omodei}, {Orlando}, {Ormes}, {Paneque}, {Panetta},
  {Parent}, {Pepe}, {Pesce-Rollins}, {Piron}, {Porter}, {Rain{\`o}}, {Razzano},
  {Reimer}, {Reimer}, {Reposeur}, {Ritz}, {Rodriguez}, {Romani}, {Ryde},
  {Sadrozinski}, {Sambruna}, {Sanchez}, {Sander}, {Sato}, {Parkinson},
  {Sgr{\`o}}, {Smith}, {Smith}, {Spandre}, {Spinelli}, {Starck}, {Strickman},
  {Strong}, {Suson}, {Tajima}, {Takahashi}, {Takahashi}, {Tanaka}, {Taylor},
  {Thayer}, {Thompson}, {Torres}, {Tosti}, {Uchiyama}, {Usher}, {Vilchez},
  {Vitale}, {Waite}, {Wood}, {Ylinen}, {Ziegler}, {Aller}, {Aller},
  {Kellermann}, {Kovalev}, {Kovalev}, {Lister}, \& {Pushkarev}}]{abdo09}
{Abdo}, A.~A., {Ackermann}, M., {Ajello}, M., {Asano}, K., {Baldini}, L.,
  {Ballet}, J., {Barbiellini}, G., {Bastieri}, D., {Baughman}, B.~M.,
  {Bechtol}, K., {Bellazzini}, R., {Blandford}, R.~D., {Bloom}, E.~D.,
  {Bonamente}, E., {Borgland}, A.~W., {Bregeon}, J., {Brez}, A., {Brigida}, M.,
  {Bruel}, P., {Burnett}, T.~H., {Caliandro}, G.~A., {Cameron}, R.~A.,
  {Caraveo}, P.~A., {Casandjian}, J.~M., {Cavazzuti}, E., {Cecchi}, C.,
  {Celotti}, A., {Chekhtman}, A., {Cheung}, C.~C., {Chiang}, J., {Ciprini}, S.,
  {Claus}, R., {Cohen-Tanugi}, J., {Colafrancesco}, S., {Cominsky}, L.~R.,
  {Conrad}, J., {Costamante}, L., {Dermer}, C.~D., {de Angelis}, A., {de
  Palma}, F., {Digel}, S.~W., {Donato}, D., {do Couto e Silva}, E., {Drell},
  P.~S., {Dubois}, R., {Dumora}, D., {Farnier}, C., {Favuzzi}, C., {Finke}, J.,
  {Focke}, W.~B., {Frailis}, M., {Fukazawa}, Y., {Funk}, S., {Fusco}, P.,
  {Gargano}, F., {Georganopoulos}, M., {Germani}, S., {Giebels}, B.,
  {Giglietto}, N., {Giordano}, F., {Glanzman}, T., {Grenier}, I.~A., {Grondin},
  M.-H., {Grove}, J.~E., {Guillemot}, L., {Guiriec}, S., {Hanabata}, Y.,
  {Harding}, A.~K., {Hartman}, R.~C., {Hayashida}, M., {Hays}, E., {Hughes},
  R.~E., {J{\'o}hannesson}, G., {Johnson}, A.~S., {Johnson}, R.~P., {Johnson},
  W.~N., {Kadler}, M., {Kamae}, T., {Kanai}, Y., {Katagiri}, H., {Kataoka}, J.,
  {Kawai}, N., {Kerr}, M., {Kn{\"o}dlseder}, J., {Kuehn}, F., {Kuss}, M.,
  {Latronico}, L., {Lemoine-Goumard}, M., {Longo}, F., {Loparco}, F., {Lott},
  B., {Lovellette}, M.~N., {Lubrano}, P., {Madejski}, G.~M., {Makeev}, A.,
  {Mazziotta}, M.~N., {McEnery}, J.~E., {Meurer}, C., {Michelson}, P.~F.,
  {Mitthumsiri}, W., {Mizuno}, T., {Moiseev}, A.~A., {Monte}, C., {Monzani},
  M.~E., {Morselli}, A., {Moskalenko}, I.~V., {Murgia}, S., {Nakamori}, T.,
  {Nolan}, P.~L., {Norris}, J.~P., {Nuss}, E., {Ohsugi}, T., {Omodei}, N.,
  {Orlando}, E., {Ormes}, J.~F., {Paneque}, D., {Panetta}, J.~H., {Parent}, D.,
  {Pepe}, M., {Pesce-Rollins}, M., {Piron}, F., {Porter}, T.~A., {Rain{\`o}},
  S., {Razzano}, M., {Reimer}, A., {Reimer}, O., {Reposeur}, T., {Ritz}, S.,
  {Rodriguez}, A.~Y., {Romani}, R.~W., {Ryde}, F., {Sadrozinski}, H.~F.-W.,
  {Sambruna}, R., {Sanchez}, D., {Sander}, A., {Sato}, R., {Parkinson},
  P.~M.~S., {Sgr{\`o}}, C., {Smith}, D.~A., {Smith}, P.~D., {Spandre}, G.,
  {Spinelli}, P., {Starck}, J.-L., {Strickman}, M.~S., {Strong}, A.~W.,
  {Suson}, D.~J., {Tajima}, H., {Takahashi}, H., {Takahashi}, T., {Tanaka}, T.,
  {Taylor}, G.~B., {Thayer}, J.~G., {Thompson}, D.~J., {Torres}, D.~F.,
  {Tosti}, G., {Uchiyama}, Y., {Usher}, T.~L., {Vilchez}, N., {Vitale}, V.,
  {Waite}, A.~P., {Wood}, K.~S., {Ylinen}, T., {Ziegler}, M., {Aller}, H.~D.,
  {Aller}, M.~F., {Kellermann}, K.~I., {Kovalev}, Y.~Y., {Kovalev}, Y.~A.,
  {Lister}, M.~L., \& {Pushkarev}, A.~B. 2009, \apj, 699, 31

\bibitem[{{Abdo} {et~al.}(2010{\natexlab{b}}){Abdo}, {Ackermann}, {Ajello},
  {Atwood}, {Axelsson}, {Baldini}, {Ballet}, {Barbiellini}, {Bastieri},
  {Bechtol}, {Bellazzini}, {Berenji}, {Blandford}, {Bloom}, {Bonamente},
  {Borgland}, {Bouvier}, {Bregeon}, {Brez}, {Brigida}, {Bruel}, {Burnett},
  {Buson}, {Caliandro}, {Cameron}, {Caraveo}, {Carrigan}, {Casandjian},
  {Cavazzuti}, {Cecchi}, {{\c C}elik}, {Charles}, {Chekhtman}, {Cheung},
  {Chiang}, {Ciprini}, {Claus}, {Cohen-Tanugi}, {Conrad}, {Cutini}, {Dermer},
  {de Angelis}, {de Palma}, {Digel}, {Silva}, {Drell}, {Dubois}, {Dumora},
  {Farnier}, {Favuzzi}, {Fegan}, {Focke}, {Fortin}, {Frailis}, {Fukazawa},
  {Funk}, {Fusco}, {Gargano}, {Gasparrini}, {Gehrels}, {Germani}, {Giebels},
  {Giglietto}, {Giommi}, {Giordano}, {Glanzman}, {Godfrey}, {Grenier},
  {Grondin}, {Grove}, {Guillemot}, {Guiriec}, {Harding}, {Hartman},
  {Hayashida}, {Hays}, {Healey}, {Horan}, {Hughes}, {Jackson},
  {J{\'o}hannesson}, {Johnson}, {Johnson}, {Kamae}, {Katagiri}, {Kataoka},
  {Kawai}, {Kerr}, {Kn{\"o}dlseder}, {Kuss}, {Lande}, {Latronico},
  {Lemoine-Goumard}, {Longo}, {Loparco}, {Lott}, {Lovellette}, {Lubrano},
  {Madejski}, {Makeev}, {Mazziotta}, {McConville}, {McEnery}, {Meurer},
  {Michelson}, {Mitthumsiri}, {Mizuno}, {Moiseev}, {Monte}, {Monzani},
  {Morselli}, {Moskalenko}, {Murgia}, {Nolan}, {Norris}, {Nuss}, {Ohsugi},
  {Omodei}, {Orlando}, {Ormes}, {Paneque}, {Panetta}, {Parent}, {Pelassa},
  {Pepe}, {Persic}, {Pesce-Rollins}, {Piron}, {Porter}, {Rain{\`o}}, {Rando},
  {Razzano}, {Reimer}, {Reimer}, {Reposeur}, {Ritz}, {Rochester}, {Rodriguez},
  {Romani}, {Roth}, {Ryde}, {Sadrozinski}, {Sanchez}, {Sander}, {Saz
  Parkinson}, {Scargle}, {Sgr{\`o}}, {Siskind}, {Smith}, {Smith}, {Spandre},
  {Spinelli}, {Strickman}, {Suson}, {Tajima}, {Takahashi}, {Takahashi},
  {Tanaka}, {Thayer}, {Thayer}, {Thompson}, {Tibaldo}, {Torres}, {Tosti},
  {Tramacere}, {Uchiyama}, {Usher}, {Vasileiou}, {Vilchez}, {Villata},
  {Vitale}, {Waite}, {Wang}, {Winer}, {Wood}, {Ylinen}, \& {Ziegler}}]{abdo10a}
{Abdo}, A.~A., {Ackermann}, M., {Ajello}, M., {Atwood}, W.~B., {Axelsson}, M.,
  {Baldini}, L., {Ballet}, J., {Barbiellini}, G., {Bastieri}, D., {Bechtol},
  K., {Bellazzini}, R., {Berenji}, B., {Blandford}, R.~D., {Bloom}, E.~D.,
  {Bonamente}, E., {Borgland}, A.~W., {Bouvier}, A., {Bregeon}, J., {Brez}, A.,
  {Brigida}, M., {Bruel}, P., {Burnett}, T.~H., {Buson}, S., {Caliandro},
  G.~A., {Cameron}, R.~A., {Caraveo}, P.~A., {Carrigan}, S., {Casandjian},
  J.~M., {Cavazzuti}, E., {Cecchi}, C., {{\c C}elik}, {\"O}., {Charles}, E.,
  {Chekhtman}, A., {Cheung}, C.~C., {Chiang}, J., {Ciprini}, S., {Claus}, R.,
  {Cohen-Tanugi}, J., {Conrad}, J., {Cutini}, S., {Dermer}, C.~D., {de
  Angelis}, A., {de Palma}, F., {Digel}, S.~W., {Silva}, E.~d.~C.~e., {Drell},
  P.~S., {Dubois}, R., {Dumora}, D., {Farnier}, C., {Favuzzi}, C., {Fegan},
  S.~J., {Focke}, W.~B., {Fortin}, P., {Frailis}, M., {Fukazawa}, Y., {Funk},
  S., {Fusco}, P., {Gargano}, F., {Gasparrini}, D., {Gehrels}, N., {Germani},
  S., {Giebels}, B., {Giglietto}, N., {Giommi}, P., {Giordano}, F., {Glanzman},
  T., {Godfrey}, G., {Grenier}, I.~A., {Grondin}, M.-H., {Grove}, J.~E.,
  {Guillemot}, L., {Guiriec}, S., {Harding}, A.~K., {Hartman}, R.~C.,
  {Hayashida}, M., {Hays}, E., {Healey}, S.~E., {Horan}, D., {Hughes}, R.~E.,
  {Jackson}, M.~S., {J{\'o}hannesson}, G., {Johnson}, A.~S., {Johnson}, W.~N.,
  {Kamae}, T., {Katagiri}, H., {Kataoka}, J., {Kawai}, N., {Kerr}, M.,
  {Kn{\"o}dlseder}, J., {Kuss}, M., {Lande}, J., {Latronico}, L.,
  {Lemoine-Goumard}, M., {Longo}, F., {Loparco}, F., {Lott}, B., {Lovellette},
  M.~N., {Lubrano}, P., {Madejski}, G.~M., {Makeev}, A., {Mazziotta}, M.~N.,
  {McConville}, W., {McEnery}, J.~E., {Meurer}, C., {Michelson}, P.~F.,
  {Mitthumsiri}, W., {Mizuno}, T., {Moiseev}, A.~A., {Monte}, C., {Monzani},
  M.~E., {Morselli}, A., {Moskalenko}, I.~V., {Murgia}, S., {Nolan}, P.~L.,
  {Norris}, J.~P., {Nuss}, E., {Ohsugi}, T., {Omodei}, N., {Orlando}, E.,
  {Ormes}, J.~F., {Paneque}, D., {Panetta}, J.~H., {Parent}, D., {Pelassa}, V.,
  {Pepe}, M., {Persic}, M., {Pesce-Rollins}, M., {Piron}, F., {Porter}, T.~A.,
  {Rain{\`o}}, S., {Rando}, R., {Razzano}, M., {Reimer}, A., {Reimer}, O.,
  {Reposeur}, T., {Ritz}, S., {Rochester}, L.~S., {Rodriguez}, A.~Y., {Romani},
  R.~W., {Roth}, M., {Ryde}, F., {Sadrozinski}, H.~F.-W., {Sanchez}, D.,
  {Sander}, A., {Saz Parkinson}, P.~M., {Scargle}, J.~D., {Sgr{\`o}}, C.,
  {Siskind}, E.~J., {Smith}, D.~A., {Smith}, P.~D., {Spandre}, G., {Spinelli},
  P., {Strickman}, M.~S., {Suson}, D.~J., {Tajima}, H., {Takahashi}, H.,
  {Takahashi}, T., {Tanaka}, T., {Thayer}, J.~B., {Thayer}, J.~G., {Thompson},
  D.~J., {Tibaldo}, L., {Torres}, D.~F., {Tosti}, G., {Tramacere}, A.,
  {Uchiyama}, Y., {Usher}, T.~L., {Vasileiou}, V., {Vilchez}, N., {Villata},
  M., {Vitale}, V., {Waite}, A.~P., {Wang}, P., {Winer}, B.~L., {Wood}, K.~S.,
  {Ylinen}, T., \& {Ziegler}, M. 2010{\natexlab{b}}, \apj, 710, 1271

\bibitem[{{Abdo} {et~al.}(2010{\natexlab{c}}){Abdo}, {Ackermann}, {Ajello},
  {Atwood}, {Axelsson}, {Baldini}, {Ballet}, {Barbiellini}, {Bastieri},
  {Bechtol}, {Bellazzini}, {Berenji}, {Bloom}, {Bonamente}, {Borgland},
  {Bregeon}, {Brez}, {Brigida}, {Bruel}, {Burnett}, {Caliandro}, {Cameron},
  {Caraveo}, {Casandjian}, {Cavazzuti}, {Cecchi}, {{\c C}elik}, {Charles},
  {Chekhtman}, {Cheung}, {Chiang}, {Ciprini}, {Claus}, {Cohen-Tanugi},
  {Conrad}, {Dermer}, {de Angelis}, {de Palma}, {Digel}, {Silva}, {Drell},
  {Drlica-Wagner}, {Dubois}, {Dumora}, {Farnier}, {Favuzzi}, {Fegan}, {Focke},
  {Foschini}, {Frailis}, {Fukazawa}, {Funk}, {Fusco}, {Gargano}, {Gasparrini},
  {Gehrels}, {Germani}, {Giebels}, {Giglietto}, {Giordano}, {Glanzman},
  {Godfrey}, {Grenier}, {Grondin}, {Grove}, {Guillemot}, {Guiriec}, {Hanabata},
  {Harding}, {Hayashida}, {Hays}, {Hughes}, {J{\'o}hannesson}, {Johnson},
  {Johnson}, {Johnson}, {Kamae}, {Katagiri}, {Kataoka}, {Kawai}, {Kerr},
  {Kn{\"o}dlseder}, {Kocian}, {Kuss}, {Lande}, {Latronico}, {Lemoine-Goumard},
  {Longo}, {Loparco}, {Lott}, {Lovellette}, {Lubrano}, {Madejski}, {Makeev},
  {Mazziotta}, {McConville}, {McEnery}, {Meurer}, {Michelson}, {Mitthumsiri},
  {Mizuno}, {Moiseev}, {Monte}, {Monzani}, {Morselli}, {Moskalenko}, {Murgia},
  {Nakamori}, {Nolan}, {Norris}, {Nuss}, {Ohsugi}, {Omodei}, {Orlando},
  {Ormes}, {Ozaki}, {Paneque}, {Panetta}, {Parent}, {Pelassa}, {Pepe},
  {Pesce-Rollins}, {Piron}, {Porter}, {Rain{\`o}}, {Rando}, {Razzano},
  {Reimer}, {Reimer}, {Reposeur}, {Ritz}, {Rodriguez}, {Romani}, {Roth},
  {Ryde}, {Sadrozinski}, {Sander}, {Saz Parkinson}, {Scargle}, {Sellerholm},
  {Sgr{\`o}}, {Shaw}, {Smith}, {Smith}, {Spandre}, {Spinelli}, {Strickman},
  {Strong}, {Suson}, {Takahashi}, {Tanaka}, {Thayer}, {Thayer}, {Thompson},
  {Tibaldo}, {Tibolla}, {Torres}, {Tosti}, {Tramacere}, {Uchiyama}, {Usher},
  {Vasileiou}, {Vilchez}, {Vitale}, {Waite}, {Wang}, {Winer}, {Wood}, {Ylinen},
  {Ziegler}, \& {Fermi LAT Collaboration}}]{abdo10b}
{Abdo}, A.~A., {Ackermann}, M., {Ajello}, M., {Atwood}, W.~B., {Axelsson}, M.,
  {Baldini}, L., {Ballet}, J., {Barbiellini}, G., {Bastieri}, D., {Bechtol},
  K., {Bellazzini}, R., {Berenji}, B., {Bloom}, E.~D., {Bonamente}, E.,
  {Borgland}, A.~W., {Bregeon}, J., {Brez}, A., {Brigida}, M., {Bruel}, P.,
  {Burnett}, T.~H., {Caliandro}, G.~A., {Cameron}, R.~A., {Caraveo}, P.~A.,
  {Casandjian}, J.~M., {Cavazzuti}, E., {Cecchi}, C., {{\c C}elik}, {\"O}.,
  {Charles}, E., {Chekhtman}, A., {Cheung}, C.~C., {Chiang}, J., {Ciprini}, S.,
  {Claus}, R., {Cohen-Tanugi}, J., {Conrad}, J., {Dermer}, C.~D., {de Angelis},
  A., {de Palma}, F., {Digel}, S.~W., {Silva}, E.~d.~C.~e., {Drell}, P.~S.,
  {Drlica-Wagner}, A., {Dubois}, R., {Dumora}, D., {Farnier}, C., {Favuzzi},
  C., {Fegan}, S.~J., {Focke}, W.~B., {Foschini}, L., {Frailis}, M.,
  {Fukazawa}, Y., {Funk}, S., {Fusco}, P., {Gargano}, F., {Gasparrini}, D.,
  {Gehrels}, N., {Germani}, S., {Giebels}, B., {Giglietto}, N., {Giordano}, F.,
  {Glanzman}, T., {Godfrey}, G., {Grenier}, I.~A., {Grondin}, M.-H., {Grove},
  J.~E., {Guillemot}, L., {Guiriec}, S., {Hanabata}, Y., {Harding}, A.~K.,
  {Hayashida}, M., {Hays}, E., {Hughes}, R.~E., {J{\'o}hannesson}, G.,
  {Johnson}, A.~S., {Johnson}, R.~P., {Johnson}, W.~N., {Kamae}, T.,
  {Katagiri}, H., {Kataoka}, J., {Kawai}, N., {Kerr}, M., {Kn{\"o}dlseder}, J.,
  {Kocian}, M.~L., {Kuss}, M., {Lande}, J., {Latronico}, L., {Lemoine-Goumard},
  M., {Longo}, F., {Loparco}, F., {Lott}, B., {Lovellette}, M.~N., {Lubrano},
  P., {Madejski}, G.~M., {Makeev}, A., {Mazziotta}, M.~N., {McConville}, W.,
  {McEnery}, J.~E., {Meurer}, C., {Michelson}, P.~F., {Mitthumsiri}, W.,
  {Mizuno}, T., {Moiseev}, A.~A., {Monte}, C., {Monzani}, M.~E., {Morselli},
  A., {Moskalenko}, I.~V., {Murgia}, S., {Nakamori}, T., {Nolan}, P.~L.,
  {Norris}, J.~P., {Nuss}, E., {Ohsugi}, T., {Omodei}, N., {Orlando}, E.,
  {Ormes}, J.~F., {Ozaki}, M., {Paneque}, D., {Panetta}, J.~H., {Parent}, D.,
  {Pelassa}, V., {Pepe}, M., {Pesce-Rollins}, M., {Piron}, F., {Porter}, T.~A.,
  {Rain{\`o}}, S., {Rando}, R., {Razzano}, M., {Reimer}, A., {Reimer}, O.,
  {Reposeur}, T., {Ritz}, S., {Rodriguez}, A.~Y., {Romani}, R.~W., {Roth}, M.,
  {Ryde}, F., {Sadrozinski}, H.~F.-W., {Sander}, A., {Saz Parkinson}, P.~M.,
  {Scargle}, J.~D., {Sellerholm}, A., {Sgr{\`o}}, C., {Shaw}, M.~S., {Smith},
  D.~A., {Smith}, P.~D., {Spandre}, G., {Spinelli}, P., {Strickman}, M.~S.,
  {Strong}, A.~W., {Suson}, D.~J., {Takahashi}, H., {Tanaka}, T., {Thayer},
  J.~B., {Thayer}, J.~G., {Thompson}, D.~J., {Tibaldo}, L., {Tibolla}, O.,
  {Torres}, D.~F., {Tosti}, G., {Tramacere}, A., {Uchiyama}, Y., {Usher},
  T.~L., {Vasileiou}, V., {Vilchez}, N., {Vitale}, V., {Waite}, A.~P., {Wang},
  P., {Winer}, B.~L., {Wood}, K.~S., {Ylinen}, T., {Ziegler}, M., \& {Fermi LAT
  Collaboration}. 2010{\natexlab{c}}, \apjl, 709, L152

\bibitem[{{Abdo} {et~al.}(2010{\natexlab{d}}){Abdo}, {Ackermann}, {Ajello},
  {Atwood}, {Baldini}, {Ballet}, {Barbiellini}, {Bastieri}, {Baughman},
  {Bechtol}, {Bellazzini}, {Berenji}, {Blandford}, {Bloom}, {Bonamente},
  {Borgland}, {Bouvier}, {Brandt}, {Bregeon}, {Brez}, {Brigida}, {Bruel},
  {Buehler}, {Buson}, {Caliandro}, {Cameron}, {Cannon}, {Caraveo}, {Carrigan},
  {Casandjian}, {Cavazzuti}, {Cecchi}, {{\c C}elik}, {Charles}, {Chekhtman},
  {Cheung}, {Chiang}, {Ciprini}, {Claus}, {Cohen-Tanugi}, {Colafrancesco},
  {Cominsky}, {Conrad}, {Costamante}, {Davis}, {Dermer}, {de Angelis}, {de
  Palma}, {Silva}, {Drell}, {Dubois}, {Dumora}, {Falcone}, {Farnier},
  {Favuzzi}, {Fegan}, {Finke}, {Focke}, {Fortin}, {Frailis}, {Fukazawa},
  {Funk}, {Fusco}, {Gargano}, {Gasparrini}, {Gehrels}, {Georganopoulos},
  {Germani}, {Giebels}, {Giglietto}, {Giommi}, {Giordano}, {Giroletti},
  {Glanzman}, {Godfrey}, {Grandi}, {Grenier}, {Grondin}, {Grove}, {Guillemot},
  {Guiriec}, {Hadasch}, {Harding}, {Hase}, {Hayashida}, {Hays}, {Horan},
  {Hughes}, {Itoh}, {Jackson}, {J{\'o}hannesson}, {Johnson}, {Johnson},
  {Johnson}, {Kadler}, {Kamae}, {Katagiri}, {Kataoka}, {Kawai}, {Kishishita},
  {Kn{\"o}dlseder}, {Kuss}, {Lande}, {Latronico}, {Lee}, {Lemoine-Goumard},
  {Llena Garde}, {Longo}, {Loparco}, {Lott}, {Lovellette}, {Lubrano}, {Makeev},
  {Mazziotta}, {McConville}, {McEnery}, {Michelson}, {Mitthumsiri}, {Mizuno},
  {Moiseev}, {Monte}, {Monzani}, {Morselli}, {Moskalenko}, {Murgia},
  {M{\"u}ller}, {Nakamori}, {Naumann-Godo}, {Nolan}, {Norris}, {Nuss}, {Ohno},
  {Ohsugi}, {Ojha}, {Okumura}, {Omodei}, {Orlando}, {Ormes}, {Ozaki}, {Pagani},
  {Paneque}, {Panetta}, {Parent}, {Pelassa}, {Pepe}, {Pesce-Rollins}, {Piron},
  {Pl{\"o}tz}, {Porter}, {Rain{\`o}}, {Rando}, {Razzano}, {Razzaque}, {Reimer},
  {Reimer}, {Reposeur}, {Ripken}, {Ritz}, {Rodriguez}, {Roth}, {Ryde},
  {Sadrozinski}, {Sanchez}, {Sander}, {Scargle}, {Sgr{\`o}}, {Siskind},
  {Smith}, {Spandre}, {Spinelli}, {Starck}, {Stawarz}, {Strickman}, {Suson},
  {Tajima}, {Takahashi}, {Takahashi}, {Tanaka}, {Thayer}, {Thayer}, {Thompson},
  {Tibaldo}, {Torres}, {Tosti}, {Tramacere}, {Uchiyama}, {Usher},
  {Vandenbroucke}, {Vasileiou}, {Vilchez}, {Vitale}, {Waite}, {Wang}, {Winer},
  {Wood}, {Yang}, {Ylinen}, \& {Ziegler}}]{abdo10}
{Abdo}, A.~A., {Ackermann}, M., {Ajello}, M., {Atwood}, W.~B., {Baldini}, L.,
  {Ballet}, J., {Barbiellini}, G., {Bastieri}, D., {Baughman}, B.~M.,
  {Bechtol}, K., {Bellazzini}, R., {Berenji}, B., {Blandford}, R.~D., {Bloom},
  E.~D., {Bonamente}, E., {Borgland}, A.~W., {Bouvier}, A., {Brandt}, T.~J.,
  {Bregeon}, J., {Brez}, A., {Brigida}, M., {Bruel}, P., {Buehler}, R.,
  {Buson}, S., {Caliandro}, G.~A., {Cameron}, R.~A., {Cannon}, A., {Caraveo},
  P.~A., {Carrigan}, S., {Casandjian}, J.~M., {Cavazzuti}, E., {Cecchi}, C.,
  {{\c C}elik}, {\"O}., {Charles}, E., {Chekhtman}, A., {Cheung}, C.~C.,
  {Chiang}, J., {Ciprini}, S., {Claus}, R., {Cohen-Tanugi}, J.,
  {Colafrancesco}, S., {Cominsky}, L.~R., {Conrad}, J., {Costamante}, L.,
  {Davis}, D.~S., {Dermer}, C.~D., {de Angelis}, A., {de Palma}, F., {Silva},
  E.~d.~C.~e., {Drell}, P.~S., {Dubois}, R., {Dumora}, D., {Falcone}, A.,
  {Farnier}, C., {Favuzzi}, C., {Fegan}, S.~J., {Finke}, J., {Focke}, W.~B.,
  {Fortin}, P., {Frailis}, M., {Fukazawa}, Y., {Funk}, S., {Fusco}, P.,
  {Gargano}, F., {Gasparrini}, D., {Gehrels}, N., {Georganopoulos}, M.,
  {Germani}, S., {Giebels}, B., {Giglietto}, N., {Giommi}, P., {Giordano}, F.,
  {Giroletti}, M., {Glanzman}, T., {Godfrey}, G., {Grandi}, P., {Grenier},
  I.~A., {Grondin}, M.-H., {Grove}, J.~E., {Guillemot}, L., {Guiriec}, S.,
  {Hadasch}, D., {Harding}, A.~K., {Hase}, H., {Hayashida}, M., {Hays}, E.,
  {Horan}, D., {Hughes}, R.~E., {Itoh}, R., {Jackson}, M.~S.,
  {J{\'o}hannesson}, G., {Johnson}, A.~S., {Johnson}, T.~J., {Johnson}, W.~N.,
  {Kadler}, M., {Kamae}, T., {Katagiri}, H., {Kataoka}, J., {Kawai}, N.,
  {Kishishita}, T., {Kn{\"o}dlseder}, J., {Kuss}, M., {Lande}, J., {Latronico},
  L., {Lee}, S.-H., {Lemoine-Goumard}, M., {Llena Garde}, M., {Longo}, F.,
  {Loparco}, F., {Lott}, B., {Lovellette}, M.~N., {Lubrano}, P., {Makeev}, A.,
  {Mazziotta}, M.~N., {McConville}, W., {McEnery}, J.~E., {Michelson}, P.~F.,
  {Mitthumsiri}, W., {Mizuno}, T., {Moiseev}, A.~A., {Monte}, C., {Monzani},
  M.~E., {Morselli}, A., {Moskalenko}, I.~V., {Murgia}, S., {M{\"u}ller}, C.,
  {Nakamori}, T., {Naumann-Godo}, M., {Nolan}, P.~L., {Norris}, J.~P., {Nuss},
  E., {Ohno}, M., {Ohsugi}, T., {Ojha}, R., {Okumura}, A., {Omodei}, N.,
  {Orlando}, E., {Ormes}, J.~F., {Ozaki}, M., {Pagani}, C., {Paneque}, D.,
  {Panetta}, J.~H., {Parent}, D., {Pelassa}, V., {Pepe}, M., {Pesce-Rollins},
  M., {Piron}, F., {Pl{\"o}tz}, C., {Porter}, T.~A., {Rain{\`o}}, S., {Rando},
  R., {Razzano}, M., {Razzaque}, S., {Reimer}, A., {Reimer}, O., {Reposeur},
  T., {Ripken}, J., {Ritz}, S., {Rodriguez}, A.~Y., {Roth}, M., {Ryde}, F.,
  {Sadrozinski}, H.~F.-W., {Sanchez}, D., {Sander}, A., {Scargle}, J.~D.,
  {Sgr{\`o}}, C., {Siskind}, E.~J., {Smith}, P.~D., {Spandre}, G., {Spinelli},
  P., {Starck}, J.-L., {Stawarz}, L., {Strickman}, M.~S., {Suson}, D.~J.,
  {Tajima}, H., {Takahashi}, H., {Takahashi}, T., {Tanaka}, T., {Thayer},
  J.~B., {Thayer}, J.~G., {Thompson}, D.~J., {Tibaldo}, L., {Torres}, D.~F.,
  {Tosti}, G., {Tramacere}, A., {Uchiyama}, Y., {Usher}, T.~L.,
  {Vandenbroucke}, J., {Vasileiou}, V., {Vilchez}, N., {Vitale}, V., {Waite},
  A.~P., {Wang}, P., {Winer}, B.~L., {Wood}, K.~S., {Yang}, Z., {Ylinen}, T.,
  \& {Ziegler}, M. 2010{\natexlab{d}}, \apj, 719, 1433

\bibitem[{{Abdo} {et~al.}(2010{\natexlab{e}}){Abdo}, {Ackermann}, {Ajello},
  {Atwood}, {Baldini}, {Ballet}, {Barbiellini}, {Bastieri}, {Baughman},
  {Bechtol}, {Bellazzini}, {Berenji}, {Blandford}, {Bloom}, {Bonamente},
  {Borgland}, {Bregeon}, {Brez}, {Brigida}, {Bruel}, {Burnett}, {Buson},
  {Caliandro}, {Cameron}, {Caraveo}, {Casandjian}, {Cavazzuti}, {Cecchi}, {{\c
  C}elik}, {Charles}, {Chekhtman}, {Cheung}, {Chiang}, {Ciprini}, {Claus},
  {Cohen-Tanugi}, {Cominsky}, {Conrad}, {Cutini}, {Dermer}, {de Angelis}, {de
  Palma}, {Digel}, {di Bernardo}, {do Couto e Silva}, {Drell}, {Drlica-Wagner},
  {Dubois}, {Dumora}, {Farnier}, {Favuzzi}, {Fegan}, {Focke}, {Fortin},
  {Frailis}, {Fukazawa}, {Funk}, {Fusco}, {Gaggero}, {Gargano}, {Gasparrini},
  {Gehrels}, {Germani}, {Giebels}, {Giglietto}, {Giommi}, {Giordano},
  {Glanzman}, {Godfrey}, {Grenier}, {Grondin}, {Grove}, {Guillemot}, {Guiriec},
  {Gustafsson}, {Hanabata}, {Harding}, {Hayashida}, {Hughes}, {Itoh},
  {Jackson}, {J{\'o}hannesson}, {Johnson}, {Johnson}, {Johnson}, {Johnson},
  {Kamae}, {Katagiri}, {Kataoka}, {Kawai}, {Kerr}, {Kn{\"o}dlseder}, {Kocian},
  {Kuehn}, {Kuss}, {Lande}, {Latronico}, {Lemoine-Goumard}, {Longo}, {Loparco},
  {Lott}, {Lovellette}, {Lubrano}, {Madejski}, {Makeev}, {Mazziotta},
  {McConville}, {McEnery}, {Meurer}, {Michelson}, {Mitthumsiri}, {Mizuno},
  {Moiseev}, {Monte}, {Monzani}, {Morselli}, {Moskalenko}, {Murgia}, {Nolan},
  {Norris}, {Nuss}, {Ohsugi}, {Omodei}, {Orlando}, {Ormes}, {Paneque},
  {Panetta}, {Parent}, {Pelassa}, {Pepe}, {Pesce-Rollins}, {Piron}, {Porter},
  {Rain{\`o}}, {Rando}, {Razzano}, {Reimer}, {Reimer}, {Reposeur}, {Ritz},
  {Rochester}, {Rodriguez}, {Roth}, {Ryde}, {Sadrozinski}, {Sanchez}, {Sander},
  {Parkinson}, {Scargle}, {Sellerholm}, {Sgr{\`o}}, {Shaw}, {Siskind}, {Smith},
  {Smith}, {Spandre}, {Spinelli}, {Starck}, {Strickman}, {Strong}, {Suson},
  {Tajima}, {Takahashi}, {Takahashi}, {Tanaka}, {Thayer}, {Thayer}, {Thompson},
  {Tibaldo}, {Torres}, {Tosti}, {Tramacere}, {Uchiyama}, {Usher}, {Vasileiou},
  {Vilchez}, {Vitale}, {Waite}, {Wang}, {Winer}, {Wood}, {Ylinen}, {Ziegler},
  \& {Fermi LAT Collaboration}}]{abdo10d}
{Abdo}, A.~A., {Ackermann}, M., {Ajello}, M., {Atwood}, W.~B., {Baldini}, L.,
  {Ballet}, J., {Barbiellini}, G., {Bastieri}, D., {Baughman}, B.~M.,
  {Bechtol}, K., {Bellazzini}, R., {Berenji}, B., {Blandford}, R.~D., {Bloom},
  E.~D., {Bonamente}, E., {Borgland}, A.~W., {Bregeon}, J., {Brez}, A.,
  {Brigida}, M., {Bruel}, P., {Burnett}, T.~H., {Buson}, S., {Caliandro},
  G.~A., {Cameron}, R.~A., {Caraveo}, P.~A., {Casandjian}, J.~M., {Cavazzuti},
  E., {Cecchi}, C., {{\c C}elik}, {\"O}., {Charles}, E., {Chekhtman}, A.,
  {Cheung}, C.~C., {Chiang}, J., {Ciprini}, S., {Claus}, R., {Cohen-Tanugi},
  J., {Cominsky}, L.~R., {Conrad}, J., {Cutini}, S., {Dermer}, C.~D., {de
  Angelis}, A., {de Palma}, F., {Digel}, S.~W., {di Bernardo}, G., {do Couto e
  Silva}, E., {Drell}, P.~S., {Drlica-Wagner}, A., {Dubois}, R., {Dumora}, D.,
  {Farnier}, C., {Favuzzi}, C., {Fegan}, S.~J., {Focke}, W.~B., {Fortin}, P.,
  {Frailis}, M., {Fukazawa}, Y., {Funk}, S., {Fusco}, P., {Gaggero}, D.,
  {Gargano}, F., {Gasparrini}, D., {Gehrels}, N., {Germani}, S., {Giebels}, B.,
  {Giglietto}, N., {Giommi}, P., {Giordano}, F., {Glanzman}, T., {Godfrey}, G.,
  {Grenier}, I.~A., {Grondin}, M.-H., {Grove}, J.~E., {Guillemot}, L.,
  {Guiriec}, S., {Gustafsson}, M., {Hanabata}, Y., {Harding}, A.~K.,
  {Hayashida}, M., {Hughes}, R.~E., {Itoh}, R., {Jackson}, M.~S.,
  {J{\'o}hannesson}, G., {Johnson}, A.~S., {Johnson}, R.~P., {Johnson}, T.~J.,
  {Johnson}, W.~N., {Kamae}, T., {Katagiri}, H., {Kataoka}, J., {Kawai}, N.,
  {Kerr}, M., {Kn{\"o}dlseder}, J., {Kocian}, M.~L., {Kuehn}, F., {Kuss}, M.,
  {Lande}, J., {Latronico}, L., {Lemoine-Goumard}, M., {Longo}, F., {Loparco},
  F., {Lott}, B., {Lovellette}, M.~N., {Lubrano}, P., {Madejski}, G.~M.,
  {Makeev}, A., {Mazziotta}, M.~N., {McConville}, W., {McEnery}, J.~E.,
  {Meurer}, C., {Michelson}, P.~F., {Mitthumsiri}, W., {Mizuno}, T., {Moiseev},
  A.~A., {Monte}, C., {Monzani}, M.~E., {Morselli}, A., {Moskalenko}, I.~V.,
  {Murgia}, S., {Nolan}, P.~L., {Norris}, J.~P., {Nuss}, E., {Ohsugi}, T.,
  {Omodei}, N., {Orlando}, E., {Ormes}, J.~F., {Paneque}, D., {Panetta}, J.~H.,
  {Parent}, D., {Pelassa}, V., {Pepe}, M., {Pesce-Rollins}, M., {Piron}, F.,
  {Porter}, T.~A., {Rain{\`o}}, S., {Rando}, R., {Razzano}, M., {Reimer}, A.,
  {Reimer}, O., {Reposeur}, T., {Ritz}, S., {Rochester}, L.~S., {Rodriguez},
  A.~Y., {Roth}, M., {Ryde}, F., {Sadrozinski}, H.~F.-W., {Sanchez}, D.,
  {Sander}, A., {Parkinson}, P.~M.~S., {Scargle}, J.~D., {Sellerholm}, A.,
  {Sgr{\`o}}, C., {Shaw}, M.~S., {Siskind}, E.~J., {Smith}, D.~A., {Smith},
  P.~D., {Spandre}, G., {Spinelli}, P., {Starck}, J.-L., {Strickman}, M.~S.,
  {Strong}, A.~W., {Suson}, D.~J., {Tajima}, H., {Takahashi}, H., {Takahashi},
  T., {Tanaka}, T., {Thayer}, J.~B., {Thayer}, J.~G., {Thompson}, D.~J.,
  {Tibaldo}, L., {Torres}, D.~F., {Tosti}, G., {Tramacere}, A., {Uchiyama}, Y.,
  {Usher}, T.~L., {Vasileiou}, V., {Vilchez}, N., {Vitale}, V., {Waite}, A.~P.,
  {Wang}, P., {Winer}, B.~L., {Wood}, K.~S., {Ylinen}, T., {Ziegler}, M., \&
  {Fermi LAT Collaboration}. 2010{\natexlab{e}}, Physical Review Letters, 104,
  101101

\bibitem[{{Abdo} {et~al.}(2011{\natexlab{b}}){Abdo}, {Ackermann}, {Ajello},
  {Baldini}, {Ballet}, {Barbiellini}, {Bastieri}, {Bechtol}, {Bellazzini},
  {Berenji}, \& et~al.}]{abdo11d}
{Abdo}, A.~A., {Ackermann}, M., {Ajello}, M., {Baldini}, L., {Ballet}, J.,
  {Barbiellini}, G., {Bastieri}, D., {Bechtol}, K., {Bellazzini}, R.,
  {Berenji}, B., \& et~al. 2011{\natexlab{b}}, \apj, 731, 77

\bibitem[{{Abdo} {et~al.}(2011{\natexlab{c}}){Abdo}, {Ackermann}, {Ajello},
  {Baldini}, {Ballet}, {Barbiellini}, {Bastieri}, {Bechtol}, {Bellazzini},
  {Berenji}, \& et~al.}]{abdo11a}
---. 2011{\natexlab{c}}, \apj, 736, 131

\bibitem[{{Abdo} {et~al.}(2011{\natexlab{d}}){Abdo}, {Ackermann}, {Ajello},
  {Baldini}, {Ballet}, {Barbiellini}, {Bastieri}, {Bechtol}, {Bellazzini},
  {Berenji}, \& et~al.}]{abdo11c}
---. 2011{\natexlab{d}}, \apj, 726, 43

\bibitem[{{Abramowski} {et~al.}(2012){Abramowski}, {Acero}, {Aharonian},
  {Akhperjanian}, {Anton}, {Balzer}, {Barnacka}, {Becherini}, {Becker},
  {Bernl{\"o}hr}, {Birsin}, {Biteau}, {Bochow}, {Boisson}, {Bolmont}, {Bordas},
  {Brucker}, {Brun}, {Brun}, {Bulik}, {B{\"u}sching}, {Carrigan}, {Casanova},
  {Cerruti}, {Chadwick}, {Charbonnier}, {Chaves}, {Cheesebrough}, {Cologna},
  {Conrad}, {Couturier}, {Dalton}, {Daniel}, {Davids}, {Degrange}, {Deil},
  {Dickinson}, {Djannati-Ata{\"\i}}, {Domainko}, {O'C.~Drury}, {Dubus},
  {Dutson}, {Dyks}, {Dyrda}, {Egberts}, {Eger}, {Espigat}, {Fallon}, {Fegan},
  {Feinstein}, {Fernandes}, {Fiasson}, {Fontaine}, {F{\"o}rster},
  {F{\"u}{\ss}ling}, {Gajdus}, {Gallant}, {Garrigoux}, {Gast}, {G{\'e}rard},
  {Giebels}, {Glicenstein}, {Gl{\"u}ck}, {G{\"o}ring}, {Grondin},
  {H{\"a}ffner}, {Hague}, {Hahn}, {Hampf}, {Harris}, {Hauser}, {Heinz},
  {Heinzelmann}, {Henri}, {Hermann}, {Hillert}, {Hinton}, {Hofmann},
  {Hofverberg}, {Holler}, {Horns}, {Jacholkowska}, {Jahn}, {Jamrozy}, {Jung},
  {Kastendieck}, {Katarzy{\'n}ski}, {Katz}, {Kaufmann}, {Kh{\'e}lifi},
  {Klochkov}, {Klu{\'z}niak}, {Kneiske}, {Komin}, {Kosack}, {Kossakowski},
  {Krayzel}, {Laffon}, {Lamanna}, {Lenain}, {Lennarz}, {Lohse}, {Lopatin},
  {Lu}, {Marandon}, {Marcowith}, {Masbou}, {Maurin}, {Maxted}, {Mayer},
  {McComb}, {Medina}, {M{\'e}hault}, {Moderski}, {Mohamed}, {Moulin},
  {Naumann}, {Naumann-Godo}, {de Naurois}, {Nedbal}, {Nekrassov}, {Nguyen},
  {Nicholas}, {Niemiec}, {Nolan}, {Ohm}, {de O{\~n}a Wilhelmi}, {Opitz},
  {Ostrowski}, {Oya}, {Panter}, {Paz Arribas}, {Pekeur}, {Pelletier}, {Perez},
  {Petrucci}, {Peyaud}, {Pita}, {P{\"u}hlhofer}, {Punch}, {Quirrenbach},
  {Raue}, {Reimer}, {Reimer}, {Renaud}, {de los Reyes}, {Rieger}, {Ripken},
  {Rob}, {Rosier-Lees}, {Rowell}, {Rudak}, {Rulten}, {Sahakian}, {Sanchez},
  {Santangelo}, {Schlickeiser}, {Schulz}, {Schwanke}, {Schwarzburg},
  {Schwemmer}, {Sheidaei}, {Skilton}, {Sol}, {Spengler}, {Stawarz},
  {Steenkamp}, {Stegmann}, {Stinzing}, {Stycz}, {Sushch}, {Szostek},
  {Tavernet}, {Terrier}, {Tluczykont}, {Valerius}, {van Eldik}, {Vasileiadis},
  {Venter}, {Viana}, {Vincent}, {V{\"o}lk}, {Volpe}, {Vorobiov}, {Vorster},
  {Wagner}, {Ward}, {White}, {Wierzcholska}, {Zacharias}, {Zajczyk},
  {Zdziarski}, {Zech}, \& {Zechlin}}]{abramowski12}
{Abramowski}, A., {Acero}, F., {Aharonian}, F., {Akhperjanian}, A.~G., {Anton},
  G., {Balzer}, A., {Barnacka}, A., {Becherini}, Y., {Becker}, J.,
  {Bernl{\"o}hr}, K., {Birsin}, E., {Biteau}, J., {Bochow}, A., {Boisson}, C.,
  {Bolmont}, J., {Bordas}, P., {Brucker}, J., {Brun}, F., {Brun}, P., {Bulik},
  T., {B{\"u}sching}, I., {Carrigan}, S., {Casanova}, S., {Cerruti}, M.,
  {Chadwick}, P.~M., {Charbonnier}, A., {Chaves}, R.~C.~G., {Cheesebrough}, A.,
  {Cologna}, G., {Conrad}, J., {Couturier}, C., {Dalton}, M., {Daniel}, M.~K.,
  {Davids}, I.~D., {Degrange}, B., {Deil}, C., {Dickinson}, H.~J.,
  {Djannati-Ata{\"\i}}, A., {Domainko}, W., {O'C.~Drury}, L., {Dubus}, G.,
  {Dutson}, K., {Dyks}, J., {Dyrda}, M., {Egberts}, K., {Eger}, P., {Espigat},
  P., {Fallon}, L., {Fegan}, S., {Feinstein}, F., {Fernandes}, M.~V.,
  {Fiasson}, A., {Fontaine}, G., {F{\"o}rster}, A., {F{\"u}{\ss}ling}, M.,
  {Gajdus}, M., {Gallant}, Y.~A., {Garrigoux}, T., {Gast}, H., {G{\'e}rard},
  L., {Giebels}, B., {Glicenstein}, J.~F., {Gl{\"u}ck}, B., {G{\"o}ring}, D.,
  {Grondin}, M.-H., {H{\"a}ffner}, S., {Hague}, J.~D., {Hahn}, J., {Hampf}, D.,
  {Harris}, J., {Hauser}, M., {Heinz}, S., {Heinzelmann}, G., {Henri}, G.,
  {Hermann}, G., {Hillert}, A., {Hinton}, J.~A., {Hofmann}, W., {Hofverberg},
  P., {Holler}, M., {Horns}, D., {Jacholkowska}, A., {Jahn}, C., {Jamrozy}, M.,
  {Jung}, I., {Kastendieck}, M.~A., {Katarzy{\'n}ski}, K., {Katz}, U.,
  {Kaufmann}, S., {Kh{\'e}lifi}, B., {Klochkov}, D., {Klu{\'z}niak}, W.,
  {Kneiske}, T., {Komin}, N., {Kosack}, K., {Kossakowski}, R., {Krayzel}, F.,
  {Laffon}, H., {Lamanna}, G., {Lenain}, J.-P., {Lennarz}, D., {Lohse}, T.,
  {Lopatin}, A., {Lu}, C.-C., {Marandon}, V., {Marcowith}, A., {Masbou}, J.,
  {Maurin}, G., {Maxted}, N., {Mayer}, M., {McComb}, T.~J.~L., {Medina}, M.~C.,
  {M{\'e}hault}, J., {Moderski}, R., {Mohamed}, M., {Moulin}, E., {Naumann},
  C.~L., {Naumann-Godo}, M., {de Naurois}, M., {Nedbal}, D., {Nekrassov}, D.,
  {Nguyen}, N., {Nicholas}, B., {Niemiec}, J., {Nolan}, S.~J., {Ohm}, S., {de
  O{\~n}a Wilhelmi}, E., {Opitz}, B., {Ostrowski}, M., {Oya}, I., {Panter}, M.,
  {Paz Arribas}, M., {Pekeur}, N.~W., {Pelletier}, G., {Perez}, J., {Petrucci},
  P.-O., {Peyaud}, B., {Pita}, S., {P{\"u}hlhofer}, G., {Punch}, M.,
  {Quirrenbach}, A., {Raue}, M., {Reimer}, A., {Reimer}, O., {Renaud}, M., {de
  los Reyes}, R., {Rieger}, F., {Ripken}, J., {Rob}, L., {Rosier-Lees}, S.,
  {Rowell}, G., {Rudak}, B., {Rulten}, C.~B., {Sahakian}, V., {Sanchez}, D.~A.,
  {Santangelo}, A., {Schlickeiser}, R., {Schulz}, A., {Schwanke}, U.,
  {Schwarzburg}, S., {Schwemmer}, S., {Sheidaei}, F., {Skilton}, J.~L., {Sol},
  H., {Spengler}, G., {Stawarz}, {\L}., {Steenkamp}, R., {Stegmann}, C.,
  {Stinzing}, F., {Stycz}, K., {Sushch}, I., {Szostek}, A., {Tavernet}, J.-P.,
  {Terrier}, R., {Tluczykont}, M., {Valerius}, K., {van Eldik}, C.,
  {Vasileiadis}, G., {Venter}, C., {Viana}, A., {Vincent}, P., {V{\"o}lk},
  H.~J., {Volpe}, F., {Vorobiov}, S., {Vorster}, M., {Wagner}, S.~J., {Ward},
  M., {White}, R., {Wierzcholska}, A., {Zacharias}, M., {Zajczyk}, A.,
  {Zdziarski}, A.~A., {Zech}, A., \& {Zechlin}, H.-S. 2012, ArXiv e-prints

\bibitem[{{Abramowski} {et~al.}(2011){Abramowski}, {Acero}, {Aharonian},
  {Akhperjanian}, {Anton}, {Barnacka}, {Barres de Almeida}, {Bazer-Bachi},
  {Becherini}, \& et~al.}]{abramowski11}
{Abramowski}, A., {Acero}, F., {Aharonian}, F., {Akhperjanian}, A.~G., {Anton},
  G., {Barnacka}, A., {Barres de Almeida}, U., {Bazer-Bachi}, A.~R.,
  {Becherini}, Y., \& et~al. 2011, \aap, 533, A110

\bibitem[{{Acciari} {et~al.}(2009{\natexlab{a}}){Acciari}, {Aliu}, {Arlen},
  {Bautista}, {Beilicke}, {Benbow}, {B{\"o}ttcher}, {Bradbury}, {Buckley},
  {Bugaev}, {Butt}, {Byrum}, {Cannon}, {Celik}, {Cesarini}, {Chow}, {Ciupik},
  {Cogan}, {Colin}, {Cui}, {Dickherber}, {Duke}, {Ergin}, {Falcone}, {Fegan},
  {Finley}, {Finnegan}, {Fortin}, {Fortson}, {Furniss}, {Gall}, {Gibbs},
  {Gillanders}, {Grube}, {Guenette}, {Gyuk}, {Hanna}, {Hays}, {Holder},
  {Horan}, {Hui}, {Humensky}, {Imran}, {Kaaret}, {Karlsson}, {Kertzman},
  {Kieda}, {Kildea}, {Konopelko}, {Krawczynski}, {Krennrich}, {Lang},
  {LeBohec}, {Maier}, {McCann}, {McCutcheon}, {Millis}, {Moriarty},
  {Mukherjee}, {Nagai}, {Ong}, {Otte}, {Pandel}, {Perkins}, {Petry}, {Pohl},
  {Quinn}, {Ragan}, {Reyes}, {Reynolds}, {Roache}, {Rose}, {Schroedter},
  {Sembroski}, {Smith}, {Steele}, {Swordy}, {Theiling}, {Toner}, {Valcarcel},
  {Varlotta}, {Vassiliev}, {Wagner}, {Wakely}, {Ward}, {Weekes}, {Weinstein},
  {White}, {Williams}, {Wissel}, {Wood}, \& {Zitzer}}]{acciari09b}
{Acciari}, V., {Aliu}, E., {Arlen}, T., {Bautista}, M., {Beilicke}, M.,
  {Benbow}, W., {B{\"o}ttcher}, M., {Bradbury}, S.~M., {Buckley}, J.~H.,
  {Bugaev}, V., {Butt}, Y., {Byrum}, K., {Cannon}, A., {Celik}, O., {Cesarini},
  A., {Chow}, Y.~C., {Ciupik}, L., {Cogan}, P., {Colin}, P., {Cui}, W.,
  {Dickherber}, R., {Duke}, C., {Ergin}, T., {Falcone}, A., {Fegan}, S.~J.,
  {Finley}, J.~P., {Finnegan}, G., {Fortin}, P., {Fortson}, L., {Furniss}, A.,
  {Gall}, D., {Gibbs}, K., {Gillanders}, G.~H., {Grube}, J., {Guenette}, R.,
  {Gyuk}, G., {Hanna}, D., {Hays}, E., {Holder}, J., {Horan}, D., {Hui}, C.~M.,
  {Humensky}, T.~B., {Imran}, A., {Kaaret}, P., {Karlsson}, N., {Kertzman}, M.,
  {Kieda}, D., {Kildea}, J., {Konopelko}, A., {Krawczynski}, H., {Krennrich},
  F., {Lang}, M.~J., {LeBohec}, S., {Maier}, G., {McCann}, A., {McCutcheon},
  M., {Millis}, J., {Moriarty}, P., {Mukherjee}, R., {Nagai}, T., {Ong}, R.~A.,
  {Otte}, A.~N., {Pandel}, D., {Perkins}, J.~S., {Petry}, D., {Pohl}, M.,
  {Quinn}, J., {Ragan}, K., {Reyes}, L.~C., {Reynolds}, P.~T., {Roache}, E.,
  {Rose}, J., {Schroedter}, M., {Sembroski}, G.~H., {Smith}, A.~W., {Steele},
  D., {Swordy}, S.~P., {Theiling}, M., {Toner}, J.~A., {Valcarcel}, L.,
  {Varlotta}, A., {Vassiliev}, V.~V., {Wagner}, R.~G., {Wakely}, S.~P., {Ward},
  J.~E., {Weekes}, T.~C., {Weinstein}, A., {White}, R.~J., {Williams}, D.~A.,
  {Wissel}, S., {Wood}, M., \& {Zitzer}, B. 2009{\natexlab{a}}, \apjl, 690,
  L126

\bibitem[{{Acciari} {et~al.}(2009{\natexlab{b}}){Acciari}, {Aliu}, {Arlen},
  {Aune}, {Bautista}, {Beilicke}, {Benbow}, {Boltuch}, {Bradbury}, {Buckley},
  {Bugaev}, {Byrum}, {Cannon}, {Celik}, {Cesarini}, {Chow}, {Ciupik}, {Cogan},
  {Colin}, {Cui}, {Dickherber}, {Duke}, {Fegan}, {Finley}, {Finnegan},
  {Fortin}, {Fortson}, {Furniss}, {Galante}, {Gall}, {Gibbs}, {Gillanders},
  {Godambe}, {Grube}, {Guenette}, {Gyuk}, {Hanna}, {Holder}, {Horan}, {Hui},
  {Humensky}, {Imran}, {Kaaret}, {Karlsson}, {Kertzman}, {Kieda}, {Kildea},
  {Konopelko}, {Krawczynski}, {Krennrich}, {Lang}, {Lebohec}, {Maier},
  {McArthur}, {McCann}, {McCutcheon}, {Millis}, {Moriarty}, {Mukherjee},
  {Nagai}, {Ong}, {Otte}, {Pandel}, {Perkins}, {Pizlo}, {Pohl}, {Quinn},
  {Ragan}, {Reyes}, {Reynolds}, {Roache}, {Rose}, {Schroedter}, {Sembroski},
  {Smith}, {Steele}, {Swordy}, {Theiling}, {Thibadeau}, {Varlotta},
  {Vassiliev}, {Vincent}, {Wagner}, {Wakely}, {Ward}, {Weekes}, {Weinstein},
  {Weisgarber}, {Williams}, {Wissel}, {Wood}, \& {Zitzer}}]{acciari09d}
{Acciari}, V.~A., {Aliu}, E., {Arlen}, T., {Aune}, T., {Bautista}, M.,
  {Beilicke}, M., {Benbow}, W., {Boltuch}, D., {Bradbury}, S.~M., {Buckley},
  J.~H., {Bugaev}, V., {Byrum}, K., {Cannon}, A., {Celik}, O., {Cesarini}, A.,
  {Chow}, Y.~C., {Ciupik}, L., {Cogan}, P., {Colin}, P., {Cui}, W.,
  {Dickherber}, R., {Duke}, C., {Fegan}, S.~J., {Finley}, J.~P., {Finnegan},
  G., {Fortin}, P., {Fortson}, L., {Furniss}, A., {Galante}, N., {Gall}, D.,
  {Gibbs}, K., {Gillanders}, G.~H., {Godambe}, S., {Grube}, J., {Guenette}, R.,
  {Gyuk}, G., {Hanna}, D., {Holder}, J., {Horan}, D., {Hui}, C.~M., {Humensky},
  T.~B., {Imran}, A., {Kaaret}, P., {Karlsson}, N., {Kertzman}, M., {Kieda},
  D., {Kildea}, J., {Konopelko}, A., {Krawczynski}, H., {Krennrich}, F.,
  {Lang}, M.~J., {Lebohec}, S., {Maier}, G., {McArthur}, S., {McCann}, A.,
  {McCutcheon}, M., {Millis}, J., {Moriarty}, P., {Mukherjee}, R., {Nagai}, T.,
  {Ong}, R.~A., {Otte}, A.~N., {Pandel}, D., {Perkins}, J.~S., {Pizlo}, F.,
  {Pohl}, M., {Quinn}, J., {Ragan}, K., {Reyes}, L.~C., {Reynolds}, P.~T.,
  {Roache}, E., {Rose}, H.~J., {Schroedter}, M., {Sembroski}, G.~H., {Smith},
  A.~W., {Steele}, D., {Swordy}, S.~P., {Theiling}, M., {Thibadeau}, S.,
  {Varlotta}, A., {Vassiliev}, V.~V., {Vincent}, S., {Wagner}, R.~G., {Wakely},
  S.~P., {Ward}, J.~E., {Weekes}, T.~C., {Weinstein}, A., {Weisgarber}, T.,
  {Williams}, D.~A., {Wissel}, S., {Wood}, M., \& {Zitzer}, B.
  2009{\natexlab{b}}, \nat, 462, 770

\bibitem[{{Acciari} {et~al.}(2010{\natexlab{a}}){Acciari}, {Aliu}, {Arlen},
  {Aune}, {Bautista}, {Beilicke}, {Benbow}, {B{\"o}ttcher}, {Boltuch},
  {Bradbury}, \& et~al.}]{acciari10b}
{Acciari}, V.~A., {Aliu}, E., {Arlen}, T., {Aune}, T., {Bautista}, M.,
  {Beilicke}, M., {Benbow}, W., {B{\"o}ttcher}, M., {Boltuch}, D., {Bradbury},
  S.~M., \& et~al. 2010{\natexlab{a}}, \apjl, 715, L49

\bibitem[{{Acciari} {et~al.}(2010{\natexlab{b}}){Acciari}, {Aliu}, {Arlen},
  {Aune}, {Beilicke}, {Benbow}, {Boltuch}, {Bradbury}, {Buckley}, {Bugaev},
  {Byrum}, {Cannon}, {Cesarini}, {Chow}, {Ciupik}, {Cogan}, {Cui},
  {Dickherber}, {Duke}, {Finley}, {Finnegan}, {Fortin}, {Fortson}, {Furniss},
  {Galante}, {Gall}, {Gillanders}, {Godambe}, {Grube}, {Guenette}, {Gyuk},
  {Hanna}, {Holder}, {Hui}, {Humensky}, {Imran}, {Kaaret}, {Karlsson},
  {Kertzman}, {Kieda}, {Konopelko}, {Krawczynski}, {Krennrich}, {Lang},
  {LeBohec}, {Maier}, {McArthur}, {McCann}, {McCutcheon}, {Millis}, {Moriarty},
  {Ong}, {Otte}, {Pandel}, {Perkins}, {Pichel}, {Pohl}, {Quinn}, {Ragan},
  {Reyes}, {Reynolds}, {Roache}, {Rose}, {Rovero}, {Schroedter}, {Sembroski},
  {Senturk}, {Smith}, {Steele}, {Swordy}, {Theiling}, {Thibadeau}, {Varlotta},
  {Vincent}, {Wagner}, {Wakely}, {Ward}, {Weekes}, {Weinstein}, {Weisgarber},
  {Williams}, {Wissel}, {Wood}, {Zitzer}, {Harris}, \& {Massaro}}]{acciari10a}
{Acciari}, V.~A., {Aliu}, E., {Arlen}, T., {Aune}, T., {Beilicke}, M.,
  {Benbow}, W., {Boltuch}, D., {Bradbury}, S.~M., {Buckley}, J.~H., {Bugaev},
  V., {Byrum}, K., {Cannon}, A., {Cesarini}, A., {Chow}, Y.~C., {Ciupik}, L.,
  {Cogan}, P., {Cui}, W., {Dickherber}, R., {Duke}, C., {Finley}, J.~P.,
  {Finnegan}, G., {Fortin}, P., {Fortson}, L., {Furniss}, A., {Galante}, N.,
  {Gall}, D., {Gillanders}, G.~H., {Godambe}, S., {Grube}, J., {Guenette}, R.,
  {Gyuk}, G., {Hanna}, D., {Holder}, J., {Hui}, C.~M., {Humensky}, T.~B.,
  {Imran}, A., {Kaaret}, P., {Karlsson}, N., {Kertzman}, M., {Kieda}, D.,
  {Konopelko}, A., {Krawczynski}, H., {Krennrich}, F., {Lang}, M.~J.,
  {LeBohec}, S., {Maier}, G., {McArthur}, S., {McCann}, A., {McCutcheon}, M.,
  {Millis}, J., {Moriarty}, P., {Ong}, R.~A., {Otte}, A.~N., {Pandel}, D.,
  {Perkins}, J.~S., {Pichel}, A., {Pohl}, M., {Quinn}, J., {Ragan}, K.,
  {Reyes}, L.~C., {Reynolds}, P.~T., {Roache}, E., {Rose}, H.~J., {Rovero},
  A.~C., {Schroedter}, M., {Sembroski}, G.~H., {Senturk}, G.~D., {Smith},
  A.~W., {Steele}, D., {Swordy}, S.~P., {Theiling}, M., {Thibadeau}, S.,
  {Varlotta}, A., {Vincent}, S., {Wagner}, R.~G., {Wakely}, S.~P., {Ward},
  J.~E., {Weekes}, T.~C., {Weinstein}, A., {Weisgarber}, T., {Williams}, D.~A.,
  {Wissel}, S., {Wood}, M., {Zitzer}, B., {Harris}, D.~E., \& {Massaro}, F.
  2010{\natexlab{b}}, \apj, 716, 819

\bibitem[{{Acciari} {et~al.}(2011{\natexlab{a}}){Acciari}, {Aliu}, {Arlen},
  {Aune}, {Beilicke}, {Benbow}, {Boltuch}, {Bradbury}, {Buckley}, {Bugaev},
  {Byrum}, {Cannon}, {Cesarini}, {Ciupik}, {Cui}, {Dickherber}, {Duke},
  {Falcone}, {Finley}, {Finnegan}, {Fortson}, {Furniss}, {Galante}, {Gall},
  {Gillanders}, {Godambe}, {Grube}, {Guenette}, {Gyuk}, {Hanna}, {Holder},
  {Hui}, {Humensky}, {Imran}, {Kaaret}, {Karlsson}, {Kertzman}, {Kieda},
  {Konopelko}, {Krawczynski}, {Krennrich}, {Lang}, {Maier}, {McArthur},
  {McCutcheon}, {Moriarty}, {Ong}, {Otte}, {Ouellette}, {Pandel}, {Perkins},
  {Pichel}, {Pohl}, {Quinn}, {Ragan}, {Reyes}, {Reynolds}, {Roache}, {Rose},
  {Rovero}, {Schroedter}, {Sembroski}, {Senturk}, {Steele}, {Swordy},
  {Theiling}, {Thibadeau}, {Varlotta}, {Vassiliev}, {Vincent}, {Wagner},
  {Wakely}, {Ward}, {Weekes}, {Weinstein}, {Weisgarber}, {Williams}, {Wissel},
  {Wood}, {Zitzer}, {Garson}, {Lee}, {Sadun}, {Carini}, {Barnaby}, {Cook},
  {Maune}, {Pease}, {Smith}, {Walters}, {Berdyugin}, {Lindfors}, {Nilsson},
  {Pasanen}, {Sainio}, {Sillanpaa}, {Takalo}, {Villforth}, {Montaruli},
  {Baker}, {Lahteenmaki}, {Tornikoski}, {Hovatta}, {Nieppola}, {Aller}, \&
  {Aller}}]{acciari11b}
{Acciari}, V.~A., {Aliu}, E., {Arlen}, T., {Aune}, T., {Beilicke}, M.,
  {Benbow}, W., {Boltuch}, D., {Bradbury}, S.~M., {Buckley}, J.~H., {Bugaev},
  V., {Byrum}, K., {Cannon}, A., {Cesarini}, A., {Ciupik}, L., {Cui}, W.,
  {Dickherber}, R., {Duke}, C., {Falcone}, A., {Finley}, J.~P., {Finnegan}, G.,
  {Fortson}, L., {Furniss}, A., {Galante}, N., {Gall}, D., {Gillanders}, G.~H.,
  {Godambe}, S., {Grube}, J., {Guenette}, R., {Gyuk}, G., {Hanna}, D.,
  {Holder}, J., {Hui}, C.~M., {Humensky}, T.~B., {Imran}, A., {Kaaret}, P.,
  {Karlsson}, N., {Kertzman}, M., {Kieda}, D., {Konopelko}, A., {Krawczynski},
  H., {Krennrich}, F., {Lang}, M.~J., {Maier}, G., {McArthur}, S.,
  {McCutcheon}, M., {Moriarty}, P., {Ong}, R.~A., {Otte}, A.~N., {Ouellette},
  M., {Pandel}, D., {Perkins}, J.~S., {Pichel}, A., {Pohl}, M., {Quinn}, J.,
  {Ragan}, K., {Reyes}, L.~C., {Reynolds}, P.~T., {Roache}, E., {Rose}, H.~J.,
  {Rovero}, A.~C., {Schroedter}, M., {Sembroski}, G.~H., {Senturk}, G.~D.,
  {Steele}, D., {Swordy}, S.~P., {Theiling}, M., {Thibadeau}, S., {Varlotta},
  A., {Vassiliev}, V.~V., {Vincent}, S., {Wagner}, R.~G., {Wakely}, S.~P.,
  {Ward}, J.~E., {Weekes}, T.~C., {Weinstein}, A., {Weisgarber}, T.,
  {Williams}, D.~A., {Wissel}, S., {Wood}, M., {Zitzer}, B., {Garson}, III, A.,
  {Lee}, K., {Sadun}, A.~C., {Carini}, M., {Barnaby}, D., {Cook}, K., {Maune},
  J., {Pease}, A., {Smith}, S., {Walters}, R., {Berdyugin}, A., {Lindfors}, E.,
  {Nilsson}, K., {Pasanen}, M., {Sainio}, J., {Sillanpaa}, A., {Takalo}, L.~O.,
  {Villforth}, C., {Montaruli}, T., {Baker}, M., {Lahteenmaki}, A.,
  {Tornikoski}, M., {Hovatta}, T., {Nieppola}, E., {Aller}, H.~D., \& {Aller},
  M.~F. 2011{\natexlab{a}}, \apj, 738, 25

\bibitem[{{Acciari} {et~al.}(2011{\natexlab{b}}){Acciari}, {Aliu}, {Arlen},
  {Aune}, {Beilicke}, {Benbow}, {Boltuch}, {Bugaev}, {Cannon}, {Ciupik},
  {Cogan}, {Colin}, {Dickherber}, {Falcone}, {Fegan}, {Finley}, {Fortin},
  {Fortson}, {Furniss}, {Gall}, {Gillanders}, {Grube}, {Guenette}, {Gyuk},
  {Hanna}, {Holder}, {Horan}, {Hui}, {Humensky}, {Imran}, {Kaaret}, {Karlsson},
  {Kertzman}, {Kieda}, {Kildea}, {Konopelko}, {Krawczynski}, {Krennrich},
  {Lang}, {LeBohec}, {Maier}, {Moriarty}, {Mukherjee}, {Ong}, {Otte}, {Pandel},
  {Perkins}, {Pichel}, {Pohl}, {Quinn}, {Ragan}, {Reynolds}, {Rose},
  {Schroedter}, {Sembroski}, {Smith}, {Steele}, {Swordy}, {Theiling}, {Toner},
  {Varlotta}, {Vassiliev}, {Vincent}, {Wagner}, {Wakely}, {Ward}, {Weekes},
  {Weinstein}, {Weisgarber}, {Williams}, {Wissel}, {Wood}, \&
  {Zitzer}}]{acciari11}
{Acciari}, V.~A., {Aliu}, E., {Arlen}, T., {Aune}, T., {Beilicke}, M.,
  {Benbow}, W., {Boltuch}, D., {Bugaev}, V., {Cannon}, A., {Ciupik}, L.,
  {Cogan}, P., {Colin}, P., {Dickherber}, R., {Falcone}, A., {Fegan}, S.~J.,
  {Finley}, J.~P., {Fortin}, P., {Fortson}, L.~F., {Furniss}, A., {Gall}, D.,
  {Gillanders}, G.~H., {Grube}, J., {Guenette}, R., {Gyuk}, G., {Hanna}, D.,
  {Holder}, J., {Horan}, D., {Hui}, C.~M., {Humensky}, T.~B., {Imran}, A.,
  {Kaaret}, P., {Karlsson}, N., {Kertzman}, M., {Kieda}, D., {Kildea}, J.,
  {Konopelko}, A., {Krawczynski}, H., {Krennrich}, F., {Lang}, M.~J.,
  {LeBohec}, S., {Maier}, G., {Moriarty}, P., {Mukherjee}, R., {Ong}, R.~A.,
  {Otte}, A.~N., {Pandel}, D., {Perkins}, J.~S., {Pichel}, A., {Pohl}, M.,
  {Quinn}, J., {Ragan}, K., {Reynolds}, P.~T., {Rose}, H.~J., {Schroedter}, M.,
  {Sembroski}, G.~H., {Smith}, A.~W., {Steele}, D., {Swordy}, S.~P.,
  {Theiling}, M., {Toner}, J.~A., {Varlotta}, A., {Vassiliev}, V.~V.,
  {Vincent}, S., {Wagner}, R., {Wakely}, S.~P., {Ward}, J.~E., {Weekes}, T.~C.,
  {Weinstein}, A., {Weisgarber}, T., {Williams}, D.~A., {Wissel}, S., {Wood},
  M., \& {Zitzer}, B. 2011{\natexlab{b}}, \apj, 738, 169

\bibitem[{{Acciari} {et~al.}(2009{\natexlab{c}}){Acciari}, {Aliu}, {Arlen},
  {Bautista}, {Beilicke}, {Benbow}, {Bradbury}, {Buckley}, {Bugaev}, {Butt}, \&
  et~al.}]{acciari09a}
{Acciari}, V.~A., {Aliu}, E., {Arlen}, T., {Bautista}, M., {Beilicke}, M.,
  {Benbow}, W., {Bradbury}, S.~M., {Buckley}, J.~H., {Bugaev}, V., {Butt}, Y.,
  \& et~al. 2009{\natexlab{c}}, Science, 325, 444

\bibitem[{{Acciari} {et~al.}(2009{\natexlab{d}}){Acciari}, {Aliu}, {Arlen},
  {Beilicke}, {Benbow}, {B{\"o}ttcher}, {Bradbury}, {Buckley}, {Bugaev},
  {Butt}, {Byrum}, {Cannon}, {Celik}, {Cesarini}, {Chow}, {Ciupik}, {Cogan},
  {Cui}, {Daniel}, {Dickherber}, {Ergin}, {Falcone}, {Fegan}, {Finley},
  {Fortin}, {Fortson}, {Furniss}, {Gall}, {Gibbs}, {Gillanders}, {Godambe},
  {Grube}, {Guenette}, {Gyuk}, {Hanna}, {Hays}, {Holder}, {Horan}, {Hui},
  {Humensky}, {Imran}, {Kaaret}, {Karlsson}, {Kertzman}, {Kieda}, {Kildea},
  {Konopelko}, {Krawczynski}, {Krennrich}, {Lang}, {LeBohec}, {Maier},
  {McCann}, {McCutcheon}, {Millis}, {Moriarty}, {Mukherjee}, {Nagai}, {Ong},
  {Otte}, {Pandel}, {Perkins}, {Petry}, {Pizlo}, {Pohl}, {Quinn}, {Ragan},
  {Reyes}, {Reynolds}, {Roache}, {Rose}, {Schroedter}, {Sembroski}, {Smith},
  {Steele}, {Swordy}, {Theiling}, {Toner}, {Varlotta}, {Vassiliev}, {Wagner},
  {Wakely}, {Ward}, {Weekes}, {Weinstein}, {Williams}, {Wissel}, {Wood}, \&
  {Zitzer}}]{acciari09c}
{Acciari}, V.~A., {Aliu}, E., {Arlen}, T., {Beilicke}, M., {Benbow}, W.,
  {B{\"o}ttcher}, M., {Bradbury}, S.~M., {Buckley}, J.~H., {Bugaev}, V.,
  {Butt}, Y., {Byrum}, K., {Cannon}, A., {Celik}, O., {Cesarini}, A., {Chow},
  Y.~C., {Ciupik}, L., {Cogan}, P., {Cui}, W., {Daniel}, M.~K., {Dickherber},
  R., {Ergin}, T., {Falcone}, A., {Fegan}, S.~J., {Finley}, J.~P., {Fortin},
  P., {Fortson}, L., {Furniss}, A., {Gall}, D., {Gibbs}, K., {Gillanders},
  G.~H., {Godambe}, S., {Grube}, J., {Guenette}, R., {Gyuk}, G., {Hanna}, D.,
  {Hays}, E., {Holder}, J., {Horan}, D., {Hui}, C.~M., {Humensky}, T.~B.,
  {Imran}, A., {Kaaret}, P., {Karlsson}, N., {Kertzman}, M., {Kieda}, D.,
  {Kildea}, J., {Konopelko}, A., {Krawczynski}, H., {Krennrich}, F., {Lang},
  M.~J., {LeBohec}, S., {Maier}, G., {McCann}, A., {McCutcheon}, M., {Millis},
  J., {Moriarty}, P., {Mukherjee}, R., {Nagai}, T., {Ong}, R.~A., {Otte},
  A.~N., {Pandel}, D., {Perkins}, J.~S., {Petry}, D., {Pizlo}, F., {Pohl}, M.,
  {Quinn}, J., {Ragan}, K., {Reyes}, L.~C., {Reynolds}, P.~T., {Roache}, E.,
  {Rose}, H.~J., {Schroedter}, M., {Sembroski}, G.~H., {Smith}, A.~W.,
  {Steele}, D., {Swordy}, S.~P., {Theiling}, M., {Toner}, J.~A., {Varlotta},
  A., {Vassiliev}, V.~V., {Wagner}, R.~G., {Wakely}, S.~P., {Ward}, J.~E.,
  {Weekes}, T.~C., {Weinstein}, A., {Williams}, D.~A., {Wissel}, S., {Wood},
  M., \& {Zitzer}, B. 2009{\natexlab{d}}, \apjl, 693, L104

\bibitem[{{Acciari} {et~al.}(2009{\natexlab{e}}){Acciari}, {Aliu}, {Aune},
  {Beilicke}, {Benbow}, {B{\"o}ttcher}, {Boltuch}, {Buckley}, {Bradbury},
  {Bugaev}, {Byrum}, {Cannon}, {Cesarini}, {Ciupik}, {Cogan}, {Cui},
  {Dickherber}, {Duke}, {Falcone}, {Finley}, {Fortin}, {Fortson}, {Furniss},
  {Galante}, {Gall}, {Gibbs}, {Gillanders}, {Grube}, {Guenette}, {Gyuk},
  {Hanna}, {Holder}, {Hui}, {Humensky}, {Kaaret}, {Karlsson}, {Kertzman},
  {Kieda}, {Konopelko}, {Krawczynski}, {Krennrich}, {Lang}, {Le Bohec},
  {Maier}, {McArthur}, {McCann}, {McCutcheon}, {Millis}, {Moriarty}, {Ong},
  {Otte}, {Pandel}, {Perkins}, {Pichel}, {Pohl}, {Quinn}, {Ragan}, {Reyes},
  {Reynolds}, {Roache}, {Rose}, {Sembroski}, {Smith}, {Steele}, {Theiling},
  {Thibadeau}, {Varlotta}, {Vassiliev}, {Vincent}, {Wakely}, {Ward}, {Weekes},
  {Weinstein}, {Weisgarber}, {Williams}, {Wissel}, {Wood}, {Pian},
  {Vercellone}, {Donnarumma}, {D'Ammando}, {Bulgarelli}, {Chen}, {Giuliani},
  {Longo}, {Pacciani}, {Pucella}, {Vittorini}, {Tavani}, {Argan},
  {Barbiellini}, {Caraveo}, {Cattaneo}, {Cocco}, {Costa}, {Del Monte}, {De
  Paris}, {Di Cocco}, {Evangelista}, {Feroci}, {Fiorini}, {Froysland},
  {Frutti}, {Fuschino}, {Galli}, {Gianotti}, {Labanti}, {Lapshov},
  {Lazzarotto}, {Lipari}, {Marisaldi}, {Mastropietro}, {Mereghetti}, {Morelli},
  {Morselli}, {Pellizzoni}, {Perotti}, {Piano}, {Picozza}, {Pilia},
  {Porrovecchio}, {Prest}, {Rapisarda}, {Rappoldi}, {Rubini}, {Sabatini},
  {Soffitta}, {Trifoglio}, {Trois}, {Vallazza}, {Zambra}, {Zanello}, {Pittori},
  {Santolamazza}, {Verrecchia}, {Giommi}, {Colafrancesco}, {Salotti},
  {Villata}, {Raiteri}, {Aller}, {Aller}, {Arkharov}, {Efimova}, {Larionov},
  {Leto}, {Ligustri}, {Lindfors}, {Pasanen}, {Kurtanidze}, {Tetradze},
  {Lahteenmaki}, {Kotiranta}, {Cucchiara}, {Romano}, {Nesci}, {Pursimo},
  {Heidt}, {Benitez}, {Hiriart}, {Nilsson}, {Berdyugin}, {Mujica}, {Dultzin},
  {Lopez}, {Mommert}, {Sorcia}, \& {de la Calle Perez}}]{acciari09e}
{Acciari}, V.~A., {Aliu}, E., {Aune}, T., {Beilicke}, M., {Benbow}, W.,
  {B{\"o}ttcher}, M., {Boltuch}, D., {Buckley}, J.~H., {Bradbury}, S.~M.,
  {Bugaev}, V., {Byrum}, K., {Cannon}, A., {Cesarini}, A., {Ciupik}, L.,
  {Cogan}, P., {Cui}, W., {Dickherber}, R., {Duke}, C., {Falcone}, A.,
  {Finley}, J.~P., {Fortin}, P., {Fortson}, L., {Furniss}, A., {Galante}, N.,
  {Gall}, D., {Gibbs}, K., {Gillanders}, G.~H., {Grube}, J., {Guenette}, R.,
  {Gyuk}, G., {Hanna}, D., {Holder}, J., {Hui}, C.~M., {Humensky}, T.~B.,
  {Kaaret}, P., {Karlsson}, N., {Kertzman}, M., {Kieda}, D., {Konopelko}, A.,
  {Krawczynski}, H., {Krennrich}, F., {Lang}, M.~J., {Le Bohec}, S., {Maier},
  G., {McArthur}, S., {McCann}, A., {McCutcheon}, M., {Millis}, J., {Moriarty},
  P., {Ong}, R.~A., {Otte}, A.~N., {Pandel}, D., {Perkins}, J.~S., {Pichel},
  A., {Pohl}, M., {Quinn}, J., {Ragan}, K., {Reyes}, L.~C., {Reynolds}, P.~T.,
  {Roache}, E., {Rose}, H.~J., {Sembroski}, G.~H., {Smith}, A.~W., {Steele},
  D., {Theiling}, M., {Thibadeau}, S., {Varlotta}, A., {Vassiliev}, V.~V.,
  {Vincent}, S., {Wakely}, S.~P., {Ward}, J.~E., {Weekes}, T.~C., {Weinstein},
  A., {Weisgarber}, T., {Williams}, D.~A., {Wissel}, S., {Wood}, M., {Pian},
  E., {Vercellone}, S., {Donnarumma}, I., {D'Ammando}, F., {Bulgarelli}, A.,
  {Chen}, A.~W., {Giuliani}, A., {Longo}, F., {Pacciani}, L., {Pucella}, G.,
  {Vittorini}, V., {Tavani}, M., {Argan}, A., {Barbiellini}, G., {Caraveo}, P.,
  {Cattaneo}, P.~W., {Cocco}, V., {Costa}, E., {Del Monte}, E., {De Paris}, G.,
  {Di Cocco}, G., {Evangelista}, Y., {Feroci}, M., {Fiorini}, M., {Froysland},
  T., {Frutti}, M., {Fuschino}, F., {Galli}, M., {Gianotti}, F., {Labanti}, C.,
  {Lapshov}, I., {Lazzarotto}, F., {Lipari}, P., {Marisaldi}, M.,
  {Mastropietro}, M., {Mereghetti}, S., {Morelli}, E., {Morselli}, A.,
  {Pellizzoni}, A., {Perotti}, F., {Piano}, G., {Picozza}, P., {Pilia}, M.,
  {Porrovecchio}, G., {Prest}, M., {Rapisarda}, M., {Rappoldi}, A., {Rubini},
  A., {Sabatini}, S., {Soffitta}, P., {Trifoglio}, M., {Trois}, A., {Vallazza},
  E., {Zambra}, A., {Zanello}, D., {Pittori}, C., {Santolamazza}, P.,
  {Verrecchia}, F., {Giommi}, P., {Colafrancesco}, S., {Salotti}, L.,
  {Villata}, M., {Raiteri}, C.~M., {Aller}, H.~D., {Aller}, M.~F., {Arkharov},
  A.~A., {Efimova}, N.~V., {Larionov}, V.~M., {Leto}, P., {Ligustri}, R.,
  {Lindfors}, E., {Pasanen}, M., {Kurtanidze}, O.~M., {Tetradze}, S.~D.,
  {Lahteenmaki}, A., {Kotiranta}, M., {Cucchiara}, A., {Romano}, P., {Nesci},
  R., {Pursimo}, T., {Heidt}, J., {Benitez}, E., {Hiriart}, D., {Nilsson}, K.,
  {Berdyugin}, A., {Mujica}, R., {Dultzin}, D., {Lopez}, J.~M., {Mommert}, M.,
  {Sorcia}, M., \& {de la Calle Perez}, I. 2009{\natexlab{e}}, \apj, 707, 612

\bibitem[{{Acciari} {et~al.}(2010{\natexlab{c}}){Acciari}, {Aliu}, {Beilicke},
  {Benbow}, {Boltuch}, {B{\"o}ttcher}, {Bradbury}, {Bugaev}, {Byrum},
  {Cesarini}, {Ciupik}, {Cogan}, {Cui}, {Dickherber}, {Duke}, {Falcone},
  {Finley}, {Finnegan}, {Fortson}, {Furniss}, {Galante}, {Gall}, {Gibbs},
  {Guenette}, {Gillanders}, {Godambe}, {Grube}, {Hanna}, {Hui}, {Humensky},
  {Imran}, {Kaaret}, {Karlsson}, {Kertzman}, {Kieda}, {Krawczynski},
  {Krennrich}, {Lang}, {LeBohec}, {Maier}, {McArthur}, {McCann}, {Moriarty},
  {Nagai}, {Ong}, {Otte}, {Pandel}, {Perkins}, {Pichel}, {Pohl}, {Quinn},
  {Ragan}, {Reyes}, {Reynolds}, {Roache}, {Rose}, {Schroedter}, {Sembroski},
  {Smith}, {Steele}, {Swordy}, {Theiling}, {Thibadeau}, {Vassiliev}, {Vincent},
  {Wakely}, {Weekes}, {Weinstein}, {Weisgarber}, {Williams}, \& {VERITAS
  Collaboration}}]{acciari10c}
{Acciari}, V.~A., {Aliu}, E., {Beilicke}, M., {Benbow}, W., {Boltuch}, D.,
  {B{\"o}ttcher}, M., {Bradbury}, S.~M., {Bugaev}, V., {Byrum}, K., {Cesarini},
  A., {Ciupik}, L., {Cogan}, P., {Cui}, W., {Dickherber}, R., {Duke}, C.,
  {Falcone}, A., {Finley}, J.~P., {Finnegan}, G., {Fortson}, L., {Furniss}, A.,
  {Galante}, N., {Gall}, D., {Gibbs}, K., {Guenette}, R., {Gillanders}, G.~H.,
  {Godambe}, S., {Grube}, J., {Hanna}, D., {Hui}, C.~M., {Humensky}, T.~B.,
  {Imran}, A., {Kaaret}, P., {Karlsson}, N., {Kertzman}, M., {Kieda}, D.,
  {Krawczynski}, H., {Krennrich}, F., {Lang}, M.~J., {LeBohec}, S., {Maier},
  G., {McArthur}, S., {McCann}, A., {Moriarty}, P., {Nagai}, T., {Ong}, R.~A.,
  {Otte}, A.~N., {Pandel}, D., {Perkins}, J.~S., {Pichel}, A., {Pohl}, M.,
  {Quinn}, J., {Ragan}, K., {Reyes}, L.~C., {Reynolds}, P.~T., {Roache}, E.,
  {Rose}, H.~J., {Schroedter}, M., {Sembroski}, G.~H., {Smith}, A.~W.,
  {Steele}, D., {Swordy}, S.~P., {Theiling}, M., {Thibadeau}, S., {Vassiliev},
  V.~V., {Vincent}, S., {Wakely}, S.~P., {Weekes}, T.~C., {Weinstein}, A.,
  {Weisgarber}, T., {Williams}, D.~A., \& {VERITAS Collaboration}.
  2010{\natexlab{c}}, \apjl, 709, L163

\bibitem[{{Acciari} {et~al.}(2008){Acciari}, {Aliu}, {Beilicke}, {Benbow},
  {B{\"o}ttcher}, {Bradbury}, {Buckley}, {Bugaev}, {Butt}, {Celik}, {Cesarini},
  {Ciupik}, {Chow}, {Cogan}, {Colin}, {Cui}, {Daniel}, {Ergin}, {Falcone},
  {Fegan}, {Finley}, {Finnegan}, {Fortin}, {Fortson}, {Furniss}, {Gall},
  {Gillanders}, {Grube}, {Guenette}, {Gyuk}, {Hanna}, {Hays}, {Holder},
  {Horan}, {Hui}, {Humensky}, {Imran}, {Kaaret}, {Karlsson}, {Kertzman},
  {Kieda}, {Konopelko}, {Krawczynski}, {Krennrich}, {Lang}, {LeBohec}, {Lee},
  {Maier}, {McCann}, {McCutcheon}, {Moriarty}, {Mukherjee}, {Nagai}, {Niemiec},
  {Ong}, {Pandel}, {Perkins}, {Petry}, {Pohl}, {Quinn}, {Ragan}, {Reyes},
  {Reynolds}, {Roache}, {Rose}, {Schroedter}, {Sembroski}, {Smith}, {Steele},
  {Swordy}, {Toner}, {Vassiliev}, {Wagner}, {Wakely}, {Ward}, {Weekes},
  {Weinstein}, {White}, {Williams}, {Wissel}, {Wood}, \& {Zitzer}}]{acciari08}
{Acciari}, V.~A., {Aliu}, E., {Beilicke}, M., {Benbow}, W., {B{\"o}ttcher}, M.,
  {Bradbury}, S.~M., {Buckley}, J.~H., {Bugaev}, V., {Butt}, Y., {Celik}, O.,
  {Cesarini}, A., {Ciupik}, L., {Chow}, Y.~C.~K., {Cogan}, P., {Colin}, P.,
  {Cui}, W., {Daniel}, M.~K., {Ergin}, T., {Falcone}, A.~D., {Fegan}, S.~J.,
  {Finley}, J.~P., {Finnegan}, G., {Fortin}, P., {Fortson}, L.~F., {Furniss},
  A., {Gall}, D., {Gillanders}, G.~H., {Grube}, J., {Guenette}, R., {Gyuk}, G.,
  {Hanna}, D., {Hays}, E., {Holder}, J., {Horan}, D., {Hui}, C.~M., {Humensky},
  T.~B., {Imran}, A., {Kaaret}, P., {Karlsson}, N., {Kertzman}, M., {Kieda},
  D.~B., {Konopelko}, A., {Krawczynski}, H., {Krennrich}, F., {Lang}, M.~J.,
  {LeBohec}, S., {Lee}, K., {Maier}, G., {McCann}, A., {McCutcheon}, M.,
  {Moriarty}, P., {Mukherjee}, R., {Nagai}, T., {Niemiec}, J., {Ong}, R.~A.,
  {Pandel}, D., {Perkins}, J.~S., {Petry}, D., {Pohl}, M., {Quinn}, J.,
  {Ragan}, K., {Reyes}, L.~C., {Reynolds}, P.~T., {Roache}, E., {Rose}, H.~J.,
  {Schroedter}, M., {Sembroski}, G.~H., {Smith}, A.~W., {Steele}, D., {Swordy},
  S.~P., {Toner}, J.~A., {Vassiliev}, V.~V., {Wagner}, R., {Wakely}, S.~P.,
  {Ward}, J.~E., {Weekes}, T.~C., {Weinstein}, A., {White}, R.~J., {Williams},
  D.~A., {Wissel}, S.~A., {Wood}, M., \& {Zitzer}, B. 2008, \apjl, 684, L73

\bibitem[{{Acero} {et~al.}(2009){Acero}, {Aharonian}, {Akhperjanian}, {Anton},
  {Barres de Almeida}, {Bazer-Bachi}, {Becherini}, {Behera}, {Bernl{\"o}hr},
  {Bochow}, {Boisson}, {Bolmont}, {Borrel}, {Brucker}, {Brun}, {Brun},
  {B{\"u}hler}, {Bulik}, {B{\"u}sching}, {Boutelier}, {Chadwick},
  {Charbonnier}, {Chaves}, {Cheesebrough}, {Chounet}, {Clapson}, {Coignet},
  {Dalton}, {Daniel}, {Davids}, {Degrange}, {Deil}, {Dickinson},
  {Djannati-Ata{\"\i}}, {Domainko}, {Drury}, {Dubois}, {Dubus}, {Dyks},
  {Dyrda}, {Egberts}, {Emmanoulopoulos}, {Espigat}, {Farnier}, {Fegan},
  {Feinstein}, {Fiasson}, {F{\"o}rster}, {Fontaine}, {F{\"u}{\ss}ling},
  {Gabici}, {Gallant}, {G{\'e}rard}, {Gerbig}, {Giebels}, {Glicenstein},
  {Gl{\"u}ck}, {Goret}, {G{\"o}ring}, {Hauser}, {Hauser}, {Heinz},
  {Heinzelmann}, {Henri}, {Hermann}, {Hinton}, {Hoffmann}, {Hofmann},
  {Hofverberg}, {Hoppe}, {Horns}, {Jacholkowska}, {de Jager}, {Jahn}, {Jung},
  {Katarzy{\'n}ski}, {Katz}, {Kaufmann}, {Kerschhaggl}, {Khangulyan},
  {Kh{\'e}lifi}, {Keogh}, {Klochkov}, {Klu{\'z}niak}, {Kneiske}, {Komin},
  {Kosack}, {Kossakowski}, {Lamanna}, {Lenain}, {Lohse}, {Marandon},
  {Martineau-Huynh}, {Marcowith}, {Masbou}, {Maurin}, {McComb}, {Medina},
  {M{\'e}hault}, {Moderski}, {Moulin}, {Naumann-Godo}, {de Naurois}, {Nedbal},
  {Nekrassov}, {Nicholas}, {Niemiec}, {Nolan}, {Ohm}, {Olive}, {Wilhelmi},
  {Orford}, {Ostrowski}, {Panter}, {Arribas}, {Pedaletti}, {Pelletier},
  {Petrucci}, {Pita}, {P{\"u}hlhofer}, {Punch}, {Quirrenbach}, {Raubenheimer},
  {Raue}, {Rayner}, {Reimer}, {Renaud}, {Rieger}, {Ripken}, {Rob},
  {Rosier-Lees}, {Rowell}, {Rudak}, {Rulten}, {Ruppel}, {Sahakian},
  {Santangelo}, {Schlickeiser}, {Sch{\"o}ck}, {Schwanke}, {Schwarzburg},
  {Schwemmer}, {Shalchi}, {Sikora}, {Skilton}, {Sol}, {Stawarz}, {Steenkamp},
  {Stegmann}, {Stinzing}, {Superina}, {Szostek}, {Tam}, {Tavernet}, {Terrier},
  {Tibolla}, {Tluczykont}, {van Eldik}, {Vasileiadis}, {Venter}, {Venter},
  {Vialle}, {Vincent}, {Vivier}, {V{\"o}lk}, {Volpe}, {Wagner}, {Ward},
  {Zdziarski}, \& {Zech}}]{acero09}
{Acero}, F., {Aharonian}, F., {Akhperjanian}, A.~G., {Anton}, G., {Barres de
  Almeida}, U., {Bazer-Bachi}, A.~R., {Becherini}, Y., {Behera}, B.,
  {Bernl{\"o}hr}, K., {Bochow}, A., {Boisson}, C., {Bolmont}, J., {Borrel}, V.,
  {Brucker}, J., {Brun}, F., {Brun}, P., {B{\"u}hler}, R., {Bulik}, T.,
  {B{\"u}sching}, I., {Boutelier}, T., {Chadwick}, P.~M., {Charbonnier}, A.,
  {Chaves}, R.~C.~G., {Cheesebrough}, A., {Chounet}, L.-M., {Clapson}, A.~C.,
  {Coignet}, G., {Dalton}, M., {Daniel}, M.~K., {Davids}, I.~D., {Degrange},
  B., {Deil}, C., {Dickinson}, H.~J., {Djannati-Ata{\"\i}}, A., {Domainko}, W.,
  {Drury}, L.~O.~., {Dubois}, F., {Dubus}, G., {Dyks}, J., {Dyrda}, M.,
  {Egberts}, K., {Emmanoulopoulos}, D., {Espigat}, P., {Farnier}, C., {Fegan},
  S., {Feinstein}, F., {Fiasson}, A., {F{\"o}rster}, A., {Fontaine}, G.,
  {F{\"u}{\ss}ling}, M., {Gabici}, S., {Gallant}, Y.~A., {G{\'e}rard}, L.,
  {Gerbig}, D., {Giebels}, B., {Glicenstein}, J.~F., {Gl{\"u}ck}, B., {Goret},
  P., {G{\"o}ring}, D., {Hauser}, D., {Hauser}, M., {Heinz}, S., {Heinzelmann},
  G., {Henri}, G., {Hermann}, G., {Hinton}, J.~A., {Hoffmann}, A., {Hofmann},
  W., {Hofverberg}, P., {Hoppe}, S., {Horns}, D., {Jacholkowska}, A., {de
  Jager}, O.~C., {Jahn}, C., {Jung}, I., {Katarzy{\'n}ski}, K., {Katz}, U.,
  {Kaufmann}, S., {Kerschhaggl}, M., {Khangulyan}, D., {Kh{\'e}lifi}, B.,
  {Keogh}, D., {Klochkov}, D., {Klu{\'z}niak}, W., {Kneiske}, T., {Komin}, N.,
  {Kosack}, K., {Kossakowski}, R., {Lamanna}, G., {Lenain}, J.-P., {Lohse}, T.,
  {Marandon}, V., {Martineau-Huynh}, O., {Marcowith}, A., {Masbou}, J.,
  {Maurin}, D., {McComb}, T.~J.~L., {Medina}, M.~C., {M{\'e}hault}, J.,
  {Moderski}, R., {Moulin}, E., {Naumann-Godo}, M., {de Naurois}, M., {Nedbal},
  D., {Nekrassov}, D., {Nicholas}, B., {Niemiec}, J., {Nolan}, S.~J., {Ohm},
  S., {Olive}, J.-F., {Wilhelmi}, E.~d.~O., {Orford}, K.~J., {Ostrowski}, M.,
  {Panter}, M., {Arribas}, M.~P., {Pedaletti}, G., {Pelletier}, G., {Petrucci},
  P.-O., {Pita}, S., {P{\"u}hlhofer}, G., {Punch}, M., {Quirrenbach}, A.,
  {Raubenheimer}, B.~C., {Raue}, M., {Rayner}, S.~M., {Reimer}, O., {Renaud},
  M., {Rieger}, F., {Ripken}, J., {Rob}, L., {Rosier-Lees}, S., {Rowell}, G.,
  {Rudak}, B., {Rulten}, C.~B., {Ruppel}, J., {Sahakian}, V., {Santangelo}, A.,
  {Schlickeiser}, R., {Sch{\"o}ck}, F.~M., {Schwanke}, U., {Schwarzburg}, S.,
  {Schwemmer}, S., {Shalchi}, A., {Sikora}, M., {Skilton}, J.~L., {Sol}, H.,
  {Stawarz}, {\L}., {Steenkamp}, R., {Stegmann}, C., {Stinzing}, F.,
  {Superina}, G., {Szostek}, A., {Tam}, P.~H., {Tavernet}, J.-P., {Terrier},
  R., {Tibolla}, O., {Tluczykont}, M., {van Eldik}, C., {Vasileiadis}, G.,
  {Venter}, C., {Venter}, L., {Vialle}, J.~P., {Vincent}, P., {Vivier}, M.,
  {V{\"o}lk}, H.~J., {Volpe}, F., {Wagner}, S.~J., {Ward}, M., {Zdziarski},
  A.~A., \& {Zech}, A. 2009, Science, 326, 1080

\bibitem[{{Ackermann} {et~al.}(2011){Ackermann}, {Ajello}, {Allafort},
  {Antolini}, {Atwood}, {Axelsson}, {Baldini}, {Ballet}, {Barbiellini},
  {Bastieri}, {Bechtol}, {Bellazzini}, {Berenji}, {Blandford}, {Bloom},
  {Bonamente}, {Borgland}, {Bottacini}, {Bouvier}, {Bregeon}, {Brigida},
  {Bruel}, {Buehler}, {Burnett}, {Buson}, {Caliandro}, {Cameron}, {Caraveo},
  {Casandjian}, {Cavazzuti}, {Cecchi}, {Charles}, {Cheung}, {Chiang},
  {Ciprini}, {Claus}, {Cohen-Tanugi}, {Conrad}, {Costamante}, {Cutini}, {de
  Angelis}, {de Palma}, {Dermer}, {Digel}, {Silva}, {Drell}, {Dubois},
  {Escande}, {Favuzzi}, {Fegan}, {Ferrara}, {Finke}, {Focke}, {Fortin},
  {Frailis}, {Fukazawa}, {Funk}, {Fusco}, {Gargano}, {Gasparrini}, {Gehrels},
  {Germani}, {Giebels}, {Giglietto}, {Giommi}, {Giordano}, {Giroletti},
  {Glanzman}, {Godfrey}, {Grenier}, {Grove}, {Guiriec}, {Gustafsson},
  {Hadasch}, {Hayashida}, {Hays}, {Healey}, {Horan}, {Hou}, {Hughes},
  {Iafrate}, {J{\'o}hannesson}, {Johnson}, {Johnson}, {Kamae}, {Katagiri},
  {Kataoka}, {Kn{\"o}dlseder}, {Kuss}, {Lande}, {Larsson}, {Latronico},
  {Longo}, {Loparco}, {Lott}, {Lovellette}, {Lubrano}, {Madejski}, {Mazziotta},
  {McConville}, {McEnery}, {Michelson}, {Mitthumsiri}, {Mizuno}, {Moiseev},
  {Monte}, {Monzani}, {Moretti}, {Morselli}, {Moskalenko}, {Murgia},
  {Nakamori}, {Naumann-Godo}, {Nolan}, {Norris}, {Nuss}, {Ohno}, {Ohsugi},
  {Okumura}, {Omodei}, {Orienti}, {Orlando}, {Ormes}, {Ozaki}, {Paneque},
  {Parent}, {Pesce-Rollins}, {Pierbattista}, {Piranomonte}, {Piron}, {Pivato},
  {Porter}, {Rain{\`o}}, {Rando}, {Razzano}, {Razzaque}, {Reimer}, {Reimer},
  {Ritz}, {Rochester}, {Romani}, {Roth}, {Sanchez}, {Sbarra}, {Scargle},
  {Schalk}, {Sgr{\`o}}, {Shaw}, {Siskind}, {Spandre}, {Spinelli}, {Strong},
  {Suson}, {Tajima}, {Takahashi}, {Takahashi}, {Tanaka}, {Thayer}, {Thayer},
  {Thompson}, {Tibaldo}, {Tinivella}, {Torres}, {Tosti}, {Troja}, {Uchiyama},
  {Vandenbroucke}, {Vasileiou}, {Vianello}, {Vitale}, {Waite}, {Wallace},
  {Wang}, {Winer}, {Wood}, {Wood}, \& {Zimmer}}]{ackermann11}
{Ackermann}, M., {Ajello}, M., {Allafort}, A., {Antolini}, E., {Atwood}, W.~B.,
  {Axelsson}, M., {Baldini}, L., {Ballet}, J., {Barbiellini}, G., {Bastieri},
  D., {Bechtol}, K., {Bellazzini}, R., {Berenji}, B., {Blandford}, R.~D.,
  {Bloom}, E.~D., {Bonamente}, E., {Borgland}, A.~W., {Bottacini}, E.,
  {Bouvier}, A., {Bregeon}, J., {Brigida}, M., {Bruel}, P., {Buehler}, R.,
  {Burnett}, T.~H., {Buson}, S., {Caliandro}, G.~A., {Cameron}, R.~A.,
  {Caraveo}, P.~A., {Casandjian}, J.~M., {Cavazzuti}, E., {Cecchi}, C.,
  {Charles}, E., {Cheung}, C.~C., {Chiang}, J., {Ciprini}, S., {Claus}, R.,
  {Cohen-Tanugi}, J., {Conrad}, J., {Costamante}, L., {Cutini}, S., {de
  Angelis}, A., {de Palma}, F., {Dermer}, C.~D., {Digel}, S.~W., {Silva},
  E.~d.~C.~e., {Drell}, P.~S., {Dubois}, R., {Escande}, L., {Favuzzi}, C.,
  {Fegan}, S.~J., {Ferrara}, E.~C., {Finke}, J., {Focke}, W.~B., {Fortin}, P.,
  {Frailis}, M., {Fukazawa}, Y., {Funk}, S., {Fusco}, P., {Gargano}, F.,
  {Gasparrini}, D., {Gehrels}, N., {Germani}, S., {Giebels}, B., {Giglietto},
  N., {Giommi}, P., {Giordano}, F., {Giroletti}, M., {Glanzman}, T., {Godfrey},
  G., {Grenier}, I.~A., {Grove}, J.~E., {Guiriec}, S., {Gustafsson}, M.,
  {Hadasch}, D., {Hayashida}, M., {Hays}, E., {Healey}, S.~E., {Horan}, D.,
  {Hou}, X., {Hughes}, R.~E., {Iafrate}, G., {J{\'o}hannesson}, G., {Johnson},
  A.~S., {Johnson}, W.~N., {Kamae}, T., {Katagiri}, H., {Kataoka}, J.,
  {Kn{\"o}dlseder}, J., {Kuss}, M., {Lande}, J., {Larsson}, S., {Latronico},
  L., {Longo}, F., {Loparco}, F., {Lott}, B., {Lovellette}, M.~N., {Lubrano},
  P., {Madejski}, G.~M., {Mazziotta}, M.~N., {McConville}, W., {McEnery},
  J.~E., {Michelson}, P.~F., {Mitthumsiri}, W., {Mizuno}, T., {Moiseev}, A.~A.,
  {Monte}, C., {Monzani}, M.~E., {Moretti}, E., {Morselli}, A., {Moskalenko},
  I.~V., {Murgia}, S., {Nakamori}, T., {Naumann-Godo}, M., {Nolan}, P.~L.,
  {Norris}, J.~P., {Nuss}, E., {Ohno}, M., {Ohsugi}, T., {Okumura}, A.,
  {Omodei}, N., {Orienti}, M., {Orlando}, E., {Ormes}, J.~F., {Ozaki}, M.,
  {Paneque}, D., {Parent}, D., {Pesce-Rollins}, M., {Pierbattista}, M.,
  {Piranomonte}, S., {Piron}, F., {Pivato}, G., {Porter}, T.~A., {Rain{\`o}},
  S., {Rando}, R., {Razzano}, M., {Razzaque}, S., {Reimer}, A., {Reimer}, O.,
  {Ritz}, S., {Rochester}, L.~S., {Romani}, R.~W., {Roth}, M., {Sanchez},
  D.~A., {Sbarra}, C., {Scargle}, J.~D., {Schalk}, T.~L., {Sgr{\`o}}, C.,
  {Shaw}, M.~S., {Siskind}, E.~J., {Spandre}, G., {Spinelli}, P., {Strong},
  A.~W., {Suson}, D.~J., {Tajima}, H., {Takahashi}, H., {Takahashi}, T.,
  {Tanaka}, T., {Thayer}, J.~G., {Thayer}, J.~B., {Thompson}, D.~J., {Tibaldo},
  L., {Tinivella}, M., {Torres}, D.~F., {Tosti}, G., {Troja}, E., {Uchiyama},
  Y., {Vandenbroucke}, J., {Vasileiou}, V., {Vianello}, G., {Vitale}, V.,
  {Waite}, A.~P., {Wallace}, E., {Wang}, P., {Winer}, B.~L., {Wood}, D.~L.,
  {Wood}, K.~S., \& {Zimmer}, S. 2011, \apj, 743, 171

\bibitem[{{Aharonian} {et~al.}(2009){Aharonian}, {Akhperjanian}, {Anton}, {de
  Almeida}, {Bazer-Bachi}, {Becherini}, {Behera}, {Benbow}, {Bernl{\"o}hr},
  {Boisson}, {Bochow}, {Borrel}, {Brion}, {Brucker}, {Brun}, {B{\"u}hler},
  {Bulik}, {B{\"u}sching}, {Boutelier}, {Chadwick}, {Charbonnier}, {Chaves},
  {Cheesebrough}, {Chounet}, {Clapson}, {Coignet}, {Dalton}, {Daniel},
  {Davids}, {Degrange}, {Deil}, {Dickinson}, {Djannati-Ata{\"\i}}, {Domainko},
  {Drury}, {Dubois}, {Dubus}, {Dyks}, {Dyrda}, {Egberts}, {Emmanoulopoulos},
  {Espigat}, {Farnier}, {Feinstein}, {Fiasson}, {F{\"o}rster}, {Fontaine},
  {F{\"u}{\ss}ling}, {Gabici}, {Gallant}, {G{\'e}rard}, {Giebels},
  {Glicenstein}, {Gl{\"u}ck}, {Goret}, {G{\"o}hring}, {Hauser}, {Hauser},
  {Heinz}, {Heinzelmann}, {Henri}, {Hermann}, {Hinton}, {Hoffmann}, {Hofmann},
  {Holleran}, {Hoppe}, {Horns}, {Jacholkowska}, {de Jager}, {Jahn}, {Jung},
  {Katarzy{\'n}ski}, {Katz}, {Kaufmann}, {Kendziorra}, {Kerschhaggl},
  {Khangulyan}, {Kh{\'e}lifi}, {Keogh}, {Klu{\'z}niak}, {Kneiske}, {Komin},
  {Kosack}, {Lamanna}, {Latham}, {Lenain}, {Lohse}, {Marandon}, {Martin},
  {Martineau-Huynh}, {Marcowith}, {Maurin}, {McComb}, {Medina}, {Moderski},
  {Moulin}, {Naumann-Godo}, {de Naurois}, {Nedbal}, {Nekrassov}, {Niemiec},
  {Nolan}, {Ohm}, {Olive}, {de O{\~n}a Wilhelmi}, {Orford}, {Ostrowski},
  {Panter}, {Arribas}, {Pedaletti}, {Pelletier}, {Petrucci}, {Pita},
  {P{\"u}hlhofer}, {Punch}, {Quirrenbach}, {Raubenheimer}, {Raue}, {Rayner},
  {Renaud}, {Rieger}, {Ripken}, {Rob}, {Rosier-Lees}, {Rowell}, {Rudak},
  {Rulten}, {Ruppel}, {Sahakian}, {Santangelo}, {Schlickeiser}, {Sch{\"o}ck},
  {Schr{\"o}der}, {Schwanke}, {Schwarzburg}, {Schwemmer}, {Shalchi}, {Sikora},
  {Skilton}, {Sol}, {Spangler}, {Stawarz}, {Steenkamp}, {Stegmann}, {Superina},
  {Szostek}, {Tam}, {Tavernet}, {Terrier}, {Tibolla}, {Tluczykont}, {van
  Eldik}, {Vasileiadis}, {Venter}, {Venter}, {Vialle}, {Vincent}, {Vink},
  {Vivier}, {V{\"o}lk}, {Volpe}, {Wagner}, {Ward}, {Zdziarski}, \&
  {Zech}}]{aharonian09}
{Aharonian}, F., {Akhperjanian}, A.~G., {Anton}, G., {de Almeida}, U.~B.,
  {Bazer-Bachi}, A.~R., {Becherini}, Y., {Behera}, B., {Benbow}, W.,
  {Bernl{\"o}hr}, K., {Boisson}, C., {Bochow}, A., {Borrel}, V., {Brion}, E.,
  {Brucker}, J., {Brun}, P., {B{\"u}hler}, R., {Bulik}, T., {B{\"u}sching}, I.,
  {Boutelier}, T., {Chadwick}, P.~M., {Charbonnier}, A., {Chaves}, R.~C.~G.,
  {Cheesebrough}, A., {Chounet}, L.-M., {Clapson}, A.~C., {Coignet}, G.,
  {Dalton}, M., {Daniel}, M.~K., {Davids}, I.~D., {Degrange}, B., {Deil}, C.,
  {Dickinson}, H.~J., {Djannati-Ata{\"\i}}, A., {Domainko}, W., {Drury}, L.~O.,
  {Dubois}, F., {Dubus}, G., {Dyks}, J., {Dyrda}, M., {Egberts}, K.,
  {Emmanoulopoulos}, D., {Espigat}, P., {Farnier}, C., {Feinstein}, F.,
  {Fiasson}, A., {F{\"o}rster}, A., {Fontaine}, G., {F{\"u}{\ss}ling}, M.,
  {Gabici}, S., {Gallant}, Y.~A., {G{\'e}rard}, L., {Giebels}, B.,
  {Glicenstein}, J.-F., {Gl{\"u}ck}, B., {Goret}, P., {G{\"o}hring}, D.,
  {Hauser}, D., {Hauser}, M., {Heinz}, S., {Heinzelmann}, G., {Henri}, G.,
  {Hermann}, G., {Hinton}, J.~A., {Hoffmann}, A., {Hofmann}, W., {Holleran},
  M., {Hoppe}, S., {Horns}, D., {Jacholkowska}, A., {de Jager}, O.~C., {Jahn},
  C., {Jung}, I., {Katarzy{\'n}ski}, K., {Katz}, U., {Kaufmann}, S.,
  {Kendziorra}, E., {Kerschhaggl}, M., {Khangulyan}, D., {Kh{\'e}lifi}, B.,
  {Keogh}, D., {Klu{\'z}niak}, W., {Kneiske}, T., {Komin}, N., {Kosack}, K.,
  {Lamanna}, G., {Latham}, I.~J., {Lenain}, J.-P., {Lohse}, T., {Marandon}, V.,
  {Martin}, J.~M., {Martineau-Huynh}, O., {Marcowith}, A., {Maurin}, D.,
  {McComb}, T.~J.~L., {Medina}, M.~C., {Moderski}, R., {Moulin}, E.,
  {Naumann-Godo}, M., {de Naurois}, M., {Nedbal}, D., {Nekrassov}, D.,
  {Niemiec}, J., {Nolan}, S.~J., {Ohm}, S., {Olive}, J.-F., {de O{\~n}a
  Wilhelmi}, E., {Orford}, K.~J., {Ostrowski}, M., {Panter}, M., {Arribas},
  M.~P., {Pedaletti}, G., {Pelletier}, G., {Petrucci}, P.-O., {Pita}, S.,
  {P{\"u}hlhofer}, G., {Punch}, M., {Quirrenbach}, A., {Raubenheimer}, B.~C.,
  {Raue}, M., {Rayner}, S.~M., {Renaud}, M., {Rieger}, F., {Ripken}, J., {Rob},
  L., {Rosier-Lees}, S., {Rowell}, G., {Rudak}, B., {Rulten}, C.~B., {Ruppel},
  J., {Sahakian}, V., {Santangelo}, A., {Schlickeiser}, R., {Sch{\"o}ck},
  F.~M., {Schr{\"o}der}, R., {Schwanke}, U., {Schwarzburg}, S., {Schwemmer},
  S., {Shalchi}, A., {Sikora}, M., {Skilton}, J.~L., {Sol}, H., {Spangler}, D.,
  {Stawarz}, {\L}., {Steenkamp}, R., {Stegmann}, C., {Superina}, G., {Szostek},
  A., {Tam}, P.~H., {Tavernet}, J.-P., {Terrier}, R., {Tibolla}, O.,
  {Tluczykont}, M., {van Eldik}, C., {Vasileiadis}, G., {Venter}, C., {Venter},
  L., {Vialle}, J.~P., {Vincent}, P., {Vink}, J., {Vivier}, M., {V{\"o}lk},
  H.~J., {Volpe}, F., {Wagner}, S.~J., {Ward}, M., {Zdziarski}, A.~A., \&
  {Zech}, A. 2009, \apjl, 695, L40

\bibitem[{{Aharonian} {et~al.}(2005{\natexlab{a}}){Aharonian}, {Akhperjanian},
  {Aye}, {Bazer-Bachi}, {Beilicke}, {Benbow}, {Berge}, {Berghaus},
  {Bernl{\"o}hr}, {Boisson}, {Bolz}, {Braun}, {Breitling}, {Brown}, {Bussons
  Gordo}, {Chadwick}, {Chounet}, {Cornils}, {Costamante}, {Degrange},
  {Djannati-Ata{\"\i}}, {O'C.~Drury}, {Dubus}, {Emmanoulopoulos}, {Espigat},
  {Feinstein}, {Fleury}, {Fontaine}, {Fuchs}, {Funk}, {Gallant}, {Giebels},
  {Gillessen}, {Glicenstein}, {Goret}, {Hadjichristidis}, {Hauser},
  {Heinzelmann}, {Henri}, {Hermann}, {Hinton}, {Hofmann}, {Holleran}, {Horns},
  {de Jager}, {Kh{\'e}lifi}, {Komin}, {Konopelko}, {Latham}, {Le Gallou},
  {Lemi{\`e}re}, {Lemoine-Goumard}, {Leroy}, {Lohse}, {Martineau-Huynh},
  {Marcowith}, {Masterson}, {McComb}, {de Naurois}, {Nolan}, {Noutsos},
  {Orford}, {Osborne}, {Ouchrif}, {Panter}, {Pelletier}, {Pita},
  {P{\"u}hlhofer}, {Punch}, {Raubenheimer}, {Raue}, {Raux}, {Rayner},
  {Redondo}, {Reimer}, {Reimer}, {Ripken}, {Rob}, {Rolland}, {Rowell},
  {Sahakian}, {Saug{\'e}}, {Schlenker}, {Schlickeiser}, {Schuster}, {Schwanke},
  {Siewert}, {Sol}, {Steenkamp}, {Stegmann}, {Tavernet}, {Terrier},
  {Th{\'e}oret}, {Tluczykont}, {Vasileiadis}, {Venter}, {Vincent}, {V{\"o}lk},
  \& {Wagner}}]{aharonian05b}
{Aharonian}, F., {Akhperjanian}, A.~G., {Aye}, K.-M., {Bazer-Bachi}, A.~R.,
  {Beilicke}, M., {Benbow}, W., {Berge}, D., {Berghaus}, P., {Bernl{\"o}hr},
  K., {Boisson}, C., {Bolz}, O., {Braun}, I., {Breitling}, F., {Brown}, A.~M.,
  {Bussons Gordo}, J., {Chadwick}, P.~M., {Chounet}, L.-M., {Cornils}, R.,
  {Costamante}, L., {Degrange}, B., {Djannati-Ata{\"\i}}, A., {O'C.~Drury}, L.,
  {Dubus}, G., {Emmanoulopoulos}, D., {Espigat}, P., {Feinstein}, F., {Fleury},
  P., {Fontaine}, G., {Fuchs}, Y., {Funk}, S., {Gallant}, Y.~A., {Giebels}, B.,
  {Gillessen}, S., {Glicenstein}, J.~F., {Goret}, P., {Hadjichristidis}, C.,
  {Hauser}, M., {Heinzelmann}, G., {Henri}, G., {Hermann}, G., {Hinton}, J.~A.,
  {Hofmann}, W., {Holleran}, M., {Horns}, D., {de Jager}, O.~C., {Kh{\'e}lifi},
  B., {Komin}, N., {Konopelko}, A., {Latham}, I.~J., {Le Gallou}, R.,
  {Lemi{\`e}re}, A., {Lemoine-Goumard}, M., {Leroy}, N., {Lohse}, T.,
  {Martineau-Huynh}, O., {Marcowith}, A., {Masterson}, C., {McComb}, T.~J.~L.,
  {de Naurois}, M., {Nolan}, S.~J., {Noutsos}, A., {Orford}, K.~J., {Osborne},
  J.~L., {Ouchrif}, M., {Panter}, M., {Pelletier}, G., {Pita}, S.,
  {P{\"u}hlhofer}, G., {Punch}, M., {Raubenheimer}, B.~C., {Raue}, M., {Raux},
  J., {Rayner}, S.~M., {Redondo}, I., {Reimer}, A., {Reimer}, O., {Ripken}, J.,
  {Rob}, L., {Rolland}, L., {Rowell}, G., {Sahakian}, V., {Saug{\'e}}, L.,
  {Schlenker}, S., {Schlickeiser}, R., {Schuster}, C., {Schwanke}, U.,
  {Siewert}, M., {Sol}, H., {Steenkamp}, R., {Stegmann}, C., {Tavernet}, J.-P.,
  {Terrier}, R., {Th{\'e}oret}, C.~G., {Tluczykont}, M., {Vasileiadis}, G.,
  {Venter}, C., {Vincent}, P., {V{\"o}lk}, H.~J., \& {Wagner}, S.~J.
  2005{\natexlab{a}}, \aap, 436, L17

\bibitem[{{Aharonian} {et~al.}(2005{\natexlab{b}}){Aharonian}, {Akhperjanian},
  {Aye}, {Bazer-Bachi}, {Beilicke}, {Benbow}, {Berge}, {Berghaus},
  {Bernl{\"o}hr}, {Bolz}, {Boisson}, {Borgmeier}, {Breitling}, {Brown},
  {Bussons Gordo}, {Chadwick}, {Chitnis}, {Chounet}, {Cornils}, {Costamante},
  {Degrange}, {Djannati-Ata{\"\i}}, {Drury}, {Ergin}, {Espigat}, {Feinstein},
  {Fleury}, {Fontaine}, {Funk}, {Gallant}, {Giebels}, {Gillessen}, {Goret},
  {Guy}, {Hadjichristidis}, {Hauser}, {Heinzelmann}, {Henri}, {Hermann},
  {Hinton}, {Hofmann}, {Holleran}, {Horns}, {de Jager}, {Jung I.},
  {Kh{\'e}lifi}, {Komin}, {Konopelko}, {Latham}, {Le Gallou}, {Lemoine},
  {Lemi{\`e}re}, {Leroy}, {Lohse}, {Marcowith}, {Masterson}, {McComb}, {de
  Naurois}, {Nolan}, {Noutsos}, {Orford}, {Osborne}, {Ouchrif}, {Panter},
  {Pelletier}, {Pita}, {Pohl}, {P{\"u}hlhofer}, {Punch}, {Raubenheimer},
  {Raue}, {Raux}, {Rayner}, {Redondo}, {Reimer}, {Reimer}, {Ripken}, {Rivoal},
  {Rob}, {Rolland}, {Rowell}, {Sahakian}, {Saug{\'e}}, {Schlenker},
  {Schlickeiser}, {Schuster}, {Schwanke}, {Siewert}, {Sol}, {Steenkamp},
  {Stegmann}, {Tavernet}, {Th{\'e}oret}, {Tluczykont}, {van der Walt},
  {Vasileiadis}, {Vincent}, {Visser}, {V{\"o}lk}, \& {Wagner}}]{aharonian05a}
{Aharonian}, F., {Akhperjanian}, A.~G., {Aye}, K.-M., {Bazer-Bachi}, A.~R.,
  {Beilicke}, M., {Benbow}, W., {Berge}, D., {Berghaus}, P., {Bernl{\"o}hr},
  K., {Bolz}, O., {Boisson}, C., {Borgmeier}, C., {Breitling}, F., {Brown},
  A.~M., {Bussons Gordo}, J., {Chadwick}, P.~M., {Chitnis}, V.~R., {Chounet},
  L.-M., {Cornils}, R., {Costamante}, L., {Degrange}, B., {Djannati-Ata{\"\i}},
  A., {Drury}, L.~O., {Ergin}, T., {Espigat}, P., {Feinstein}, F., {Fleury},
  P., {Fontaine}, G., {Funk}, S., {Gallant}, Y.~A., {Giebels}, B., {Gillessen},
  S., {Goret}, P., {Guy}, J., {Hadjichristidis}, C., {Hauser}, M.,
  {Heinzelmann}, G., {Henri}, G., {Hermann}, G., {Hinton}, J.~A., {Hofmann},
  W., {Holleran}, M., {Horns}, D., {de Jager}, O.~C., {Jung I.}, {Kh{\'e}lifi},
  B., {Komin}, N., {Konopelko}, A., {Latham}, I.~J., {Le Gallou}, R.,
  {Lemoine}, M., {Lemi{\`e}re}, A., {Leroy}, N., {Lohse}, T., {Marcowith}, A.,
  {Masterson}, C., {McComb}, T.~J.~L., {de Naurois}, M., {Nolan}, S.~J.,
  {Noutsos}, A., {Orford}, K.~J., {Osborne}, J.~L., {Ouchrif}, M., {Panter},
  M., {Pelletier}, G., {Pita}, S., {Pohl}, M., {P{\"u}hlhofer}, G., {Punch},
  M., {Raubenheimer}, B.~C., {Raue}, M., {Raux}, J., {Rayner}, S.~M.,
  {Redondo}, I., {Reimer}, A., {Reimer}, O., {Ripken}, J., {Rivoal}, M., {Rob},
  L., {Rolland}, L., {Rowell}, G., {Sahakian}, V., {Saug{\'e}}, L.,
  {Schlenker}, S., {Schlickeiser}, R., {Schuster}, C., {Schwanke}, U.,
  {Siewert}, M., {Sol}, H., {Steenkamp}, R., {Stegmann}, C., {Tavernet}, J.-P.,
  {Th{\'e}oret}, C.~G., {Tluczykont}, M., {van der Walt}, D.~J., {Vasileiadis},
  G., {Vincent}, P., {Visser}, B., {V{\"o}lk}, H.~J., \& {Wagner}, S.~J.
  2005{\natexlab{b}}, \aap, 430, 865

\bibitem[{{Aharonian} {et~al.}(2007{\natexlab{a}}){Aharonian}, {Akhperjanian},
  {Barres de Almeida}, {Bazer-Bachi}, {Behera}, {Beilicke}, {Benbow},
  {Bernl{\"o}hr}, {Boisson}, {Bolz}, {Borrel}, {Braun}, {Brion}, {Brown},
  {B{\"u}hler}, {Bulik}, {B{\"u}sching}, {Boutelier}, {Carrigan}, {Chadwick},
  {Chounet}, {Clapson}, {Coignet}, {Cornils}, {Costamante}, {Dalton},
  {Degrange}, {Dickinson}, {Djannati-Ata{\"\i}}, {Domainko}, {O'C.~Drury},
  {Dubois}, {Dubus}, {Dyks}, {Egberts}, {Emmanoulopoulos}, {Espigat},
  {Farnier}, {Feinstein}, {Fiasson}, {F{\"o}rster}, {Fontaine}, {Funk},
  {F{\"u}{\ss}ling}, {Gallant}, {Giebels}, {Glicenstein}, {Gl{\"u}ck}, {Goret},
  {Hadjichristidis}, {Hauser}, {Hauser}, {Heinzelmann}, {Henri}, {Hermann},
  {Hinton}, {Hoffmann}, {Hofmann}, {Holleran}, {Hoppe}, {Horns},
  {Jacholkowska}, {de Jager}, {Jung}, {Katarzy{\'n}ski}, {Kendziorra},
  {Kerschhaggl}, {Kh{\'e}lifi}, {Keogh}, {Komin}, {Kosack}, {Lamanna},
  {Latham}, {Lemi{\`e}re}, {Lemoine-Goumard}, {Lenain}, {Lohse}, {Martin},
  {Martineau-Huynh}, {Marcowith}, {Masterson}, {Maurin}, {Maurin}, {McComb},
  {Moderski}, {Moulin}, {de Naurois}, {Nedbal}, {Nolan}, {Ohm}, {Olive}, {de
  O{\~n}a Wilhelmi}, {Orford}, {Osborne}, {Ostrowski}, {Panter}, {Pedaletti},
  {Pelletier}, {Petrucci}, {Pita}, {P{\"u}hlhofer}, {Punch}, {Ranchon},
  {Raubenheimer}, {Raue}, {Rayner}, {Renaud}, {Ripken}, {Rob}, {Rolland},
  {Rosier-Lees}, {Rowell}, {Rudak}, {Ruppel}, {Sahakian}, {Santangelo},
  {Schlickeiser}, {Sch{\"o}ck}, {Schr{\"o}der}, {Schwanke}, {Schwarzburg},
  {Schwemmer}, {Shalchi}, {Sol}, {Spangler}, {Stawarz}, {Steenkamp},
  {Stegmann}, {Superina}, {Tam}, {Tavernet}, {Terrier}, {van Eldik},
  {Vasileiadis}, {Venter}, {Vialle}, {Vincent}, {Vivier}, {V{\"o}lk}, {Volpe},
  {Wagner}, {Ward}, {Zdziarski}, \& {Zech}}]{aharonian07}
{Aharonian}, F., {Akhperjanian}, A.~G., {Barres de Almeida}, U., {Bazer-Bachi},
  A.~R., {Behera}, B., {Beilicke}, M., {Benbow}, W., {Bernl{\"o}hr}, K.,
  {Boisson}, C., {Bolz}, O., {Borrel}, V., {Braun}, I., {Brion}, E., {Brown},
  A.~M., {B{\"u}hler}, R., {Bulik}, T., {B{\"u}sching}, I., {Boutelier}, T.,
  {Carrigan}, S., {Chadwick}, P.~M., {Chounet}, L.-M., {Clapson}, A.~C.,
  {Coignet}, G., {Cornils}, R., {Costamante}, L., {Dalton}, M., {Degrange}, B.,
  {Dickinson}, H.~J., {Djannati-Ata{\"\i}}, A., {Domainko}, W., {O'C.~Drury},
  L., {Dubois}, F., {Dubus}, G., {Dyks}, J., {Egberts}, K., {Emmanoulopoulos},
  D., {Espigat}, P., {Farnier}, C., {Feinstein}, F., {Fiasson}, A.,
  {F{\"o}rster}, A., {Fontaine}, G., {Funk}, S., {F{\"u}{\ss}ling}, M.,
  {Gallant}, Y.~A., {Giebels}, B., {Glicenstein}, J.~F., {Gl{\"u}ck}, B.,
  {Goret}, P., {Hadjichristidis}, C., {Hauser}, D., {Hauser}, M.,
  {Heinzelmann}, G., {Henri}, G., {Hermann}, G., {Hinton}, J.~A., {Hoffmann},
  A., {Hofmann}, W., {Holleran}, M., {Hoppe}, S., {Horns}, D., {Jacholkowska},
  A., {de Jager}, O.~C., {Jung}, I., {Katarzy{\'n}ski}, K., {Kendziorra}, E.,
  {Kerschhaggl}, M., {Kh{\'e}lifi}, B., {Keogh}, D., {Komin}, N., {Kosack}, K.,
  {Lamanna}, G., {Latham}, I.~J., {Lemi{\`e}re}, A., {Lemoine-Goumard}, M.,
  {Lenain}, J.-P., {Lohse}, T., {Martin}, J.~M., {Martineau-Huynh}, O.,
  {Marcowith}, A., {Masterson}, C., {Maurin}, D., {Maurin}, G., {McComb},
  T.~J.~L., {Moderski}, R., {Moulin}, E., {de Naurois}, M., {Nedbal}, D.,
  {Nolan}, S.~J., {Ohm}, S., {Olive}, J.-P., {de O{\~n}a Wilhelmi}, E.,
  {Orford}, K.~J., {Osborne}, J.~L., {Ostrowski}, M., {Panter}, M.,
  {Pedaletti}, G., {Pelletier}, G., {Petrucci}, P.-O., {Pita}, S.,
  {P{\"u}hlhofer}, G., {Punch}, M., {Ranchon}, S., {Raubenheimer}, B.~C.,
  {Raue}, M., {Rayner}, S.~M., {Renaud}, M., {Ripken}, J., {Rob}, L.,
  {Rolland}, L., {Rosier-Lees}, S., {Rowell}, G., {Rudak}, B., {Ruppel}, J.,
  {Sahakian}, V., {Santangelo}, A., {Schlickeiser}, R., {Sch{\"o}ck}, F.,
  {Schr{\"o}der}, R., {Schwanke}, U., {Schwarzburg}, S., {Schwemmer}, S.,
  {Shalchi}, A., {Sol}, H., {Spangler}, D., {Stawarz}, {\L}., {Steenkamp}, R.,
  {Stegmann}, C., {Superina}, G., {Tam}, P.~H., {Tavernet}, J.-P., {Terrier},
  R., {van Eldik}, C., {Vasileiadis}, G., {Venter}, C., {Vialle}, J.~P.,
  {Vincent}, P., {Vivier}, M., {V{\"o}lk}, H.~J., {Volpe}, F., {Wagner}, S.~J.,
  {Ward}, M., {Zdziarski}, A.~A., \& {Zech}, A. 2007{\natexlab{a}}, \aap, 475,
  L9

\bibitem[{{Aharonian} {et~al.}(1999){Aharonian}, {Akhperjanian}, {Barrio},
  {Bernl{\"o}hr}, {Bojahr}, {Calle}, {Contreras}, {Cortina}, {Daum}, {Deckers},
  {Denninghoff}, {Fonseca}, {Gebauer}, {Gonzalez}, {Heinzelmann}, {Hemberger},
  {Hermann}, {Hess}, {Heusler}, {Hofmann}, {Hohl}, {Horns}, {Ibarra},
  {Kankanyan}, {Kestel}, {Kirstein}, {K{\"o}hler}, {Kornmayer}, {Kranich},
  {Krawczynski}, {Lampeitl}, {Lindner}, {Lorenz}, {Magnussen}, {Meyer},
  {Mirzoyan}, {Moralejo}, {Padilla}, {Panter}, {Petry}, {Plaga},
  {Plyasheshnikov}, {Prahl}, {P{\"u}hlhofer}, {Rauterberg}, {Renault}, {Rhode},
  {R{\"o}hring}, {Sahakian}, {Samorski}, {Schmele}, {Schr{\"o}der}, {Stamm},
  {Wiebel-Sooth}, {Wiedner}, {Willmer}, {Wirth}, \& {Wittek}}]{aharonian99b}
{Aharonian}, F., {Akhperjanian}, A.~G., {Barrio}, J.~A., {Bernl{\"o}hr}, K.,
  {Bojahr}, H., {Calle}, I., {Contreras}, J.~L., {Cortina}, J., {Daum}, A.,
  {Deckers}, T., {Denninghoff}, S., {Fonseca}, V., {Gebauer}, J., {Gonzalez},
  J.~C., {Heinzelmann}, G., {Hemberger}, M., {Hermann}, G., {Hess}, M.,
  {Heusler}, A., {Hofmann}, W., {Hohl}, H., {Horns}, D., {Ibarra}, A.,
  {Kankanyan}, R., {Kestel}, M., {Kirstein}, O., {K{\"o}hler}, C., {Kornmayer},
  H., {Kranich}, D., {Krawczynski}, H., {Lampeitl}, H., {Lindner}, A.,
  {Lorenz}, E., {Magnussen}, N., {Meyer}, H., {Mirzoyan}, R., {Moralejo}, A.,
  {Padilla}, L., {Panter}, M., {Petry}, D., {Plaga}, R., {Plyasheshnikov}, A.,
  {Prahl}, J., {P{\"u}hlhofer}, G., {Rauterberg}, G., {Renault}, C., {Rhode},
  W., {R{\"o}hring}, A., {Sahakian}, V., {Samorski}, M., {Schmele}, D.,
  {Schr{\"o}der}, F., {Stamm}, W., {Wiebel-Sooth}, B., {Wiedner}, C.,
  {Willmer}, M., {Wirth}, H., \& {Wittek}, W. 1999, \aap, 349, 29

\bibitem[{{Aharonian} {et~al.}(2007{\natexlab{b}}){Aharonian}, {Akhperjanian},
  {Bazer-Bachi}, {Behera}, {Beilicke}, {Benbow}, {Berge}, {Bernl{\"o}hr},
  {Boisson}, {Bolz}, {Borrel}, {Boutelier}, {Braun}, {Brion}, {Brown},
  {B{\"u}hler}, {B{\"u}sching}, {Bulik}, {Carrigan}, {Chadwick}, {Clapson},
  {Chounet}, {Coignet}, {Cornils}, {Costamante}, {Degrange}, {Dickinson},
  {Djannati-Ata{\"\i}}, {Domainko}, {Drury}, {Dubus}, {Dyks}, {Egberts},
  {Emmanoulopoulos}, {Espigat}, {Farnier}, {Feinstein}, {Fiasson},
  {F{\"o}rster}, {Fontaine}, {Funk}, {Funk}, {F{\"u}{\ss}ling}, {Gallant},
  {Giebels}, {Glicenstein}, {Gl{\"u}ck}, {Goret}, {Hadjichristidis}, {Hauser},
  {Hauser}, {Heinzelmann}, {Henri}, {Hermann}, {Hinton}, {Hoffmann}, {Hofmann},
  {Holleran}, {Hoppe}, {Horns}, {Jacholkowska}, {de Jager}, {Kendziorra},
  {Kerschhaggl}, {Kh{\'e}lifi}, {Komin}, {Kosack}, {Lamanna}, {Latham}, {Le
  Gallou}, {Lemi{\`e}re}, {Lemoine-Goumard}, {Lenain}, {Lohse}, {Martin},
  {Martineau-Huynh}, {Marcowith}, {Masterson}, {Maurin}, {McComb}, {Moderski},
  {Moulin}, {de Naurois}, {Nedbal}, {Nolan}, {Olive}, {Orford}, {Osborne},
  {Ostrowski}, {Panter}, {Pedaletti}, {Pelletier}, {Petrucci}, {Pita},
  {P{\"u}hlhofer}, {Punch}, {Ranchon}, {Raubenheimer}, {Raue}, {Rayner},
  {Renaud}, {Ripken}, {Rob}, {Rolland}, {Rosier-Lees}, {Rowell}, {Rudak},
  {Ruppel}, {Sahakian}, {Santangelo}, {Saug{\'e}}, {Schlenker}, {Schlickeiser},
  {Schr{\"o}der}, {Schwanke}, {Schwarzburg}, {Schwemmer}, {Shalchi}, {Sol},
  {Spangler}, {Stawarz}, {Steenkamp}, {Stegmann}, {Superina}, {Tam},
  {Tavernet}, {Terrier}, {van Eldik}, {Vasileiadis}, {Venter}, {Vialle},
  {Vincent}, {Vivier}, {V{\"o}lk}, {Volpe}, {Wagner}, {Ward}, \&
  {Zdziarski}}]{aharonian07b}
{Aharonian}, F., {Akhperjanian}, A.~G., {Bazer-Bachi}, A.~R., {Behera}, B.,
  {Beilicke}, M., {Benbow}, W., {Berge}, D., {Bernl{\"o}hr}, K., {Boisson}, C.,
  {Bolz}, O., {Borrel}, V., {Boutelier}, T., {Braun}, I., {Brion}, E., {Brown},
  A.~M., {B{\"u}hler}, R., {B{\"u}sching}, I., {Bulik}, T., {Carrigan}, S.,
  {Chadwick}, P.~M., {Clapson}, A.~C., {Chounet}, L.-M., {Coignet}, G.,
  {Cornils}, R., {Costamante}, L., {Degrange}, B., {Dickinson}, H.~J.,
  {Djannati-Ata{\"\i}}, A., {Domainko}, W., {Drury}, L.~O., {Dubus}, G.,
  {Dyks}, J., {Egberts}, K., {Emmanoulopoulos}, D., {Espigat}, P., {Farnier},
  C., {Feinstein}, F., {Fiasson}, A., {F{\"o}rster}, A., {Fontaine}, G.,
  {Funk}, S., {Funk}, S., {F{\"u}{\ss}ling}, M., {Gallant}, Y.~A., {Giebels},
  B., {Glicenstein}, J.~F., {Gl{\"u}ck}, B., {Goret}, P., {Hadjichristidis},
  C., {Hauser}, D., {Hauser}, M., {Heinzelmann}, G., {Henri}, G., {Hermann},
  G., {Hinton}, J.~A., {Hoffmann}, A., {Hofmann}, W., {Holleran}, M., {Hoppe},
  S., {Horns}, D., {Jacholkowska}, A., {de Jager}, O.~C., {Kendziorra}, E.,
  {Kerschhaggl}, M., {Kh{\'e}lifi}, B., {Komin}, N., {Kosack}, K., {Lamanna},
  G., {Latham}, I.~J., {Le Gallou}, R., {Lemi{\`e}re}, A., {Lemoine-Goumard},
  M., {Lenain}, J.-P., {Lohse}, T., {Martin}, J.~M., {Martineau-Huynh}, O.,
  {Marcowith}, A., {Masterson}, C., {Maurin}, G., {McComb}, T.~J.~L.,
  {Moderski}, R., {Moulin}, E., {de Naurois}, M., {Nedbal}, D., {Nolan}, S.~J.,
  {Olive}, J.-P., {Orford}, K.~J., {Osborne}, J.~L., {Ostrowski}, M., {Panter},
  M., {Pedaletti}, G., {Pelletier}, G., {Petrucci}, P.-O., {Pita}, S.,
  {P{\"u}hlhofer}, G., {Punch}, M., {Ranchon}, S., {Raubenheimer}, B.~C.,
  {Raue}, M., {Rayner}, S.~M., {Renaud}, M., {Ripken}, J., {Rob}, L.,
  {Rolland}, L., {Rosier-Lees}, S., {Rowell}, G., {Rudak}, B., {Ruppel}, J.,
  {Sahakian}, V., {Santangelo}, A., {Saug{\'e}}, L., {Schlenker}, S.,
  {Schlickeiser}, R., {Schr{\"o}der}, R., {Schwanke}, U., {Schwarzburg}, S.,
  {Schwemmer}, S., {Shalchi}, A., {Sol}, H., {Spangler}, D., {Stawarz}, {\L}.,
  {Steenkamp}, R., {Stegmann}, C., {Superina}, G., {Tam}, P.~H., {Tavernet},
  J.-P., {Terrier}, R., {van Eldik}, C., {Vasileiadis}, G., {Venter}, C.,
  {Vialle}, J.~P., {Vincent}, P., {Vivier}, M., {V{\"o}lk}, H.~J., {Volpe}, F.,
  {Wagner}, S.~J., {Ward}, M., \& {Zdziarski}, A.~A. 2007{\natexlab{b}}, \apjl,
  664, L71

\bibitem[{{Aharonian} {et~al.}(2006{\natexlab{a}}){Aharonian}, {Akhperjanian},
  {Bazer-Bachi}, {Beilicke}, {Benbow}, {Berge}, {Bernl{\"o}hr}, {Boisson},
  {Bolz}, {Borrel}, {Braun}, {Breitling}, {Brown}, {Chadwick}, {Chounet},
  {Cornils}, {Costamante}, {Degrange}, {Dickinson}, {Djannati-Ata{\"\i}},
  {Drury}, {Dubus}, {Emmanoulopoulos}, {Espigat}, {Feinstein}, {Fontaine},
  {Fuchs}, {Funk}, {Gallant}, {Giebels}, {Gillessen}, {Glicenstein}, {Goret},
  {Hadjichristidis}, {Hauser}, {Hauser}, {Heinzelmann}, {Henri}, {Hermann},
  {Hinton}, {Hofmann}, {Holleran}, {Horns}, {Jacholkowska}, {de Jager},
  {Kh{\'e}lifi}, {Klages}, {Komin}, {Konopelko}, {Latham}, {Le Gallou},
  {Lemi{\`e}re}, {Lemoine-Goumard}, {Leroy}, {Lohse}, {Martin},
  {Martineau-Huynh}, {Marcowith}, {Masterson}, {McComb}, {de Naurois}, {Nolan},
  {Noutsos}, {Orford}, {Osborne}, {Ouchrif}, {Panter}, {Pelletier}, {Pita},
  {P{\"u}hlhofer}, {Punch}, {Raubenheimer}, {Raue}, {Raux}, {Rayner}, {Reimer},
  {Reimer}, {Ripken}, {Rob}, {Rolland}, {Rowell}, {Sahakian}, {Saug{\'e}},
  {Schlenker}, {Schlickeiser}, {Schuster}, {Schwanke}, {Siewert}, {Sol},
  {Spangler}, {Steenkamp}, {Stegmann}, {Tavernet}, {Terrier}, {Th{\'e}oret},
  {Tluczykont}, {van Eldik}, {Vasileiadis}, {Venter}, {Vincent}, {V{\"o}lk}, \&
  {Wagner}}]{aharonian06}
{Aharonian}, F., {Akhperjanian}, A.~G., {Bazer-Bachi}, A.~R., {Beilicke}, M.,
  {Benbow}, W., {Berge}, D., {Bernl{\"o}hr}, K., {Boisson}, C., {Bolz}, O.,
  {Borrel}, V., {Braun}, I., {Breitling}, F., {Brown}, A.~M., {Chadwick},
  P.~M., {Chounet}, L.-M., {Cornils}, R., {Costamante}, L., {Degrange}, B.,
  {Dickinson}, H.~J., {Djannati-Ata{\"\i}}, A., {Drury}, L.~O., {Dubus}, G.,
  {Emmanoulopoulos}, D., {Espigat}, P., {Feinstein}, F., {Fontaine}, G.,
  {Fuchs}, Y., {Funk}, S., {Gallant}, Y.~A., {Giebels}, B., {Gillessen}, S.,
  {Glicenstein}, J.~F., {Goret}, P., {Hadjichristidis}, C., {Hauser}, D.,
  {Hauser}, M., {Heinzelmann}, G., {Henri}, G., {Hermann}, G., {Hinton}, J.~A.,
  {Hofmann}, W., {Holleran}, M., {Horns}, D., {Jacholkowska}, A., {de Jager},
  O.~C., {Kh{\'e}lifi}, B., {Klages}, S., {Komin}, N., {Konopelko}, A.,
  {Latham}, I.~J., {Le Gallou}, R., {Lemi{\`e}re}, A., {Lemoine-Goumard}, M.,
  {Leroy}, N., {Lohse}, T., {Martin}, J.~M., {Martineau-Huynh}, O.,
  {Marcowith}, A., {Masterson}, C., {McComb}, T.~J.~L., {de Naurois}, M.,
  {Nolan}, S.~J., {Noutsos}, A., {Orford}, K.~J., {Osborne}, J.~L., {Ouchrif},
  M., {Panter}, M., {Pelletier}, G., {Pita}, S., {P{\"u}hlhofer}, G., {Punch},
  M., {Raubenheimer}, B.~C., {Raue}, M., {Raux}, J., {Rayner}, S.~M., {Reimer},
  A., {Reimer}, O., {Ripken}, J., {Rob}, L., {Rolland}, L., {Rowell}, G.,
  {Sahakian}, V., {Saug{\'e}}, L., {Schlenker}, S., {Schlickeiser}, R.,
  {Schuster}, C., {Schwanke}, U., {Siewert}, M., {Sol}, H., {Spangler}, D.,
  {Steenkamp}, R., {Stegmann}, C., {Tavernet}, J.-P., {Terrier}, R.,
  {Th{\'e}oret}, C.~G., {Tluczykont}, M., {van Eldik}, C., {Vasileiadis}, G.,
  {Venter}, C., {Vincent}, P., {V{\"o}lk}, H.~J., \& {Wagner}, S.~J.
  2006{\natexlab{a}}, \nat, 440, 1018

\bibitem[{{Aharonian} {et~al.}(2006{\natexlab{b}}){Aharonian}, {Akhperjanian},
  {Bazer-Bachi}, {Beilicke}, {Benbow}, {Berge}, {Bernl{\"o}hr}, {Boisson},
  {Bolz}, {Borrel}, {Braun}, {Brown}, {B{\"u}hler}, {B{\"u}sching}, {Carrigan},
  {Chadwick}, {Chounet}, {Coignet}, {Cornils}, {Costamante}, {Degrange},
  {Dickinson}, {Djannati-Ata{\"\i}}, {Drury}, {Dubus}, {Egberts},
  {Emmanoulopoulos}, {Espigat}, {Feinstein}, {Ferrero}, {Fiasson}, {Fontaine},
  {Funk}, {Funk}, {F{\"u}{\ss}ling}, {Gallant}, {Giebels}, {Glicenstein},
  {Goret}, {Hadjichristidis}, {Hauser}, {Hauser}, {Heinzelmann}, {Henri},
  {Hermann}, {Hinton}, {Hoffmann}, {Hofmann}, {Holleran}, {Hoppe}, {Horns},
  {Jacholkowska}, {de Jager}, {Kendziorra}, {Kerschhaggl}, {Kh{\'e}lifi},
  {Komin}, {Konopelko}, {Kosack}, {Lamanna}, {Latham}, {Le Gallou},
  {Lemi{\`e}re}, {Lemoine-Goumard}, {Lenain}, {Lohse}, {Martin},
  {Martineau-Huynh}, {Marcowith}, {Masterson}, {Maurin}, {McComb}, {Moulin},
  {de Naurois}, {Nedbal}, {Nolan}, {Noutsos}, {Orford}, {Osborne}, {Ouchrif},
  {Panter}, {Pelletier}, {Pita}, {P{\"u}hlhofer}, {Punch}, {Ranchon},
  {Raubenheimer}, {Raue}, {Rayner}, {Reimer}, {Ripken}, {Rob}, {Rolland},
  {Rosier-Lees}, {Rowell}, {Sahakian}, {Santangelo}, {Saug{\'e}}, {Schlenker},
  {Schlickeiser}, {Schr{\"o}der}, {Schwanke}, {Schwarzburg}, {Schwemmer},
  {Shalchi}, {Sol}, {Spangler}, {Spanier}, {Steenkamp}, {Stegmann}, {Superina},
  {Tam}, {Tavernet}, {Terrier}, {Tluczykont}, {van Eldik}, {Vasileiadis},
  {Venter}, {Vialle}, {Vincent}, {V{\"o}lk}, {Wagner}, \&
  {Ward}}]{aharonian06a}
{Aharonian}, F., {Akhperjanian}, A.~G., {Bazer-Bachi}, A.~R., {Beilicke}, M.,
  {Benbow}, W., {Berge}, D., {Bernl{\"o}hr}, K., {Boisson}, C., {Bolz}, O.,
  {Borrel}, V., {Braun}, I., {Brown}, A.~M., {B{\"u}hler}, R., {B{\"u}sching},
  I., {Carrigan}, S., {Chadwick}, P.~M., {Chounet}, L.-M., {Coignet}, G.,
  {Cornils}, R., {Costamante}, L., {Degrange}, B., {Dickinson}, H.~J.,
  {Djannati-Ata{\"\i}}, A., {Drury}, L.~O., {Dubus}, G., {Egberts}, K.,
  {Emmanoulopoulos}, D., {Espigat}, P., {Feinstein}, F., {Ferrero}, E.,
  {Fiasson}, A., {Fontaine}, G., {Funk}, S., {Funk}, S., {F{\"u}{\ss}ling}, M.,
  {Gallant}, Y.~A., {Giebels}, B., {Glicenstein}, J.~F., {Goret}, P.,
  {Hadjichristidis}, C., {Hauser}, D., {Hauser}, M., {Heinzelmann}, G.,
  {Henri}, G., {Hermann}, G., {Hinton}, J.~A., {Hoffmann}, A., {Hofmann}, W.,
  {Holleran}, M., {Hoppe}, S., {Horns}, D., {Jacholkowska}, A., {de Jager},
  O.~C., {Kendziorra}, E., {Kerschhaggl}, M., {Kh{\'e}lifi}, B., {Komin}, N.,
  {Konopelko}, A., {Kosack}, K., {Lamanna}, G., {Latham}, I.~J., {Le Gallou},
  R., {Lemi{\`e}re}, A., {Lemoine-Goumard}, M., {Lenain}, J.-P., {Lohse}, T.,
  {Martin}, J.~M., {Martineau-Huynh}, O., {Marcowith}, A., {Masterson}, C.,
  {Maurin}, G., {McComb}, T.~J.~L., {Moulin}, E., {de Naurois}, M., {Nedbal},
  D., {Nolan}, S.~J., {Noutsos}, A., {Orford}, K.~J., {Osborne}, J.~L.,
  {Ouchrif}, M., {Panter}, M., {Pelletier}, G., {Pita}, S., {P{\"u}hlhofer},
  G., {Punch}, M., {Ranchon}, S., {Raubenheimer}, B.~C., {Raue}, M., {Rayner},
  S.~M., {Reimer}, A., {Ripken}, J., {Rob}, L., {Rolland}, L., {Rosier-Lees},
  S., {Rowell}, G., {Sahakian}, V., {Santangelo}, A., {Saug{\'e}}, L.,
  {Schlenker}, S., {Schlickeiser}, R., {Schr{\"o}der}, R., {Schwanke}, U.,
  {Schwarzburg}, S., {Schwemmer}, S., {Shalchi}, A., {Sol}, H., {Spangler}, D.,
  {Spanier}, F., {Steenkamp}, R., {Stegmann}, C., {Superina}, G., {Tam}, P.~H.,
  {Tavernet}, J.-P., {Terrier}, R., {Tluczykont}, M., {van Eldik}, C.,
  {Vasileiadis}, G., {Venter}, C., {Vialle}, J.~P., {Vincent}, P., {V{\"o}lk},
  H.~J., {Wagner}, S.~J., \& {Ward}, M. 2006{\natexlab{b}}, Science, 314, 1424

\bibitem[{{Aharonian} {et~al.}(2012){Aharonian}, {Essey}, {Kusenko}, \&
  {Prosekin}}]{aharonian12}
{Aharonian}, F., {Essey}, W., {Kusenko}, A., \& {Prosekin}, A. 2012, ArXiv
  e-prints

\bibitem[{{Aharonian} {et~al.}(2008){Aharonian}, {Khangulyan}, \&
  {Costamante}}]{aharonian08}
{Aharonian}, F.~A., {Khangulyan}, D., \& {Costamante}, L. 2008, \mnras, 387,
  1206

\bibitem[{{Albert} {et~al.}(2006{\natexlab{a}}){Albert}, {Aliu}, {Anderhub},
  {Antoranz}, {Armada}, {Asensio}, {Baixeras}, {Barrio}, {Bartko}, {Bastieri},
  {Becker}, {Bednarek}, {Berger}, {Bigongiari}, {Biland}, {Bisesi}, {Bock},
  {Bordas}, {Bosch-Ramon}, {Bretz}, {Britvitch}, {Camara}, {Carmona},
  {Chilingarian}, {Ciprini}, {Coarasa}, {Commichau}, {Contreras}, {Cortina},
  {Curtef}, {Danielyan}, {Dazzi}, {De Angelis}, {de los Reyes}, {De Lotto},
  {Domingo-Santamar{\'{\i}}a}, {Dorner}, {Doro}, {Errando}, {Fagiolini},
  {Ferenc}, {Fern{\'a}ndez}, {Firpo}, {Flix}, {Fonseca}, {Font}, {Fuchs},
  {Galante}, {Garczarczyk}, {Gaug}, {Giller}, {Goebel}, {Hakobyan},
  {Hayashida}, {Hengstebeck}, {H{\"o}hne}, {Hose}, {Hsu}, {Jacon}, {Kalekin},
  {Kosyra}, {Kranich}, {Laatiaoui}, {Laille}, {Lenisa}, {Liebing}, {Lindfors},
  {Lombardi}, {Longo}, {L{\'o}pez}, {L{\'o}pez}, {Lorenz}, {Majumdar},
  {Maneva}, {Mannheim}, {Mansutti}, {Mariotti}, {Mart{\'{\i}}nez}, {Mazin},
  {Merck}, {Meucci}, {Meyer}, {Miranda}, {Mirzoyan}, {Mizobuchi}, {Moralejo},
  {Nilsson}, {Ninkovic}, {O{\~n}a-Wilhelmi}, {Ordu{\~n}a}, {Otte}, {Oya},
  {Paneque}, {Paoletti}, {Paredes}, {Pasanen}, {Pascoli}, {Pauss}, {Pegna},
  {Persic}, {Peruzzo}, {Piccioli}, {Poller}, {Prandini}, {Raymers}, {Rhode},
  {Rib{\'o}}, {Rico}, {Riegel}, {Rissi}, {Robert}, {R{\"u}gamer}, {Saggion},
  {S{\'a}nchez}, {Sartori}, {Scalzotto}, {Scapin}, {Schmitt}, {Schweizer},
  {Shayduk}, {Shinozaki}, {Shore}, {Sidro}, {Sillanp{\"a}{\"a}}, {Sobczynska},
  {Stamerra}, {Stark}, {Takalo}, {Temnikov}, {Tescaro}, {Teshima}, {Tonello},
  {Torres}, {Torres}, {Turini}, {Vankov}, {Vitale}, {Wagner}, {Wibig},
  {Wittek}, {Zanin}, \& {Zapatero}}]{albert06a}
{Albert}, J., {Aliu}, E., {Anderhub}, H., {Antoranz}, P., {Armada}, A.,
  {Asensio}, M., {Baixeras}, C., {Barrio}, J.~A., {Bartko}, H., {Bastieri}, D.,
  {Becker}, J., {Bednarek}, W., {Berger}, K., {Bigongiari}, C., {Biland}, A.,
  {Bisesi}, E., {Bock}, R.~K., {Bordas}, P., {Bosch-Ramon}, V., {Bretz}, T.,
  {Britvitch}, I., {Camara}, M., {Carmona}, E., {Chilingarian}, A., {Ciprini},
  S., {Coarasa}, J.~A., {Commichau}, S., {Contreras}, J.~L., {Cortina}, J.,
  {Curtef}, V., {Danielyan}, V., {Dazzi}, F., {De Angelis}, A., {de los Reyes},
  R., {De Lotto}, B., {Domingo-Santamar{\'{\i}}a}, E., {Dorner}, D., {Doro},
  M., {Errando}, M., {Fagiolini}, M., {Ferenc}, D., {Fern{\'a}ndez}, E.,
  {Firpo}, R., {Flix}, J., {Fonseca}, M.~V., {Font}, L., {Fuchs}, M.,
  {Galante}, N., {Garczarczyk}, M., {Gaug}, M., {Giller}, M., {Goebel}, F.,
  {Hakobyan}, D., {Hayashida}, M., {Hengstebeck}, T., {H{\"o}hne}, D., {Hose},
  J., {Hsu}, C.~C., {Jacon}, P., {Kalekin}, O., {Kosyra}, R., {Kranich}, D.,
  {Laatiaoui}, M., {Laille}, A., {Lenisa}, T., {Liebing}, P., {Lindfors}, E.,
  {Lombardi}, S., {Longo}, F., {L{\'o}pez}, J., {L{\'o}pez}, M., {Lorenz}, E.,
  {Majumdar}, P., {Maneva}, G., {Mannheim}, K., {Mansutti}, O., {Mariotti}, M.,
  {Mart{\'{\i}}nez}, M., {Mazin}, D., {Merck}, C., {Meucci}, M., {Meyer}, M.,
  {Miranda}, J.~M., {Mirzoyan}, R., {Mizobuchi}, S., {Moralejo}, A., {Nilsson},
  K., {Ninkovic}, J., {O{\~n}a-Wilhelmi}, E., {Ordu{\~n}a}, R., {Otte}, N.,
  {Oya}, I., {Paneque}, D., {Paoletti}, R., {Paredes}, J.~M., {Pasanen}, M.,
  {Pascoli}, D., {Pauss}, F., {Pegna}, R., {Persic}, M., {Peruzzo}, L.,
  {Piccioli}, A., {Poller}, M., {Prandini}, E., {Raymers}, A., {Rhode}, W.,
  {Rib{\'o}}, M., {Rico}, J., {Riegel}, B., {Rissi}, M., {Robert}, A.,
  {R{\"u}gamer}, S., {Saggion}, A., {S{\'a}nchez}, A., {Sartori}, P.,
  {Scalzotto}, V., {Scapin}, V., {Schmitt}, R., {Schweizer}, T., {Shayduk}, M.,
  {Shinozaki}, K., {Shore}, S.~N., {Sidro}, N., {Sillanp{\"a}{\"a}}, A.,
  {Sobczynska}, D., {Stamerra}, A., {Stark}, L.~S., {Takalo}, L., {Temnikov},
  P., {Tescaro}, D., {Teshima}, M., {Tonello}, N., {Torres}, A., {Torres},
  D.~F., {Turini}, N., {Vankov}, H., {Vitale}, V., {Wagner}, R.~M., {Wibig},
  T., {Wittek}, W., {Zanin}, R., \& {Zapatero}, J. 2006{\natexlab{a}}, \apjl,
  648, L105

\bibitem[{{Albert} {et~al.}(2006{\natexlab{b}}){Albert}, {Aliu}, {Anderhub},
  {Antoranz}, {Armada}, {Asensio}, {Baixeras}, {Barrio}, {Bartko}, {Bastieri},
  {Bednarek}, {Berger}, {Bigongiari}, {Biland}, {Bisesi}, {Bock}, {Bretz},
  {Britvitch}, {Camara}, {Chilingarian}, {Ciprini}, {Coarasa}, {Commichau},
  {Contreras}, {Cortina}, {Danielyan}, {Dazzi}, {De Angelis}, {de los Reyes},
  {De Lotto}, {Domingo-Santamar{\'{\i}}a}, {Dorner}, {Doro}, {Errando},
  {Ferenc}, {Fern{\'a}ndez}, {Firpo}, {Flix}, {Fonseca}, {Font}, {Galante},
  {Garczarczyk}, {Gaug}, {Gebauer}, {Giannitrapani}, {Giller}, {Goebel},
  {Hakobyan}, {Hayashida}, {Hengstebeck}, {H{\"o}hne}, {Hose}, {Jacon},
  {Kalekin}, {Kranich}, {Laille}, {Lenisa}, {Liebing}, {Lindfors}, {Longo},
  {L{\'o}pez}, {L{\'o}pez}, {Lorenz}, {Lucarelli}, {Majumdar}, {Maneva},
  {Mannheim}, {Mariotti}, {Mart{\'{\i}}nez}, {Mase}, {Mazin}, {Merck}, {Merck},
  {Meucci}, {Meyer}, {Miranda}, {Mirzoyan}, {Mizobuchi}, {Moralejo}, {Nilsson},
  {O{\~n}a-Wilhelmi}, {Ordu{\~n}a}, {Otte}, {Oya}, {Paneque}, {Paoletti},
  {Pasanen}, {Pascoli}, {Pauss}, {Pavel}, {Pegna}, {Peruzzo}, {Piccioli},
  {Pin}, {Prandini}, {Rico}, {Rhode}, {Riegel}, {Rissi}, {Robert}, {Rossato},
  {R{\"u}gamer}, {Saggion}, {S{\'a}nchez}, {Sartori}, {Scalzotto}, {Schmitt},
  {Schweizer}, {Shayduk}, {Shinozaki}, {Sidro}, {Sillanp{\"a}{\"a}},
  {Sobczynska}, {Stamerra}, {Stark}, {Takalo}, {Temnikov}, {Tescaro},
  {Teshima}, {Tonello}, {Torres}, {Torres}, {Turini}, {Vankov}, {Vitale},
  {Wagner}, {Wibig}, {Wittek}, \& {Zapatero}}]{albert06b}
{Albert}, J., {Aliu}, E., {Anderhub}, H., {Antoranz}, P., {Armada}, A.,
  {Asensio}, M., {Baixeras}, C., {Barrio}, J.~A., {Bartko}, H., {Bastieri}, D.,
  {Bednarek}, W., {Berger}, K., {Bigongiari}, C., {Biland}, A., {Bisesi}, E.,
  {Bock}, R.~K., {Bretz}, T., {Britvitch}, I., {Camara}, M., {Chilingarian},
  A., {Ciprini}, S., {Coarasa}, J.~A., {Commichau}, S., {Contreras}, J.~L.,
  {Cortina}, J., {Danielyan}, V., {Dazzi}, F., {De Angelis}, A., {de los
  Reyes}, R., {De Lotto}, B., {Domingo-Santamar{\'{\i}}a}, E., {Dorner}, D.,
  {Doro}, M., {Errando}, M., {Ferenc}, D., {Fern{\'a}ndez}, E., {Firpo}, R.,
  {Flix}, J., {Fonseca}, M.~V., {Font}, L., {Galante}, N., {Garczarczyk}, M.,
  {Gaug}, M., {Gebauer}, J., {Giannitrapani}, R., {Giller}, M., {Goebel}, F.,
  {Hakobyan}, D., {Hayashida}, M., {Hengstebeck}, T., {H{\"o}hne}, D., {Hose},
  J., {Jacon}, P., {Kalekin}, O., {Kranich}, D., {Laille}, A., {Lenisa}, T.,
  {Liebing}, P., {Lindfors}, E., {Longo}, F., {L{\'o}pez}, J., {L{\'o}pez}, M.,
  {Lorenz}, E., {Lucarelli}, F., {Majumdar}, P., {Maneva}, G., {Mannheim}, K.,
  {Mariotti}, M., {Mart{\'{\i}}nez}, M., {Mase}, K., {Mazin}, D., {Merck}, C.,
  {Merck}, M., {Meucci}, M., {Meyer}, M., {Miranda}, J.~M., {Mirzoyan}, R.,
  {Mizobuchi}, S., {Moralejo}, A., {Nilsson}, K., {O{\~n}a-Wilhelmi}, E.,
  {Ordu{\~n}a}, R., {Otte}, N., {Oya}, I., {Paneque}, D., {Paoletti}, R.,
  {Pasanen}, M., {Pascoli}, D., {Pauss}, F., {Pavel}, N., {Pegna}, R.,
  {Peruzzo}, L., {Piccioli}, A., {Pin}, M., {Prandini}, E., {Rico}, J.,
  {Rhode}, W., {Riegel}, B., {Rissi}, M., {Robert}, A., {Rossato}, G.,
  {R{\"u}gamer}, S., {Saggion}, A., {S{\'a}nchez}, A., {Sartori}, P.,
  {Scalzotto}, V., {Schmitt}, R., {Schweizer}, T., {Shayduk}, M., {Shinozaki},
  K., {Sidro}, N., {Sillanp{\"a}{\"a}}, A., {Sobczynska}, D., {Stamerra}, A.,
  {Stark}, L., {Takalo}, L., {Temnikov}, P., {Tescaro}, D., {Teshima}, M.,
  {Tonello}, N., {Torres}, A., {Torres}, D.~F., {Turini}, N., {Vankov}, H.,
  {Vitale}, V., {Wagner}, R.~M., {Wibig}, T., {Wittek}, W., \& {Zapatero}, J.
  2006{\natexlab{b}}, \apj, 639, 761

\bibitem[{{Albert} {et~al.}(2007{\natexlab{a}}){Albert}, {Aliu}, {Anderhub},
  {Antoranz}, {Armada}, {Baixeras}, {Barrio}, {Bartko}, {Bastieri}, {Becker},
  {Bednarek}, {Berger}, {Bigongiari}, {Biland}, {Bock}, {Bordas},
  {Bosch-Ramon}, {Bretz}, {Britvitch}, {Camara}, {Carmona}, {Chilingarian},
  {Coarasa}, {Commichau}, {Contreras}, {Cortina}, {Costado}, {Curtef},
  {Danielyan}, {Dazzi}, {De Angelis}, {Delgado}, {de los Reyes}, {De Lotto},
  {Domingo-Santamar{\'{\i}}a}, {Dorner}, {Doro}, {Errando}, {Fagiolini},
  {Ferenc}, {Fern{\'a}ndez}, {Firpo}, {Flix}, {Fonseca}, {Font}, {Fuchs},
  {Galante}, {Garc{\'{\i}}a-L{\'o}pez}, {Garczarczyk}, {Gaug}, {Giller},
  {Goebel}, {Hakobyan}, {Hayashida}, {Hengstebeck}, {Herrero}, {H{\"o}hne},
  {Hose}, {Hsu}, {Jacon}, {Jogler}, {Kosyra}, {Kranich}, {Kritzer}, {Laille},
  {Lindfors}, {Lombardi}, {Longo}, {L{\'o}pez}, {L{\'o}pez}, {Lorenz},
  {Majumdar}, {Maneva}, {Mannheim}, {Mansutti}, {Mariotti}, {Mart{\'{\i}}nez},
  {Mazin}, {Merck}, {Meucci}, {Meyer}, {Miranda}, {Mirzoyan}, {Mizobuchi},
  {Moralejo}, {Nilsson}, {Ninkovic}, {O{\~n}a-Wilhelmi}, {Otte}, {Oya},
  {Paneque}, {Panniello}, {Paoletti}, {Paredes}, {Pasanen}, {Pascoli}, {Pauss},
  {Pegna}, {Persic}, {Peruzzo}, {Piccioli}, {Poller}, {Prandini}, {Puchades},
  {Raymers}, {Rhode}, {Rib{\'o}}, {Rico}, {Rissi}, {Robert}, {R{\"u}gamer},
  {Saggion}, {S{\'a}nchez}, {Sartori}, {Scalzotto}, {Scapin}, {Schmitt},
  {Schweizer}, {Shayduk}, {Shinozaki}, {Shore}, {Sidro}, {Sillanp{\"a}{\"a}},
  {Sobczynska}, {Stamerra}, {Stark}, {Takalo}, {Temnikov}, {Tescaro},
  {Teshima}, {Tonello}, {Torres}, {Turini}, {Vankov}, {Vitale}, {Wagner},
  {Wibig}, {Wittek}, {Zandanel}, {Zanin}, \& {Zapatero}}]{albert07}
{Albert}, J., {Aliu}, E., {Anderhub}, H., {Antoranz}, P., {Armada}, A.,
  {Baixeras}, C., {Barrio}, J.~A., {Bartko}, H., {Bastieri}, D., {Becker},
  J.~K., {Bednarek}, W., {Berger}, K., {Bigongiari}, C., {Biland}, A., {Bock},
  R.~K., {Bordas}, P., {Bosch-Ramon}, V., {Bretz}, T., {Britvitch}, I.,
  {Camara}, M., {Carmona}, E., {Chilingarian}, A., {Coarasa}, J.~A.,
  {Commichau}, S., {Contreras}, J.~L., {Cortina}, J., {Costado}, M.~T.,
  {Curtef}, V., {Danielyan}, V., {Dazzi}, F., {De Angelis}, A., {Delgado}, C.,
  {de los Reyes}, R., {De Lotto}, B., {Domingo-Santamar{\'{\i}}a}, E.,
  {Dorner}, D., {Doro}, M., {Errando}, M., {Fagiolini}, M., {Ferenc}, D.,
  {Fern{\'a}ndez}, E., {Firpo}, R., {Flix}, J., {Fonseca}, M.~V., {Font}, L.,
  {Fuchs}, M., {Galante}, N., {Garc{\'{\i}}a-L{\'o}pez}, R., {Garczarczyk}, M.,
  {Gaug}, M., {Giller}, M., {Goebel}, F., {Hakobyan}, D., {Hayashida}, M.,
  {Hengstebeck}, T., {Herrero}, A., {H{\"o}hne}, D., {Hose}, J., {Hsu}, C.~C.,
  {Jacon}, P., {Jogler}, T., {Kosyra}, R., {Kranich}, D., {Kritzer}, R.,
  {Laille}, A., {Lindfors}, E., {Lombardi}, S., {Longo}, F., {L{\'o}pez}, J.,
  {L{\'o}pez}, M., {Lorenz}, E., {Majumdar}, P., {Maneva}, G., {Mannheim}, K.,
  {Mansutti}, O., {Mariotti}, M., {Mart{\'{\i}}nez}, M., {Mazin}, D., {Merck},
  C., {Meucci}, M., {Meyer}, M., {Miranda}, J.~M., {Mirzoyan}, R., {Mizobuchi},
  S., {Moralejo}, A., {Nilsson}, K., {Ninkovic}, J., {O{\~n}a-Wilhelmi}, E.,
  {Otte}, N., {Oya}, I., {Paneque}, D., {Panniello}, M., {Paoletti}, R.,
  {Paredes}, J.~M., {Pasanen}, M., {Pascoli}, D., {Pauss}, F., {Pegna}, R.,
  {Persic}, M., {Peruzzo}, L., {Piccioli}, A., {Poller}, M., {Prandini}, E.,
  {Puchades}, N., {Raymers}, A., {Rhode}, W., {Rib{\'o}}, M., {Rico}, J.,
  {Rissi}, M., {Robert}, A., {R{\"u}gamer}, S., {Saggion}, A., {S{\'a}nchez},
  A., {Sartori}, P., {Scalzotto}, V., {Scapin}, V., {Schmitt}, R., {Schweizer},
  T., {Shayduk}, M., {Shinozaki}, K., {Shore}, S.~N., {Sidro}, N.,
  {Sillanp{\"a}{\"a}}, A., {Sobczynska}, D., {Stamerra}, A., {Stark}, L.~S.,
  {Takalo}, L., {Temnikov}, P., {Tescaro}, D., {Teshima}, M., {Tonello}, N.,
  {Torres}, D.~F., {Turini}, N., {Vankov}, H., {Vitale}, V., {Wagner}, R.~M.,
  {Wibig}, T., {Wittek}, W., {Zandanel}, F., {Zanin}, R., \& {Zapatero}, J.
  2007{\natexlab{a}}, \apjl, 666, L17

\bibitem[{{Albert} {et~al.}(2007{\natexlab{b}}){Albert}, {Aliu}, {Anderhub},
  {Antoranz}, {Armada}, {Baixeras}, {Barrio}, {Bartko}, {Bastieri}, {Becker},
  {Bednarek}, {Berger}, {Bigongiari}, {Biland}, {Bock}, {Bordas},
  {Bosch-Ramon}, {Bretz}, {Britvitch}, {Camara}, {Carmona}, {Chilingarian},
  {Coarasa}, {Commichau}, {Contreras}, {Cortina}, {Costado}, {Curtef},
  {Danielyan}, {Dazzi}, {De Angelis}, {Delgado}, {de los Reyes}, {De Lotto},
  {Domingo-Santamar{\'{\i}}a}, {Dorner}, {Doro}, {Errando}, {Fagiolini},
  {Ferenc}, {Fern{\'a}ndez}, {Firpo}, {Flix}, {Fonseca}, {Font}, {Fuchs},
  {Galante}, {Garc{\'{\i}}a-L{\'o}pez}, {Garczarczyk}, {Gaug}, {Giller},
  {Goebel}, {Hakobyan}, {Hayashida}, {Hengstebeck}, {Herrero}, {H{\"o}hne},
  {Hose}, {Hsu}, {Jacon}, {Jogler}, {Kosyra}, {Kranich}, {Kritzer}, {Laille},
  {Lindfors}, {Lombardi}, {Longo}, {L{\'o}pez}, {L{\'o}pez}, {Lorenz},
  {Majumdar}, {Maneva}, {Mannheim}, {Mansutti}, {Mariotti}, {Mart{\'{\i}}nez},
  {Mazin}, {Merck}, {Meucci}, {Meyer}, {Miranda}, {Mirzoyan}, {Mizobuchi},
  {Moralejo}, {Nieto}, {Nilsson}, {Ninkovic}, {O{\~n}a-Wilhelmi}, {Otte},
  {Oya}, {Paneque}, {Panniello}, {Paoletti}, {Paredes}, {Pasanen}, {Pascoli},
  {Pauss}, {Pegna}, {Perlman}, {Persic}, {Peruzzo}, {Piccioli}, {Prandini},
  {Puchades}, {Raymers}, {Rhode}, {Rib{\'o}}, {Rico}, {Rissi}, {Robert},
  {R{\"u}gamer}, {Saggion}, {Saito}, {S{\'a}nchez}, {Sartori}, {Scalzotto},
  {Scapin}, {Schmitt}, {Schweizer}, {Shayduk}, {Shinozaki}, {Shore}, {Sidro},
  {Sillanp{\"a}{\"a}}, {Sobczynska}, {Stamerra}, {Stark}, {Takalo},
  {Tavecchio}, {Temnikov}, {Tescaro}, {Teshima}, {Torres}, {Turini}, {Vankov},
  {Vitale}, {Wagner}, {Wibig}, {Wittek}, {Zandanel}, {Zanin}, \&
  {Zapatero}}]{albert07b}
{Albert}, J., {Aliu}, E., {Anderhub}, H., {Antoranz}, P., {Armada}, A.,
  {Baixeras}, C., {Barrio}, J.~A., {Bartko}, H., {Bastieri}, D., {Becker},
  J.~K., {Bednarek}, W., {Berger}, K., {Bigongiari}, C., {Biland}, A., {Bock},
  R.~K., {Bordas}, P., {Bosch-Ramon}, V., {Bretz}, T., {Britvitch}, I.,
  {Camara}, M., {Carmona}, E., {Chilingarian}, A., {Coarasa}, J.~A.,
  {Commichau}, S., {Contreras}, J.~L., {Cortina}, J., {Costado}, M.~T.,
  {Curtef}, V., {Danielyan}, V., {Dazzi}, F., {De Angelis}, A., {Delgado}, C.,
  {de los Reyes}, R., {De Lotto}, B., {Domingo-Santamar{\'{\i}}a}, E.,
  {Dorner}, D., {Doro}, M., {Errando}, M., {Fagiolini}, M., {Ferenc}, D.,
  {Fern{\'a}ndez}, E., {Firpo}, R., {Flix}, J., {Fonseca}, M.~V., {Font}, L.,
  {Fuchs}, M., {Galante}, N., {Garc{\'{\i}}a-L{\'o}pez}, R.~J., {Garczarczyk},
  M., {Gaug}, M., {Giller}, M., {Goebel}, F., {Hakobyan}, D., {Hayashida}, M.,
  {Hengstebeck}, T., {Herrero}, A., {H{\"o}hne}, D., {Hose}, J., {Hsu}, C.~C.,
  {Jacon}, P., {Jogler}, T., {Kosyra}, R., {Kranich}, D., {Kritzer}, R.,
  {Laille}, A., {Lindfors}, E., {Lombardi}, S., {Longo}, F., {L{\'o}pez}, J.,
  {L{\'o}pez}, M., {Lorenz}, E., {Majumdar}, P., {Maneva}, G., {Mannheim}, K.,
  {Mansutti}, O., {Mariotti}, M., {Mart{\'{\i}}nez}, M., {Mazin}, D., {Merck},
  C., {Meucci}, M., {Meyer}, M., {Miranda}, J.~M., {Mirzoyan}, R., {Mizobuchi},
  S., {Moralejo}, A., {Nieto}, D., {Nilsson}, K., {Ninkovic}, J.,
  {O{\~n}a-Wilhelmi}, E., {Otte}, N., {Oya}, I., {Paneque}, D., {Panniello},
  M., {Paoletti}, R., {Paredes}, J.~M., {Pasanen}, M., {Pascoli}, D., {Pauss},
  F., {Pegna}, R., {Perlman}, E., {Persic}, M., {Peruzzo}, L., {Piccioli}, A.,
  {Prandini}, E., {Puchades}, N., {Raymers}, A., {Rhode}, W., {Rib{\'o}}, M.,
  {Rico}, J., {Rissi}, M., {Robert}, A., {R{\"u}gamer}, S., {Saggion}, A.,
  {Saito}, T., {S{\'a}nchez}, A., {Sartori}, P., {Scalzotto}, V., {Scapin}, V.,
  {Schmitt}, R., {Schweizer}, T., {Shayduk}, M., {Shinozaki}, K., {Shore},
  S.~N., {Sidro}, N., {Sillanp{\"a}{\"a}}, A., {Sobczynska}, D., {Stamerra},
  A., {Stark}, L.~S., {Takalo}, L., {Tavecchio}, F., {Temnikov}, P., {Tescaro},
  D., {Teshima}, M., {Torres}, D.~F., {Turini}, N., {Vankov}, H., {Vitale}, V.,
  {Wagner}, R.~M., {Wibig}, T., {Wittek}, W., {Zandanel}, F., {Zanin}, R., \&
  {Zapatero}, J. 2007{\natexlab{b}}, \apjl, 667, L21

\bibitem[{{Aleksic} {et~al.}(2011){Aleksic}, {Alvarez}, {Antonelli},
  {Antoranz}, {Asensio}, {Backes}, {Barrio}, {Bastieri}, {Becerra Gonzalez},
  {Bednarek}, {Berdyugin}, {Berger}, {Bernardini}, {Biland}, {Blanch}, {Bock},
  {Boller}, {Bonnoli}, {Borla Tridon}, {Braun}, {Bretz}, {Canellas}, {Carmona},
  {Carosi}, {Colin}, {Colombo}, {Contreras}, {Cortina}, {Cossio}, {Covino},
  {Dazzi}, {De Angelis}, {De Caneva}, {De Cea del Pozo}, {De Lotto}, {Delgado
  Mendez}, {Diago Ortega}, {Doert}, {Dominguez1}, {Dominis Prester}, {Dorner},
  {Doro}, {Elsaesser}, {Ferenc}, {Fonseca}, {Font}, {Fruck}, {Garcia Lopez},
  {Garczarczyk}, {Garrido}, {Giavitto}, {Godinovic}, {Hadasch}, {H{\"a}fner},
  {Herrero}, {Hildebrand}, {H{\"o}hne-M{\"o}nch}, {Hose}, {Hrupec}, {Huber},
  {Jogler}, {Kellermann}, {Klepser}, {Krahenbuh}, {Krause}, {La Barbera},
  {Lelas}, {Leonardo}, {Lindfors}, {Lombardi}, {Lopez}, {Lopez}, {Lorenz},
  {Makariev}, {Maneva}, {Mankuzhiyil}, {Mannheim}, {Maraschi}, {Mariotti},
  {Martinez}, {Mazin}, {Meucci}, {Miranda}, {Mirzoyan}, {Miyamoto}, {Moldon},
  {Moralejo}, {Munar-Adrover}, {Nieto}, {Nilsson}, {Orito}, {Oya}, {Paneque},
  {Paoletti}, {Pardo}, {Paredes}, {Partini}, {Pasanen}, {Pauss},
  {Perez-Torres}, {Persic}, {Peruzzo}, {Pilia}, {Pochon}, {Prada}, {Prada
  Moroni}, {Prandini}, {Puljak}, {Reichardt}, {Reinthal}, {Rhode}, {Ribo},
  {Rico}, {R{\"u}gamer}, {Saggion}, {Saito}, {Saito}, {Salvati}, {Satalecka},
  {Scalzotto}, {Scapin}, {Schultz}, {Schweizer}, {Shayduk}, {Shore},
  {Sillanpaa}, {Sitarek}, {Sobczynska}, {Spanier}, {Spiro}, {Stamerra},
  {Steinke}, {Storz}, {Strah}, {Suric}, {Takalo}, {Takami}, {Tavecchio},
  {Temnikov}, {Terzic}, {Tescaro}, {Teshima}, {Tibolla}, {Torres}, {Treves},
  {Uellenbeck}, {Vankov}, {Vogler}, {Wagner}, {Weitzel}, {Zabalza}, {Zandanel},
  \& {Zanin}}]{aleksic11c}
{Aleksic}, J., {Alvarez}, E.~A., {Antonelli}, L.~A., {Antoranz}, P., {Asensio},
  M., {Backes}, M., {Barrio}, J.~A., {Bastieri}, D., {Becerra Gonzalez}, J.,
  {Bednarek}, W., {Berdyugin}, A., {Berger}, K., {Bernardini}, E., {Biland},
  A., {Blanch}, O., {Bock}, R.~K., {Boller}, A., {Bonnoli}, G., {Borla Tridon},
  D., {Braun}, I., {Bretz}, T., {Canellas}, A., {Carmona}, E., {Carosi}, A.,
  {Colin}, P., {Colombo}, E., {Contreras}, J.~L., {Cortina}, J., {Cossio}, L.,
  {Covino}, S., {Dazzi}, F., {De Angelis}, A., {De Caneva}, G., {De Cea del
  Pozo}, E., {De Lotto}, B., {Delgado Mendez}, C., {Diago Ortega}, A., {Doert},
  M., {Dominguez1}, A., {Dominis Prester}, D., {Dorner}, D., {Doro}, M.,
  {Elsaesser}, D., {Ferenc}, D., {Fonseca}, M.~V., {Font}, L., {Fruck}, C.,
  {Garcia Lopez}, R.~J., {Garczarczyk}, M., {Garrido}, D., {Giavitto}, G.,
  {Godinovic}, N., {Hadasch}, D., {H{\"a}fner}, D., {Herrero}, A.,
  {Hildebrand}, D., {H{\"o}hne-M{\"o}nch}, D., {Hose}, J., {Hrupec}, D.,
  {Huber}, B., {Jogler}, T., {Kellermann}, H., {Klepser}, S., {Krahenbuh}, T.,
  {Krause}, J., {La Barbera}, A., {Lelas}, D., {Leonardo}, E., {Lindfors}, E.,
  {Lombardi}, S., {Lopez}, A., {Lopez}, M., {Lorenz}, E., {Makariev}, M.,
  {Maneva}, G., {Mankuzhiyil}, N., {Mannheim}, K., {Maraschi}, L., {Mariotti},
  M., {Martinez}, M., {Mazin}, D., {Meucci}, M., {Miranda}, J.~M., {Mirzoyan},
  R., {Miyamoto}, H., {Moldon}, J., {Moralejo}, A., {Munar-Adrover}, P.,
  {Nieto}, D., {Nilsson}, K., {Orito}, R., {Oya}, I., {Paneque}, D.,
  {Paoletti}, R., {Pardo}, S., {Paredes}, J.~M., {Partini}, S., {Pasanen}, M.,
  {Pauss}, F., {Perez-Torres}, M.~A., {Persic}, M., {Peruzzo}, L., {Pilia}, M.,
  {Pochon}, J., {Prada}, F., {Prada Moroni}, P.~G., {Prandini}, E., {Puljak},
  I., {Reichardt}, I., {Reinthal}, R., {Rhode}, W., {Ribo}, M., {Rico}, J.,
  {R{\"u}gamer}, S., {Saggion}, A., {Saito}, K., {Saito}, T.~Y., {Salvati}, M.,
  {Satalecka}, K., {Scalzotto}, V., {Scapin}, V., {Schultz}, C., {Schweizer},
  T., {Shayduk}, M., {Shore}, S.~N., {Sillanpaa}, A., {Sitarek}, J.,
  {Sobczynska}, D., {Spanier}, F., {Spiro}, S., {Stamerra}, A., {Steinke}, B.,
  {Storz}, J., {Strah}, N., {Suric}, T., {Takalo}, L., {Takami}, H.,
  {Tavecchio}, F., {Temnikov}, P., {Terzic}, T., {Tescaro}, D., {Teshima}, M.,
  {Tibolla}, O., {Torres}, D.~F., {Treves}, A., {Uellenbeck}, M., {Vankov}, H.,
  {Vogler}, P., {Wagner}, R.~M., {Weitzel}, Q., {Zabalza}, V., {Zandanel}, F.,
  \& {Zanin}, R. 2011, ArXiv: 1589

\bibitem[{{Aleksi{\'c}} {et~al.}(2010{\natexlab{a}}){Aleksi{\'c}}, {Anderhub},
  {Antonelli}, {Antoranz}, {Backes}, {Baixeras}, {Balestra}, {Barrio},
  {Bastieri}, {Becerra Gonz{\'a}lez}, {Becker}, {Bednarek}, {Berdyugin},
  {Berger}, {Bernardini}, {Biland}, {Bock}, {Bonnoli}, {Bordas}, {Borla
  Tridon}, {Bosch-Ramon}, {Bose}, {Braun}, {Bretz}, {Britzger}, {Camara},
  {Carmona}, {Carosi}, {Colin}, {Commichau}, {Contreras}, {Cortina}, {Costado},
  {Covino}, {Dazzi}, {de Angelis}, {de Cea Del Pozo}, {de Los Reyes}, {de
  Lotto}, {de Maria}, {de Sabata}, {Delgado Mendez}, {Doert},
  {Dom{\'{\i}}nguez}, {Dominis Prester}, {Dorner}, {Doro}, {Elsaesser},
  {Errando}, {Ferenc}, {Fonseca}, {Font}, {Garc{\'{\i}}a L{\'o}pez},
  {Garczarczyk}, {Gaug}, {Godinovic}, {Hadasch}, {Herrero}, {Hildebrand},
  {H{\"o}hne-M{\"o}nch}, {Hose}, {Hrupec}, {Hsu}, {Jogler}, {Klepser},
  {Kr{\"a}henb{\"u}hl}, {Kranich}, {La Barbera}, {Laille}, {Leonardo},
  {Lindfors}, {Lombardi}, {Longo}, {L{\'o}pez}, {Lorenz}, {Majumdar}, {Maneva},
  {Mankuzhiyil}, {Mannheim}, {Maraschi}, {Mariotti}, {Mart{\'{\i}}nez},
  {Mazin}, {Meucci}, {Miranda}, {Mirzoyan}, {Miyamoto}, {Mold{\'o}n}, {Moles},
  {Moralejo}, {Nieto}, {Nilsson}, {Ninkovic}, {Orito}, {Oya}, {Paoletti},
  {Paredes}, {Partini}, {Pasanen}, {Pascoli}, {Pauss}, {Pegna}, {Perez-Torres},
  {Persic}, {Peruzzo}, {Prada}, {Prandini}, {Puchades}, {Puljak}, {Reichardt},
  {Rhode}, {Rib{\'o}}, {Rico}, {Rissi}, {R{\"u}gamer}, {Saggion}, {Saito},
  {Salvati}, {S{\'a}nchez-Conde}, {Satalecka}, {Scalzotto}, {Scapin},
  {Schweizer}, {Shayduk}, {Shore}, {Sierpowska-Bartosik}, {Sillanp{\"a}{\"a}},
  {Sitarek}, {Sobczynska}, {Spanier}, {Spiro}, {Stamerra}, {Steinke}, {Strah},
  {Struebig}, {Suric}, {Takalo}, {Tavecchio}, {Temnikov}, {Tescaro}, {Teshima},
  {Torres}, {Vankov}, {Wagner}, {Zabalza}, {Zandanel}, {Zanin}, \& {MAGIC
  Collaboration}}]{aleksic10b}
{Aleksi{\'c}}, J., {Anderhub}, H., {Antonelli}, L.~A., {Antoranz}, P.,
  {Backes}, M., {Baixeras}, C., {Balestra}, S., {Barrio}, J.~A., {Bastieri},
  D., {Becerra Gonz{\'a}lez}, J., {Becker}, J.~K., {Bednarek}, W., {Berdyugin},
  A., {Berger}, K., {Bernardini}, E., {Biland}, A., {Bock}, R.~K., {Bonnoli},
  G., {Bordas}, P., {Borla Tridon}, D., {Bosch-Ramon}, V., {Bose}, D., {Braun},
  I., {Bretz}, T., {Britzger}, D., {Camara}, M., {Carmona}, E., {Carosi}, A.,
  {Colin}, P., {Commichau}, S., {Contreras}, J.~L., {Cortina}, J., {Costado},
  M.~T., {Covino}, S., {Dazzi}, F., {de Angelis}, A., {de Cea Del Pozo}, E.,
  {de Los Reyes}, R., {de Lotto}, B., {de Maria}, M., {de Sabata}, F., {Delgado
  Mendez}, C., {Doert}, M., {Dom{\'{\i}}nguez}, A., {Dominis Prester}, D.,
  {Dorner}, D., {Doro}, M., {Elsaesser}, D., {Errando}, M., {Ferenc}, D.,
  {Fonseca}, M.~V., {Font}, L., {Garc{\'{\i}}a L{\'o}pez}, R.~J.,
  {Garczarczyk}, M., {Gaug}, M., {Godinovic}, N., {Hadasch}, D., {Herrero}, A.,
  {Hildebrand}, D., {H{\"o}hne-M{\"o}nch}, D., {Hose}, J., {Hrupec}, D., {Hsu},
  C.~C., {Jogler}, T., {Klepser}, S., {Kr{\"a}henb{\"u}hl}, T., {Kranich}, D.,
  {La Barbera}, A., {Laille}, A., {Leonardo}, E., {Lindfors}, E., {Lombardi},
  S., {Longo}, F., {L{\'o}pez}, M., {Lorenz}, E., {Majumdar}, P., {Maneva}, G.,
  {Mankuzhiyil}, N., {Mannheim}, K., {Maraschi}, L., {Mariotti}, M.,
  {Mart{\'{\i}}nez}, M., {Mazin}, D., {Meucci}, M., {Miranda}, J.~M.,
  {Mirzoyan}, R., {Miyamoto}, H., {Mold{\'o}n}, J., {Moles}, M., {Moralejo},
  A., {Nieto}, D., {Nilsson}, K., {Ninkovic}, J., {Orito}, R., {Oya}, I.,
  {Paoletti}, R., {Paredes}, J.~M., {Partini}, S., {Pasanen}, M., {Pascoli},
  D., {Pauss}, F., {Pegna}, R.~G., {Perez-Torres}, M.~A., {Persic}, M.,
  {Peruzzo}, L., {Prada}, F., {Prandini}, E., {Puchades}, N., {Puljak}, I.,
  {Reichardt}, I., {Rhode}, W., {Rib{\'o}}, M., {Rico}, J., {Rissi}, M.,
  {R{\"u}gamer}, S., {Saggion}, A., {Saito}, T.~Y., {Salvati}, M.,
  {S{\'a}nchez-Conde}, M., {Satalecka}, K., {Scalzotto}, V., {Scapin}, V.,
  {Schweizer}, T., {Shayduk}, M., {Shore}, S.~N., {Sierpowska-Bartosik}, A.,
  {Sillanp{\"a}{\"a}}, A., {Sitarek}, J., {Sobczynska}, D., {Spanier}, F.,
  {Spiro}, S., {Stamerra}, A., {Steinke}, B., {Strah}, N., {Struebig}, J.~C.,
  {Suric}, T., {Takalo}, L., {Tavecchio}, F., {Temnikov}, P., {Tescaro}, D.,
  {Teshima}, M., {Torres}, D.~F., {Vankov}, H., {Wagner}, R.~M., {Zabalza}, V.,
  {Zandanel}, F., {Zanin}, R., \& {MAGIC Collaboration}. 2010{\natexlab{a}},
  \aap, 519, A32

\bibitem[{{Aleksi{\'c}} {et~al.}(2010{\natexlab{b}}){Aleksi{\'c}}, {Antonelli},
  {Antoranz}, {Backes}, {Barrio}, {Bastieri}, {Becerra Gonz{\'a}lez},
  {Bednarek}, {Berdyugin}, {Berger}, {Bernardini}, {Biland}, {Blanch}, {Bock},
  {Boller}, {Bonnoli}, {Bordas}, {Borla Tridon}, {Bosch-Ramon}, {Bose},
  {Braun}, {Bretz}, {Camara}, {Ca{\~n}ellas}, {Carmona}, {Carosi}, {Colin},
  {Colombo}, {Contreras}, {Cortina}, {Cossio}, {Covino}, {Dazzi}, {De Angelis},
  {De Cea del Pozo}, {De Lotto}, {De Maria}, {De Sabata}, {Delgado Mendez},
  {Diago Ortega}, {Doert}, {Dom{\'{\i}}nguez}, {Dominis Prester}, {Dorner},
  {Doro}, {Elsaesser}, {Errando}, {Ferenc}, {Fonseca}, {Font}, {Garc{\'{\i}}a
  L{\'o}pez}, {Garczarczyk}, {Giavitto}, {Godinovi{\'c}}, {Hadasch}, {Herrero},
  {Hildebrand}, {H{\"o}hne-M{\"o}nch}, {Hose}, {Hrupec}, {Jogler}, {Klepser},
  {Kr{\"a}henb{\"u}hl}, {Kranich}, {Krause}, {La Barbera}, {Leonardo},
  {Lindfors}, {Lombardi}, {Longo}, {L{\'o}pez}, {Lorenz}, {Majumdar},
  {Makariev}, {Maneva}, {Mankuzhiyil}, {Mannheim}, {Maraschi}, {Mariotti},
  {Mart{\'{\i}}nez}, {Mazin}, {Meucci}, {Miranda}, {Mirzoyan}, {Miyamoto},
  {Mold{\'o}n}, {Moralejo}, {Nieto}, {Nilsson}, {Orito}, {Oya}, {Paoletti},
  {Paredes}, {Partini}, {Pasanen}, {Pauss}, {Pegna}, {Perez-Torres}, {Persic},
  {Peruzzo}, {Pochon}, {Prada}, {Prada Moroni}, {Prandini}, {Puchades},
  {Puljak}, {Reichardt}, {Reinthal}, {Rhode}, {Rib{\'o}}, {Rico},
  {R{\"u}gamer}, {Saggion}, {Saito}, {Saito}, {Salvati}, {S{\'a}nchez-Conde},
  {Satalecka}, {Scalzotto}, {Scapin}, {Schultz}, {Schweizer}, {Shayduk},
  {Shore}, {Sierpowska-Bartosik}, {Sillanp{\"a}{\"a}}, {Sitarek}, {Sobczynska},
  {Spanier}, {Spiro}, {Stamerra}, {Steinke}, {Storz}, {Strah}, {Struebig},
  {Suric}, {Takalo}, {Tavecchio}, {Temnikov}, {Terzi{\'c}}, {Tescaro},
  {Teshima}, {Torres}, {Vankov}, {Wagner}, {Weitzel}, {Zabalza}, {Zandanel},
  {Zanin}, {Neronov}, {Pfrommer}, {Pinzke}, {Semikoz}, \& {MAGIC
  Collaboration}}]{aleksic10}
{Aleksi{\'c}}, J., {Antonelli}, L.~A., {Antoranz}, P., {Backes}, M., {Barrio},
  J.~A., {Bastieri}, D., {Becerra Gonz{\'a}lez}, J., {Bednarek}, W.,
  {Berdyugin}, A., {Berger}, K., {Bernardini}, E., {Biland}, A., {Blanch}, O.,
  {Bock}, R.~K., {Boller}, A., {Bonnoli}, G., {Bordas}, P., {Borla Tridon}, D.,
  {Bosch-Ramon}, V., {Bose}, D., {Braun}, I., {Bretz}, T., {Camara}, M.,
  {Ca{\~n}ellas}, A., {Carmona}, E., {Carosi}, A., {Colin}, P., {Colombo}, E.,
  {Contreras}, J.~L., {Cortina}, J., {Cossio}, L., {Covino}, S., {Dazzi}, F.,
  {De Angelis}, A., {De Cea del Pozo}, E., {De Lotto}, B., {De Maria}, M., {De
  Sabata}, F., {Delgado Mendez}, C., {Diago Ortega}, A., {Doert}, M.,
  {Dom{\'{\i}}nguez}, A., {Dominis Prester}, D., {Dorner}, D., {Doro}, M.,
  {Elsaesser}, D., {Errando}, M., {Ferenc}, D., {Fonseca}, M.~V., {Font}, L.,
  {Garc{\'{\i}}a L{\'o}pez}, R.~J., {Garczarczyk}, M., {Giavitto}, G.,
  {Godinovi{\'c}}, N., {Hadasch}, D., {Herrero}, A., {Hildebrand}, D.,
  {H{\"o}hne-M{\"o}nch}, D., {Hose}, J., {Hrupec}, D., {Jogler}, T., {Klepser},
  S., {Kr{\"a}henb{\"u}hl}, T., {Kranich}, D., {Krause}, J., {La Barbera}, A.,
  {Leonardo}, E., {Lindfors}, E., {Lombardi}, S., {Longo}, F., {L{\'o}pez}, M.,
  {Lorenz}, E., {Majumdar}, P., {Makariev}, M., {Maneva}, G., {Mankuzhiyil},
  N., {Mannheim}, K., {Maraschi}, L., {Mariotti}, M., {Mart{\'{\i}}nez}, M.,
  {Mazin}, D., {Meucci}, M., {Miranda}, J.~M., {Mirzoyan}, R., {Miyamoto}, H.,
  {Mold{\'o}n}, J., {Moralejo}, A., {Nieto}, D., {Nilsson}, K., {Orito}, R.,
  {Oya}, I., {Paoletti}, R., {Paredes}, J.~M., {Partini}, S., {Pasanen}, M.,
  {Pauss}, F., {Pegna}, R.~G., {Perez-Torres}, M.~A., {Persic}, M., {Peruzzo},
  L., {Pochon}, J., {Prada}, F., {Prada Moroni}, P.~G., {Prandini}, E.,
  {Puchades}, N., {Puljak}, I., {Reichardt}, I., {Reinthal}, R., {Rhode}, W.,
  {Rib{\'o}}, M., {Rico}, J., {R{\"u}gamer}, S., {Saggion}, A., {Saito}, K.,
  {Saito}, T.~Y., {Salvati}, M., {S{\'a}nchez-Conde}, M., {Satalecka}, K.,
  {Scalzotto}, V., {Scapin}, V., {Schultz}, C., {Schweizer}, T., {Shayduk}, M.,
  {Shore}, S.~N., {Sierpowska-Bartosik}, A., {Sillanp{\"a}{\"a}}, A.,
  {Sitarek}, J., {Sobczynska}, D., {Spanier}, F., {Spiro}, S., {Stamerra}, A.,
  {Steinke}, B., {Storz}, J., {Strah}, N., {Struebig}, J.~C., {Suric}, T.,
  {Takalo}, L., {Tavecchio}, F., {Temnikov}, P., {Terzi{\'c}}, T., {Tescaro},
  D., {Teshima}, M., {Torres}, D.~F., {Vankov}, H., {Wagner}, R.~M., {Weitzel},
  Q., {Zabalza}, V., {Zandanel}, F., {Zanin}, R., {Neronov}, A., {Pfrommer},
  C., {Pinzke}, A., {Semikoz}, D.~V., \& {MAGIC Collaboration}.
  2010{\natexlab{b}}, \apjl, 723, L207

\bibitem[{{Aleksi{\'c}} {et~al.}(2011{\natexlab{a}}){Aleksi{\'c}}, {Antonelli},
  {Antoranz}, {Backes}, {Barrio}, {Bastieri}, {Becerra Gonz{\'a}lez},
  {Bednarek}, {Berdyugin}, {Berger}, {Bernardini}, {Biland}, {Blanch}, {Bock},
  {Boller}, {Bonnoli}, {Borla Tridon}, {Braun}, {Bretz}, {Ca{\~n}ellas},
  {Carmona}, {Carosi}, {Colin}, {Colombo}, {Contreras}, {Cortina}, {Cossio},
  {Covino}, {Dazzi}, {De Angelis}, {De Cea del Pozo}, {De Lotto}, {Delgado
  Mendez}, {Diago Ortega}, {Doert}, {Dom{\'{\i}}nguez}, {Dominis Prester},
  {Dorner}, {Doro}, {Elsaesser}, {Ferenc}, {Fonseca}, {Font}, {Fruck},
  {Garc{\'{\i}}a L{\'o}pez}, {Garczarczyk}, {Garrido}, {Giavitto},
  {Godinovi{\'c}}, {Hadasch}, {H{\"a}fner}, {Herrero}, {Hildebrand},
  {H{\"o}hne-M{\"o}nch}, {Hose}, {Hrupec}, {Huber}, {Jogler}, {Klepser},
  {Kr{\"a}henb{\"u}hl}, {Krause}, {La Barbera}, {Lelas}, {Leonardo},
  {Lindfors}, {Lombardi}, {L{\'o}pez}, {Lorenz}, {Makariev}, {Maneva},
  {Mankuzhiyil}, {Mannheim}, {Maraschi}, {Mariotti}, {Mart{\'{\i}}nez},
  {Mazin}, {Meucci}, {Miranda}, {Mirzoyan}, {Miyamoto}, {Mold{\'o}n},
  {Moralejo}, {Nieto}, {Nilsson}, {Orito}, {Oya}, {Paneque}, {Paoletti},
  {Pardo}, {Paredes}, {Partini}, {Pasanen}, {Pauss}, {Perez-Torres}, {Persic},
  {Peruzzo}, {Pilia}, {Pochon}, {Prada}, {Prada Moroni}, {Prandini}, {Puljak},
  {Reichardt}, {Reinthal}, {Rhode}, {Rib{\'o}}, {Rico}, {R{\"u}gamer},
  {Saggion}, {Saito}, {Saito}, {Salvati}, {Satalecka}, {Scalzotto}, {Scapin},
  {Schultz}, {Schweizer}, {Shayduk}, {Shore}, {Sillanp{\"a}{\"a}}, {Sitarek},
  {Sobczynska}, {Spanier}, {Spiro}, {Stamerra}, {Steinke}, {Storz}, {Strah},
  {Suri{\'c}}, {Takalo}, {Tavecchio}, {Temnikov}, {Terzi{\'c}}, {Tescaro},
  {Teshima}, {Thom}, {Tibolla}, {Torres}, {Treves}, {Vankov}, {Vogler},
  {Wagner}, {Weitzel}, {Zabalza}, {Zandanel}, {Zanin}, {MAGIC Collaboration},
  {Tanaka}, {Wood}, \& {Buson}}]{aleksic11}
{Aleksi{\'c}}, J., {Antonelli}, L.~A., {Antoranz}, P., {Backes}, M., {Barrio},
  J.~A., {Bastieri}, D., {Becerra Gonz{\'a}lez}, J., {Bednarek}, W.,
  {Berdyugin}, A., {Berger}, K., {Bernardini}, E., {Biland}, A., {Blanch}, O.,
  {Bock}, R.~K., {Boller}, A., {Bonnoli}, G., {Borla Tridon}, D., {Braun}, I.,
  {Bretz}, T., {Ca{\~n}ellas}, A., {Carmona}, E., {Carosi}, A., {Colin}, P.,
  {Colombo}, E., {Contreras}, J.~L., {Cortina}, J., {Cossio}, L., {Covino}, S.,
  {Dazzi}, F., {De Angelis}, A., {De Cea del Pozo}, E., {De Lotto}, B.,
  {Delgado Mendez}, C., {Diago Ortega}, A., {Doert}, M., {Dom{\'{\i}}nguez},
  A., {Dominis Prester}, D., {Dorner}, D., {Doro}, M., {Elsaesser}, D.,
  {Ferenc}, D., {Fonseca}, M.~V., {Font}, L., {Fruck}, C., {Garc{\'{\i}}a
  L{\'o}pez}, R.~J., {Garczarczyk}, M., {Garrido}, D., {Giavitto}, G.,
  {Godinovi{\'c}}, N., {Hadasch}, D., {H{\"a}fner}, D., {Herrero}, A.,
  {Hildebrand}, D., {H{\"o}hne-M{\"o}nch}, D., {Hose}, J., {Hrupec}, D.,
  {Huber}, B., {Jogler}, T., {Klepser}, S., {Kr{\"a}henb{\"u}hl}, T., {Krause},
  J., {La Barbera}, A., {Lelas}, D., {Leonardo}, E., {Lindfors}, E.,
  {Lombardi}, S., {L{\'o}pez}, M., {Lorenz}, E., {Makariev}, M., {Maneva}, G.,
  {Mankuzhiyil}, N., {Mannheim}, K., {Maraschi}, L., {Mariotti}, M.,
  {Mart{\'{\i}}nez}, M., {Mazin}, D., {Meucci}, M., {Miranda}, J.~M.,
  {Mirzoyan}, R., {Miyamoto}, H., {Mold{\'o}n}, J., {Moralejo}, A., {Nieto},
  D., {Nilsson}, K., {Orito}, R., {Oya}, I., {Paneque}, D., {Paoletti}, R.,
  {Pardo}, S., {Paredes}, J.~M., {Partini}, S., {Pasanen}, M., {Pauss}, F.,
  {Perez-Torres}, M.~A., {Persic}, M., {Peruzzo}, L., {Pilia}, M., {Pochon},
  J., {Prada}, F., {Prada Moroni}, P.~G., {Prandini}, E., {Puljak}, I.,
  {Reichardt}, I., {Reinthal}, R., {Rhode}, W., {Rib{\'o}}, M., {Rico}, J.,
  {R{\"u}gamer}, S., {Saggion}, A., {Saito}, K., {Saito}, T.~Y., {Salvati}, M.,
  {Satalecka}, K., {Scalzotto}, V., {Scapin}, V., {Schultz}, C., {Schweizer},
  T., {Shayduk}, M., {Shore}, S.~N., {Sillanp{\"a}{\"a}}, A., {Sitarek}, J.,
  {Sobczynska}, D., {Spanier}, F., {Spiro}, S., {Stamerra}, A., {Steinke}, B.,
  {Storz}, J., {Strah}, N., {Suri{\'c}}, T., {Takalo}, L., {Tavecchio}, F.,
  {Temnikov}, P., {Terzi{\'c}}, T., {Tescaro}, D., {Teshima}, M., {Thom}, M.,
  {Tibolla}, O., {Torres}, D.~F., {Treves}, A., {Vankov}, H., {Vogler}, P.,
  {Wagner}, R.~M., {Weitzel}, Q., {Zabalza}, V., {Zandanel}, F., {Zanin}, R.,
  {MAGIC Collaboration}, {Tanaka}, Y.~T., {Wood}, D.~L., \& {Buson}, S.
  2011{\natexlab{a}}, \apjl, 730, L8

\bibitem[{{Aleksi{\'c}} {et~al.}(2011{\natexlab{b}}){Aleksi{\'c}}, {Antonelli},
  {Antoranz}, {Backes}, {Barrio}, {Bastieri}, {Becerra Gonz{\'a}lez},
  {Bednarek}, {Berdyugin}, {Berger}, {Bernardini}, {Biland}, {Blanch}, {Bock},
  {Boller}, {Bonnoli}, {Borla Tridon}, {Braun}, {Bretz}, {Ca{\~n}ellas},
  {Carmona}, {Carosi}, {Colin}, {Colombo}, {Contreras}, {Cortina}, {Cossio},
  {Covino}, {Dazzi}, {de Angelis}, {de Cea Del Pozo}, {de Lotto}, {Delgado
  Mendez}, {Diago Ortega}, {Doert}, {Dom{\'{\i}}nguez}, {Dominis Prester},
  {Dorner}, {Doro}, {Elsaesser}, {Ferenc}, {Fonseca}, {Font}, {Fruck},
  {Garc{\'{\i}}a L{\'o}pez}, {Garczarczyk}, {Garrido}, {Giavitto},
  {Godinovi{\'c}}, {Hadasch}, {H{\"a}fner}, {Herrero}, {Hildebrand}, {Hose},
  {Hrupec}, {Huber}, {Jogler}, {Klepser}, {Kr{\"a}henb{\"u}hl}, {Krause}, {La
  Barbera}, {Lelas}, {Leonardo}, {Lindfors}, {Lombardi}, {L{\'o}pez}, {Lorenz},
  {Majumdar}, {Makariev}, {Maneva}, {Mankuzhiyil}, {Mannheim}, {Maraschi},
  {Mariotti}, {Mart{\'{\i}}nez}, {Mazin}, {Meucci}, {Miranda}, {Mirzoyan},
  {Miyamoto}, {Mold{\'o}n}, {Moralejo}, {Nieto}, {Nilsson}, {Orito}, {Oya},
  {Paoletti}, {Pardo}, {Paredes}, {Partini}, {Pasanen}, {Pauss},
  {Perez-Torres}, {Persic}, {Peruzzo}, {Pilia}, {Pochon}, {Prada}, {Prada
  Moroni}, {Prandini}, {Puljak}, {Reichardt}, {Reinthal}, {Rhode}, {Rib{\'o}},
  {Rico}, {R{\"u}gamer}, {R{\"u}ger}, {Saggion}, {Saito}, {Saito}, {Salvati},
  {Satalecka}, {Scalzotto}, {Scapin}, {Schultz}, {Schweizer}, {Shayduk},
  {Shore}, {Sillanp{\"a}{\"a}}, {Sitarek}, {Sobczynska}, {Spanier}, {Spiro},
  {Stamerra}, {Steinke}, {Storz}, {Strah}, {Suri{\'c}}, {Takalo}, {Tavecchio},
  {Temnikov}, {Terzi{\'c}}, {Tescaro}, {Teshima}, {Thom}, {Tibolla}, {Torres},
  {Treves}, {Vankov}, {Vogler}, {Wagner}, {Weitzel}, {Zabalza}, {Zandanel}, \&
  {Zanin}}]{aleksic11b}
{Aleksi{\'c}}, J., {Antonelli}, L.~A., {Antoranz}, P., {Backes}, M., {Barrio},
  J.~A., {Bastieri}, D., {Becerra Gonz{\'a}lez}, J., {Bednarek}, W.,
  {Berdyugin}, A., {Berger}, K., {Bernardini}, E., {Biland}, A., {Blanch}, O.,
  {Bock}, R.~K., {Boller}, A., {Bonnoli}, G., {Borla Tridon}, D., {Braun}, I.,
  {Bretz}, T., {Ca{\~n}ellas}, A., {Carmona}, E., {Carosi}, A., {Colin}, P.,
  {Colombo}, E., {Contreras}, J.~L., {Cortina}, J., {Cossio}, L., {Covino}, S.,
  {Dazzi}, F., {de Angelis}, A., {de Cea Del Pozo}, E., {de Lotto}, B.,
  {Delgado Mendez}, C., {Diago Ortega}, A., {Doert}, M., {Dom{\'{\i}}nguez},
  A., {Dominis Prester}, D., {Dorner}, D., {Doro}, M., {Elsaesser}, D.,
  {Ferenc}, D., {Fonseca}, M.~V., {Font}, L., {Fruck}, C., {Garc{\'{\i}}a
  L{\'o}pez}, R.~J., {Garczarczyk}, M., {Garrido}, D., {Giavitto}, G.,
  {Godinovi{\'c}}, N., {Hadasch}, D., {H{\"a}fner}, D., {Herrero}, A.,
  {Hildebrand}, D., {Hose}, J., {Hrupec}, D., {Huber}, B., {Jogler}, T.,
  {Klepser}, S., {Kr{\"a}henb{\"u}hl}, T., {Krause}, J., {La Barbera}, A.,
  {Lelas}, D., {Leonardo}, E., {Lindfors}, E., {Lombardi}, S., {L{\'o}pez}, M.,
  {Lorenz}, E., {Majumdar}, P., {Makariev}, M., {Maneva}, G., {Mankuzhiyil},
  N., {Mannheim}, K., {Maraschi}, L., {Mariotti}, M., {Mart{\'{\i}}nez}, M.,
  {Mazin}, D., {Meucci}, M., {Miranda}, J.~M., {Mirzoyan}, R., {Miyamoto}, H.,
  {Mold{\'o}n}, J., {Moralejo}, A., {Nieto}, D., {Nilsson}, K., {Orito}, R.,
  {Oya}, I., {Paoletti}, R., {Pardo}, S., {Paredes}, J.~M., {Partini}, S.,
  {Pasanen}, M., {Pauss}, F., {Perez-Torres}, M.~A., {Persic}, M., {Peruzzo},
  L., {Pilia}, M., {Pochon}, J., {Prada}, F., {Prada Moroni}, P.~G.,
  {Prandini}, E., {Puljak}, I., {Reichardt}, I., {Reinthal}, R., {Rhode}, W.,
  {Rib{\'o}}, M., {Rico}, J., {R{\"u}gamer}, S., {R{\"u}ger}, M., {Saggion},
  A., {Saito}, K., {Saito}, T.~Y., {Salvati}, M., {Satalecka}, K., {Scalzotto},
  V., {Scapin}, V., {Schultz}, C., {Schweizer}, T., {Shayduk}, M., {Shore},
  S.~N., {Sillanp{\"a}{\"a}}, A., {Sitarek}, J., {Sobczynska}, D., {Spanier},
  F., {Spiro}, S., {Stamerra}, A., {Steinke}, B., {Storz}, J., {Strah}, N.,
  {Suri{\'c}}, T., {Takalo}, L., {Tavecchio}, F., {Temnikov}, P., {Terzi{\'c}},
  T., {Tescaro}, D., {Teshima}, M., {Thom}, M., {Tibolla}, O., {Torres}, D.~F.,
  {Treves}, A., {Vankov}, H., {Vogler}, P., {Wagner}, R.~M., {Weitzel}, Q.,
  {Zabalza}, V., {Zandanel}, F., \& {Zanin}, R. 2011{\natexlab{b}}, \aap, 530,
  A4

\bibitem[{{Anderhub} {et~al.}(2009){Anderhub}, {Antonelli}, {Antoranz},
  {Backes}, {Baixeras}, {Balestra}, {Barrio}, {Bastieri}, {Becerra
  Gonz{\'a}lez}, {Becker}, {Bednarek}, {Berdyugin}, {Berger}, {Bernardini},
  {Biland}, {Bock}, {Bonnoli}, {Bordas}, {Borla Tridon}, {Bosch-Ramon}, {Bose},
  {Braun}, {Bretz}, {Britzger}, {Camara}, {Carmona}, {Carosi}, {Colin},
  {Commichau}, {Contreras}, {Cortina}, {Costado}, {Covino}, {Dazzi}, {De
  Angelis}, {de Cea del Pozo}, {De los Reyes}, {De Lotto}, {De Maria}, {De
  Sabata}, {Delgado Mendez}, {Dom{\'{\i}}nguez}, {Dominis Prester}, {Dorner},
  {Doro}, {Elsaesser}, {Errando}, {Ferenc}, {Fern{\'a}ndez}, {Firpo},
  {Fonseca}, {Font}, {Galante}, {Garc{\'{\i}}a L{\'o}pez}, {Garczarczyk},
  {Gaug}, {Godinovic}, {Goebel}, {Hadasch}, {Herrero}, {Hildebrand},
  {H{\"o}hne-M{\"o}nch}, {Hose}, {Hrupec}, {Hsu}, {Jogler}, {Klepser},
  {Kranich}, {La Barbera}, {Laille}, {Leonardo}, {Lindfors}, {Lombardi},
  {Longo}, {L{\'o}pez}, {Lorenz}, {Majumdar}, {Maneva}, {Mankuzhiyil},
  {Mannheim}, {Maraschi}, {Mariotti}, {Mart{\'{\i}}nez}, {Mazin}, {Meucci},
  {Miranda}, {Mirzoyan}, {Miyamoto}, {Mold{\'o}n}, {Moles}, {Moralejo},
  {Nieto}, {Nilsson}, {Ninkovic}, {Orito}, {Oya}, {Paoletti}, {Paredes},
  {Pasanen}, {Pascoli}, {Pauss}, {Pegna}, {Perez-Torres}, {Persic}, {Peruzzo},
  {Prada}, {Prandini}, {Puchades}, {Puljak}, {Reichardt}, {Rhode}, {Rib{\'o}},
  {Rico}, {Rissi}, {Robert}, {R{\"u}gamer}, {Saggion}, {Sainio}, {Saito},
  {Salvati}, {S{\'a}nchez-Conde}, {Satalecka}, {Scalzotto}, {Scapin},
  {Schweizer}, {Shayduk}, {Shore}, {Sierpowska-Bartosik}, {Sillanp{\"a}{\"a}},
  {Sitarek}, {Sobczynska}, {Spanier}, {Spiro}, {Stamerra}, {Stark}, {Suric},
  {Takalo}, {Tavecchio}, {Temnikov}, {Tescaro}, {Teshima}, {Torres}, {Turini},
  {Vankov}, {Wagner}, {Villforth}, {Zabalza}, {Zandanel}, {Zanin}, \&
  {Zapatero}}]{anderhub09}
{Anderhub}, H., {Antonelli}, L.~A., {Antoranz}, P., {Backes}, M., {Baixeras},
  C., {Balestra}, S., {Barrio}, J.~A., {Bastieri}, D., {Becerra Gonz{\'a}lez},
  J., {Becker}, J.~K., {Bednarek}, W., {Berdyugin}, A., {Berger}, K.,
  {Bernardini}, E., {Biland}, A., {Bock}, R.~K., {Bonnoli}, G., {Bordas}, P.,
  {Borla Tridon}, D., {Bosch-Ramon}, V., {Bose}, D., {Braun}, I., {Bretz}, T.,
  {Britzger}, D., {Camara}, M., {Carmona}, E., {Carosi}, A., {Colin}, P.,
  {Commichau}, S., {Contreras}, J.~L., {Cortina}, J., {Costado}, M.~T.,
  {Covino}, S., {Dazzi}, F., {De Angelis}, A., {de Cea del Pozo}, E., {De los
  Reyes}, R., {De Lotto}, B., {De Maria}, M., {De Sabata}, F., {Delgado
  Mendez}, C., {Dom{\'{\i}}nguez}, A., {Dominis Prester}, D., {Dorner}, D.,
  {Doro}, M., {Elsaesser}, D., {Errando}, M., {Ferenc}, D., {Fern{\'a}ndez},
  E., {Firpo}, R., {Fonseca}, M.~V., {Font}, L., {Galante}, N., {Garc{\'{\i}}a
  L{\'o}pez}, R.~J., {Garczarczyk}, M., {Gaug}, M., {Godinovic}, N., {Goebel},
  F., {Hadasch}, D., {Herrero}, A., {Hildebrand}, D., {H{\"o}hne-M{\"o}nch},
  D., {Hose}, J., {Hrupec}, D., {Hsu}, C.~C., {Jogler}, T., {Klepser}, S.,
  {Kranich}, D., {La Barbera}, A., {Laille}, A., {Leonardo}, E., {Lindfors},
  E., {Lombardi}, S., {Longo}, F., {L{\'o}pez}, M., {Lorenz}, E., {Majumdar},
  P., {Maneva}, G., {Mankuzhiyil}, N., {Mannheim}, K., {Maraschi}, L.,
  {Mariotti}, M., {Mart{\'{\i}}nez}, M., {Mazin}, D., {Meucci}, M., {Miranda},
  J.~M., {Mirzoyan}, R., {Miyamoto}, H., {Mold{\'o}n}, J., {Moles}, M.,
  {Moralejo}, A., {Nieto}, D., {Nilsson}, K., {Ninkovic}, J., {Orito}, R.,
  {Oya}, I., {Paoletti}, R., {Paredes}, J.~M., {Pasanen}, M., {Pascoli}, D.,
  {Pauss}, F., {Pegna}, R.~G., {Perez-Torres}, M.~A., {Persic}, M., {Peruzzo},
  L., {Prada}, F., {Prandini}, E., {Puchades}, N., {Puljak}, I., {Reichardt},
  I., {Rhode}, W., {Rib{\'o}}, M., {Rico}, J., {Rissi}, M., {Robert}, A.,
  {R{\"u}gamer}, S., {Saggion}, A., {Sainio}, J., {Saito}, T.~Y., {Salvati},
  M., {S{\'a}nchez-Conde}, M., {Satalecka}, K., {Scalzotto}, V., {Scapin}, V.,
  {Schweizer}, T., {Shayduk}, M., {Shore}, S.~N., {Sierpowska-Bartosik}, A.,
  {Sillanp{\"a}{\"a}}, A., {Sitarek}, J., {Sobczynska}, D., {Spanier}, F.,
  {Spiro}, S., {Stamerra}, A., {Stark}, L.~S., {Suric}, T., {Takalo}, L.,
  {Tavecchio}, F., {Temnikov}, P., {Tescaro}, D., {Teshima}, M., {Torres},
  D.~F., {Turini}, N., {Vankov}, H., {Wagner}, R.~M., {Villforth}, C.,
  {Zabalza}, V., {Zandanel}, F., {Zanin}, R., \& {Zapatero}, J. 2009, \apjl,
  704, L129

\bibitem[{{Arendt} \& {Dwek}(2003)}]{arendt03}
{Arendt}, R.~G. \& {Dwek}, E. 2003, \apj, 585, 305

\bibitem[{{Audouze} \& {Tinsley}(1976)}]{audouze76}
{Audouze}, J. \& {Tinsley}, B.~M. 1976, \araa, 14, 43

\bibitem[{{Ballantyne} {et~al.}(2006){Ballantyne}, {Shi}, {Rieke}, {Donley},
  {Papovich}, \& {Rigby}}]{ballantyne06}
{Ballantyne}, D.~R., {Shi}, Y., {Rieke}, G.~H., {Donley}, J.~L., {Papovich},
  C., \& {Rigby}, J.~R. 2006, \apj, 653, 1070

\bibitem[{{Beacom}(2010)}]{beacom10}
{Beacom}, J.~F. 2010, Annual Review of Nuclear and Particle Science, 60, 439

\bibitem[{{Beacom} \& {Vagins}(2004)}]{beacom04}
{Beacom}, J.~F. \& {Vagins}, M.~R. 2004, Physical Review Letters, 93, 171101

\bibitem[{{Benbow}(2011)}]{benbow11}
{Benbow}, W. 2011, arXiv: 1110.0040, L3003

\bibitem[{{Benbow, W. et al.}(2011)}]{benbow11b}
{Benbow, W. et al.} 2011, ArXiv: 1110.0040

\bibitem[{{Berger} {et~al.}(2011{\natexlab{a}}){Berger}, {Dominis Prester},
  {Tavecchio}, {Terzi{\'c}}, \& {for the MAGIC Collaboration}}]{berger11a}
{Berger}, K., {Dominis Prester}, D., {Tavecchio}, F., {Terzi{\'c}}, T., \& {for
  the MAGIC Collaboration}. 2011{\natexlab{a}}, ArXiv: 1109.5879

\bibitem[{{Berger} {et~al.}(2011{\natexlab{b}}){Berger}, {Giavitto},
  {Lindfors}, {Takalo}, {Paneque}, {Stamerra}, \& {On behalf of the MAGIC
  Collaboration}}]{berger11b}
{Berger}, K., {Giavitto}, G., {Lindfors}, E., {Takalo}, L., {Paneque}, D.,
  {Stamerra}, A., \& {On behalf of the MAGIC Collaboration}.
  2011{\natexlab{b}}, ArXiv: 1110.6368

\bibitem[{{Bernstein}(2007)}]{bernstein07}
{Bernstein}, R.~A. 2007, \apj, 666, 663

\bibitem[{{Bernstein} {et~al.}(2002){Bernstein}, {Freedman}, \&
  {Madore}}]{bernstein02}
{Bernstein}, R.~A., {Freedman}, W.~L., \& {Madore}, B.~F. 2002, \apj, 571, 56

\bibitem[{{Berta} {et~al.}(2010){Berta}, {Magnelli}, {Lutz}, {Altieri},
  {Aussel}, {Andreani}, {Bauer}, {Bongiovanni}, {Cava}, {Cepa}, {Cimatti},
  {Daddi}, {Dominguez}, {Elbaz}, {Feuchtgruber}, {F{\"o}rster Schreiber},
  {Genzel}, {Gruppioni}, {Katterloher}, {Magdis}, {Maiolino}, {Nordon},
  {P{\'e}rez Garc{\'{\i}}a}, {Poglitsch}, {Popesso}, {Pozzi}, {Riguccini},
  {Rodighiero}, {Saintonge}, {Santini}, {Sanchez-Portal}, {Shao}, {Sturm},
  {Tacconi}, {Valtchanov}, {Wetzstein}, \& {Wieprecht}}]{berta10}
{Berta}, S., {Magnelli}, B., {Lutz}, D., {Altieri}, B., {Aussel}, H.,
  {Andreani}, P., {Bauer}, O., {Bongiovanni}, A., {Cava}, A., {Cepa}, J.,
  {Cimatti}, A., {Daddi}, E., {Dominguez}, H., {Elbaz}, D., {Feuchtgruber}, H.,
  {F{\"o}rster Schreiber}, N.~M., {Genzel}, R., {Gruppioni}, C., {Katterloher},
  R., {Magdis}, G., {Maiolino}, R., {Nordon}, R., {P{\'e}rez Garc{\'{\i}}a},
  A.~M., {Poglitsch}, A., {Popesso}, P., {Pozzi}, F., {Riguccini}, L.,
  {Rodighiero}, G., {Saintonge}, A., {Santini}, P., {Sanchez-Portal}, M.,
  {Shao}, L., {Sturm}, E., {Tacconi}, L.~J., {Valtchanov}, I., {Wetzstein}, M.,
  \& {Wieprecht}, E. 2010, \aap, 518, L30+

\bibitem[{{Berta} {et~al.}(2011){Berta}, {Magnelli}, {Nordon}, {Lutz}, {Wuyts},
  {Altieri}, {Andreani}, {Aussel}, {Casta{\~n}eda}, {Cepa}, {Cimatti}, {Daddi},
  {Elbaz}, {F{\"o}rster Schreiber}, {Genzel}, {Le Floc'h}, {Maiolino},
  {P{\'e}rez-Fournon}, {Poglitsch}, {Popesso}, {Pozzi}, {Riguccini},
  {Rodighiero}, {Sanchez-Portal}, {Sturm}, {Tacconi}, \&
  {Valtchanov}}]{berta11}
{Berta}, S., {Magnelli}, B., {Nordon}, R., {Lutz}, D., {Wuyts}, S., {Altieri},
  B., {Andreani}, P., {Aussel}, H., {Casta{\~n}eda}, H., {Cepa}, J., {Cimatti},
  A., {Daddi}, E., {Elbaz}, D., {F{\"o}rster Schreiber}, N.~M., {Genzel}, R.,
  {Le Floc'h}, E., {Maiolino}, R., {P{\'e}rez-Fournon}, I., {Poglitsch}, A.,
  {Popesso}, P., {Pozzi}, F., {Riguccini}, L., {Rodighiero}, G.,
  {Sanchez-Portal}, M., {Sturm}, E., {Tacconi}, L.~J., \& {Valtchanov}, I.
  2011, \aap, 532, A49+

\bibitem[{{B{\'e}thermin} {et~al.}(2010{\natexlab{a}}){B{\'e}thermin}, {Dole},
  {Beelen}, \& {Aussel}}]{bethermin10a}
{B{\'e}thermin}, M., {Dole}, H., {Beelen}, A., \& {Aussel}, H.
  2010{\natexlab{a}}, \aap, 512, A78+

\bibitem[{{B{\'e}thermin} {et~al.}(2010{\natexlab{b}}){B{\'e}thermin}, {Dole},
  {Cousin}, \& {Bavouzet}}]{bethermin10b}
{B{\'e}thermin}, M., {Dole}, H., {Cousin}, M., \& {Bavouzet}, N.
  2010{\natexlab{b}}, \aap, 516, A43+

\bibitem[{{B{\'e}thermin} {et~al.}(2012){B{\'e}thermin}, {Le Floc'h}, {Ilbert},
  {Conley}, {Lagache}, {Amblard}, {Arumugam}, {Aussel}, {Berta}, {Bock},
  {Boselli}, {Buat}, {Casey}, {Castro-Rodr{\'{\i}}guez}, {Cava}, {Clements},
  {Cooray}, {Dowell}, {Eales}, {Farrah}, {Franceschini}, {Glenn}, {Griffin},
  {Hatziminaoglou}, {Heinis}, {Ibar}, {Ivison}, {Kartaltepe}, {Levenson},
  {Magdis}, {Marchetti}, {Marsden}, {Nguyen}, {O'Halloran}, {Oliver}, {Omont},
  {Page}, {Panuzzo}, {Papageorgiou}, {Pearson}, {P{\'e}rez-Fournon}, {Pohlen},
  {Rigopoulou}, {Roseboom}, {Rowan-Robinson}, {Salvato}, {Schulz}, {Scott},
  {Seymour}, {Shupe}, {Smith}, {Symeonidis}, {Trichas}, {Tugwell}, {Vaccari},
  {Valtchanov}, {Vieira}, {Viero}, {Wang}, {Xu}, \& {Zemcov}}]{bethermin12}
{B{\'e}thermin}, M., {Le Floc'h}, E., {Ilbert}, O., {Conley}, A., {Lagache},
  G., {Amblard}, A., {Arumugam}, V., {Aussel}, H., {Berta}, S., {Bock}, J.,
  {Boselli}, A., {Buat}, V., {Casey}, C.~M., {Castro-Rodr{\'{\i}}guez}, N.,
  {Cava}, A., {Clements}, D.~L., {Cooray}, A., {Dowell}, C.~D., {Eales}, S.,
  {Farrah}, D., {Franceschini}, A., {Glenn}, J., {Griffin}, M.,
  {Hatziminaoglou}, E., {Heinis}, S., {Ibar}, E., {Ivison}, R.~J.,
  {Kartaltepe}, J.~S., {Levenson}, L., {Magdis}, G., {Marchetti}, L.,
  {Marsden}, G., {Nguyen}, H.~T., {O'Halloran}, B., {Oliver}, S.~J., {Omont},
  A., {Page}, M.~J., {Panuzzo}, P., {Papageorgiou}, A., {Pearson}, C.~P.,
  {P{\'e}rez-Fournon}, I., {Pohlen}, M., {Rigopoulou}, D., {Roseboom}, I.~G.,
  {Rowan-Robinson}, M., {Salvato}, M., {Schulz}, B., {Scott}, D., {Seymour},
  N., {Shupe}, D.~L., {Smith}, A.~J., {Symeonidis}, M., {Trichas}, M.,
  {Tugwell}, K.~E., {Vaccari}, M., {Valtchanov}, I., {Vieira}, J.~D., {Viero},
  M., {Wang}, L., {Xu}, C.~K., \& {Zemcov}, M. 2012, \aap, 542, A58

\bibitem[{{Biller} {et~al.}(1995){Biller}, {Akerlof}, {Buckley}, {Cawley},
  {Chantell}, {Fegan}, {Fennell}, {Gaidos}, {Hillas}, {Kerrick}, {Lamb},
  {Lewis}, {Meyer}, {Mohanty}, {O'Flaherty}, {Punch}, {Reynolds}, {Rose},
  {Rovero}, {Schubnell}, {Sembroski}, {Weekes}, \& {Wilson}}]{biller95}
{Biller}, S.~D., {Akerlof}, C.~W., {Buckley}, J., {Cawley}, M.~F., {Chantell},
  M., {Fegan}, D.~J., {Fennell}, S., {Gaidos}, J.~A., {Hillas}, A.~M.,
  {Kerrick}, A.~D., {Lamb}, R.~C., {Lewis}, D.~A., {Meyer}, D.~I., {Mohanty},
  G., {O'Flaherty}, K.~S., {Punch}, M., {Reynolds}, P.~T., {Rose}, H.~J.,
  {Rovero}, A.~C., {Schubnell}, M.~S., {Sembroski}, G., {Weekes}, T.~C., \&
  {Wilson}, C. 1995, \apj, 445, 227

\bibitem[{Biller {et~al.}(1998)Biller, Buckley, Burdett, Bussons~Gordo,
  Carter-Lewis, Fegan, Finley, Gaidos, Hillas, Krennrich, Lamb, Lessard,
  McEnery, Mohanty, Quinn, Rodgers, Rose, Samuelson, Sembroski, Skelton,
  Weekes, \& Zweerink}]{biller98}
Biller, S.~D., Buckley, J., Burdett, A., Bussons~Gordo, J., Carter-Lewis,
  D.~A., Fegan, D.~J., Finley, J., Gaidos, J.~A., Hillas, A.~M., Krennrich, F.,
  Lamb, R.~C., Lessard, R., McEnery, J.~E., Mohanty, G., Quinn, J., Rodgers,
  A.~J., Rose, H.~J., Samuelson, F., Sembroski, G., Skelton, P., Weekes, T.~C.,
  \& Zweerink, J. 1998, Phys. Rev. Lett., 80, 2992

\bibitem[{{Boettcher}(2010)}]{boettcher10}
{Boettcher}, M. 2010, ArXiv: 1006.5048

\bibitem[{{B{\"o}ttcher} {et~al.}(2008){B{\"o}ttcher}, {Dermer}, \&
  {Finke}}]{bottcher08}
{B{\"o}ttcher}, M., {Dermer}, C.~D., \& {Finke}, J.~D. 2008, \apjl, 679, L9

\bibitem[{{Bromm} \& {Loeb}(2002)}]{bromm02}
{Bromm}, V. \& {Loeb}, A. 2002, \apj, 575, 111

\bibitem[{{Bruzual} \& {Charlot}(2003)}]{bruzual03}
{Bruzual}, G. \& {Charlot}, S. 2003, \mnras, 344, 1000

\bibitem[{{{\c S}ent{\"u}rk} {et~al.}(2011){{\c S}ent{\"u}rk}, {for the VERITAS
  Collaboration}, {Fortin}, {Horan}, \& {for the Fermi-LAT
  Collaboration}}]{senturk11}
{{\c S}ent{\"u}rk}, G.~D., {for the VERITAS Collaboration}, {Fortin}, P.,
  {Horan}, D., \& {for the Fermi-LAT Collaboration}. 2011, ArXiv: 1109.6035

\bibitem[{{Cambr{\'e}sy} {et~al.}(2001){Cambr{\'e}sy}, {Reach}, {Beichman}, \&
  {Jarrett}}]{cambresy01}
{Cambr{\'e}sy}, L., {Reach}, W.~T., {Beichman}, C.~A., \& {Jarrett}, T.~H.
  2001, \apj, 555, 563

\bibitem[{{Catanese} {et~al.}(1997){Catanese}, {Bradbury}, {Breslin},
  {Buckley}, {Carter-Lewis}, {Cawley}, {Dermer}, {Fegan}, {Finley}, {Gaidos},
  {Hillas}, {Johnson}, {Krennrich}, {Lamb}, {Lessard}, {Macomb}, {McEnery},
  {Moriarty}, {Quinn}, {Rodgers}, {Rose}, {Samuelson}, {Sembroski},
  {Srinivasan}, {Weekes}, \& {Zweerink}}]{catanese97}
{Catanese}, M., {Bradbury}, S.~M., {Breslin}, A.~C., {Buckley}, J.~H.,
  {Carter-Lewis}, D.~A., {Cawley}, M.~F., {Dermer}, C.~D., {Fegan}, D.~J.,
  {Finley}, J.~P., {Gaidos}, J.~A., {Hillas}, A.~M., {Johnson}, W.~N.,
  {Krennrich}, F., {Lamb}, R.~C., {Lessard}, R.~W., {Macomb}, D.~J., {McEnery},
  J.~E., {Moriarty}, P., {Quinn}, J., {Rodgers}, A.~J., {Rose}, H.~J.,
  {Samuelson}, F.~W., {Sembroski}, G.~H., {Srinivasan}, R., {Weekes}, T.~C., \&
  {Zweerink}, J. 1997, \apjl, 487, L143

\bibitem[{{Cerruti, M. for the H.E.S.S. Collaboration }(2011)}]{cerruti11}
{Cerruti, M. for the H.E.S.S. Collaboration }. 2011, ArXiv; 1110.2119

\bibitem[{{Chary} {et~al.}(2004){Chary}, {Casertano}, {Dickinson}, {Ferguson},
  {Eisenhardt}, {Elbaz}, {Grogin}, {Moustakas}, {Reach}, \& {Yan}}]{chary04}
{Chary}, R., {Casertano}, S., {Dickinson}, M.~E., {Ferguson}, H.~C.,
  {Eisenhardt}, P.~R.~M., {Elbaz}, D., {Grogin}, N.~A., {Moustakas}, L.~A.,
  {Reach}, W.~T., \& {Yan}, H. 2004, \apjs, 154, 80

\bibitem[{{Chary} \& {Elbaz}(2001)}]{chary01}
{Chary}, R. \& {Elbaz}, D. 2001, \apj, 556, 562

\bibitem[{{Colin} {et~al.}(2011){Colin}, {Becerra Gonzalez}, {Lindfors},
  {Lombardi}, {Sitarek}, {Stamera}, \& {for the MAGIC Collaboration}}]{colin11}
{Colin}, P., {Becerra Gonzalez}, J., {Lindfors}, E., {Lombardi}, S., {Sitarek},
  J., {Stamera}, A., \& {for the MAGIC Collaboration}. 2011, ArXiv: 1110.0236

\bibitem[{{Condon} {et~al.}(1991){Condon}, {Anderson}, \& {Helou}}]{condon91}
{Condon}, J.~J., {Anderson}, M.~L., \& {Helou}, G. 1991, \apj, 376, 95

\bibitem[{{Coppin} {et~al.}(2006){Coppin}, {Chapin}, {Mortier}, {Scott},
  {Borys}, {Dunlop}, {Halpern}, {Hughes}, {Pope}, {Scott}, {Serjeant}, {Wagg},
  {Alexander}, {Almaini}, {Aretxaga}, {Babbedge}, {Best}, {Blain}, {Chapman},
  {Clements}, {Crawford}, {Dunne}, {Eales}, {Edge}, {Farrah}, {Gazta{\~n}aga},
  {Gear}, {Granato}, {Greve}, {Fox}, {Ivison}, {Jarvis}, {Jenness}, {Lacey},
  {Lepage}, {Mann}, {Marsden}, {Martinez-Sansigre}, {Oliver}, {Page},
  {Peacock}, {Pearson}, {Percival}, {Priddey}, {Rawlings}, {Rowan-Robinson},
  {Savage}, {Seigar}, {Sekiguchi}, {Silva}, {Simpson}, {Smail}, {Stevens},
  {Takagi}, {Vaccari}, {van Kampen}, \& {Willott}}]{coppin06}
{Coppin}, K., {Chapin}, E.~L., {Mortier}, A.~M.~J., {Scott}, S.~E., {Borys},
  C., {Dunlop}, J.~S., {Halpern}, M., {Hughes}, D.~H., {Pope}, A., {Scott}, D.,
  {Serjeant}, S., {Wagg}, J., {Alexander}, D.~M., {Almaini}, O., {Aretxaga},
  I., {Babbedge}, T., {Best}, P.~N., {Blain}, A., {Chapman}, S., {Clements},
  D.~L., {Crawford}, M., {Dunne}, L., {Eales}, S.~A., {Edge}, A.~C., {Farrah},
  D., {Gazta{\~n}aga}, E., {Gear}, W.~K., {Granato}, G.~L., {Greve}, T.~R.,
  {Fox}, M., {Ivison}, R.~J., {Jarvis}, M.~J., {Jenness}, T., {Lacey}, C.,
  {Lepage}, K., {Mann}, R.~G., {Marsden}, G., {Martinez-Sansigre}, A.,
  {Oliver}, S., {Page}, M.~J., {Peacock}, J.~A., {Pearson}, C.~P., {Percival},
  W.~J., {Priddey}, R.~S., {Rawlings}, S., {Rowan-Robinson}, M., {Savage},
  R.~S., {Seigar}, M., {Sekiguchi}, K., {Silva}, L., {Simpson}, C., {Smail},
  I., {Stevens}, J.~A., {Takagi}, T., {Vaccari}, M., {van Kampen}, E., \&
  {Willott}, C.~J. 2006, \mnras, 372, 1621

\bibitem[{{Dermer}(2012)}]{dermer12a}
{Dermer}, C.~D. 2012, ArXiv e-prints

\bibitem[{{Djannati-Atai } {et~al.}(1999){Djannati-Atai }, {Piron}, {Barrau},
  {Iacoucci}, {Punch}, {Tavernet}, {Bazer-Bachi}, {Cabot}, {Chounet},
  {Debiais}, {Degrange}, {Dezalay}, {Dumora}, {Espigat}, {Fabre}, {Fleury},
  {Fontaine}, {Ghesqui{\`e}re}, {Goret}, {Gouiffes}, {Grenier}, {Le Bohec},
  {Malet}, {Meynadier}, {Mohanty}, {Nuss}, {Par{\'e}}, {Qu{\'e}bert}, {Ragan},
  {Renault}, {Rivoal}, {Rob}, {Schahmaneche}, \& {Smith}}]{djannati-atai99}
{Djannati-Atai }, A., {Piron}, F., {Barrau}, A., {Iacoucci}, L., {Punch}, M.,
  {Tavernet}, J.-P., {Bazer-Bachi}, R., {Cabot}, H., {Chounet}, L.-M.,
  {Debiais}, G., {Degrange}, B., {Dezalay}, J.-P., {Dumora}, D., {Espigat}, P.,
  {Fabre}, B., {Fleury}, P., {Fontaine}, G., {Ghesqui{\`e}re}, C., {Goret}, P.,
  {Gouiffes}, C., {Grenier}, I.~A., {Le Bohec}, S., {Malet}, I., {Meynadier},
  C., {Mohanty}, G., {Nuss}, E., {Par{\'e}}, E., {Qu{\'e}bert}, J., {Ragan},
  K., {Renault}, C., {Rivoal}, M., {Rob}, L., {Schahmaneche}, K., \& {Smith},
  D.~A. 1999, \aap, 350, 17

\bibitem[{{Dole} {et~al.}(2006){Dole}, {Lagache}, {Puget}, {Caputi},
  {Fern{\'a}ndez-Conde}, {Le Floc'h}, {Papovich}, {P{\'e}rez-Gonz{\'a}lez},
  {Rieke}, \& {Blaylock}}]{dole06}
{Dole}, H., {Lagache}, G., {Puget}, J.-L., {Caputi}, K.~I.,
  {Fern{\'a}ndez-Conde}, N., {Le Floc'h}, E., {Papovich}, C.,
  {P{\'e}rez-Gonz{\'a}lez}, P.~G., {Rieke}, G.~H., \& {Blaylock}, M. 2006,
  \aap, 451, 417

\bibitem[{{Dole} {et~al.}(2004){Dole}, {Rieke}, {Lagache}, {Puget},
  {Alonso-Herrero}, {Bai}, {Blaylock}, {Egami}, {Engelbracht}, {Gordon},
  {Hines}, {Kelly}, {Le Floc'h}, {Misselt}, {Morrison}, {Muzerolle},
  {Papovich}, {Perez-Gonzalez}, {Rieke}, {Rigby}, {Neugebauer}, {Stansberry},
  {Su}, {Young}, {Beichman}, \& {Richards}}]{dole04}
{Dole}, H., {Rieke}, G.~H., {Lagache}, G., {Puget}, J., {Alonso-Herrero}, A.,
  {Bai}, L., {Blaylock}, M., {Egami}, E., {Engelbracht}, C.~W., {Gordon},
  K.~D., {Hines}, D.~C., {Kelly}, D.~M., {Le Floc'h}, E., {Misselt}, K.~A.,
  {Morrison}, J.~E., {Muzerolle}, J., {Papovich}, C., {Perez-Gonzalez}, P.~G.,
  {Rieke}, M.~J., {Rigby}, J.~R., {Neugebauer}, G., {Stansberry}, J.~A., {Su},
  K.~Y.~L., {Young}, E.~T., {Beichman}, C.~A., \& {Richards}, P.~L. 2004, ArXiv
  Astrophysics e-prints

\bibitem[{{Dom{\'{\i}}nguez} {et~al.}(2011){Dom{\'{\i}}nguez}, {Primack},
  {Rosario}, {Prada}, {Gilmore}, {Faber}, {Koo}, {Somerville},
  {P{\'e}rez-Torres}, {P{\'e}rez-Gonz{\'a}lez}, {Huang}, {Davis},
  {Guhathakurta}, {Barmby}, {Conselice}, {Lozano}, {Newman}, \&
  {Cooper}}]{dominguez11}
{Dom{\'{\i}}nguez}, A., {Primack}, J.~R., {Rosario}, D.~J., {Prada}, F.,
  {Gilmore}, R.~C., {Faber}, S.~M., {Koo}, D.~C., {Somerville}, R.~S.,
  {P{\'e}rez-Torres}, M.~A., {P{\'e}rez-Gonz{\'a}lez}, P., {Huang}, J.-S.,
  {Davis}, M., {Guhathakurta}, P., {Barmby}, P., {Conselice}, C.~J., {Lozano},
  M., {Newman}, J.~A., \& {Cooper}, M.~C. 2011, \mnras, 410, 2556

\bibitem[{{Donnarumma} {et~al.}(2009){Donnarumma}, {Pucella}, {Vittorini},
  {D'Ammando}, {Vercellone}, {Raiteri}, {Villata}, {Perri}, {Chen}, {Smart},
  {Kataoka}, {Kawai}, {Mori}, {Tosti}, {Impiombato}, {Takahashi}, {Sato},
  {Tavani}, {Bulgarelli}, {Chen}, {Giuliani}, {Longo}, {Pacciani}, {Argan},
  {Barbiellini}, {Boffelli}, {Caraveo}, {Cattaneo}, {Cocco}, {Contessi},
  {Costa}, {Del Monte}, {De Paris}, {Di Cocco}, {Evangelista}, {Feroci},
  {Ferrari}, {Fiorini}, {Froysland}, {Frutti}, {Fuschino}, {Galli}, {Gianotti},
  {Labanti}, {Lapshov}, {Lazzarotto}, {Lipari}, {Marisaldi}, {Mastropietro},
  {Mereghetti}, {Morelli}, {Moretti}, {Morselli}, {Pellizzoni}, {Perotti},
  {Piano}, {Picozza}, {Pilia}, {Porrovecchio}, {Prest}, {Rapisarda},
  {Rappoldi}, {Rubini}, {Sabatini}, {Scalise}, {Soffitta}, {Striani},
  {Trifoglio}, {Trois}, {Vallazza}, {Zambra}, {Zanello}, {Pittori},
  {Santolamazza}, {Verrecchia}, {Giommi}, {Antonelli}, {Colafrancesco}, \&
  {Salotti}}]{donnarumma09}
{Donnarumma}, I., {Pucella}, G., {Vittorini}, V., {D'Ammando}, F.,
  {Vercellone}, S., {Raiteri}, C.~M., {Villata}, M., {Perri}, M., {Chen},
  W.~P., {Smart}, R.~L., {Kataoka}, J., {Kawai}, N., {Mori}, Y., {Tosti}, G.,
  {Impiombato}, D., {Takahashi}, T., {Sato}, R., {Tavani}, M., {Bulgarelli},
  A., {Chen}, A.~W., {Giuliani}, A., {Longo}, F., {Pacciani}, L., {Argan}, A.,
  {Barbiellini}, G., {Boffelli}, F., {Caraveo}, P., {Cattaneo}, P.~W., {Cocco},
  V., {Contessi}, T., {Costa}, E., {Del Monte}, E., {De Paris}, G., {Di Cocco},
  G., {Evangelista}, Y., {Feroci}, M., {Ferrari}, A., {Fiorini}, M.,
  {Froysland}, T., {Frutti}, M., {Fuschino}, F., {Galli}, M., {Gianotti}, F.,
  {Labanti}, C., {Lapshov}, I., {Lazzarotto}, F., {Lipari}, P., {Marisaldi},
  M., {Mastropietro}, M., {Mereghetti}, S., {Morelli}, E., {Moretti}, E.,
  {Morselli}, A., {Pellizzoni}, A., {Perotti}, F., {Piano}, G., {Picozza}, P.,
  {Pilia}, M., {Porrovecchio}, G., {Prest}, M., {Rapisarda}, M., {Rappoldi},
  A., {Rubini}, A., {Sabatini}, S., {Scalise}, E., {Soffitta}, P., {Striani},
  E., {Trifoglio}, M., {Trois}, A., {Vallazza}, E., {Zambra}, A., {Zanello},
  D., {Pittori}, C., {Santolamazza}, P., {Verrecchia}, F., {Giommi}, P.,
  {Antonelli}, L.~A., {Colafrancesco}, S., \& {Salotti}, L. 2009, \apj, 707,
  1115

\bibitem[{{Draper} \& {Ballantyne}(2009)}]{draper09}
{Draper}, A.~R. \& {Ballantyne}, D.~R. 2009, \apj, 707, 778

\bibitem[{{Dube} {et~al.}(1979){Dube}, {Wickes}, \& {Wilkinson}}]{dube79}
{Dube}, R.~R., {Wickes}, W.~C., \& {Wilkinson}, D.~T. 1979, \apj, 232, 333

\bibitem[{{Dunne} {et~al.}(2000){Dunne}, {Eales}, {Edmunds}, {Ivison},
  {Alexander}, \& {Clements}}]{dunne00}
{Dunne}, L., {Eales}, S., {Edmunds}, M., {Ivison}, R., {Alexander}, P., \&
  {Clements}, D.~L. 2000, \mnras, 315, 115

\bibitem[{{Dwek} \& {Arendt}(1998)}]{dwek98b}
{Dwek}, E. \& {Arendt}, R.~G. 1998, \apjl, 508, L9

\bibitem[{{Dwek} {et~al.}(1998){Dwek}, {Arendt}, {Hauser}, {Fixsen}, {Kelsall},
  {Leisawitz}, {Pei}, {Wright}, {Mather}, {Moseley}, {Odegard}, {Shafer},
  {Silverberg}, \& {Weiland}}]{dwek98a}
{Dwek}, E., {Arendt}, R.~G., {Hauser}, M.~G., {Fixsen}, D., {Kelsall}, T.,
  {Leisawitz}, D., {Pei}, Y.~C., {Wright}, E.~L., {Mather}, J.~C., {Moseley},
  S.~H., {Odegard}, N., {Shafer}, R., {Silverberg}, R.~F., \& {Weiland}, J.~L.
  1998, \apj, 508, 106

\bibitem[{{Dwek} {et~al.}(2005{\natexlab{a}}){Dwek}, {Arendt}, \&
  {Krennrich}}]{dwek05c}
{Dwek}, E., {Arendt}, R.~G., \& {Krennrich}, F. 2005{\natexlab{a}}, \apj, 635,
  784

\bibitem[{{Dwek} \& {Barker}(2002)}]{dwek02}
{Dwek}, E. \& {Barker}, M.~K. 2002, \apj, 575, 7

\bibitem[{{Dwek} \& {Krennrich}(2005)}]{dwek05a}
{Dwek}, E. \& {Krennrich}, F. 2005, \apj, 618, 657

\bibitem[{{Dwek} {et~al.}(2005{\natexlab{b}}){Dwek}, {Krennrich}, \&
  {Arendt}}]{dwek05b}
{Dwek}, E., {Krennrich}, F., \& {Arendt}, R.~G. 2005{\natexlab{b}}, \apj, 634,
  155

\bibitem[{{Dwek} \& {Slavin}(1994)}]{dwek94}
{Dwek}, E. \& {Slavin}, J. 1994, \apj, 436, 696

\bibitem[{{Dwek} {et~al.}(2011){Dwek}, {Staguhn}, {Arendt}, {Capak}, {Kovacs},
  {Benford}, {Fixsen}, {Karim}, {Leclercq}, {Maher}, {Moseley}, {Schinnerer},
  \& {Sharp}}]{dwek11b}
{Dwek}, E., {Staguhn}, J.~G., {Arendt}, R.~G., {Capak}, P.~L., {Kovacs}, A.,
  {Benford}, D.~J., {Fixsen}, D., {Karim}, A., {Leclercq}, S., {Maher}, S.~F.,
  {Moseley}, S.~H., {Schinnerer}, E., \& {Sharp}, E.~H. 2011, \apj, 738, 36

\bibitem[{{Elbaz} {et~al.}(2002){Elbaz}, {Cesarsky}, {Chanial}, {Aussel},
  {Franceschini}, {Fadda}, \& {Chary}}]{elbaz02}
{Elbaz}, D., {Cesarsky}, C.~J., {Chanial}, P., {Aussel}, H., {Franceschini},
  A., {Fadda}, D., \& {Chary}, R.~R. 2002, \aap, 384, 848

\bibitem[{{Errando, M. et al.}(2011)}]{errando11}
{Errando, M. et al.} 2011, ArXiv: 1111.1209

\bibitem[{{Fazio} {et~al.}(2004){Fazio}, {Ashby}, {Barmby}, {Hora}, {Huang},
  {Pahre}, {Wang}, {Willner}, {Arendt}, {Moseley}, {Brodwin}, {Eisenhardt},
  {Stern}, {Tollestrup}, \& {Wright}}]{fazio04}
{Fazio}, G.~G., {Ashby}, M.~L.~N., {Barmby}, P., {Hora}, J.~L., {Huang}, J.-S.,
  {Pahre}, M.~A., {Wang}, Z., {Willner}, S.~P., {Arendt}, R.~G., {Moseley},
  S.~H., {Brodwin}, M., {Eisenhardt}, P., {Stern}, D., {Tollestrup}, E.~V., \&
  {Wright}, E.~L. 2004, \apjs, 154, 39

\bibitem[{{Finkbeiner} {et~al.}(2000){Finkbeiner}, {Davis}, \&
  {Schlegel}}]{finkbeiner00}
{Finkbeiner}, D.~P., {Davis}, M., \& {Schlegel}, D.~J. 2000, \apj, 544, 81

\bibitem[{{Finke} \& {Razzaque}(2009)}]{finke09}
{Finke}, J.~D. \& {Razzaque}, S. 2009, \apj, 698, 1761

\bibitem[{{Finke} {et~al.}(2010){Finke}, {Razzaque}, \& {Dermer}}]{finke10}
{Finke}, J.~D., {Razzaque}, S., \& {Dermer}, C.~D. 2010, \apj, 712, 238

\bibitem[{{Fioc} \& {Rocca-Volmerange}(1997)}]{fioc97}
{Fioc}, M. \& {Rocca-Volmerange}, B. 1997, \aap, 326, 950

\bibitem[{{Fixsen} {et~al.}(1998){Fixsen}, {Dwek}, {Mather}, {Bennett}, \&
  {Shafer}}]{fixsen98}
{Fixsen}, D.~J., {Dwek}, E., {Mather}, J.~C., {Bennett}, C.~L., \& {Shafer},
  R.~A. 1998, \apj, 508, 123

\bibitem[{{Fontanot} \& {Somerville}(2011)}]{fontanot11}
{Fontanot}, F. \& {Somerville}, R.~S. 2011, \mnras, 416, 2962

\bibitem[{{Fontanot} {et~al.}(2009){Fontanot}, {Somerville}, {Silva}, {Monaco},
  \& {Skibba}}]{fontanot09}
{Fontanot}, F., {Somerville}, R.~S., {Silva}, L., {Monaco}, P., \& {Skibba}, R.
  2009, \mnras, 392, 553

\bibitem[{{Fossati} {et~al.}(2008){Fossati}, {Buckley}, {Bond}, {Bradbury},
  {Carter-Lewis}, {Chow}, {Cui}, {Falcone}, {Finley}, {Gaidos}, {Grube},
  {Holder}, {Horan}, {Horns}, {Jordan}, {Kieda}, {Kildea}, {Krawczynski},
  {Krennrich}, {Lang}, {LeBohec}, {Lee}, {Moriarty}, {Ong}, {Petry}, {Quinn},
  {Sembroski}, {Wakely}, \& {Weekes}}]{fossati08}
{Fossati}, G., {Buckley}, J.~H., {Bond}, I.~H., {Bradbury}, S.~M.,
  {Carter-Lewis}, D.~A., {Chow}, Y.~C.~K., {Cui}, W., {Falcone}, A.~D.,
  {Finley}, J.~P., {Gaidos}, J.~A., {Grube}, J., {Holder}, J., {Horan}, D.,
  {Horns}, D., {Jordan}, M.~M., {Kieda}, D.~B., {Kildea}, J., {Krawczynski},
  H., {Krennrich}, F., {Lang}, M.~J., {LeBohec}, S., {Lee}, K., {Moriarty}, P.,
  {Ong}, R.~A., {Petry}, D., {Quinn}, J., {Sembroski}, G.~H., {Wakely}, S.~P.,
  \& {Weekes}, T.~C. 2008, \apj, 677, 906

\bibitem[{{Fossati} {et~al.}(1998){Fossati}, {Maraschi}, {Celotti}, {Comastri},
  \& {Ghisellini}}]{fossati98}
{Fossati}, G., {Maraschi}, L., {Celotti}, A., {Comastri}, A., \& {Ghisellini},
  G. 1998, \mnras, 299, 433

\bibitem[{{Franceschini} {et~al.}(2002){Franceschini}, {Fadda}, {Cesarsky},
  {Elbaz}, {Flores}, \& {Granato}}]{franceschini02}
{Franceschini}, A., {Fadda}, D., {Cesarsky}, C.~J., {Elbaz}, D., {Flores}, H.,
  \& {Granato}, G.~L. 2002, \apj, 568, 470

\bibitem[{{Franceschini} {et~al.}(2008){Franceschini}, {Rodighiero}, \&
  {Vaccari}}]{franceschini08}
{Franceschini}, A., {Rodighiero}, G., \& {Vaccari}, M. 2008, \aap, 487, 837

\bibitem[{{Frayer} {et~al.}(2006){Frayer}, {Huynh}, {Chary}, {Dickinson},
  {Elbaz}, {Fadda}, {Surace}, {Teplitz}, {Yan}, \& {Mobasher}}]{frayer06}
{Frayer}, D.~T., {Huynh}, M.~T., {Chary}, R., {Dickinson}, M., {Elbaz}, D.,
  {Fadda}, D., {Surace}, J.~A., {Teplitz}, H.~I., {Yan}, L., \& {Mobasher}, B.
  2006, \apjl, 647, L9

\bibitem[{{Fumagalli} {et~al.}(2012){Fumagalli}, {Furniss}, {O'Meara},
  {Prochaska}, {Williams}, \& {Farina}}]{fumagalli12}
{Fumagalli}, M., {Furniss}, A., {O'Meara}, J.~M., {Prochaska}, J.~X.,
  {Williams}, D.~A., \& {Farina}, E.~P. 2012, ArXiv e-prints

\bibitem[{{Funk} {et~al.}(1998){Funk}, {Magnussen}, {Meyer}, {Rhode},
  {Westerhoff}, \& {Wiebel-Sooth}}]{funk98}
{Funk}, B., {Magnussen}, N., {Meyer}, H., {Rhode}, W., {Westerhoff}, S., \&
  {Wiebel-Sooth}, B. 1998, Astroparticle Physics, 9, 97

\bibitem[{{Gaidos} {et~al.}(1996){Gaidos}, {Akerlof}, {Biller}, {Boyle},
  {Breslin}, {Buckley}, {Carter-Lewis}, {Catanese}, {Cawley}, {Fegan},
  {Finley}, {Gordo}, {Hillas}, {Krennrich}, {Lamb}, {Lessard}, {McEnery},
  {Masterson}, {Mohanty}, {Moriarty}, {Quinn}, {Rodgers}, {Rose}, {Samuelson},
  {Schubnell}, {Sembroski}, {Srinivasan}, {Weekes}, {Wilson}, \&
  {Zweerink}}]{gaidos96}
{Gaidos}, J.~A., {Akerlof}, C.~W., {Biller}, S., {Boyle}, P.~J., {Breslin},
  A.~C., {Buckley}, J.~H., {Carter-Lewis}, D.~A., {Catanese}, M., {Cawley},
  M.~F., {Fegan}, D.~J., {Finley}, J.~P., {Gordo}, J.~B., {Hillas}, A.~M.,
  {Krennrich}, F., {Lamb}, R.~C., {Lessard}, R.~W., {McEnery}, J.~E.,
  {Masterson}, C., {Mohanty}, G., {Moriarty}, P., {Quinn}, J., {Rodgers},
  A.~J., {Rose}, H.~J., {Samuelson}, F., {Schubnell}, M.~S., {Sembroski},
  G.~H., {Srinivasan}, R., {Weekes}, T.~C., {Wilson}, C.~L., \& {Zweerink}, J.
  1996, \nat, 383, 319

\bibitem[{{Gandhi} \& {Fabian}(2003)}]{gandhi03}
{Gandhi}, P. \& {Fabian}, A.~C. 2003, \mnras, 339, 1095

\bibitem[{{Gardner} {et~al.}(2000){Gardner}, {Brown}, \&
  {Ferguson}}]{gardner00}
{Gardner}, J.~P., {Brown}, T.~M., \& {Ferguson}, H.~C. 2000, \apjl, 542, L79

\bibitem[{{Gardner} {et~al.}(2006){Gardner}, {Mather}, {Clampin}, {Doyon},
  {Greenhouse}, {Hammel}, {Hutchings}, {Jakobsen}, {Lilly}, {Long}, {Lunine},
  {McCaughrean}, {Mountain}, {Nella}, {Rieke}, {Rieke}, {Rix}, {Smith},
  {Sonneborn}, {Stiavelli}, {Stockman}, {Windhorst}, \& {Wright}}]{gardner06}
{Gardner}, J.~P., {Mather}, J.~C., {Clampin}, M., {Doyon}, R., {Greenhouse},
  M.~A., {Hammel}, H.~B., {Hutchings}, J.~B., {Jakobsen}, P., {Lilly}, S.~J.,
  {Long}, K.~S., {Lunine}, J.~I., {McCaughrean}, M.~J., {Mountain}, M.,
  {Nella}, J., {Rieke}, G.~H., {Rieke}, M.~J., {Rix}, H.-W., {Smith}, E.~P.,
  {Sonneborn}, G., {Stiavelli}, M., {Stockman}, H.~S., {Windhorst}, R.~A., \&
  {Wright}, G.~S. 2006, \ssr, 123, 485

\bibitem[{{Georganopoulos} {et~al.}(2008){Georganopoulos}, {Sambruna},
  {Kazanas}, {Cillis}, {Cheung}, {Perlman}, {Blundell}, \&
  {Davis}}]{georganopoulos08}
{Georganopoulos}, M., {Sambruna}, R.~M., {Kazanas}, D., {Cillis}, A.~N.,
  {Cheung}, C.~C., {Perlman}, E.~S., {Blundell}, K.~M., \& {Davis}, D.~S. 2008,
  \apjl, 686, L5

\bibitem[{{Georganopoulos} {et~al.}(2010){Georganopoulos}, {Sambruna},
  {Kazanas}, {Davis}, {Cillis}, {Cheung}, {Perlman}, \&
  {Blundell}}]{georganopoulos10a}
{Georganopoulos}, M., {Sambruna}, R.~M., {Kazanas}, D., {Davis}, D.~S.,
  {Cillis}, A.~N., {Cheung}, C.~C., {Perlman}, E.~S., \& {Blundell}, K.~M.
  2010, in Astronomical Society of the Pacific Conference Series, Vol. 427,
  Accretion and Ejection in AGN: a Global View, ed. L.~{Maraschi},
  G.~{Ghisellini}, R.~{Della Ceca}, \& F.~{Tavecchio}, 177

\bibitem[{{Gilmore} {et~al.}(2011){Gilmore}, {Somerville}, {Primack}, \&
  {Dom{\'{\i}}nguez}}]{gilmore11}
{Gilmore}, R.~C., {Somerville}, R.~S., {Primack}, J.~R., \& {Dom{\'{\i}}nguez},
  A. 2011, ArXiv: 1104.0671

\bibitem[{{Gordon} {et~al.}(2001){Gordon}, {Misselt}, {Witt}, \&
  {Clayton}}]{gordon01}
{Gordon}, K.~D., {Misselt}, K.~A., {Witt}, A.~N., \& {Clayton}, G.~C. 2001,
  \apj, 551, 269

\bibitem[{{Guy} {et~al.}(2000){Guy}, {Renault}, {Aharonian}, {Rivoal}, \&
  {Tavernet}}]{guy00}
{Guy}, J., {Renault}, C., {Aharonian}, F.~A., {Rivoal}, M., \& {Tavernet},
  J.-P. 2000, \aap, 359, 419

\bibitem[{{Haarsma} \& {Partridge}(1998)}]{haarsma98}
{Haarsma}, D.~B. \& {Partridge}, R.~B. 1998, \apjl, 503, L5

\bibitem[{{Haarsma} {et~al.}(2000){Haarsma}, {Partridge}, {Windhorst}, \&
  {Richards}}]{Haarsma00}
{Haarsma}, D.~B., {Partridge}, R.~B., {Windhorst}, R.~A., \& {Richards}, E.~A.
  2000, \apj, 544, 641

\bibitem[{{Hainline} {et~al.}(2009){Hainline}, {Blain}, {Smail}, {Frayer},
  {Chapman}, {Ivison}, \& {Alexander}}]{hainline09}
{Hainline}, L.~J., {Blain}, A.~W., {Smail}, I., {Frayer}, D.~T., {Chapman},
  S.~C., {Ivison}, R.~J., \& {Alexander}, D.~M. 2009, \apj, 699, 1610

\bibitem[{{Hauser} {et~al.}(1998){Hauser}, {Arendt}, {Kelsall}, {Dwek},
  {Odegard}, {Weiland}, {Freudenreich}, {Reach}, {Silverberg}, {Moseley},
  {Pei}, {Lubin}, {Mather}, {Shafer}, {Smoot}, {Weiss}, {Wilkinson}, \&
  {Wright}}]{hauser98}
{Hauser}, M.~G., {Arendt}, R.~G., {Kelsall}, T., {Dwek}, E., {Odegard}, N.,
  {Weiland}, J.~L., {Freudenreich}, H.~T., {Reach}, W.~T., {Silverberg}, R.~F.,
  {Moseley}, S.~H., {Pei}, Y.~C., {Lubin}, P., {Mather}, J.~C., {Shafer},
  R.~A., {Smoot}, G.~F., {Weiss}, R., {Wilkinson}, D.~T., \& {Wright}, E.~L.
  1998, \apj, 508, 25

\bibitem[{{Hauser} \& {Dwek}(2001)}]{hauser01}
{Hauser}, M.~G. \& {Dwek}, E. 2001, \araa, 39, 249

\bibitem[{Helgason \& Kashlinsky(2012)}]{helgason12b}
Helgason, K. \& Kashlinsky, A. 2012, The Astrophysical Journal Letters, 758,
  L13

\bibitem[{{Helgason} {et~al.}(2012){Helgason}, {Ricotti}, \&
  {Kashlinsky}}]{helgason12a}
{Helgason}, K., {Ricotti}, M., \& {Kashlinsky}, A. 2012, \apj, 752, 113

\bibitem[{{Hildebrand} {et~al.}(2011){Hildebrand}, {Lombardi}, {Colin},
  {Sitarek}, {Zandanel}, {Prada}, {for the MAGIC Collaboration}, {Pfrommer}, \&
  {Pinzke}}]{hildebrand11}
{Hildebrand}, D., {Lombardi}, S., {Colin}, P., {Sitarek}, J., {Zandanel}, F.,
  {Prada}, F., {for the MAGIC Collaboration}, {Pfrommer}, C., \& {Pinzke}, A.
  2011, ArXiv: 1110.5358

\bibitem[{{Hinshaw} {et~al.}(2009){Hinshaw}, {Weiland}, {Hill}, {Odegard},
  {Larson}, {Bennett}, {Dunkley}, {Gold}, {Greason}, {Jarosik}, {Komatsu},
  {Nolta}, {Page}, {Spergel}, {Wollack}, {Halpern}, {Kogut}, {Limon}, {Meyer},
  {Tucker}, \& {Wright}}]{hinshaw09}
{Hinshaw}, G., {Weiland}, J.~L., {Hill}, R.~S., {Odegard}, N., {Larson}, D.,
  {Bennett}, C.~L., {Dunkley}, J., {Gold}, B., {Greason}, M.~R., {Jarosik}, N.,
  {Komatsu}, E., {Nolta}, M.~R., {Page}, L., {Spergel}, D.~N., {Wollack}, E.,
  {Halpern}, M., {Kogut}, A., {Limon}, M., {Meyer}, S.~S., {Tucker}, G.~S., \&
  {Wright}, E.~L. 2009, \apjs, 180, 225

\bibitem[{{Hinton} \& {Hofmann}(2009)}]{hinton09}
{Hinton}, J.~A. \& {Hofmann}, W. 2009, \araa, 47, 523

\bibitem[{{Hopkins} \& {Beacom}(2006)}]{hopkins06}
{Hopkins}, A.~M. \& {Beacom}, J.~F. 2006, \apj, 651, 142

\bibitem[{{Hopwood} {et~al.}(2010){Hopwood}, {Serjeant}, {Negrello}, {Pearson},
  {Egami}, {Im}, {Kneib}, {Ko}, {Lee}, {Lee}, {Matsuhara}, {Nakagawa}, {Smail},
  \& {Takagi}}]{hopwood10}
{Hopwood}, R., {Serjeant}, S., {Negrello}, M., {Pearson}, C., {Egami}, E.,
  {Im}, M., {Kneib}, J.-P., {Ko}, J., {Lee}, H.~M., {Lee}, M.~G., {Matsuhara},
  H., {Nakagawa}, T., {Smail}, I., \& {Takagi}, T. 2010, \apjl, 716, L45

\bibitem[{{Horiuchi} {et~al.}(2009){Horiuchi}, {Beacom}, \&
  {Dwek}}]{horiuchi09}
{Horiuchi}, S., {Beacom}, J.~F., \& {Dwek}, E. 2009, \prd, 79, 083013

\bibitem[{{Inoue}(2011)}]{inoue11}
{Inoue}, Y. 2011, \apj, 733, 66

\bibitem[{{Inoue} \& {Ioka}(2012)}]{inoue12}
{Inoue}, Y. \& {Ioka}, K. 2012, \prd, 86, 023003

\bibitem[{{Kampert}(2012)}]{kampert12}
{Kampert}, K.-H.~f. 2012, ArXiv e-prints

\bibitem[{{Kashlinsky}(2005)}]{kashlinsky05a}
{Kashlinsky}, A. 2005, \physrep, 409, 361

\bibitem[{{Kashlinsky} {et~al.}(2012){Kashlinsky}, {Arendt}, {Ashby}, {Fazio},
  {Mather}, \& {Moseley}}]{kashlinsky12}
{Kashlinsky}, A., {Arendt}, R.~G., {Ashby}, M.~L.~N., {Fazio}, G.~G., {Mather},
  J., \& {Moseley}, S.~H. 2012, \apj, 753, 63

\bibitem[{{Kashlinsky} {et~al.}(1996){Kashlinsky}, {Mather}, {Odenwald}, \&
  {Hauser}}]{kashlinsky96}
{Kashlinsky}, A., {Mather}, J.~C., {Odenwald}, S., \& {Hauser}, M.~G. 1996,
  \apj, 470, 681

\bibitem[{{Katarzy{\'n}ski} {et~al.}(2006){Katarzy{\'n}ski}, {Ghisellini},
  {Tavecchio}, {Gracia}, \& {Maraschi}}]{katarzynski06}
{Katarzy{\'n}ski}, K., {Ghisellini}, G., {Tavecchio}, F., {Gracia}, J., \&
  {Maraschi}, L. 2006, \mnras, 368, L52

\bibitem[{{Keenan} {et~al.}(2010){Keenan}, {Barger}, {Cowie}, \&
  {Wang}}]{keenan10}
{Keenan}, R.~C., {Barger}, A.~J., {Cowie}, L.~L., \& {Wang}, W.-H. 2010, \apj,
  723, 40

\bibitem[{{Kelsall} {et~al.}(1998){Kelsall}, {Weiland}, {Franz}, {Reach},
  {Arendt}, {Dwek}, {Freudenreich}, {Hauser}, {Moseley}, {Odegard},
  {Silverberg}, \& {Wright}}]{kelsall98}
{Kelsall}, T., {Weiland}, J.~L., {Franz}, B.~A., {Reach}, W.~T., {Arendt},
  R.~G., {Dwek}, E., {Freudenreich}, H.~T., {Hauser}, M.~G., {Moseley}, S.~H.,
  {Odegard}, N.~P., {Silverberg}, R.~F., \& {Wright}, E.~L. 1998, \apj, 508, 44

\bibitem[{{Kennicutt}(1998)}]{kennicutt98b}
{Kennicutt}, Jr., R.~C. 1998, \apj, 498, 541

\bibitem[{{Krawczynski} {et~al.}(2004){Krawczynski}, {Hughes}, {Horan},
  {Aharonian}, {Aller}, {Aller}, {Boltwood}, {Buckley}, {Coppi}, {Fossati},
  {G{\"o}tting}, {Holder}, {Horns}, {Kurtanidze}, {Marscher}, {Nikolashvili},
  {Remillard}, {Sadun}, \& {Schr{\"o}der}}]{Krawczynski04}
{Krawczynski}, H., {Hughes}, S.~B., {Horan}, D., {Aharonian}, F., {Aller},
  M.~F., {Aller}, H., {Boltwood}, P., {Buckley}, J., {Coppi}, P., {Fossati},
  G., {G{\"o}tting}, N., {Holder}, J., {Horns}, D., {Kurtanidze}, O.~M.,
  {Marscher}, A.~P., {Nikolashvili}, M., {Remillard}, R.~A., {Sadun}, A., \&
  {Schr{\"o}der}, M. 2004, \apj, 601, 151

\bibitem[{{Krennrich} {et~al.}(2002){Krennrich}, {Bond}, {Bradbury}, {Buckley},
  {Carter-Lewis}, {Cui}, {de la Calle Perez}, {Fegan}, {Fegan}, {Finley},
  {Gaidos}, {Gibbs}, {Gillanders}, {Hall}, {Hillas}, {Holder}, {Horan},
  {Jordan}, {Kertzman}, {Kieda}, {Kildea}, {Knapp}, {Kosack}, {Lang},
  {LeBohec}, {Moriarty}, {M{\"u}ller}, {Ong}, {Pallassini}, {Petry}, {Quinn},
  {Reay}, {Reynolds}, {Rose}, {Sembroski}, {Sidwell}, {Stanton}, {Swordy},
  {Vassiliev}, {Wakely}, \& {Weekes}}]{krennrich02}
{Krennrich}, F., {Bond}, I.~H., {Bradbury}, S.~M., {Buckley}, J.~H.,
  {Carter-Lewis}, D.~A., {Cui}, W., {de la Calle Perez}, I., {Fegan}, D.~J.,
  {Fegan}, S.~J., {Finley}, J.~P., {Gaidos}, J.~A., {Gibbs}, K., {Gillanders},
  G.~H., {Hall}, T.~A., {Hillas}, A.~M., {Holder}, J., {Horan}, D., {Jordan},
  M., {Kertzman}, M., {Kieda}, D., {Kildea}, J., {Knapp}, J., {Kosack}, K.,
  {Lang}, M.~J., {LeBohec}, S., {Moriarty}, P., {M{\"u}ller}, D., {Ong}, R.~A.,
  {Pallassini}, R., {Petry}, D., {Quinn}, J., {Reay}, N.~W., {Reynolds}, P.~T.,
  {Rose}, H.~J., {Sembroski}, G.~H., {Sidwell}, R., {Stanton}, N., {Swordy},
  S.~P., {Vassiliev}, V.~V., {Wakely}, S.~P., \& {Weekes}, T.~C. 2002, \apjl,
  575, L9

\bibitem[{{Krennrich} {et~al.}(2008){Krennrich}, {Dwek}, \&
  {Imran}}]{krennrich08}
{Krennrich}, F., {Dwek}, E., \& {Imran}, A. 2008, \apjl, 689, L93

\bibitem[{{Kronberg}(1994)}]{kronberg94}
{Kronberg}, P.~P. 1994, Reports on Progress in Physics, 57, 325

\bibitem[{{Kronberg}(2010)}]{kronberg10}
---. 2010, ArXiv e-prints

\bibitem[{{Kusenko}(2012)}]{kusenko12}
{Kusenko}, A. 2012, ArXiv e-prints

\bibitem[{{Kutyrev} {et~al.}(2008){Kutyrev}, {Arendt}, {Dwek}, {Moseley},
  {Rapchun}, \& {Silverberg}}]{kutyrev08}
{Kutyrev}, A.~S., {Arendt}, R., {Dwek}, E., {Moseley}, S.~H., {Rapchun}, D., \&
  {Silverberg}, R.~F. 2008, in Society of Photo-Optical Instrumentation
  Engineers (SPIE) Conference Series, Vol. 7014, Society of Photo-Optical
  Instrumentation Engineers (SPIE) Conference Series

\bibitem[{{Kutyrev} {et~al.}(2004){Kutyrev}, {Bennett}, {Moseley}, {Rapchun},
  \& {Stewart}}]{kutyrev04}
{Kutyrev}, A.~S., {Bennett}, C.~L., {Moseley}, S.~H., {Rapchun}, D., \&
  {Stewart}, K.~P. 2004, in Society of Photo-Optical Instrumentation Engineers
  (SPIE) Conference Series, Vol. 5492, Society of Photo-Optical Instrumentation
  Engineers (SPIE) Conference Series, ed. {A.~F.~M.~Moorwood \& M.~Iye},
  1172--1178

\bibitem[{{Lagache} {et~al.}(2007){Lagache}, {Bavouzet}, {Fernandez-Conde},
  {Ponthieu}, {Rodet}, {Dole}, {Miville-Desch{\^e}nes}, \& {Puget}}]{lagache07}
{Lagache}, G., {Bavouzet}, N., {Fernandez-Conde}, N., {Ponthieu}, N., {Rodet},
  T., {Dole}, H., {Miville-Desch{\^e}nes}, M.-A., \& {Puget}, J.-L. 2007,
  \apjl, 665, L89

\bibitem[{{Lagache} {et~al.}(2003){Lagache}, {Dole}, \& {Puget}}]{lagache03}
{Lagache}, G., {Dole}, H., \& {Puget}, J.-L. 2003, \mnras, 338, 555

\bibitem[{{Lagache} {et~al.}(2000){Lagache}, {Haffner}, {Reynolds}, \&
  {Tufte}}]{lagache00a}
{Lagache}, G., {Haffner}, L.~M., {Reynolds}, R.~J., \& {Tufte}, S.~L. 2000,
  \aap, 354, 247

\bibitem[{{Lagache} \& {Puget}(2000)}]{lagache00b}
{Lagache}, G. \& {Puget}, J.~L. 2000, \aap, 355, 17

\bibitem[{{Lagache} {et~al.}(2005){Lagache}, {Puget}, \& {Dole}}]{lagache05}
{Lagache}, G., {Puget}, J.-L., \& {Dole}, H. 2005, \araa, 43, 727

\bibitem[{{Landt}(2012)}]{landt12}
{Landt}, H. 2012, \mnras, 423, L84

\bibitem[{{Lefa} {et~al.}(2011){Lefa}, {Rieger}, \& {Aharonian}}]{lefa11}
{Lefa}, E., {Rieger}, F.~M., \& {Aharonian}, F.~A. 2011, \apj, 740, 64

\bibitem[{{Leinert} {et~al.}(1998){Leinert}, {Bowyer}, {Haikala}, {Hanner},
  {Hauser}, {Levasseur-Regourd}, {Mann}, {Mattila}, {Reach}, {Schlosser},
  {Staude}, {Toller}, {Weiland}, {Weinberg}, \& {Witt}}]{leinert98}
{Leinert}, C., {Bowyer}, S., {Haikala}, L.~K., {Hanner}, M.~S., {Hauser},
  M.~G., {Levasseur-Regourd}, A.-C., {Mann}, I., {Mattila}, K., {Reach}, W.~T.,
  {Schlosser}, W., {Staude}, H.~J., {Toller}, G.~N., {Weiland}, J.~L.,
  {Weinberg}, J.~L., \& {Witt}, A.~N. 1998, \aaps, 127, 1

\bibitem[{{Leitherer} {et~al.}(1999){Leitherer}, {Schaerer}, {Goldader},
  {Gonz{\'a}lez Delgado}, {Robert}, {Kune}, {de Mello}, {Devost}, \&
  {Heckman}}]{leitherer99}
{Leitherer}, C., {Schaerer}, D., {Goldader}, J.~D., {Gonz{\'a}lez Delgado},
  R.~M., {Robert}, C., {Kune}, D.~F., {de Mello}, D.~F., {Devost}, D., \&
  {Heckman}, T.~M. 1999, \apjs, 123, 3

\bibitem[{{Levenson} \& {Wright}(2008)}]{levenson08}
{Levenson}, L.~R. \& {Wright}, E.~L. 2008, \apj, 683, 585

\bibitem[{{Levenson} {et~al.}(2007){Levenson}, {Wright}, \&
  {Johnson}}]{levenson07}
{Levenson}, L.~R., {Wright}, E.~L., \& {Johnson}, B.~D. 2007, \apj, 666, 34

\bibitem[{{Lin} {et~al.}(1992){Lin}, {Bertsch}, {Chiang}, {Fichtel}, {Hartman},
  {Hunter}, {Kanbach}, {Kniffen}, {Kwok}, {Mattox}, {Mayer-Hasselwander},
  {Michelson}, {von Montigny}, {Nolan}, {Pinkau}, {Schneid}, {Sreekumar}, \&
  {Thompson}}]{lin92}
{Lin}, Y.~C., {Bertsch}, D.~L., {Chiang}, J., {Fichtel}, C.~E., {Hartman},
  R.~C., {Hunter}, S.~D., {Kanbach}, G., {Kniffen}, D.~A., {Kwok}, P.~W.,
  {Mattox}, J.~R., {Mayer-Hasselwander}, H.~A., {Michelson}, P.~F., {von
  Montigny}, C., {Nolan}, P.~L., {Pinkau}, K., {Schneid}, E., {Sreekumar}, P.,
  \& {Thompson}, D.~J. 1992, \apjl, 401, L61

\bibitem[{{Lisenfeld} {et~al.}(1996){Lisenfeld}, {Voelk}, \&
  {Xu}}]{lisenfeld96}
{Lisenfeld}, U., {Voelk}, H.~J., \& {Xu}, C. 1996, \aap, 306, 677

\bibitem[{{MacMinn} \& {Primack}(1996)}]{macminn96}
{MacMinn}, D. \& {Primack}, J.~R. 1996, \ssr, 75, 413

\bibitem[{{Madau} {et~al.}(1996){Madau}, {Ferguson}, {Dickinson}, {Giavalisco},
  {Steidel}, \& {Fruchter}}]{madau96}
{Madau}, P., {Ferguson}, H.~C., {Dickinson}, M.~E., {Giavalisco}, M.,
  {Steidel}, C.~C., \& {Fruchter}, A. 1996, \mnras, 283, 1388

\bibitem[{{Madau} \& {Pozzetti}(2000)}]{madau00}
{Madau}, P. \& {Pozzetti}, L. 2000, \mnras, 312, L9

\bibitem[{{Malkov} \& {O'C Drury}(2001)}]{malkov01}
{Malkov}, M.~A. \& {O'C Drury}, L. 2001, Reports on Progress in Physics, 64,
  429

\bibitem[{{Mankuzhiyil} {et~al.}(2010){Mankuzhiyil}, {Persic}, \&
  {Tavecchio}}]{mankuzhiyil10}
{Mankuzhiyil}, N., {Persic}, M., \& {Tavecchio}, F. 2010, \apjl, 715, L16

\bibitem[{{Mannheim}(1993)}]{mannheim93}
{Mannheim}, K. 1993, \aap, 269, 67

\bibitem[{{Mannheim}(1998)}]{mannheim98}
---. 1998, Science, 279, 684

\bibitem[{{Maraston} \& {Str{\"o}mb{\"a}ck}(2011)}]{maraston11}
{Maraston}, C. \& {Str{\"o}mb{\"a}ck}, G. 2011, \mnras, 1817

\bibitem[{{Marsden} {et~al.}(2009){Marsden}, {Ade}, {Bock}, {Chapin}, {Devlin},
  {Dicker}, {Griffin}, {Gundersen}, {Halpern}, {Hargrave}, {Hughes}, {Klein},
  {Mauskopf}, {Magnelli}, {Moncelsi}, {Netterfield}, {Ngo}, {Olmi}, {Pascale},
  {Patanchon}, {Rex}, {Scott}, {Semisch}, {Thomas}, {Truch}, {Tucker},
  {Tucker}, {Viero}, \& {Wiebe}}]{marsden09}
{Marsden}, G., {Ade}, P.~A.~R., {Bock}, J.~J., {Chapin}, E.~L., {Devlin},
  M.~J., {Dicker}, S.~R., {Griffin}, M., {Gundersen}, J.~O., {Halpern}, M.,
  {Hargrave}, P.~C., {Hughes}, D.~H., {Klein}, J., {Mauskopf}, P., {Magnelli},
  B., {Moncelsi}, L., {Netterfield}, C.~B., {Ngo}, H., {Olmi}, L., {Pascale},
  E., {Patanchon}, G., {Rex}, M., {Scott}, D., {Semisch}, C., {Thomas}, N.,
  {Truch}, M.~D.~P., {Tucker}, C., {Tucker}, G.~S., {Viero}, M.~P., \& {Wiebe},
  D.~V. 2009, \apj, 707, 1729

\bibitem[{{Matsumoto}(2001)}]{matsumoto01}
{Matsumoto}, T. 2001, in IAU Symposium, Vol. 204, The Extragalactic Infrared
  Background and its Cosmological Implications, ed. {M.~Harwit \&
  M.~G.~Hauser}, 87

\bibitem[{{Matsumoto} {et~al.}(2005){Matsumoto}, {Matsuura}, {Murakami},
  {Tanaka}, {Freund}, {Lim}, {Cohen}, {Kawada}, \& {Noda}}]{matsumoto05a}
{Matsumoto}, T., {Matsuura}, S., {Murakami}, H., {Tanaka}, M., {Freund}, M.,
  {Lim}, M., {Cohen}, M., {Kawada}, M., \& {Noda}, M. 2005, \apj, 626, 31

\bibitem[{{Matsuoka} {et~al.}(2011){Matsuoka}, {Ienaka}, {Kawara}, \&
  {Oyabu}}]{matsuoka11}
{Matsuoka}, Y., {Ienaka}, N., {Kawara}, K., \& {Oyabu}, S. 2011, \apj, 736, 119

\bibitem[{{Matsuura} {et~al.}(2011){Matsuura}, {Shirahata}, {Kawada},
  {Takeuchi}, {Burgarella}, {Clements}, {Jeong}, {Hanami}, {Khan}, {Matsuhara},
  {Nakagawa}, {Oyabu}, {Pearson}, {Pollo}, {Serjeant}, {Takagi}, \&
  {White}}]{matsuura11b}
{Matsuura}, S., {Shirahata}, M., {Kawada}, M., {Takeuchi}, T.~T., {Burgarella},
  D., {Clements}, D.~L., {Jeong}, W.-S., {Hanami}, H., {Khan}, S.~A.,
  {Matsuhara}, H., {Nakagawa}, T., {Oyabu}, S., {Pearson}, C.~P., {Pollo}, A.,
  {Serjeant}, S., {Takagi}, T., \& {White}, G.~J. 2011, \apj, 737, 2

\bibitem[{{Mattila}(1990)}]{mattila90}
{Mattila}, K. 1990, in IAU Symposium, Vol. 139, The Galactic and Extragalactic
  Background Radiation, ed. {S.~Bowyer \& C.~Leinert}, 257--268

\bibitem[{{Maurer} {et~al.}(2012){Maurer}, {Raue}, {Kneiske}, {Horns},
  {Els{\"a}sser}, \& {Hauschildt}}]{maurer12}
{Maurer}, A., {Raue}, M., {Kneiske}, T., {Horns}, D., {Els{\"a}sser}, D., \&
  {Hauschildt}, P.~H. 2012, \apj, 745, 166

\bibitem[{{Mazin} \& {Raue}(2007)}]{mazin07}
{Mazin}, D. \& {Raue}, M. 2007, \aap, 471, 439

\bibitem[{{Metcalfe} {et~al.}(2003){Metcalfe}, {Kneib}, {McBreen}, {Altieri},
  {Biviano}, {Delaney}, {Elbaz}, {Kessler}, {Leech}, {Okumura}, {Ott},
  {Perez-Martinez}, {Sanchez-Fernandez}, \& {Schulz}}]{metcalfe03}
{Metcalfe}, L., {Kneib}, J.-P., {McBreen}, B., {Altieri}, B., {Biviano}, A.,
  {Delaney}, M., {Elbaz}, D., {Kessler}, M.~F., {Leech}, K., {Okumura}, K.,
  {Ott}, S., {Perez-Martinez}, R., {Sanchez-Fernandez}, C., \& {Schulz}, B.
  2003, \aap, 407, 791

\bibitem[{{Meyer} {et~al.}(2012){Meyer}, {Raue}, {Mazin}, \& {Horns}}]{meyer12}
{Meyer}, M., {Raue}, M., {Mazin}, D., \& {Horns}, D. 2012, \aap, 542, A59

\bibitem[{{Micha{\l}owski} {et~al.}(2010){Micha{\l}owski}, {Hjorth}, \&
  {Watson}}]{michalowski10c}
{Micha{\l}owski}, M., {Hjorth}, J., \& {Watson}, D. 2010, \aap, 514, A67

\bibitem[{{Milliard} {et~al.}(1992){Milliard}, {Donas}, {Laget}, {Armand}, \&
  {Vuillemin}}]{milliard92}
{Milliard}, B., {Donas}, J., {Laget}, M., {Armand}, C., \& {Vuillemin}, A.
  1992, \aap, 257, 24

\bibitem[{{Mo} {et~al.}(2010){Mo}, {van den Bosch}, \& {White}}]{mo10}
{Mo}, H., {van den Bosch}, F.~C., \& {White}, S. 2010, {Galaxy Formation and
  Evolution}

\bibitem[{{Mushotzky} {et~al.}(2000){Mushotzky}, {Cowie}, {Barger}, \&
  {Arnaud}}]{mushotzky00}
{Mushotzky}, R.~F., {Cowie}, L.~L., {Barger}, A.~J., \& {Arnaud}, K.~A. 2000,
  \nat, 404, 459

\bibitem[{{Nenkova} {et~al.}(2000){Nenkova}, {Ivezi{\'c}}, \&
  {Elitzur}}]{nenkova00}
{Nenkova}, M., {Ivezi{\'c}}, {\v Z}., \& {Elitzur}, M. 2000, Thermal Emission
  Spectroscopy and Analysis of Dust, Disks, and Regoliths, 196, 77

\bibitem[{{Neronov} {et~al.}(2010){Neronov}, {Semikoz}, \& {Vovk}}]{neronov10}
{Neronov}, A., {Semikoz}, D., \& {Vovk}, I. 2010, \aap, 519, L6

\bibitem[{{Nieppola} {et~al.}(2006){Nieppola}, {Tornikoski}, \&
  {Valtaoja}}]{nieppola06}
{Nieppola}, E., {Tornikoski}, M., \& {Valtaoja}, E. 2006, \aap, 445, 441

\bibitem[{{Odegard} {et~al.}(2007){Odegard}, {Arendt}, {Dwek}, {Haffner},
  {Hauser}, \& {Reynolds}}]{odegard07}
{Odegard}, N., {Arendt}, R.~G., {Dwek}, E., {Haffner}, L.~M., {Hauser}, M.~G.,
  \& {Reynolds}, R.~J. 2007, \apj, 667, 11

\bibitem[{{Oliver} {et~al.}(2010){Oliver}, {Wang}, {Smith}, {Altieri},
  {Amblard}, {Arumugam}, {Auld}, {Aussel}, {Babbedge}, {Blain}, {Bock},
  {Boselli}, {Buat}, {Burgarella}, {Castro-Rodr{\'{\i}}guez}, {Cava},
  {Chanial}, {Clements}, {Conley}, {Conversi}, {Cooray}, {Dowell}, {Dwek},
  {Eales}, {Elbaz}, {Fox}, {Franceschini}, {Gear}, {Glenn}, {Griffin},
  {Halpern}, {Hatziminaoglou}, {Ibar}, {Isaak}, {Ivison}, {Lagache},
  {Levenson}, {Lu}, {Madden}, {Maffei}, {Mainetti}, {Marchetti},
  {Mitchell-Wynne}, {Mortier}, {Nguyen}, {O'Halloran}, {Omont}, {Page},
  {Panuzzo}, {Papageorgiou}, {Pearson}, {P{\'e}rez-Fournon}, {Pohlen},
  {Rawlings}, {Raymond}, {Rigopoulou}, {Rizzo}, {Roseboom}, {Rowan-Robinson},
  {S{\'a}nchez Portal}, {Savage}, {Schulz}, {Scott}, {Seymour}, {Shupe},
  {Stevens}, {Symeonidis}, {Trichas}, {Tugwell}, {Vaccari}, {Valiante},
  {Valtchanov}, {Vieira}, {Vigroux}, {Ward}, {Wright}, {Xu}, \&
  {Zemcov}}]{oliver10}
{Oliver}, S.~J., {Wang}, L., {Smith}, A.~J., {Altieri}, B., {Amblard}, A.,
  {Arumugam}, V., {Auld}, R., {Aussel}, H., {Babbedge}, T., {Blain}, A.,
  {Bock}, J., {Boselli}, A., {Buat}, V., {Burgarella}, D.,
  {Castro-Rodr{\'{\i}}guez}, N., {Cava}, A., {Chanial}, P., {Clements}, D.~L.,
  {Conley}, A., {Conversi}, L., {Cooray}, A., {Dowell}, C.~D., {Dwek}, E.,
  {Eales}, S., {Elbaz}, D., {Fox}, M., {Franceschini}, A., {Gear}, W., {Glenn},
  J., {Griffin}, M., {Halpern}, M., {Hatziminaoglou}, E., {Ibar}, E., {Isaak},
  K., {Ivison}, R.~J., {Lagache}, G., {Levenson}, L., {Lu}, N., {Madden}, S.,
  {Maffei}, B., {Mainetti}, G., {Marchetti}, L., {Mitchell-Wynne}, K.,
  {Mortier}, A.~M.~J., {Nguyen}, H.~T., {O'Halloran}, B., {Omont}, A., {Page},
  M.~J., {Panuzzo}, P., {Papageorgiou}, A., {Pearson}, C.~P.,
  {P{\'e}rez-Fournon}, I., {Pohlen}, M., {Rawlings}, J.~I., {Raymond}, G.,
  {Rigopoulou}, D., {Rizzo}, D., {Roseboom}, I.~G., {Rowan-Robinson}, M.,
  {S{\'a}nchez Portal}, M., {Savage}, R., {Schulz}, B., {Scott}, D., {Seymour},
  N., {Shupe}, D.~L., {Stevens}, J.~A., {Symeonidis}, M., {Trichas}, M.,
  {Tugwell}, K.~E., {Vaccari}, M., {Valiante}, E., {Valtchanov}, I., {Vieira},
  J.~D., {Vigroux}, L., {Ward}, R., {Wright}, G., {Xu}, C.~K., \& {Zemcov}, M.
  2010, \aap, 518, L21+

\bibitem[{{Orr} {et~al.}(2011){Orr}, {Krennrich}, \& {Dwek}}]{orr11}
{Orr}, M.~R., {Krennrich}, F., \& {Dwek}, E. 2011, \apj, 733, 77

\bibitem[{{Pagel}(2001)}]{pagel01}
{Pagel}, B.~E.~J. 2001, \pasp, 113, 137

\bibitem[{{Papovich} {et~al.}(2004){Papovich}, {Dole}, {Egami}, {Le Floc'h},
  {P{\'e}rez-Gonz{\'a}lez}, {Alonso-Herrero}, {Bai}, {Beichman}, {Blaylock},
  {Engelbracht}, {Gordon}, {Hines}, {Misselt}, {Morrison}, {Mould},
  {Muzerolle}, {Neugebauer}, {Richards}, {Rieke}, {Rieke}, {Rigby}, {Su}, \&
  {Young}}]{papovich04}
{Papovich}, C., {Dole}, H., {Egami}, E., {Le Floc'h}, E.,
  {P{\'e}rez-Gonz{\'a}lez}, P.~G., {Alonso-Herrero}, A., {Bai}, L., {Beichman},
  C.~A., {Blaylock}, M., {Engelbracht}, C.~W., {Gordon}, K.~D., {Hines}, D.~C.,
  {Misselt}, K.~A., {Morrison}, J.~E., {Mould}, J., {Muzerolle}, J.,
  {Neugebauer}, G., {Richards}, P.~L., {Rieke}, G.~H., {Rieke}, M.~J., {Rigby},
  J.~R., {Su}, K.~Y.~L., \& {Young}, E.~T. 2004, \apjs, 154, 70

\bibitem[{{Pei} \& {Fall}(1995)}]{pei95}
{Pei}, Y.~C. \& {Fall}, S.~M. 1995, \apj, 454, 69

\bibitem[{{Pei} {et~al.}(1999){Pei}, {Fall}, \& {Hauser}}]{pei99}
{Pei}, Y.~C., {Fall}, S.~M., \& {Hauser}, M.~G. 1999, \apj, 522, 604

\bibitem[{{P{\'e}nin} {et~al.}(2012){P{\'e}nin}, {Lagache}, {Noriega-Crespo},
  {Grain}, {Miville-Desch{\^e}nes}, {Ponthieu}, {Martin}, {Blagrave}, \&
  {Lockman}}]{penin12}
{P{\'e}nin}, A., {Lagache}, G., {Noriega-Crespo}, A., {Grain}, J.,
  {Miville-Desch{\^e}nes}, M.-A., {Ponthieu}, N., {Martin}, P., {Blagrave}, K.,
  \& {Lockman}, F.~J. 2012, \aap, 543, A123

\bibitem[{{Petry} {et~al.}(2002){Petry}, {Bond}, {Bradbury}, {Buckley},
  {Carter-Lewis}, {Cui}, {Duke}, {de la Calle Perez}, {Falcone}, {Fegan},
  {Fegan}, {Finley}, {Gaidos}, {Gibbs}, {Gammell}, {Hall}, {Hall}, {Hillas},
  {Holder}, {Horan}, {Jordan}, {Kertzman}, {Kieda}, {Kildea}, {Knapp},
  {Kosack}, {Krennrich}, {LeBohec}, {Moriarty}, {M{\"u}ller}, {Nagai}, {Ong},
  {Page}, {Pallassini}, {Power-Mooney}, {Quinn}, {Reay}, {Reynolds}, {Rose},
  {Schroedter}, {Sembroski}, {Sidwell}, {Stanton}, {Swordy}, {Vassiliev},
  {Wakely}, {Walker}, \& {Weekes}}]{petry02}
{Petry}, D., {Bond}, I.~H., {Bradbury}, S.~M., {Buckley}, J.~H.,
  {Carter-Lewis}, D.~A., {Cui}, W., {Duke}, C., {de la Calle Perez}, I.,
  {Falcone}, A., {Fegan}, D.~J., {Fegan}, S.~J., {Finley}, J.~P., {Gaidos},
  J.~A., {Gibbs}, K., {Gammell}, S., {Hall}, J., {Hall}, T.~A., {Hillas},
  A.~M., {Holder}, J., {Horan}, D., {Jordan}, M., {Kertzman}, M., {Kieda}, D.,
  {Kildea}, J., {Knapp}, J., {Kosack}, K., {Krennrich}, F., {LeBohec}, S.,
  {Moriarty}, P., {M{\"u}ller}, D., {Nagai}, T.~N., {Ong}, R., {Page}, M.,
  {Pallassini}, R., {Power-Mooney}, B., {Quinn}, J., {Reay}, N.~W., {Reynolds},
  P.~T., {Rose}, H.~J., {Schroedter}, M., {Sembroski}, G.~H., {Sidwell}, R.,
  {Stanton}, N., {Swordy}, S.~P., {Vassiliev}, V.~V., {Wakely}, S.~P.,
  {Walker}, G., \& {Weekes}, T.~C. 2002, \apj, 580, 104

\bibitem[{{Pita} {et~al.}(2012){Pita}, {Goldoni}, {Boisson}, {Becherini},
  {G{\'e}rard}, {Lenain}, \& {Punch}}]{pita12}
{Pita}, S., {Goldoni}, P., {Boisson}, C., {Becherini}, Y., {G{\'e}rard}, L.,
  {Lenain}, J.-P., \& {Punch}, M. 2012, ArXiv e-prints

\bibitem[{{Ponente} {et~al.}(2011){Ponente}, {Ascasibar}, \&
  {Diego}}]{ponente11}
{Ponente}, P.~P., {Ascasibar}, Y., \& {Diego}, J.~M. 2011, \mnras, 418, 1467

\bibitem[{{Punch} {et~al.}(1992){Punch}, {Akerlof}, {Cawley}, {Chantell},
  {Fegan}, {Fennell}, {Gaidos}, {Hagan}, {Hillas}, {Jiang}, {Kerrick}, {Lamb},
  {Lawrence}, {Lewis}, {Meyer}, {Mohanty}, {O'Flaherty}, {Reynolds}, {Rovero},
  {Schubnell}, {Sembroski}, {Weekes}, \& {Wilson}}]{punch92}
{Punch}, M., {Akerlof}, C.~W., {Cawley}, M.~F., {Chantell}, M., {Fegan}, D.~J.,
  {Fennell}, S., {Gaidos}, J.~A., {Hagan}, J., {Hillas}, A.~M., {Jiang}, Y.,
  {Kerrick}, A.~D., {Lamb}, R.~C., {Lawrence}, M.~A., {Lewis}, D.~A., {Meyer},
  D.~I., {Mohanty}, G., {O'Flaherty}, K.~S., {Reynolds}, P.~T., {Rovero},
  A.~C., {Schubnell}, M.~S., {Sembroski}, G., {Weekes}, T.~C., \& {Wilson}, C.
  1992, \nat, 358, 477

\bibitem[{{Raue} {et~al.}(2009){Raue}, {Kneiske}, \& {Mazin}}]{raue09}
{Raue}, M., {Kneiske}, T., \& {Mazin}, D. 2009, \aap, 498, 25

\bibitem[{{Razzaque} {et~al.}(2009){Razzaque}, {Dermer}, \&
  {Finke}}]{razzaque09}
{Razzaque}, S., {Dermer}, C.~D., \& {Finke}, J.~D. 2009, \apj, 697, 483

\bibitem[{{Renault} {et~al.}(2001){Renault}, {Barrau}, {Lagache}, \&
  {Puget}}]{renault01}
{Renault}, C., {Barrau}, A., {Lagache}, G., \& {Puget}, J.-L. 2001, \aap, 371,
  771

\bibitem[{{Rodighiero} {et~al.}(2010){Rodighiero}, {Vaccari}, {Franceschini},
  {Tresse}, {Le Fevre}, {Le Brun}, {Mancini}, {Matute}, {Cimatti}, {Marchetti},
  {Ilbert}, {Arnouts}, {Bolzonella}, {Zucca}, {Bardelli}, {Lonsdale}, {Shupe},
  {Surace}, {Rowan-Robinson}, {Garilli}, {Zamorani}, {Pozzetti}, {Bondi}, {de
  la Torre}, {Vergani}, {Santini}, {Grazian}, \& {Fontana}}]{rodighiero10}
{Rodighiero}, G., {Vaccari}, M., {Franceschini}, A., {Tresse}, L., {Le Fevre},
  O., {Le Brun}, V., {Mancini}, C., {Matute}, I., {Cimatti}, A., {Marchetti},
  L., {Ilbert}, O., {Arnouts}, S., {Bolzonella}, M., {Zucca}, E., {Bardelli},
  S., {Lonsdale}, C.~J., {Shupe}, D., {Surace}, J., {Rowan-Robinson}, M.,
  {Garilli}, B., {Zamorani}, G., {Pozzetti}, L., {Bondi}, M., {de la Torre},
  S., {Vergani}, D., {Santini}, P., {Grazian}, A., \& {Fontana}, A. 2010, \aap,
  515, A8

\bibitem[{{Rowan-Robinson}(2001)}]{rowan-robinson01}
{Rowan-Robinson}, M. 2001, \nar, 45, 631

\bibitem[{{Rowan-Robinson}(2009)}]{rowan-robinson09}
---. 2009, \mnras, 394, 117

\bibitem[{{Salvaterra} \& {Ferrara}(2003)}]{salvaterra03}
{Salvaterra}, R. \& {Ferrara}, A. 2003, \mnras, 339, 973

\bibitem[{{Samuelson} {et~al.}(1998){Samuelson}, {Biller}, {Bond}, {Boyle},
  {Bradbury}, {Breslin}, {Buckley}, {Burdett}, {Buss'ons Gordo},
  {Carter-Lewis}, {Cantanese}, {Cawley}, {Fegan}, {Finley}, {Gaidos}, {Hall},
  {Hillas}, {Krennrich}, {Lamb}, {Lessard}, {McEnery}, {Masterson}, {Quinn},
  {Rodgers}, {Rose}, {Sembroski}, {Srinivasan}, {Vassiliev}, {Weekes}, \&
  {Zweerink}}]{samuelson98}
{Samuelson}, F.~W., {Biller}, S.~D., {Bond}, I.~H., {Boyle}, P.~J., {Bradbury},
  S.~M., {Breslin}, A., {Buckley}, J.~H., {Burdett}, A.~M., {Buss'ons Gordo},
  J., {Carter-Lewis}, D.~A., {Cantanese}, M., {Cawley}, M.~F., {Fegan}, D.~J.,
  {Finley}, J.~P., {Gaidos}, J.~A., {Hall}, T., {Hillas}, A.~M., {Krennrich},
  F., {Lamb}, R.~C., {Lessard}, R.~W., {McEnery}, J.~E., {Masterson}, C.,
  {Quinn}, J., {Rodgers}, A.~J., {Rose}, H.~J., {Sembroski}, G.~H.,
  {Srinivasan}, R., {Vassiliev}, V.~V., {Weekes}, T.~C., \& {Zweerink}, J.
  1998, \apjl, 501, L17

\bibitem[{{S{\'a}nchez-Conde} {et~al.}(2009){S{\'a}nchez-Conde}, {Paneque},
  {Bloom}, {Prada}, \& {Dom{\'{\i}}nguez}}]{sanchez-conde09}
{S{\'a}nchez-Conde}, M.~A., {Paneque}, D., {Bloom}, E., {Prada}, F., \&
  {Dom{\'{\i}}nguez}, A. 2009, \prd, 79, 123511

\bibitem[{{Saunders} {et~al.}(1990){Saunders}, {Rowan-Robinson}, {Lawrence},
  {Efstathiou}, {Kaiser}, {Ellis}, \& {Frenk}}]{saunders90}
{Saunders}, W., {Rowan-Robinson}, M., {Lawrence}, A., {Efstathiou}, G.,
  {Kaiser}, N., {Ellis}, R.~S., \& {Frenk}, C.~S. 1990, \mnras, 242, 318

\bibitem[{{Schlegel} {et~al.}(1998){Schlegel}, {Finkbeiner}, \&
  {Davis}}]{schlegel98}
{Schlegel}, D.~J., {Finkbeiner}, D.~P., \& {Davis}, M. 1998, \apj, 500, 525

\bibitem[{{Schroedter} {et~al.}(2005){Schroedter}, {Badran}, {Buckley},
  {Bussons Gordo}, {Carter-Lewis}, {Duke}, {Fegan}, {Fegan}, {Finley},
  {Gillanders}, {Grube}, {Horan}, {Kenny}, {Kertzman}, {Kosack}, {Krennrich},
  {Kieda}, {Kildea}, {Lang}, {Lee}, {Moriarty}, {Quinn}, {Quinn},
  {Power-Mooney}, {Sembroski}, {Wakely}, {Vassiliev}, {Weekes}, \&
  {Zweerink}}]{schroedter05}
{Schroedter}, M., {Badran}, H.~M., {Buckley}, J.~H., {Bussons Gordo}, J.,
  {Carter-Lewis}, D.~A., {Duke}, C., {Fegan}, D.~J., {Fegan}, S.~F., {Finley},
  J.~P., {Gillanders}, G.~H., {Grube}, J., {Horan}, D., {Kenny}, G.~E.,
  {Kertzman}, M., {Kosack}, K., {Krennrich}, F., {Kieda}, D.~B., {Kildea}, J.,
  {Lang}, M.~J., {Lee}, K., {Moriarty}, P., {Quinn}, J., {Quinn}, M.,
  {Power-Mooney}, B., {Sembroski}, G.~H., {Wakely}, S.~P., {Vassiliev}, V.~V.,
  {Weekes}, T.~C., \& {Zweerink}, J. 2005, \apj, 634, 947

\bibitem[{{Shang} {et~al.}(2012){Shang}, {Haiman}, {Knox}, \& {Oh}}]{shang12}
{Shang}, C., {Haiman}, Z., {Knox}, L., \& {Oh}, S.~P. 2012, \mnras, 421, 2832

\bibitem[{{Shectman}(1973)}]{shectman73}
{Shectman}, S.~A. 1973, \apj, 179, 681

\bibitem[{{Shectman}(1974)}]{shectman74}
---. 1974, \apj, 188, 233

\bibitem[{{Silva} {et~al.}(1998){Silva}, {Granato}, {Bressan}, \&
  {Danese}}]{silva98}
{Silva}, L., {Granato}, G.~L., {Bressan}, A., \& {Danese}, L. 1998, \apj, 509,
  103

\bibitem[{{Soifer} {et~al.}(2008){Soifer}, {Helou}, \& {Werner}}]{soifer08}
{Soifer}, B.~T., {Helou}, G., \& {Werner}, M. 2008, \araa, 46, 201

\bibitem[{{Soifer} \& {Neugebauer}(1991)}]{soifer91}
{Soifer}, B.~T. \& {Neugebauer}, G. 1991, \aj, 101, 354

\bibitem[{{Sokolsky} \& {for the HiRes Collaboration}(2010)}]{sokolsky10}
{Sokolsky}, P. \& {for the HiRes Collaboration}. 2010, ArXiv e-prints

\bibitem[{Somerville {et~al.}(2011)Somerville, Gilmore, Primack, \&
  Dominguez}]{somerville11}
Somerville, R.~S., Gilmore, R.~C., Primack, J.~R., \& Dominguez, A. 2011,
  arXiv: 1104.0669

\bibitem[{Stanev {et~al.}(2000)Stanev, Engel, M\"ucke, Protheroe, \&
  Rachen}]{stanev00}
Stanev, T., Engel, R., M\"ucke, A., Protheroe, R.~J., \& Rachen, J.~P. 2000,
  Phys. Rev. D, 62, 093005

\bibitem[{{Stanev} \& {Franceschini}(1998)}]{stanev98}
{Stanev}, T. \& {Franceschini}, A. 1998, \apjl, 494, L159

\bibitem[{{Stecker} {et~al.}(2007){Stecker}, {Baring}, \&
  {Summerlin}}]{stecker07}
{Stecker}, F.~W., {Baring}, M.~G., \& {Summerlin}, E.~J. 2007, \apjl, 667, L29

\bibitem[{{Stecker} \& {de Jager}(1993)}]{stecker93}
{Stecker}, F.~W. \& {de Jager}, O.~C. 1993, \apjl, 415, L71

\bibitem[{{Stecker} {et~al.}(1992){Stecker}, {de Jager}, \&
  {Salamon}}]{stecker92}
{Stecker}, F.~W., {de Jager}, O.~C., \& {Salamon}, M.~H. 1992, \apjl, 390, L49

\bibitem[{{Stecker} {et~al.}(2006){Stecker}, {Malkan}, \&
  {Scully}}]{stecker06b}
{Stecker}, F.~W., {Malkan}, M.~A., \& {Scully}, S.~T. 2006, \apj, 648, 774

\bibitem[{{Stecker} \& {Scully}(2006)}]{stecker06}
{Stecker}, F.~W. \& {Scully}, S.~T. 2006, \apjl, 652, L9

\bibitem[{{Teplitz} {et~al.}(2011){Teplitz}, {Chary}, {Elbaz}, {Dickinson},
  {Bridge}, {Colbert}, {Le Floc'h}, {Frayer}, {Howell}, {Koo}, {Papovich},
  {Phillips}, {Scarlata}, {Siana}, {Spinrad}, \& {Stern}}]{teplitz11}
{Teplitz}, H.~I., {Chary}, R., {Elbaz}, D., {Dickinson}, M., {Bridge}, C.,
  {Colbert}, J., {Le Floc'h}, E., {Frayer}, D.~T., {Howell}, J.~H., {Koo},
  D.~C., {Papovich}, C., {Phillips}, A., {Scarlata}, C., {Siana}, B.,
  {Spinrad}, H., \& {Stern}, D. 2011, \aj, 141, 1

\bibitem[{{Tinsley}(1981)}]{tinsley81}
{Tinsley}, B.~M. 1981, \apj, 250, 758

\bibitem[{{Totani} {et~al.}(2001){Totani}, {Yoshii}, {Maihara}, {Iwamuro}, \&
  {Motohara}}]{totani01}
{Totani}, T., {Yoshii}, Y., {Maihara}, T., {Iwamuro}, F., \& {Motohara}, K.
  2001, \apj, 559, 592

\bibitem[{{Treister} {et~al.}(2006){Treister}, {Urry}, {Van Duyne},
  {Dickinson}, {Chary}, {Alexander}, {Bauer}, {Natarajan}, {Lira}, \&
  {Grogin}}]{treister06a}
{Treister}, E., {Urry}, C.~M., {Van Duyne}, J., {Dickinson}, M., {Chary},
  R.-R., {Alexander}, D.~M., {Bauer}, F., {Natarajan}, P., {Lira}, P., \&
  {Grogin}, N.~A. 2006, \apj, 640, 603

\bibitem[{{Vassiliev}(2000)}]{vassiliev00}
{Vassiliev}, V.~V. 2000, Astroparticle Physics, 12, 217

\bibitem[{{V{\"o}lk} {et~al.}(1996){V{\"o}lk}, {Aharonian}, \&
  {Breitschwerdt}}]{volk96}
{V{\"o}lk}, H.~J., {Aharonian}, F.~A., \& {Breitschwerdt}, D. 1996, \ssr, 75,
  279

\bibitem[{{Volpe} {et~al.}(2011){Volpe}, {Ohm}, {Hauser}, {Kaufmann},
  {G{\'e}rard}, {Costamante}, {Fegan}, \& {Ajello}}]{volpe11}
{Volpe}, F., {Ohm}, S., {Hauser}, M., {Kaufmann}, S., {G{\'e}rard}, L.,
  {Costamante}, L., {Fegan}, S., \& {Ajello}, M. 2011, ArXiv: 1105.5114

\bibitem[{{Weekes}(2008)}]{weekes08a}
{Weekes}, T.~C. 2008, in American Institute of Physics Conference Series, Vol.
  1085, American Institute of Physics Conference Series, ed. {F.~A.~Aharonian,
  W.~Hofmann, \& F.~Rieger}, 3--17

\bibitem[{{Wehrle} {et~al.}(1998){Wehrle}, {Pian}, {Urry}, {Maraschi},
  {McHardy}, {Lawson}, {Ghisellini}, {Hartman}, {Madejski}, {Makino},
  {Marscher}, {Wagner}, {Webb}, {Aldering}, {Aller}, {Aller}, {Backman},
  {Balonek}, {Boltwood}, {Bonnell}, {Caplinger}, {Celotti}, {Collmar},
  {Dalton}, {Drucker}, {Falomo}, {Fichtel}, {Freudling}, {Gear},
  {Gonzalez-Perez}, {Hall}, {Inoue}, {Johnson}, {Kazanas}, {Kidger}, {Kii},
  {Kollgaard}, {Kondo}, {Kurfess}, {Lin}, {McCollum}, {McNaron-Brown},
  {Nagase}, {Nair}, {Penton}, {Pesce}, {Pohl}, {Raiteri}, {Renda}, {Robson},
  {Sambruna}, {Schirmer}, {Shrader}, {Sikora}, {Sillanpaeae}, {Smith},
  {Stevens}, {Stocke}, {Takalo}, {Teraesranta}, {Thompson}, {Thompson},
  {Tornikoski}, {Tosti}, {Treves}, {Turcotte}, {Unwin}, {Valtaoja}, {Villata},
  {Xu}, {Yamashita}, \& {Zook}}]{wehrle98}
{Wehrle}, A.~E., {Pian}, E., {Urry}, C.~M., {Maraschi}, L., {McHardy}, I.~M.,
  {Lawson}, A.~J., {Ghisellini}, G., {Hartman}, R.~C., {Madejski}, G.~M.,
  {Makino}, F., {Marscher}, A.~P., {Wagner}, S.~J., {Webb}, J.~R., {Aldering},
  G.~S., {Aller}, M.~F., {Aller}, H.~D., {Backman}, D.~E., {Balonek}, T.~J.,
  {Boltwood}, P., {Bonnell}, J., {Caplinger}, J., {Celotti}, A., {Collmar}, W.,
  {Dalton}, J., {Drucker}, A., {Falomo}, R., {Fichtel}, C.~E., {Freudling}, W.,
  {Gear}, W.~K., {Gonzalez-Perez}, N., {Hall}, P., {Inoue}, H., {Johnson},
  W.~N., {Kazanas}, D., {Kidger}, M.~R., {Kii}, T., {Kollgaard}, R.~I.,
  {Kondo}, Y., {Kurfess}, J., {Lin}, Y.~C., {McCollum}, B., {McNaron-Brown},
  K., {Nagase}, F., {Nair}, A.~D., {Penton}, S., {Pesce}, J.~E., {Pohl}, M.,
  {Raiteri}, C.~M., {Renda}, M., {Robson}, E.~I., {Sambruna}, R.~M.,
  {Schirmer}, A.~F., {Shrader}, C., {Sikora}, M., {Sillanpaeae}, A., {Smith},
  P.~S., {Stevens}, J.~A., {Stocke}, J., {Takalo}, L.~O., {Teraesranta}, H.,
  {Thompson}, D.~J., {Thompson}, R., {Tornikoski}, M., {Tosti}, G., {Treves},
  A., {Turcotte}, P., {Unwin}, S.~C., {Valtaoja}, E., {Villata}, M., {Xu}, W.,
  {Yamashita}, A., \& {Zook}, A. 1998, \apj, 497, 178

\bibitem[{{Wright}(2004)}]{wright04}
{Wright}, E.~L. 2004, \nar, 48, 465

\bibitem[{{Xu} {et~al.}(2005){Xu}, {Donas}, {Arnouts}, {Wyder}, {Seibert},
  {Iglesias-P{\'a}ramo}, {Blaizot}, {Small}, {Milliard}, {Schiminovich},
  {Martin}, {Barlow}, {Bianchi}, {Byun}, {Forster}, {Friedman}, {Heckman},
  {Jelinsky}, {Lee}, {Madore}, {Malina}, {Morrissey}, {Neff}, {Rich},
  {Siegmund}, {Szalay}, \& {Welsh}}]{xu05}
{Xu}, C.~K., {Donas}, J., {Arnouts}, S., {Wyder}, T.~K., {Seibert}, M.,
  {Iglesias-P{\'a}ramo}, J., {Blaizot}, J., {Small}, T., {Milliard}, B.,
  {Schiminovich}, D., {Martin}, D.~C., {Barlow}, T.~A., {Bianchi}, L., {Byun},
  Y.-I., {Forster}, K., {Friedman}, P.~G., {Heckman}, T.~M., {Jelinsky}, P.~N.,
  {Lee}, Y.-W., {Madore}, B.~F., {Malina}, R.~F., {Morrissey}, P., {Neff},
  S.~G., {Rich}, R.~M., {Siegmund}, O.~H.~W., {Szalay}, A.~S., \& {Welsh},
  B.~Y. 2005, \apjl, 619, L11

\bibitem[{{Younger} \& {Hopkins}(2011)}]{younger11}
{Younger}, J.~D. \& {Hopkins}, P.~F. 2011, \mnras, 410, 2180

\bibitem[{{Zacharopoulou} {et~al.}(2011){Zacharopoulou}, {Khangulyan},
  {Aharonian}, \& {Costamante}}]{zacharopoulou11}
{Zacharopoulou}, O., {Khangulyan}, D., {Aharonian}, F.~A., \& {Costamante}, L.
  2011, \apj, 738, 157

\bibitem[{{Zech} {et~al.}(2011){Zech}, {Behera}, {Becherini}, {Boisson},
  {Giebels}, {Hauser}, {Kastendieck}, {Kaufmann}, {Kosack}, {Lenain}, {de
  Naurois}, {Punch}, {Raue}, {Sol}, {Wagner}, \& {the
  H.~E.~S.~S.~collaboration}}]{zech11}
{Zech}, A., {Behera}, B., {Becherini}, Y., {Boisson}, C., {Giebels}, B.,
  {Hauser}, M., {Kastendieck}, M., {Kaufmann}, S., {Kosack}, K., {Lenain},
  J.~., {de Naurois}, M., {Punch}, M., {Raue}, M., {Sol}, H., {Wagner}, S., \&
  {the H.~E.~S.~S.~collaboration}. 2011, ArXiv: 1105.0840

\bibitem[{{Zemcov} {et~al.}(2010){Zemcov}, {Blain}, {Halpern}, \&
  {Levenson}}]{zemcov10}
{Zemcov}, M., {Blain}, A., {Halpern}, M., \& {Levenson}, L. 2010, \apj, 721,
  424

\end{thebibliography}

%%%%%%\bibliographystyle{elsarticle-harv}
%%%%%%\bibliography{<your-bib-database>}


% \end{thebibliography}

\end{document}